\documentclass[11pt,a4paper,twoside=semi,open=right]{scrreprt}
\usepackage{amsmath}
\pdfoutput=1
\usepackage[utf8]{inputenc}
\usepackage[english]{babel}
\usepackage{csquotes}
\usepackage{xpatch}
\usepackage{float}
\usepackage{stmaryrd}
\usepackage{ifsym}
\usepackage{tikz}
\usetikzlibrary{matrix}
\usetikzlibrary{calc}
\usepackage{textpos}
\usepackage[charter]{mathdesign}
\usepackage{graphicx}
\usepackage{hyperref}
\usepackage{booktabs,xcolor,siunitx}
\definecolor{lightgray}{gray}{0.9}
\definecolor{bluetable1}{rgb}{0.541, 0.541, 1}
\definecolor{bluetable2}{rgb}{0.541, 0.705, 1}
\definecolor{bluetable3}{rgb}{0.541, 0.882, 1}
\definecolor{bluetable4}{rgb}{0.698, 0.541, 1}
\colorlet{notgreen}{blue!50!yellow}
\usepackage{textcomp}
\usepackage{placeins} 
\usepackage{multirow}
\usepackage{pdfpages}
\usepackage{mathtools}
\usepackage{tcolorbox}
\usepackage{color,colortbl}
\usepackage{enumitem}
\usepackage{array}
\usepackage[capitalize]{cleveref}
\usepackage[singlespacing]{setspace}
\usepackage{bm}

\usepackage[automark,headsepline]{scrlayer-scrpage}

\clearpairofpagestyles
\lehead{\headmark}
\rohead{\headmark}
    \makeatletter
    \newcommand{\thickhline}{%
        \noalign {\ifnum 0=`}\fi \hrule height 1pt
        \futurelet \reserved@a \@xhline
    }
    \newcolumntype{"}{@{\vrule width 1pt}}
    \makeatother
\let\vec\bm
\allowdisplaybreaks
\begin{document}
\pagenumbering{roman}
\thispagestyle{empty}
\begin{titlepage}
	\centering
	\begin{figure}[H]
	\begin{center}
	\begin{minipage}{0.49\textwidth}
	\includegraphics[width=0.99\textwidth]{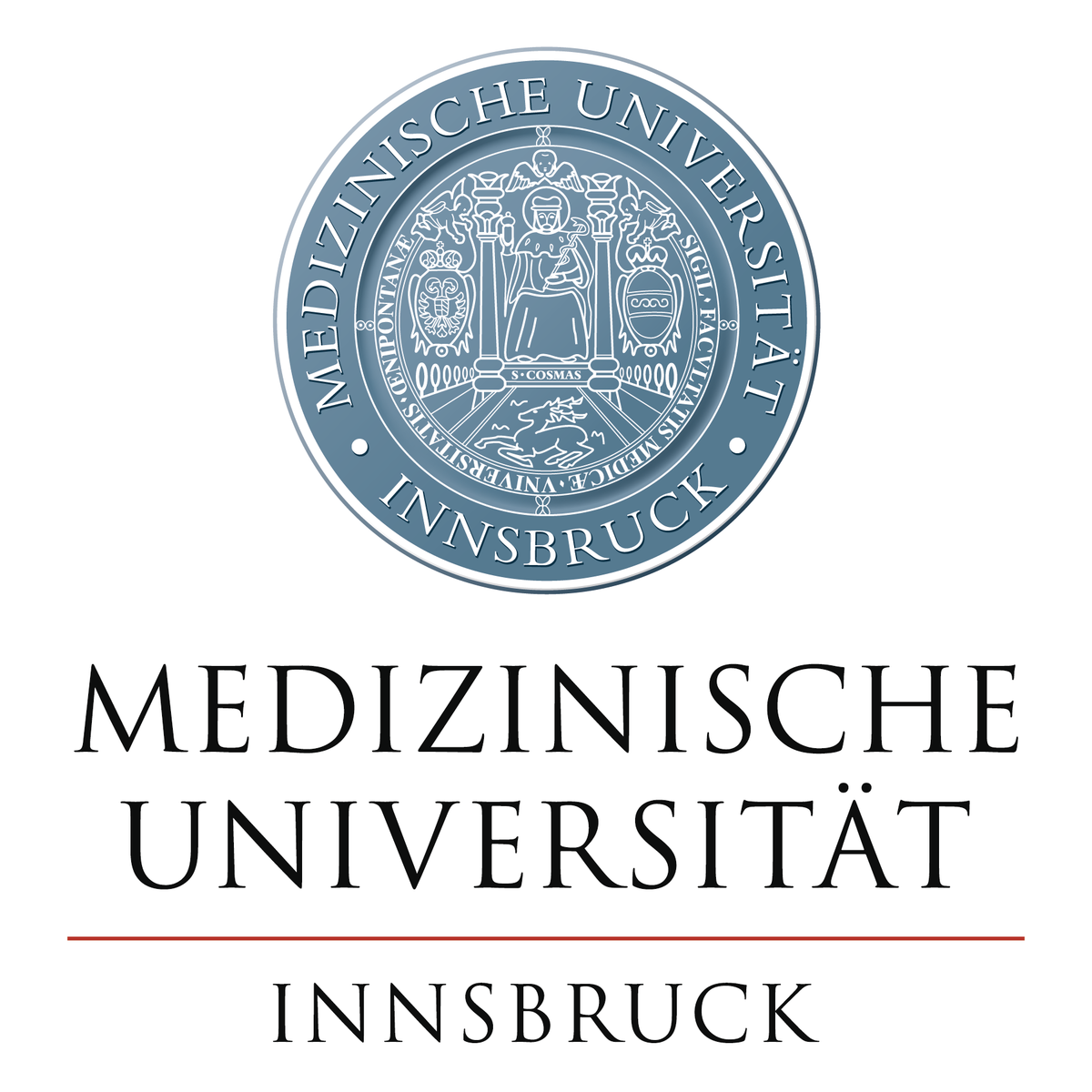}
	\end{minipage}
	\hfill
	\begin{minipage}{0.49\textwidth}
	\includegraphics[width=0.99\textwidth]{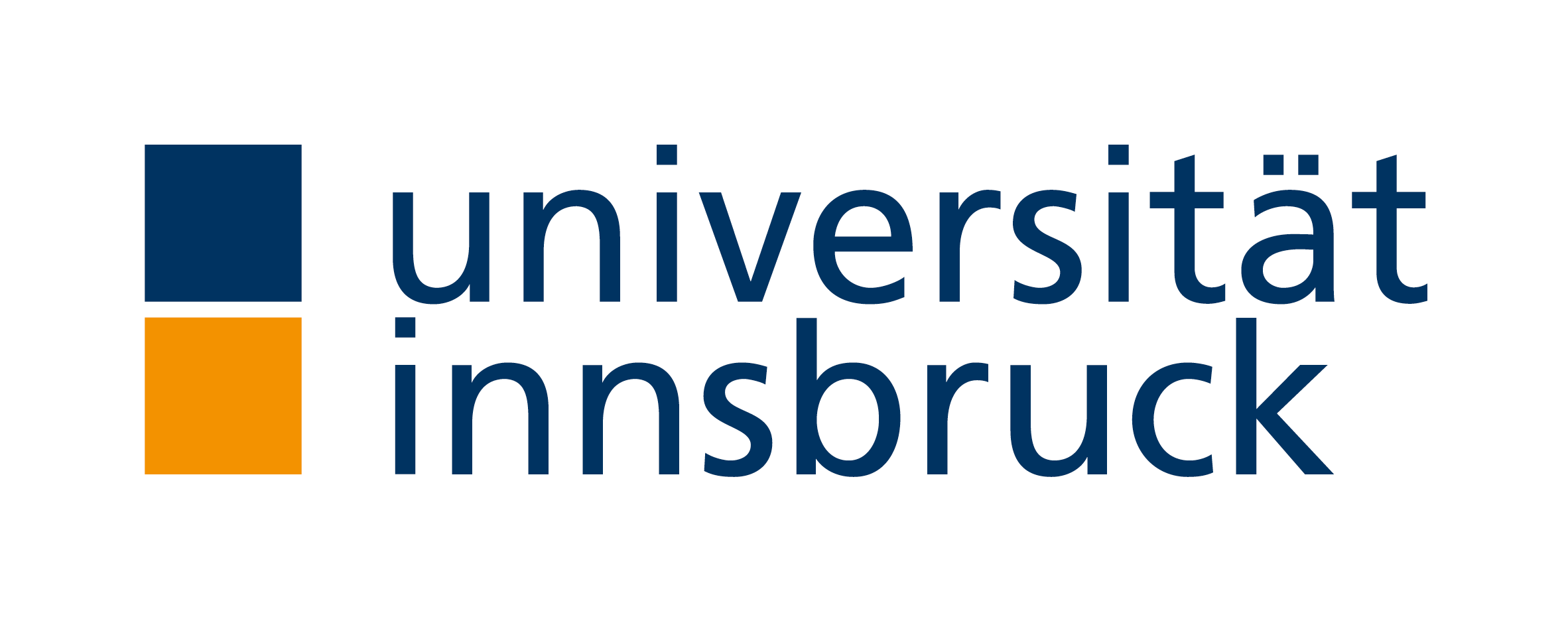}
	\end{minipage}
	\end{center}
	\end{figure}
	\hrule	
	{\huge\bfseries Accurate modeling of the fringing field effect in liquid crystal based spatial light modulators}
	\vspace{0.5cm}
	\hrule
	\makebox[\textwidth]{A thesis submitted in partial fulfillment of the requirements}\newline
	\makebox[\textwidth]{for the degree of Master of Science}\newline
	\makebox[\textwidth]{at the Leopold-Franzens University Innsbruck}
	\vfill
	{by \par}
	\vfill
	{\LARGE Simon Moser}
	\vfill
	{\Large 
	\begin{tabular}{ll}
	Supervisors:& o. Univ.-Prof. Dr. Monika Ritsch-Marte \\
	& Assoz.-Prof. Dr. Gregor Thalhammer
	\end{tabular}}
	\vfill
	{\large Division of Biomedical Physics}
	\vfill
	{\large October 2018}
\end{titlepage}
\newpage\null\thispagestyle{empty}\newpage
\chapter*{Danksagung}
An dieser Stelle möchte ich mich bei allen Leuten bedanken, die das Schreiben dieser Arbeit möglich gemacht und mich in meiner Studienzeit unterstützt haben. 

Zuerst möchte ich mich bei Monika Ritsch-Marte für die Gelegenheit bedanken, am Institut für biomedizinische Physik meine Masterarbeit schreiben zu dürfen. Darüber hinaus möchte ich Monika auch für die Betreuung und Begutachtung der Arbeit meinen Dank aussprechen. 

Ein besonderer Dank gebürt Gregor Thalhammer für die Betreuung und für die zahlreichen Ratschläge und Korrekturen, die mir bei der Erstellung meiner Masterarbeit sehr geholfen haben. Auch möchte ich mich bei Gregor für seine Zeit bedanken, die er mir oft für Fragen und Diskussionen zur Verfügung gestellt hat. 

Außerdem möchte ich mich bei meiner Familie für die Unterstützung und den Rückhalt im Studium und auch sonst bedanken.    
\clearpage
\newpage\null\thispagestyle{empty}\newpage
\chapter*{Abstract}
Liquid crystal based spatial light modulators are widely used in applied optics due to their ability to continuously modulate the phase of a light field with very high spatial resolution. A common problem in these devices is the pixel crosstalk, also called the fringing field effect, which causes the response of these devices to deviate from the ideal behavior. This fringing effect decreases the performance of the spatial light modulator and is shown to cause an asymmetry in the diffraction efficiency between positive and negative diffraction orders. We use simulations of the director distribution to reproduce diffraction efficiency measurements of binary and blazed gratings. To overcome these limitations in performance, the simulations of the director distribution in the liquid crystal layer are used to develop a fast and precise model to compute the phase response of the spatial light modulator. To compensate the fringing field effect, we implement this model in phase retrieval algorithms and calculate the phase profile corresponding to a regular spot pattern as a generic example. With this method, we are able to increase the spot uniformity significantly compared to a calculation without considering the fringing field effect. Additionally, polarization conversion efficiencies of various simple phase patterns are simulated and measured for different orientations of the spatial light modulator. We found that the polarization conversion has the the smallest effect for a setup in which the liquid crystal molecules at the alignment layer lie in the plane of incidence of the light beam.
\clearpage
\newpage\null\thispagestyle{empty}\newpage
\tableofcontents
\cleardoubleoddpage
\pagenumbering{arabic}
\chapter{Overview}

A spatial light modulator (SLM) is a device which applies a spatially varying phase or amplitude modulation to a light beam. SLMs are generally used for optical beam shaping or steering, imaging, trapping, in communication technology and adaptive optics \cite{UziEfron1994}.

Devices denoted by the term SLM can be realized in different manners, namely as digital micromirror devices (DMDs), deformable mirrors (DMs) and liquid crystal based SLMs (often referred to as  liquid crystal on silicon (LCoS) SLMs) \cite{MaurerJesacherBernetEtAl2010}. DMDs are micro-opto-electromechanical systems consisting of an array of micromirrors, which can be rotated individually to an \glqq on\grqq and \glqq off\grqq \, position, modulating the amplitude of an incoming light beam in a binary manner. DMs consist of a metal coated membrane or thin mirrors which can be deformed by a subjacent array of electrodes or mechanical actuators, respectively, providing continuous phase modulation. Whereas the functionality of DMDs and DMs is based on mechanical movement of mirrors, LCoS SLMs use electric fields to induce rotation of birefringent anisotropic liquid crystal (LC) molecules to achieve phase and/or amplitude modulation. The modulation in these LC devices can happen in a binary (ferroelectric SLMs) or in a continuous (nematic SLMs) manner. LCoS SLMs and DMDs are available with resolutions up to about $10$ megapixels and with pixel pitches in the range of about $10$ $\mu$m, whereas DMs possess a much lower actuator number ($30$ to $3000$) with pitches in the range of a few hundred $\mu$m.  DMDs as well as DMs have short (mechanical) response times ($<100$ $\mu$s), whereas LC based SLMs have longer response times in the range of $10$ ms for nematic LCs and $<1$ ms for devices using ferroelectric LCs. The total light efficiency (ratio of light intensity exiting vs. light intensity entering the device) of DMs is $\sim 100$\%, whereas LC based SLMs have efficiencies of about $20-80$\% (depending on the specific device) due to light absorption in the liquid crystal layer or at the patterned electrodes.

DMDs are often used for structured illumination microscopy (SIM), lithography, video projection systems and to correct turbid media due to their high speed and large resolution, whereas DMs are preferred in adaptive optics in astronomy, ophthalmology and microscopy to correct lower order aberrations. LC based SLMs are used for beam shaping and steering, polarization modulation and as a holographic element (e.g. in optical trapping and synthetic holography microscopy) \cite{JesacherRitsch-Marte2016,MaurerJesacherBernetEtAl2010}. A specific strength of LC based SLMs is the high achievable diffraction efficiency due to the continuous phase modulation and high resolution.

Out of the above mentioned SLM types, LC based SLMs have the slowest response time, which limits the performance. The main limitation in LC SLMs in terms of speed is the relaxation time $\tau$ of the LC molecules. After switching on the electric field, the molecules in the LC layer will reorient themselves, the angle of the long molecule axis $\varphi$ approximately following an exponential behavior in time $\varphi (t) \propto \mathrm{e}^{-t/\tau}$. Under certain circumstances the response time can be improved by overdrive switching, reducing the response time significantly to $\sim 1$ ms \cite{ThalhammerBowmanLoveEtAl2013}.

Another physical limitation of LC based SLM is the fringing field effect (also referred to as pixel crosstalk), which is caused by the interaction of the non uniform electric fields over the LC layer and the elastic forces between LC molecules. This crosstalk effect influences the phase response between adjacent pixels, and the diffraction efficiency of patterns with fine structures deviates strongly from the ideal behavior, which serves as an example for the detrimental effects of fringing. 
Therefore, in order to use the device at full capacity for high performance optical trapping or imaging, one has to understand the fringing field effect in great detail.



\section*{Operation principle of LC based SLMs}

The operating principle is based on controlling the phase shift of an incoming light beam by applying a voltage pattern on an array of electrodes across a LC layer. These arrays of pixel electrodes can provide a spatial resolution up to $1280$x$1024$ with a pixel-pitch of $10$-$20$ $\mu$m over an area of $16$x$13$ mm. 

The LCs used for this purpose are usually of the calamitic type, which are rod-shaped molecules of the size of a few nanometers. These liquid crystals exhibit a dielectric anisotropy, so the molecules possess a different polarizability along the main axes and usually the LCs are used in the nematic phase. In this phase the molecules have no positional order, but, without any external fields, the orientation of the long molecule axes strongly correlate with one another and thus can be described by a so called director $\vec{n}$. The director is simply a unit vector pointing in the direction of the long molecule axis. Another way of thinking about the orientational order in the nematic phase is through elastic interaction in LCs. The elastic energy is minimized, if the molecules are uniformly aligned. To achieve spatially dependent orientation of the molecules, one has to overcome these elastic forces.

\begin{figure}[h!]
\centering
\includegraphics[width=12cm]{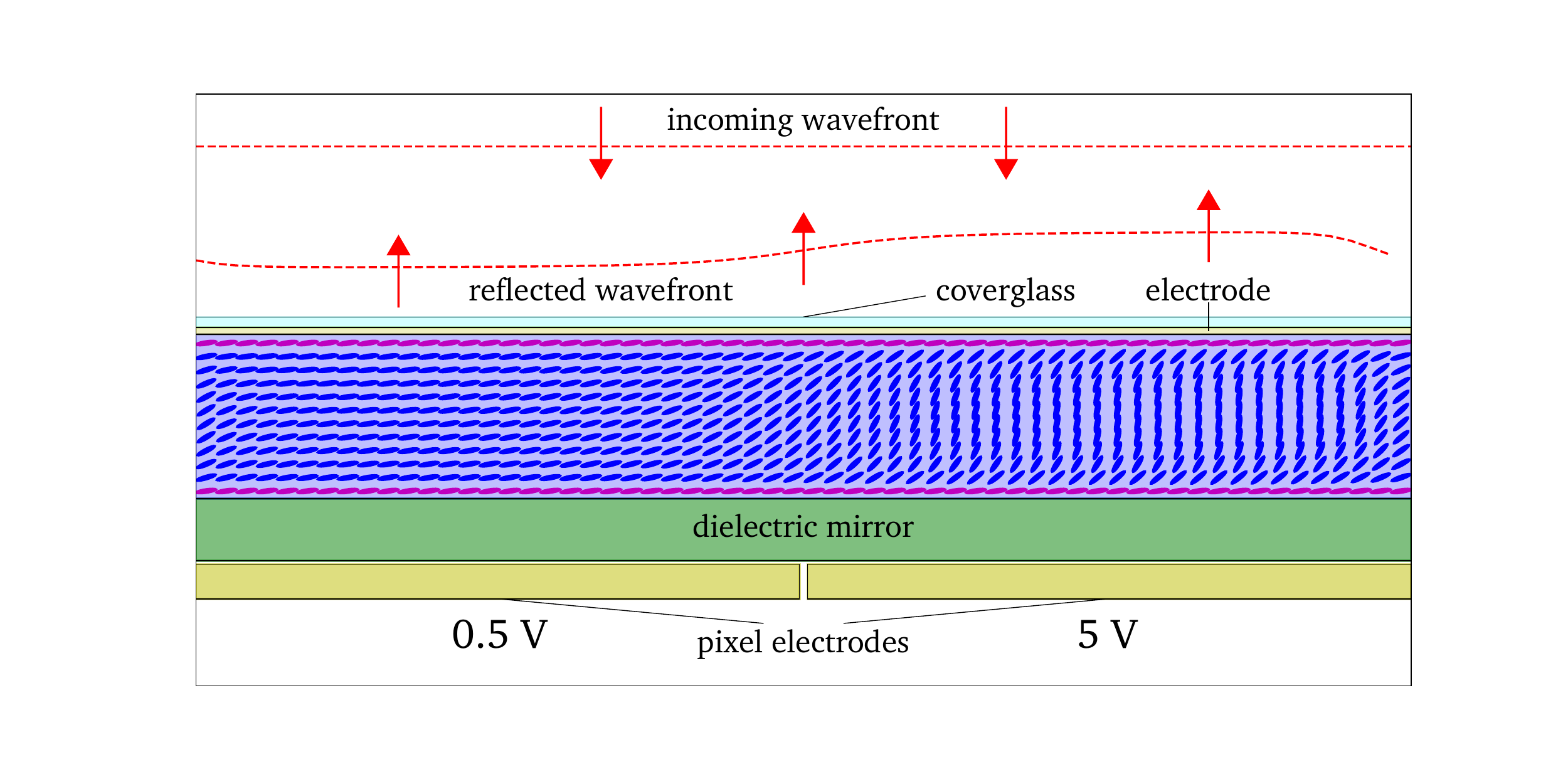}
\caption{Composition of a LCoS spatial light modulator (SLM) with dielectric mirror.}
\label{picture setup slm}
\end{figure}

The basic setup of a reflective SLM is schematically shown in  \cref{picture setup slm} for two neighboring pixels \cite{bns512}.
From top to bottom, this SLM consists of an antireflection coated coverglass with a subsequent transparent electrode. The LC layer is located between the conducting film and a dielectric mirror. Both interfaces are coated with alignment layers \cite{YangWu2014}, 
fixing the orientation of the liquid crystal molecules close to the surface (magenta colored layers in  \cref{picture setup slm}). The pixel electrodes are arranged below the dielectric mirror. 

Without any external electric field, the orientation of the molecules over the whole LC layer is defined by the orientation of the molecules at the alignment layers. In presence of a stationary external electric field of sufficient strength, the electric field exerts a torque on the molecules until the long axis of the molecules is aligned parallel to the electric field.
More precisely, the distribution of the director over the LC layer has to minimize the total free energy \cite{YangWu2014} (see  \cref{sec:theory}). 

The LC used in this SLM are positive uniaxial crystals, so the refractive index ellipsoid is defined by $n_\mathrm{e}$ and $n_\mathrm{o}$. We define an orthonormal coordinate system $(x_1,x_2,x_3)$ with axes $x_1$ and $x_2$ in the plane of the pixel electrodes and orientation of $x_1$ in the direction of the director in the alignment layers (easy axis). 
If we apply a uniform voltage pattern on the pixel electrodes, the director always lies in the $(x_1,x_3)$ plane and therefore the orientation of the liquid crystals can be described solely by the tilt angle $\theta(x_3)$ (this is a special case and these assumptions are only true if the voltage pattern only varies along the easy axis), defined as the angle between director $\vec{n}$ and the plane $(x_1,x_2)$. The refractive index for a plane wave polarized along $x_1$ with normal incidence is given by 
\begin{align}\label{equation n eff}
n(\theta) = \frac{n_\mathrm{e} n_\mathrm{o}}{\sqrt{n_\mathrm{o}^2 + (n_\mathrm{e}^2 - n_\mathrm{o}^2)\sin ^2(\theta)}}
\end{align}
where $k = 2\pi/\lambda$ is the absolute value of wave vector of the light beam. The accumulated phase shift $\Delta \phi$ of light traveling the distance $d$ two times (reflection) is then 
\begin{align}\label{equation delta phase}
\Delta \phi = 2 k \Bigg|\int_0^d \left( n(\theta(x_3)) - n(\theta_p)  \right) \mathrm{d}x_3\Bigg|.
\end{align}
Therefore, the phase shift is defined relative to the phase shift experienced if no electric field is applied. In this work, only the absolute shift $\Delta \phi$ is of interest, therefore $\Delta \phi \geq 0$. 
In this configuration, for light polarized along $x_2$ is unmodulated.

\begin{figure}[H]
\centering
\includegraphics[width=12cm]{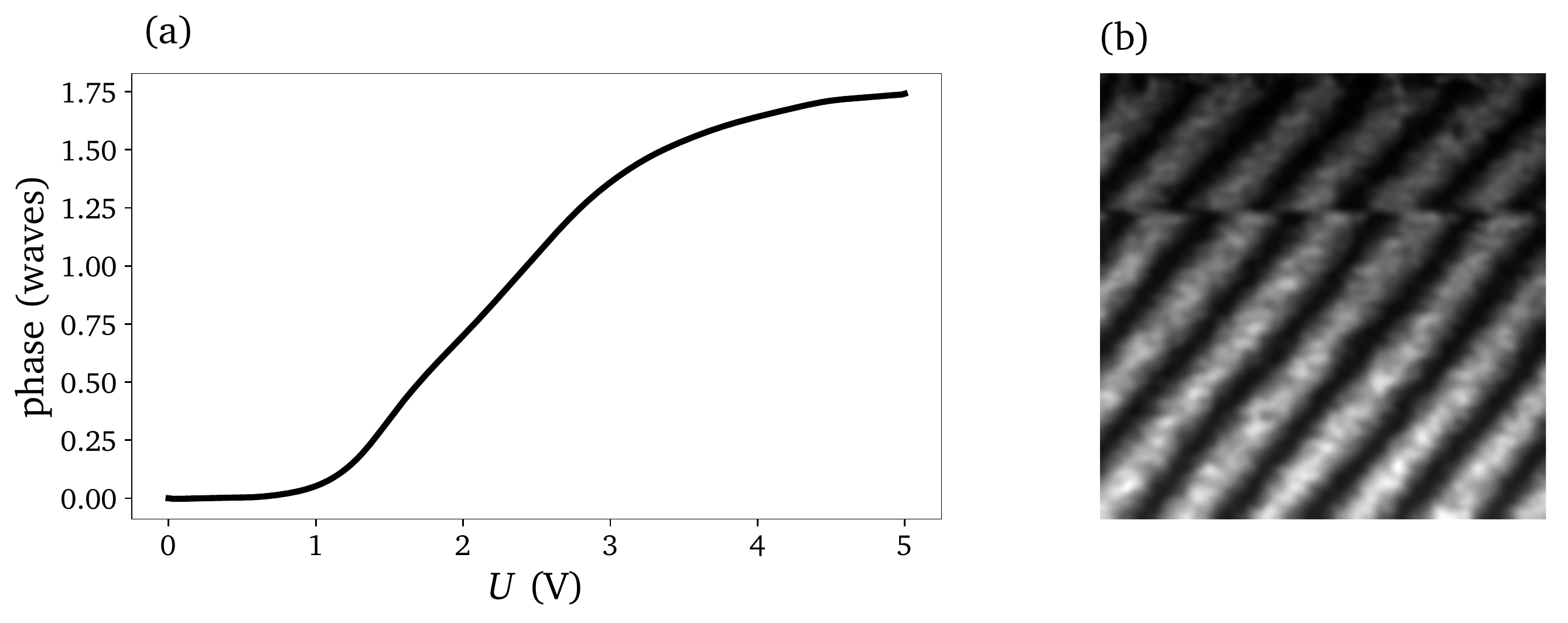}
\caption{Measured phase shift as a function of voltage for our SLM (a), interference fringes to determine the phase shift for a given voltage (b).}
\label{picture lut only measurement with pic}
\end{figure}

 \cref{picture lut only measurement with pic} (a) shows the measured phase as a function of the applied voltage at the electrodes (control voltage). To measure the phase we use a simple interferometer (see \cref{sec:comparison of experiment with simulation}). In  \cref{picture lut only measurement with pic} (b) we see interference fringes which are shifted to one another. In this case, we applied a uniform voltage pattern on the lower part, while applying no voltage on the upper part of the SLM. Through this shift between upper and lower part, we can determine the phase shift for a given voltage. For this measurement, the $512\times512$ XY Series BNS SLM was used.  

In practice, the SLM electrodes are driven by an AC voltage (for our SLM, in square-wave form). The SLM has to be driven by an AC voltage pattern to prevent charge separation due to impurity ions within the LC-cell \cite{Heilmeier1967,Murakami1997}. These transport mechanisms generally decrease the performance of the LC-device.

%
\FloatBarrier
\section*{Fringing field effect}
\begin{figure}[h!]
\centering
\includegraphics[width=12cm]{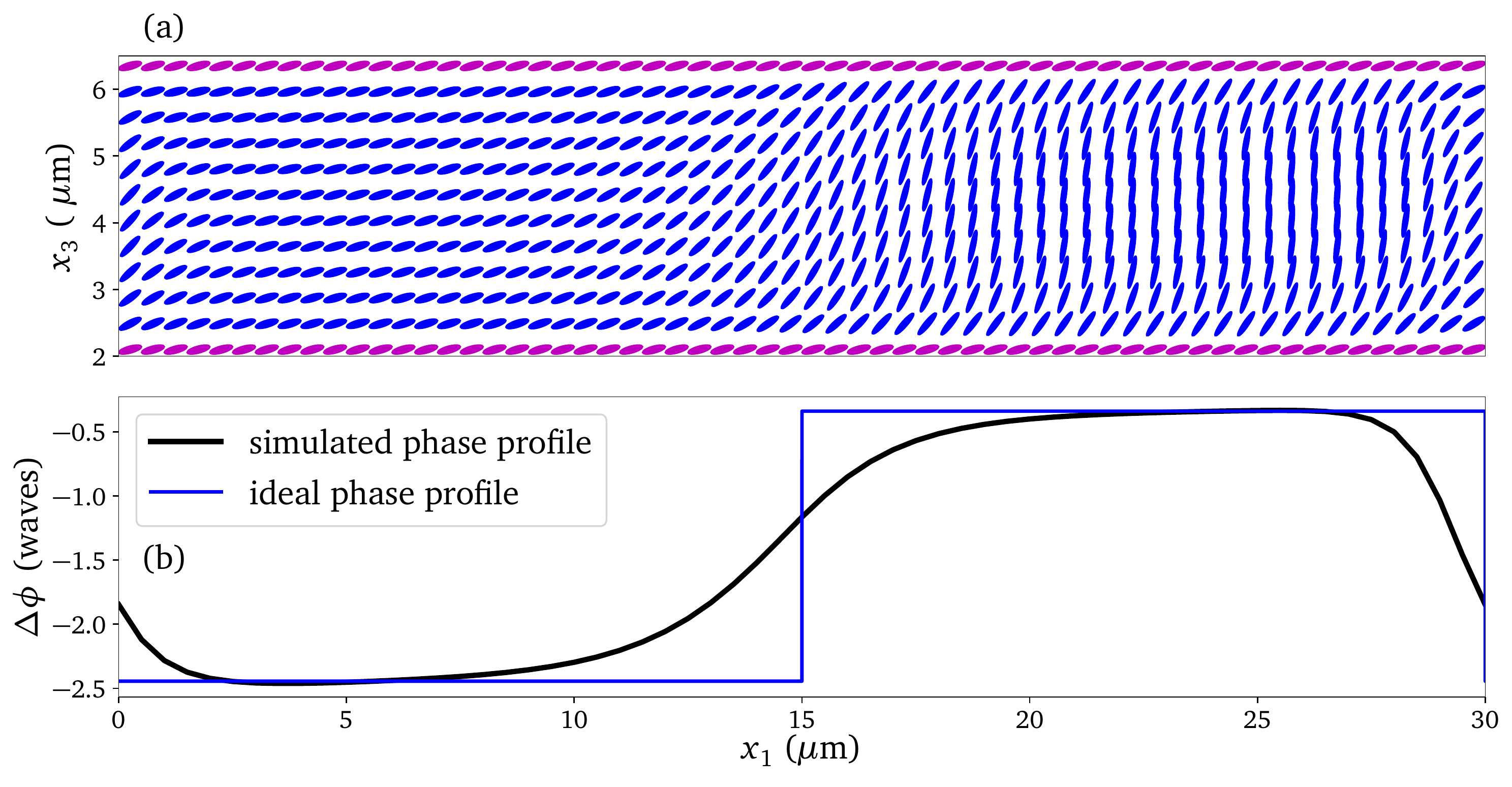}
\caption{Simulated director distribution (a) and accumulated phase shift (b).}
\label{picture fringing_lc_layer_with_phase}
\end{figure}

\begin{figure}[h!]
\centering
\includegraphics[width=12cm]{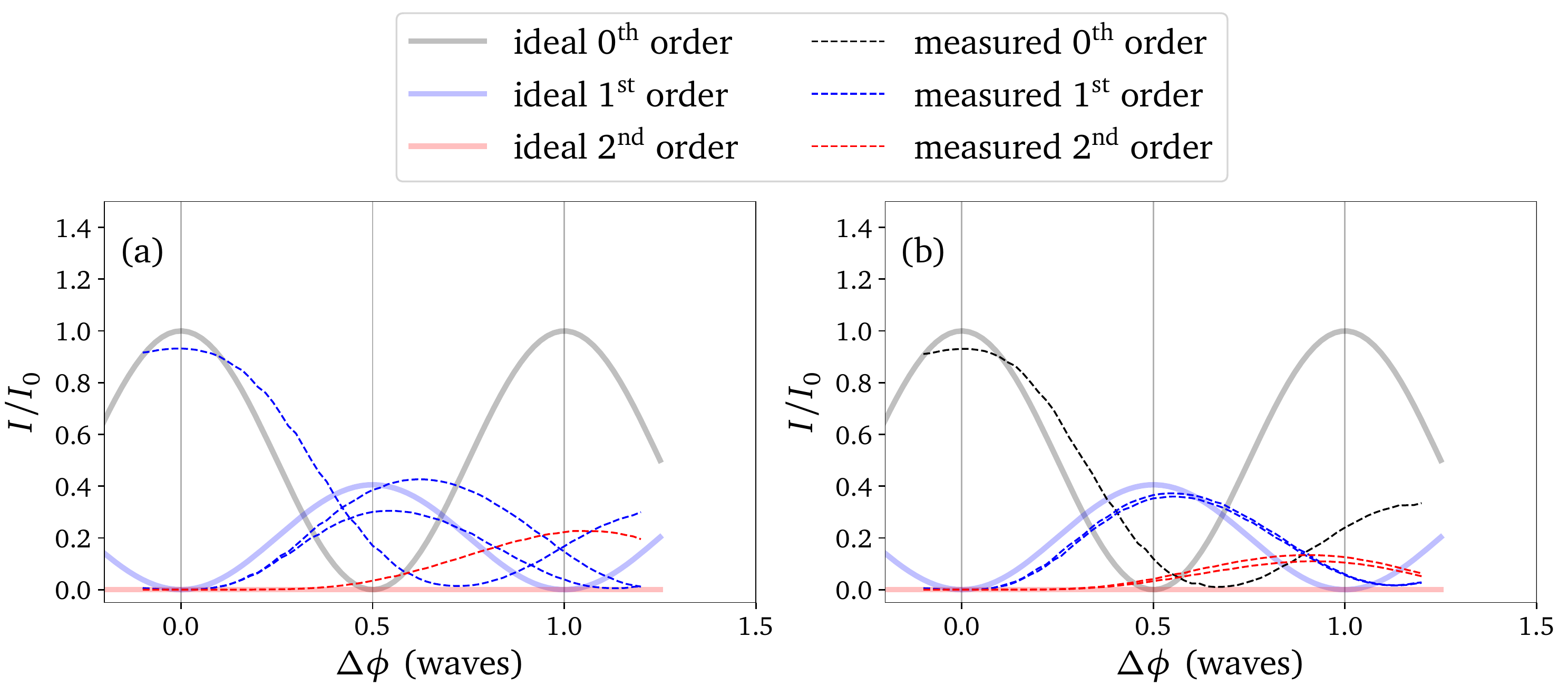}
\caption{Measured diffraction efficiencies (dashed lines) of a binary grating of period $2$ in $x_1$ (a) and in $x_2$ direction (b). The solid lines represent the diffraction efficiency for an ideal (stepwise constant) binary grating.}
\label{picture compare measurement simulation binary 2 x ideal}
\end{figure}

\begin{figure}[h!]
\centering
\includegraphics[width=12cm]{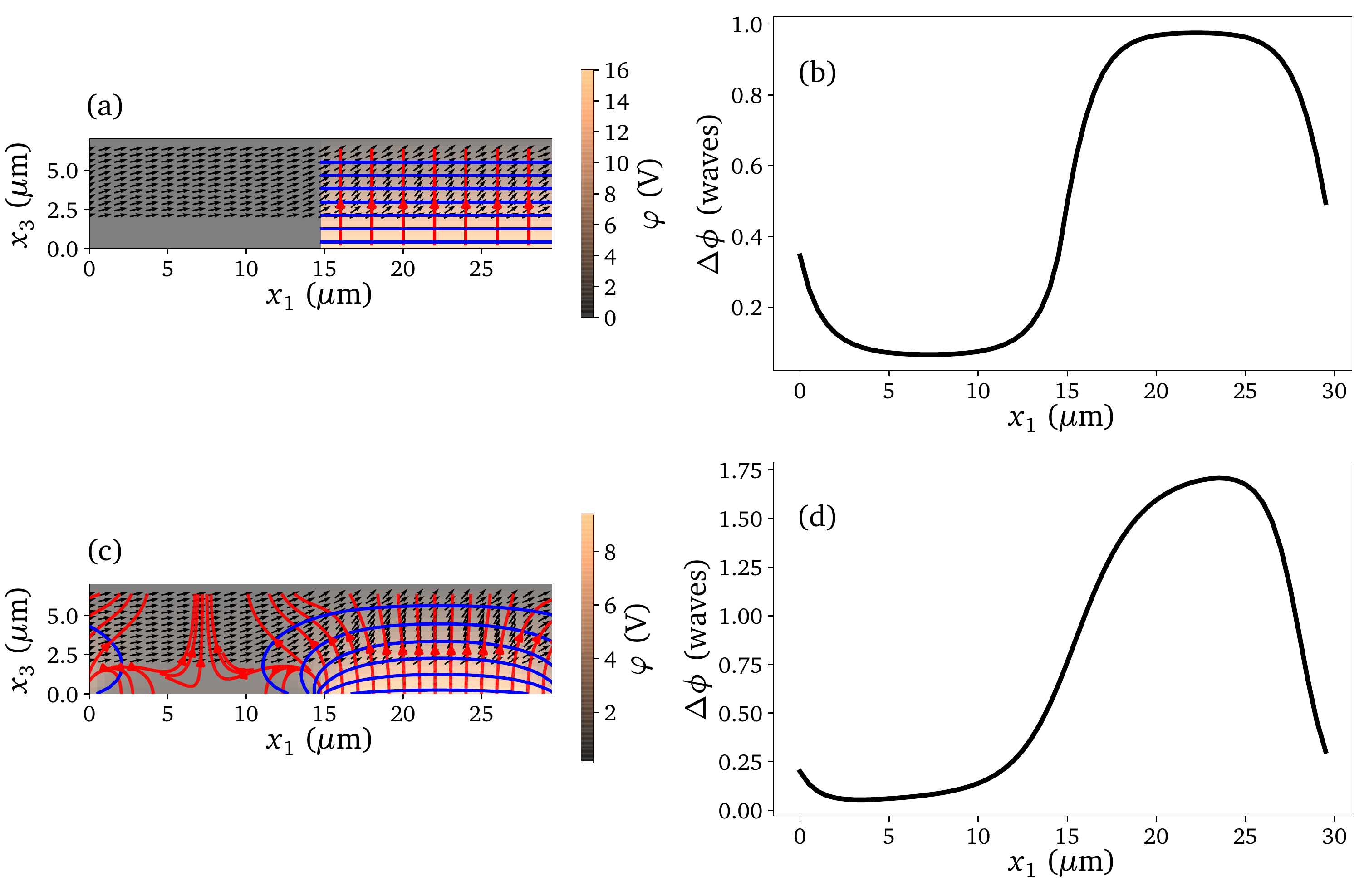}
\caption{Director distribution, potential, electric field and corresponding phase profiles in the ideal (top) and real (bottom) case.}
\label{picture fringing ideal and real pot}
\end{figure}

 \cref{picture fringing_lc_layer_with_phase} shows a simulated director distribution (a) and accumulated phase profile (b) for a binary voltage pattern. We can see, that the LC layer produces a smeared out spatial phase modulation compared to the applied voltage pattern. This crosstalk effect between pixels is generally referred to as the fringing field effect \cite{EfronApterBahat-Treidel2004}. The fringing field effect generally has two main sources:
\begin{enumerate}
\item Electric field broadening: The electric field produced by two neighboring electrodes driven with different voltages is not uniform across the LC-layer, which leads to a smoothed LC response across the LC layer.  
\item Elastic interaction of the LC: The director cannot abruptly change its orientation across the LC layer, since elastic forces between liquid crystal molecules lead to smoothed transitions between neighboring pixels with different voltages.
\end{enumerate}
These two effects influence each other. Since the LC consists of anisotropic molecules, the director $\vec{n}$ locally changes the electric field, which then retroacts again with the orientation of the director. 
In  \cref{picture compare measurement simulation binary 2 x ideal} (a) we see the measured diffraction efficiency of a binary grating applied as a voltage pattern along $x_1$ (along the easy axis) and along $x_2$ \cref{picture compare measurement simulation binary 2 x ideal} (b) (perpendicular to the easy axis). If we look at the intensities for an ideal binary grating with a phase difference $\Delta \phi=1$ wave in  \cref{picture compare measurement simulation binary 2 x ideal} we expect about $40\%$ in each of the $1^\mathrm{st}$ orders and none in the $0^\mathrm{th}$ and $2^\mathrm{nd}$. However, measurements show a different picture. In \cref{picture compare measurement simulation binary 2 x ideal} we see residual intensity in the measured $0^\mathrm{th}$ orders at $\Delta \phi = 1$ wave, which reach about $20\%$. So, the measured diffraction efficiency curves are generally broader compared to the ideal curves. Second, the minima of the $0^\mathrm{th}$ order and the maxima of the $1^\mathrm{st}$ orders do not coincide. In addition to that, the intensity of the $1^\mathrm{st}$ orders depend on the orientation of the applied grating, with an emerging asymmetry between $+1^\mathrm{st}$ and $-1^\mathrm{st}$ order for a grating along $x_1$ that does not appear along $x_2$. This asymmetry can be explained by the asymmetric fringing effect shown in \cref{picture fringing_lc_layer_with_phase}. Last, in the ideal case only odd orders ($1,3,5,...$) contribute, with intensities falling like $1/p^2$ with respect to the order $p$, while the measurements show also significant intensities in the $2^\mathrm{nd}$ orders for sufficiently large phase shifts. 
Looking at  \cref{picture compare measurement simulation binary 2 x ideal}, one could ask why the $1^\mathrm{st}$ order curves are behave differently depending on the orientation of the applied grating.  \cref{picture fringing ideal and real pot} shows a simulation of such a binary grating along $x_1$ (a,c) with the corresponding phase profiles (b,d). This picture shows the simulated director distribution (black arrows), the electric field lines (red) and the electric potential (background) with contour lines (blue). On top (a,b) we used an uniform electric field, while the bottom (c,d) director distribution has been calculated for a real electric field (without considering the effect of the dielectric medium on the electric field).

\begin{figure}[h!]
\centering
\includegraphics[width=12cm]{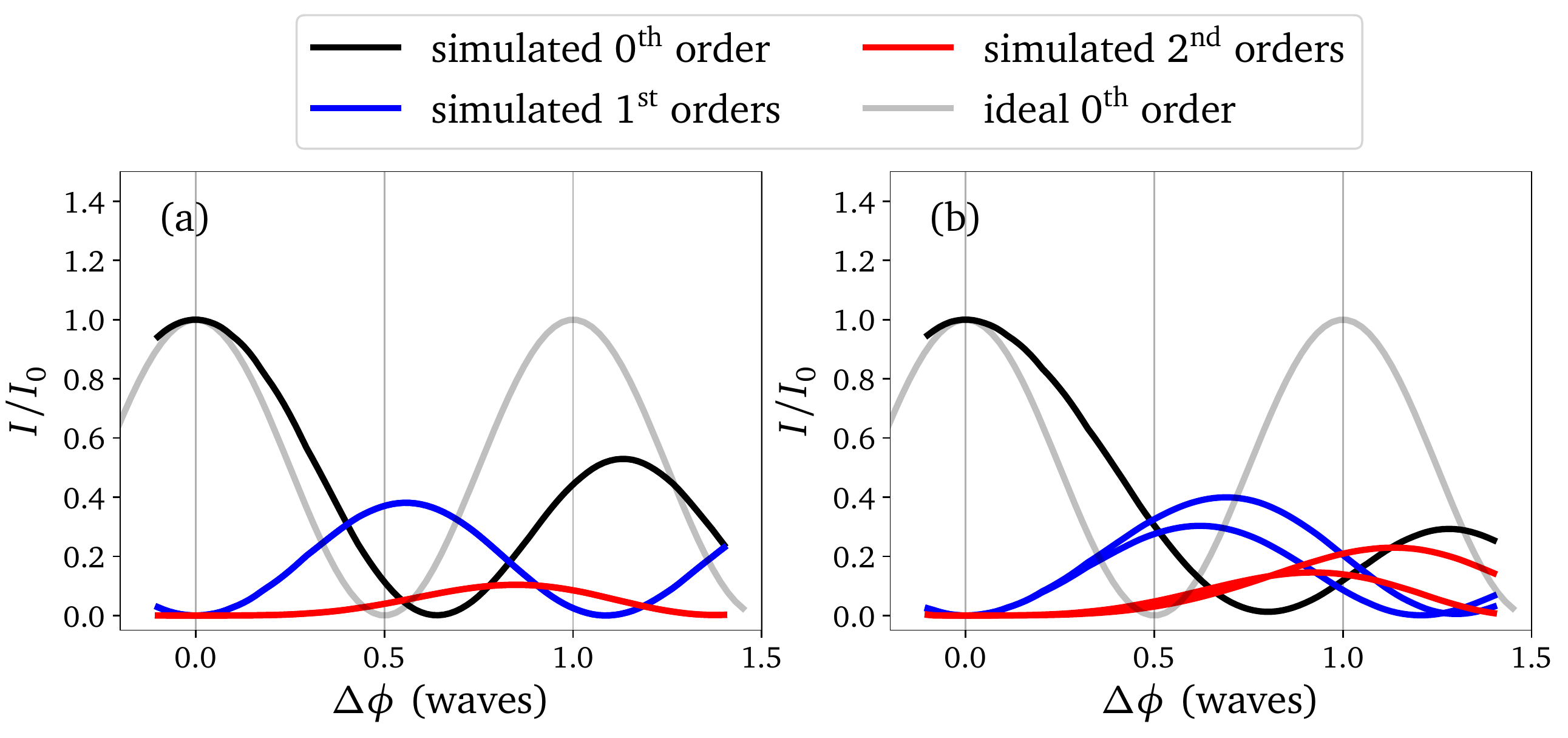}
\caption{Fringing modeled with uniform (a) and real electric field (b). The asymmetry is only visible, if the director distribution is calculated with the real electric field (b).}
\label{picture de ideal und real pot}
\end{figure}

In  \cref{picture de ideal und real pot} we see the simulated diffraction efficiencies for an uniform (a) and real (b) electric field for gratings along $x_1$. Only when modeled with a real electric field, the asymmetry in the $1^\mathrm{st}$ and $2^\mathrm{nd}$ orders starts to emerge. While the component along $x_1$ of the electric field is negative on the transition from low to high and positive on the transition from high to low, the director component along $x_1$ at the alignment layers does not change sign. Therefore the electric field tries to increase the tilt angle on the transition from low to high and decreases it from high to low. Looking at the phase profile, the transition from low to high is very smeared out, while the transition from high to low is comparatively sharp.
For a grating along $x_2$ this effect does not occur, since the director of the LC in the alignment layer only has components along $x_1$ and $x_3$. In this case, the director has a vanishing component along $x_2$ which results in symmetric transitions from low to high and vice versa.
\FloatBarrier
\section*{Compensating the fringing field effect}

\begin{figure}[h!]
\centering
\includegraphics[width=12cm]{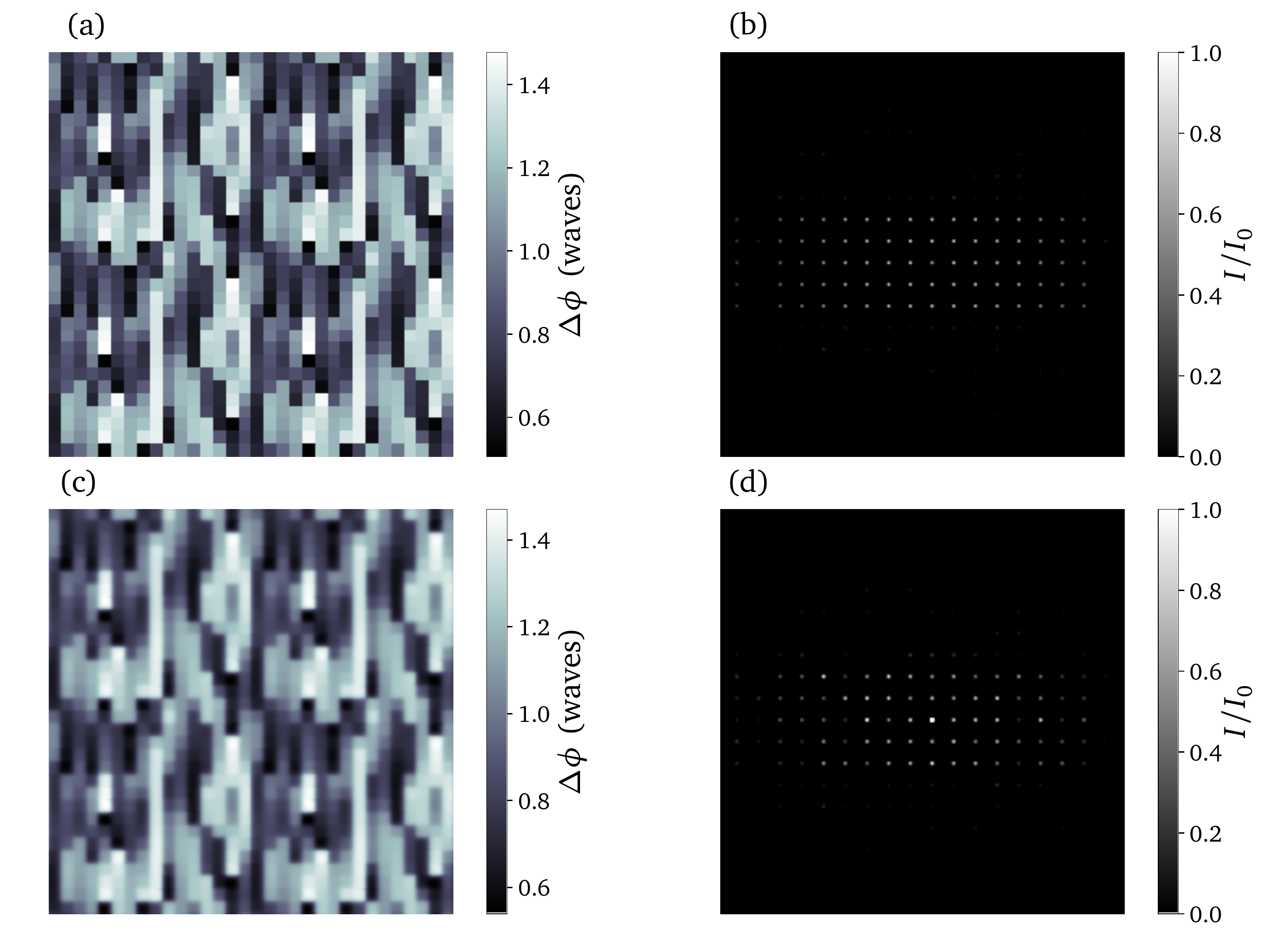}
\caption{Phase profile of ideal (a) and simulated (c) SLM response, and corresponding intensity patterns in the Fourier plane for ideal (b) and simulated (d) SLM response.}
\label{picture spot pattern ideal fringing}
\end{figure}

If the applied voltage pattern consists of small period structures and/or big phase differences between two pixels, these effects of fringing can pose a big problem. As an example for the detrimental effects of fringing, we will look at a phase pattern necessary to create a spot pattern in the Fourier plane. In  \cref{picture spot pattern ideal fringing} we see simulations for the unfringed (a) and the fringed (c) phase profiles with corresponding spot patterns (b) and (d) in the Fourier plane. We can see quite clearly, that the spot intensities vary strongly in the fringed case compared to the ideal one and the spot uniformity is reduced.

It has been shown by Persson et al. \cite{PerssonEngstroemGoksoer2012} that the effects of fringing on spot uniformity $u = 1-(I_\mathrm{max}-I_\mathrm{min})/(I_\mathrm{max}+I_\mathrm{min})$ can be reduced by modeling the real (fringed) phase profile $\phi$ through a convolution of the ideal phase profile $\phi_\mathrm{ideal}$ with a generalized Gaussian kernel $k$

\begin{align}\label{equation convolution phi_ideal kernel}
\phi = (\phi_\mathrm{ideal}*k)
\end{align}
by optimizing the kernel through comparison of the simulated and measured $1^\mathrm{st}$ diffraction order at different grating periods and orientations. They managed to increase the uniformity to $u\approx 0.9$ by including the fringing in the iterative calculation of phase patterns. 

Our goal is to develop a precise and fast model to calculate the SLM behavior for an arbitrary voltage pattern. In order to achieve that, we will use general nematic liquid crystal theory (see  \cref{sec:theory}) to model and simulate the director distribution for a given voltage pattern (see  \cref{sec:numerical implementation of LC model}) for $4$ pixels. These simulations will then be used to find the phase (and amplitude) profiles with the Berreman $4\times 4$ matrix method (see  \cref{sec:Berreman}). By varying unknown parameters (thickness of the LC-layer, birefringence etc.) we will then tailor our model to our SLM. This will happen by comparing calibration and diffraction efficiency measurements of various patterns to our simulations (see  \cref{sec:comparison of experiment with simulation}). After finding appropriate parameters to describe the SLM behavior consistently, we extend our spatially limited but very precise simulations to a much faster model by fitting our phase profiles with suitable functions (see  \cref{sec:Fringer}).

Our model will also include the effect of polarization conversion, which means a change in the polarization state of the light by passing through the LC layer. This effect can lead to a decreased contrast in the desired patterns and to deviations in the patterns themselves. 


\FloatBarrier
\chapter{Modeling the director distribution of uniaxial liquid crystals in the nematic phase}\label{sec:theory}

In this chapter we will present as a key result the differential equations used to model the liquid crystal in a SLM.


In the nematic phase, the long axes of uniaxial liquid crystal molecules possess orientational order described by the director orientation $\vec{n}$, a vector parallel to the average long LC-axis. Due to thermal fluctuations, the orientation of the molecules can deviate from the director orientation. These fluctuations are described by the nematic order parameter

\begin{align}
S = \dfrac{1}{2} \int_0^\pi (3 \cos^2 (\beta)-1) f(\beta) \mathrm{d}\beta
\end{align}
with the orientational distibution function $f(\beta)$ \cite{YangWu2014}. This order parameter $S$ can assume values from $-0.5$ (molecules lie unordered in a plane) to $1$ (perfectly ordered). 

We now want to consider spatial variations $\dfrac{\partial n_i}{\partial x_j}$ of the director $\vec{n}$. These variations are assumed to happen over a distance much larger than the size of the molecules \cite{YangWu2014}. In practice the variations happen at the scale of several microns, whereas the size of the LC-molecules is at the scale of a few nanometers. Therefore, the orientational variation can be described by a continuum theory, where deformations from the uniform state lead to an increase in the free energy $F$, similar to changes of position in solids \cite{Vertogen1988}. 
Since we assume the variations to be small, we can write the free energy density in the general form by only considering second order terms \cite{Vertogen1988}

\begin{align}
f_\mathrm{d} = f_0 + k_1 \sum_{i=1}^3 \sum_{j=1}^3 L_{ij} \dfrac{\partial n_i}{\partial x_j} + k_2 \sum_{i=1}^3 \sum_{j=1}^3\sum_{k=1}^3 L_{ijk} \dfrac{\partial ^2 n_k}{\partial x_i \partial x_j} + k_3 \sum_{i=1}^3 \sum_{j=1}^3 \sum_{k=1}^3 \sum_{l=1}^3 L_{ijkl} \dfrac{\partial n_i}{\partial x_j} \dfrac{\partial n_k}{\partial x_l}
\end{align}
with tensors $L_{ij}$, $L_{ijk}$ and $L_{ijkl}$ constructed only by the Kronecker-Delta $\delta_{ij}$, Levi-Civita Tensor $\varepsilon_{ijk}$ and $n_i$. The terms have to be invariant under transformations \cite{Gennes1994}
\begin{enumerate}
\item $\vec{n}$ $\rightarrow$ $-\vec{n}$
\item $\vec{r}=(x_1,x_2,x_3)$ $\rightarrow$ $-\vec{r}=(-x_1,-x_2,-x_3)$
\end{enumerate}
Additionally, terms of the form $\nabla \vec{g}(\vec{r})$, with $\vec{g}(\vec{r})$ being an arbitrary vector field (assuming $\vec{g}(\vec{r})$ is continuously differentiable in $V$), can be rewritten with Gauss' Theorem 
\begin{align}
\int_{V} \nabla \vec{g}(\vec{r}) \mathrm{d}\vec{r} = \int_{\partial V} \vec{g}(\vec{r}) \mathrm{d}\vec{S}.
\end{align}
These terms only describe contributions to surface energy, and not to volume energy and can therefore be neglected.

By following aforementioned conditions, the valid terms are \cite{Vertogen1988}

\begin{align}
\sum_{i=1}^3\sum_{j=1}^3 \sum_{k=1}^3 n_i n_j \dfrac{\partial n_k}{\partial x_i} \dfrac{\partial n_k}{\partial x_j} = (\vec{n} \times \nabla \times \vec{n})^2 
\end{align}
and
\begin{align}
\sum_{i=1}^3 \sum_{j=1}^3 \left(\dfrac{\partial n_j}{\partial x_i}\right)^2 = (\nabla \cdot \vec{n})^2 + (\vec{n} \cdot \nabla \times \vec{n})^2 + (\vec{n} \times \nabla \times \vec{n})^2.
\end{align}
The term $\vec{n} \cdot (\nabla \times \vec{n})$ does not satisfy condition $2$, but contributes in the case of chiral nematics, where the distortion free state also possesses a twist deformation \cite{Vertogen1988}.

\begin{figure}[h!]
\centering
\includegraphics[width=12cm]{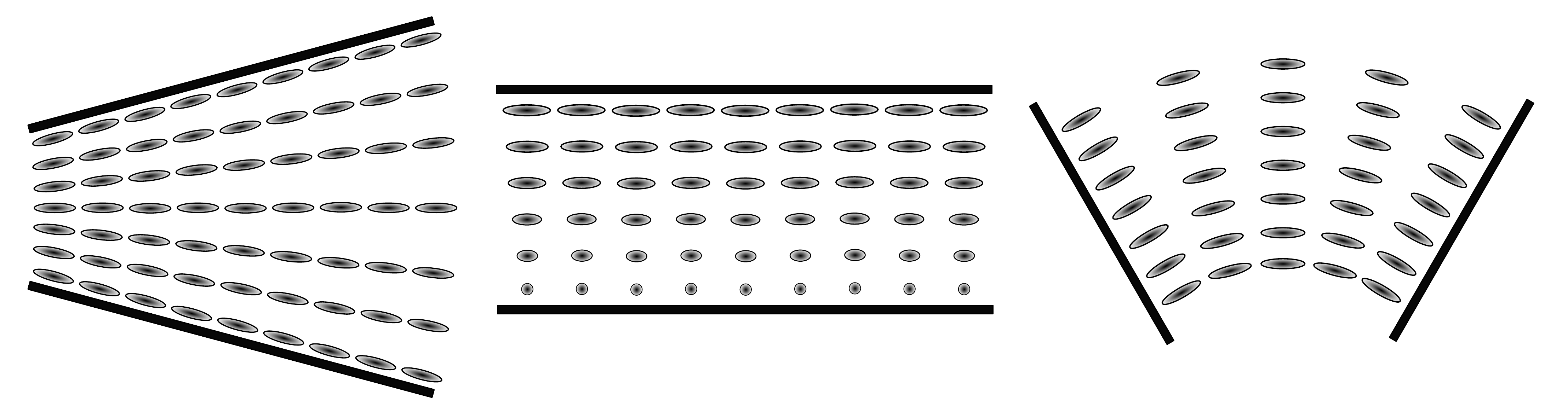}
\caption{Splay, twist and bend deformation modes of liquid crystals in the nematic phase.}
\label{picture splay twist bend deformation}
\end{figure}

By sorting the different terms we arrive at three independent terms which contribute to the so called Frank-Oseen free energy density \cite{Vertogen1988}:
\begin{align}\label{equation frank oseen free energy density compact notation}
f_\mathrm{FO} = \underbrace{ \frac{1}{2} K_{11}  (\nabla \cdot \vec{n})^2}_{\substack{\mathrm{Splay}}} + \underbrace{ \frac{1}{2} K_{22} ( \vec{n} \cdot  \nabla \times \vec{n})^2}_{\substack{\mathrm{Twist}}} + \underbrace{\frac{1}{2} K_{33} (\vec{n} \times \nabla \times \vec{n})^2}_{\substack{\mathrm{Bend}}},
\end{align}
where $K_{11}$, $K_{22}$ and $K_{33}$ denote the splay, twist and bend elastic coefficients respectively. These coefficients describe the elastic energies of the basic deformation modes of a nematic LC shown in \cref{picture splay twist bend deformation}.

\begin{figure}[h!]
\centering
\includegraphics[width=12cm]{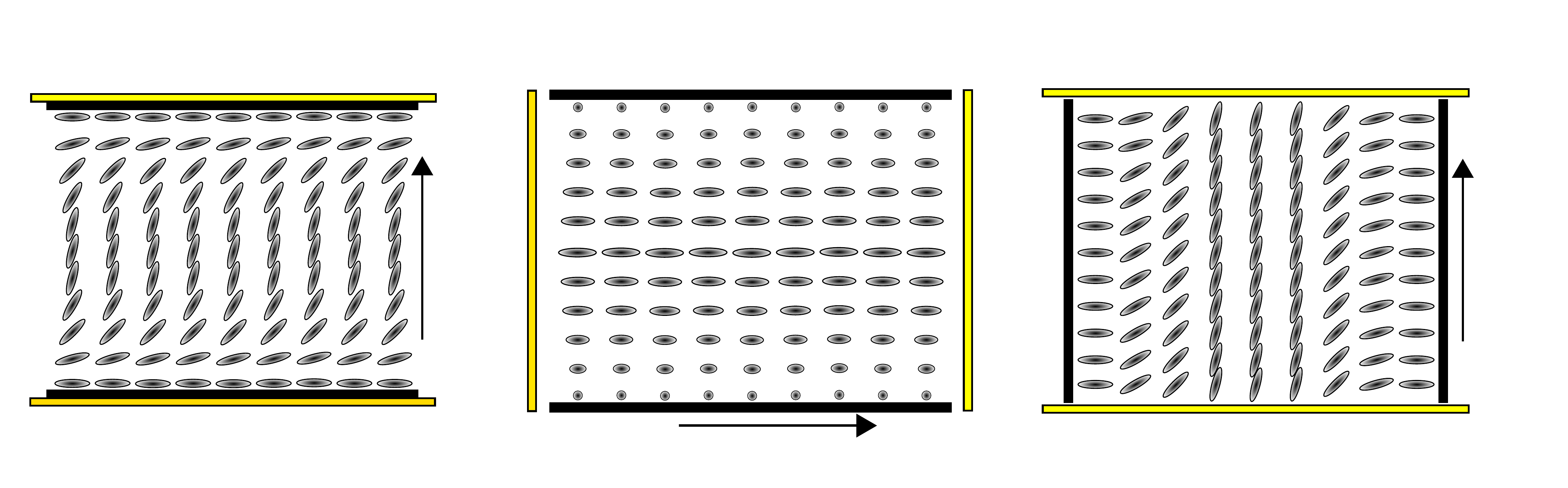}
\caption{Electric field induced elastic deformation in pure splay, twist and bend configuration. Electrodes are depicted in yellow and alignment layers in black.}
\label{picture splay twist bend}
\end{figure}

If we additionally consider an electric field, the total free energy density is \cite{YangWu2014}

\begin{align}\label{equation total free energy density compact notation}
f = \frac{1}{2} K_{11}  (\nabla \cdot \vec{n})^2 + \frac{1}{2} K_{22} ( \vec{n} \cdot  \nabla \times \vec{n})^2 + \frac{1}{2} K_{33} (\vec{n} \times \nabla \times \vec{n})^2-\frac{1}{2}\vec{D}\vec{E},
\end{align}
The last summand represents the electric energy density in a dielectric medium where $\vec{D}$ denotes the dielectric displacement field. By applying an electric field over a liquid crystal layer the molecules will reorient themselves (if the energy is sufficient) in a manner, so that the total free energy is minimized. This reorientation induced by an external electric field is called the Freedericksz Transition \cite{Chandrasekhar1992,YangWu2014}.  \cref{picture splay twist bend} shows this transition in the splay, twist and bend configuration, where the electrodes are represented in yellow and the alignment layers in black. 

\FloatBarrier
\section{Elastic deformation in the splay configuration: simple 1D model}

A SLM driven by a parallel aligned uniform voltage pattern corresponds to a Freedericksz transition in the splay configuration. In this section the basic behavior for this simple case, where the director $\vec{n}$ depends only on the $x_3$ coordinate, will be described. By setting $\vec{E} = E_0 \vec{e}_3$ and $\vec{n} = (\cos (\theta), 0 ,\sin( \theta))$ the elastic energy density yields \cite{YangWu2014}





\begin{align}\label{equation energy density splay}
f_\mathrm{splay} = \frac{1}{2}(K_{11} \cos^2 (\theta) + K_{33} \sin^2 (\theta) ) \left( \frac{\partial \theta }{\partial x_3} \right) ^2 - \frac{1}{2} \varepsilon_0 \Delta \varepsilon E_0^2.
\end{align}

A stationary solution for $\theta$ is given by minimizing the total free energy density
\begin{align}\label{equation total energy splay}
\int f_\mathrm{splay}\left(x_3,\theta (x_3),\frac{\partial \theta}{\partial x_3}(x_3)\right) \mathrm{d}x_3
\end{align}
through the Euler-Lagrange equations $\dfrac{\delta f}{\delta \theta} = 0$
\begin{align}\label{equation euler lagrange splay}
\begin{split}
\frac{\delta f}{\delta \theta} = \frac{\partial f}{\partial \theta} - \frac{\mathrm{d}}{\mathrm{d}x_3} \frac{\partial f}{\partial \left(\frac{\partial \theta}{\partial x_3}\right)} &= - (K_{33} - K_{11}) \sin (\theta) \cos(\theta) \left( \frac{\partial \theta }{\partial x_3} \right) ^2 \\
&- \big(K_{11} \cos^2 (\theta) + K_{33} \sin^2 (\theta)\big) \left( \frac{\partial^2 \theta }{\partial x_3^2} \right)  \\
&- \varepsilon_0 \Delta \varepsilon E_0^2 \sin(\theta) \cos(\theta) = 0.
\end{split}
\end{align}



We can find a stationary solution to  \cref{equation euler lagrange splay} by writing
\begin{align}\label{equation relaxation ansatz}
\gamma \frac{\partial \theta}{\partial t} = \frac{\delta f_\mathrm{splay}}{\delta \theta}
\end{align}
with the viscosity coefficient $\gamma$ (this equation does not describe the dynamics of the Freedericksz transition properly, but leads to the correct equilibrium state for $\theta$ at $\frac{\partial \theta}{\partial t} = \frac{\delta f_\mathrm{splay}}{\delta \theta} = 0$  \cite{YangWu2014}) and search for a solution for $t\rightarrow \infty$. 

We integrate \cref{equation relaxation ansatz} numerically by the iteration prescription at step $\tau$
\begin{align}
\theta^{(\tau +1)} = \theta^{(\tau)} + \alpha_\mathrm{stepsize} \Delta x_3 \left(- \frac{\delta f_\mathrm{splay}}{\delta \theta} \right)^{(\tau)}.
\end{align}
We use $30$ equidistant data points for $\theta$ along $x_3$ with parameter values $K_{33} = 19.4$ pN, $K_{11} = 9.6$ pN and $\Delta \varepsilon = 12.7$. The discrete derivatives are given by the central finite difference approximation (see \cref{sec:numerical implementation of LC model} for the appropriate numerical implementation). The electric field was set constant $E = 32\cdot U/ \Delta x_3$ over a distance of $d=4.25$ $\mathrm{\mu}$m, so $\Delta x_3 =d/31$. As boundary conditions, we chose a director with a pretilt angle of $\theta_p = 10^\circ$. The step size was set $\alpha_\mathrm{stepsize} = 10^{-4}$.

\begin{figure}[h!]
\centering
\includegraphics[width=12cm]{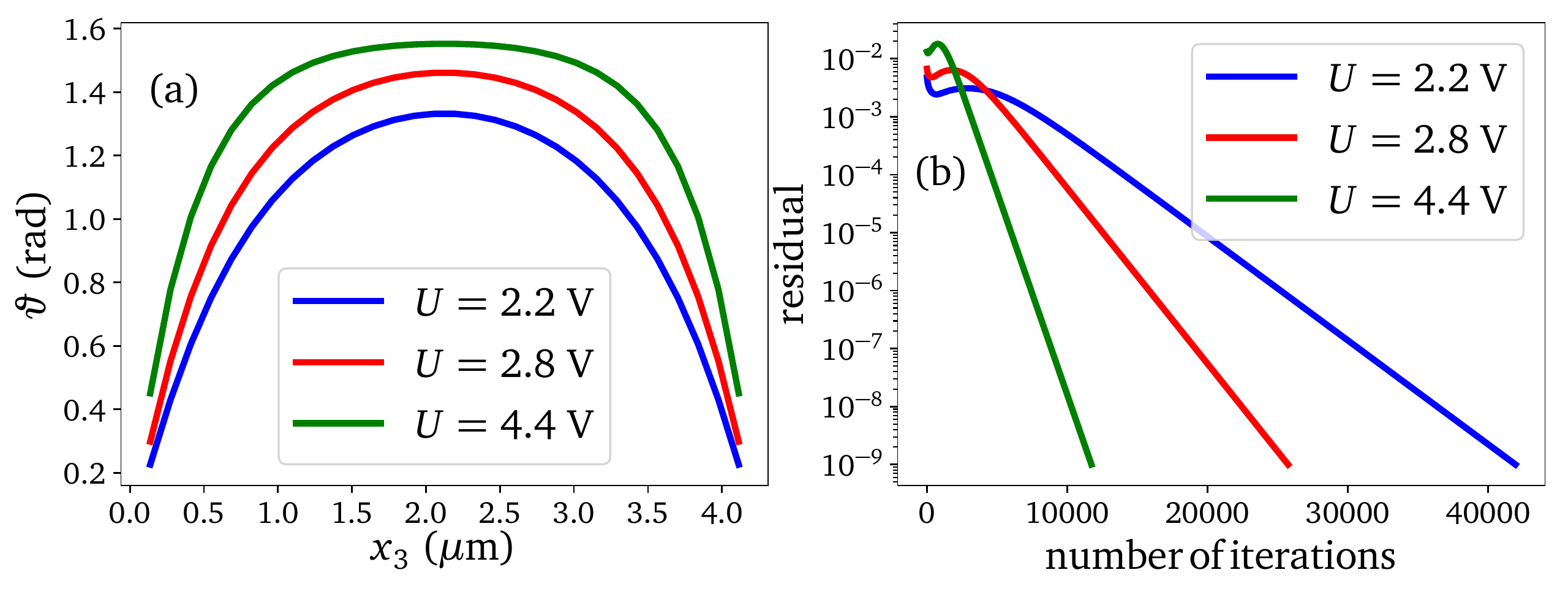}
\caption{Tilt angle $\theta$ across the LC layer at different voltages (a), residual vs. number of iterations (b).}
\label{picture Residual and number of iterations required to satisfy the abortion condition.}
\end{figure}

We stop iterating when the condition 

\begin{align}
\sum_{i=1}^{30} |\theta_i^{(\tau +1)} - \theta_i^{(\tau)}|<10^{-9}
\end{align}
is met.
 \cref{picture Residual and number of iterations required to satisfy the abortion condition.} (a) shows the solution for the tilt angle $\theta$, whereas in  \cref{picture Residual and number of iterations required to satisfy the abortion condition.} (b) we see the value of the residual as a function of the number of iterations. We see that this method has the disadvantage of needing many iterations to converge. The time needed for the blue curve in  \cref{picture Residual and number of iterations required to satisfy the abortion condition.} was $\sim 3.19$ s on an \textit{Intel\textsuperscript{\textregistered} Xeon\textsuperscript{\textregistered} CPU E5-1607 v3} @ $3.10$GHz. 

\begin{figure}[h!]
\centering
\includegraphics[width=12cm]{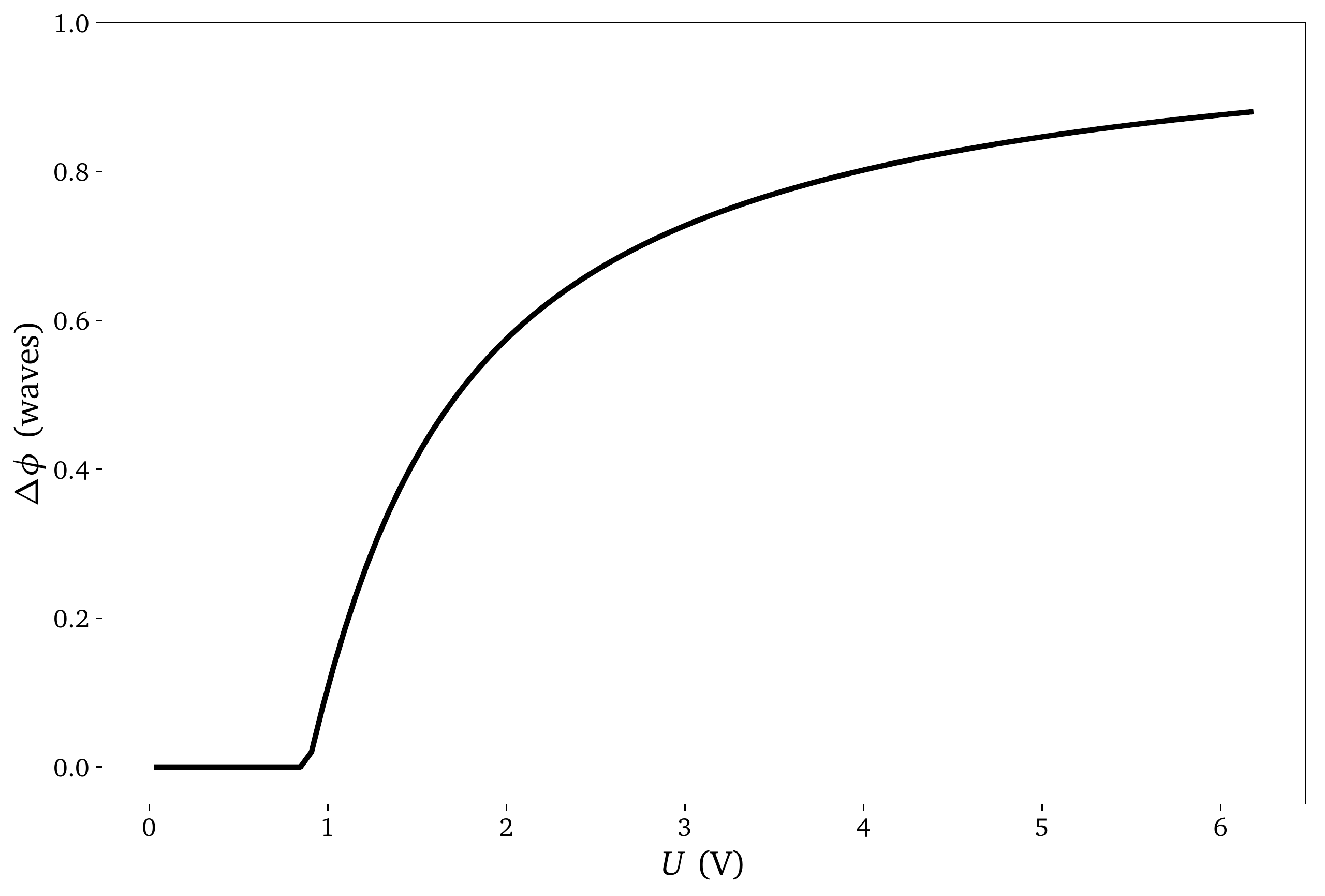}
\caption{Simulation of the accumulated phase shift as a function of the applied voltage.}
\label{picture splay lut}
\end{figure}

In  \cref{picture splay lut} we see the accumulated phase, as calculated by  \cref{equation delta phase}, over a voltage range of $0-6.25$ V. For the refractive indices the values were $n_\mathrm{e} = 2$ and $n_\mathrm{o}=1.5$.

\FloatBarrier
\section{Modeling the 3D director distribution}
\label{sec:directormodeling3D}
\FloatBarrier
\subsection{Vector representation} \label{ssec:theory vector representation}

For an arbitrary external electric field pattern, all three elastic force contributions (splay, twist and bend) have to be considered upon minimizing the total free energy. 
 \cref{equation total free energy density compact notation} written more explicitly yields \cite{YangWu2014}

\begin{align}\label{free energy density equation}
\begin{split}
f &=  K_{11} \sum_{i=1}^3 \sum_{j=1}^3 \dfrac{\partial n_i}{\partial x_i} \dfrac{\partial n_j}{\partial x_j} + \frac{1}{2} K_{22}\sum_{i=1}^3 \sum_{j=1}^3 \left( \frac{\partial n_j}{\partial x_j}\frac{\partial n_j}{\partial x_i}-\frac{\partial n_i}{\partial x_j}\frac{\partial n_j}{\partial x_i} \right) \\ 
&+\frac{1}{2}(K_{33}-K_{22})\sum_{i=1}^3 \sum_{j=1}^3 \sum_{k=1}^3  n_i n_j \frac{\partial n_k}{\partial x_i} \frac{\partial n_k}{\partial x_j} -\frac{1}{2} \varepsilon_0 \Delta \varepsilon \sum_{i=1}^3 \sum_{j=1}^3  E_iE_jn_in_j
\end{split}
\end{align}
with director components $\vec{n}=(n_1,n_2,n_3)$, Frank-Oseen elastic constants $K_{ii}$, electric field $\vec{E}$, and dielectric anisotropy $\Delta \varepsilon = \varepsilon_\mathrm{\parallel} -\varepsilon_\mathrm{\perp}$.

Similar to the $1$D case, a stationary solution for the director distribution for a given electric field is obtained by minimizing the total free energy of the system

\begin{align}
F = \int_{V} f \, \mathrm{d}\vec{r}
\end{align}
in all three dimensions. This functional can be minimized using the Euler-Lagrange equations

\begin{align}\label{euler lagrange allgemein}
-\frac{\delta f}{\delta n_i} =  \sum_{j=1}^3  \frac{\partial }{\partial x_j} \left( \frac{\partial f}{\partial n_{i,j}} \right)- \frac{\partial f}{\partial n_i}  = 0,\quad \mathrm{for} \quad i=1,2,3
\end{align}
which represent a system of coupled, nonlinear, partial differential equations.
Using  \cref{free energy density equation,euler lagrange allgemein} we get

\begin{align}\label{eulerlagrange ausgeschrieben}
\begin{split}
-\frac{\delta f}{\delta n_i} &=  (K_{11}-K_{22}) \sum_{j=1}^3 \frac{\partial^2 n_j}{\partial x_i \partial x_j} + K_{22} \sum_{j=1}^3  \frac{\partial^2 n_i}{\partial x_j^2} + \\
& (K_{33}-K_{22})  \sum_{j=1}^3 \sum_{k=1}^3 n_j n_k \frac{\partial^2 n_i}{\partial x_k \partial x_j} + \\ 
& (K_{33}-K_{22})  \sum_{j=1}^3 \sum_{k=1}^3 \bigg( n_j \frac{\partial n_i}{\partial x_k} \frac{\partial n_k}{\partial x_j} + n_k \frac{\partial n_i}{\partial x_k} \frac{\partial n_j}{\partial x_j}- n_j   \frac{\partial n_k}{\partial x_i} \frac{\partial n_k}{\partial x_j}\bigg) + \\
& \varepsilon_0 \Delta \varepsilon E_i  \sum_{j=1}^3  E_jn_j.
\end{split}
\end{align}
Explicitly written, the components  $F_i := -\dfrac{\delta f}{\delta n_i}$ become

\begin{align}\label{equation f1 explicit}
F_1=&\bigg(K_{11}+(K_{33}-K_{22})n_1^2\bigg) \frac{\partial^2 n_1}{\partial x_1^2} \notag\\
&+ \bigg(K_{22}+(K_{33}-K_{22})n_2^2\bigg) \frac{\partial^2 n_1}{\partial x_2^2} \notag\\
&+ \bigg(K_{22}+(K_{33}-K_{22})n_1^2\bigg) \frac{\partial^2 n_1}{\partial x_3^2} \notag\\
&+ (K_{11}-K_{22})\Bigg[ \Bigg.\frac{\partial^2 n_2}{\partial x_2 \partial x_1}+\frac{\partial^2 n_3}{\partial x_1 \partial x_3}\Bigg] \Bigg. \notag\\
&+ (K_{33}-K_{22})\Bigg[ \Bigg. 2n_1n_2 \frac{\partial^2 n_1}{\partial x_1 \partial x_2} + 2n_1n_3\frac{\partial^2 n_1}{\partial x_1 \partial x_3} + 2n_2n_3 \frac{\partial^2 n_1}{\partial x_2 \partial x_3}\Bigg] \Bigg. \notag\\
&+ (K_{33}-K_{22})  \Bigg[ \Bigg. n_1 \left( \frac{\partial n_1}{\partial x_1 }\right)^2 
+ n_1 \frac{\partial n_1}{\partial x_2 } \frac{\partial n_2}{\partial x_1 }
+ n_1 \frac{\partial n_1}{\partial x_3 } \frac{\partial n_3}{\partial x_1 }
+ n_2 \frac{\partial n_1}{\partial x_1 } \frac{\partial n_1}{\partial x_2 }\\
&+ 2 n_2 \frac{\partial n_1}{\partial x_2 } \frac{\partial n_2}{\partial x_2 }
+ n_2 \frac{\partial n_1}{\partial x_3 } \frac{\partial n_3}{\partial x_2 }
+ n_3 \frac{\partial n_1}{\partial x_1 } \frac{\partial n_1}{\partial x_3 }
+ n_3 \frac{\partial n_1}{\partial x_2 } \frac{\partial n_2}{\partial x_3 }
+ n_3 \frac{\partial n_1}{\partial x_3 } \frac{\partial n_3}{\partial x_3 }\notag\\
&+ n_1 \frac{\partial n_1}{\partial x_1 } \frac{\partial n_2}{\partial x_2 }
+ n_3 \frac{\partial n_1}{\partial x_3 } \frac{\partial n_2}{\partial x_2 }
+ n_1 \frac{\partial n_1}{\partial x_1 } \frac{\partial n_3}{\partial x_3 }
+ n_2 \frac{\partial n_1}{\partial x_2 } \frac{\partial n_3}{\partial x_3 } \Bigg] \Bigg.  \notag\\
&-(K_{33}-K_{22})\Bigg[ \Bigg. n_1 \left(\frac{\partial n_2}{\partial x_1 }\right)^2 
+ n_1  \left(\frac{\partial n_3}{\partial x_1 }\right)^2 \notag\\
&+ n_2 \frac{\partial n_2}{\partial x_1 } \frac{\partial n_2}{\partial x_2 } 
+ n_2 \frac{\partial n_3}{\partial x_1 }\frac{\partial n_3}{\partial x_2 }
+ n_3 \frac{\partial n_2}{\partial x_1 }\frac{\partial n_2}{\partial x_3 }
+ n_3 \frac{\partial n_3}{\partial x_1 }\frac{\partial n_3}{\partial x_3 } \Bigg] \Bigg. \notag\\
&+ \varepsilon_0(\varepsilon_\mathrm{\parallel}-\varepsilon_\mathrm{\perp})E_1 \bigg(E_1n_1+E_2n_2+E_3n_3\bigg)\notag,
\end{align}

\begin{align}\label{equation f2 explicit}
F_2=&(K_{11}-K_{22}) \bigg(\frac{\partial^2 n_1}{\partial x_1 \partial x_2}+\frac{\partial^2 n_3}{\partial x_2 \partial x_3}\bigg) \notag\\
&+ \bigg(K_{11}+(K_{33}-K_{22})n_2^2\bigg) \frac{\partial^2 n_2}{\partial x_2^2} \notag\\
&+ K_{22} \Bigg[ \Bigg.\frac{\partial^2 n_2}{\partial x_1^2} + \frac{\partial^2 n_2}{\partial x_3^2}\Bigg] \Bigg. \notag\\
&+ (K_{33}-K_{22})\Bigg[ \Bigg. n_1^2 \frac{\partial^2 n_1}{\partial x_1^2}+2n_1 n_2 \frac{\partial^2 n_2}{\partial x_1 \partial x_2} + 2 n_1 n_3 \frac{\partial^2 n_2}{\partial x_1 \partial x_3}\notag\\
&+n_2 n_3 \frac{\partial^2 n_2}{\partial x_2 \partial x_3} + n_3 ^2 \frac{\partial^2 n_2}{ \partial x_3^2}\Bigg] \Bigg. \notag\\
&+ (K_{33}-K_{22})  \Bigg[ \Bigg.  n_1 \frac{\partial n_2}{\partial x_2 } \frac{\partial n_2}{\partial x_1 }
+ n_1 \frac{\partial n_2}{\partial x_3 } \frac{\partial n_3}{\partial x_1 }
+ n_2 \frac{\partial n_2}{\partial x_1 } \frac{\partial n_1}{\partial x_2 }\\
&+ n_2 \left( \frac{\partial n_2}{\partial x_2 }\right)^2 
+ n_2 \frac{\partial n_2}{\partial x_3 } \frac{\partial n_3}{\partial x_2 }
+ n_3 \frac{\partial n_2}{\partial x_1} \frac{\partial n_1}{\partial x_3}
+ n_3 \frac{\partial n_2}{\partial x_2 } \frac{\partial n_2}{\partial x_3 }\notag\\
&+ 2 n_3 \frac{\partial n_2}{\partial x_3 } \frac{\partial n_3}{\partial x_3 }
+ n_3 \frac{\partial n_2}{\partial x_3 } \frac{\partial n_1}{\partial x_1 }
+ 2 n_1 \frac{\partial n_2}{\partial x_1 } \frac{\partial n_1}{\partial x_1 }\notag\\
&+ n_1 \frac{\partial n_2}{\partial x_1 } \frac{\partial n_3}{\partial x_3 }
+ n_2 \frac{\partial n_2}{\partial x_2 } \frac{\partial n_3}{\partial x_3 } \Bigg] \Bigg.  \notag\\
&-(K_{33}-K_{22})\Bigg[ \Bigg. + n_1 \frac{\partial n_1}{\partial x_2 } \frac{\partial n_1}{\partial x_1 }
+ n_2 \left(\frac{\partial n_1}{\partial x_2 }\right)^2 
+ n_2  \left(\frac{\partial n_3}{\partial x_2 }\right)^2 \notag\\
&+ n_1 \frac{\partial n_3}{\partial x_2 }\frac{\partial n_3}{\partial x_1 }
+ n_3 \frac{\partial n_1}{\partial x_2 }\frac{\partial n_1}{\partial x_3 }
+ n_3 \frac{\partial n_3}{\partial x_2 }\frac{\partial n_3}{\partial x_3 } \Bigg] \Bigg. \notag\\
&+ \varepsilon_0 (\varepsilon_\mathrm{\parallel}-\varepsilon_\mathrm{\perp})E_2 \bigg(E_1n_1+E_2n_2+E_3n_3\bigg) \notag
\end{align}
and
\begin{align}\label{equation f3 explicit}
F_3=&\bigg(K_{11}+(K_{33}-K_{22})n_1^2\bigg) \frac{\partial^2 n_3}{\partial x_3^2} \notag\\
&+ \bigg(K_{22}+(K_{33}-K_{22})n_1^2\bigg) \frac{\partial^2 n_3}{\partial x_1^2} \notag\\
&+ \bigg(K_{22}+(K_{33}-K_{22})n_1^2\bigg) \frac{\partial^2 n_3}{\partial x_2^2} \notag\\
&+ (K_{11}-K_{22})\Bigg[ \Bigg. \frac{\partial^2 n_1}{\partial x_3 \partial x_1}+\frac{\partial^2 n_2}{\partial x_3 \partial x_2}\Bigg] \Bigg. \notag\\
&+ (K_{33}-K_{22})\Bigg[ \Bigg. 2n_1n_2 \frac{\partial^2 n_3}{\partial x_2 \partial x_1} + 2n_1n_3\frac{\partial^2 n_3}{\partial x_1 \partial x_3} + 2n_2n_3 \frac{\partial^2 n_3}{\partial x_2 \partial x_3} \Bigg] \Bigg. \notag\\
&+ (K_{33}-K_{22})  \Bigg[ \Bigg. n_3 \left( \frac{\partial n_3}{\partial x_3 }\right)^2 
+ n_1 \frac{\partial n_3}{\partial x_1 } \frac{\partial n_1}{\partial x_1 }
+ n_1 \frac{\partial n_3}{\partial x_2 } \frac{\partial n_2}{\partial x_1 }
+ n_1 \frac{\partial n_3}{\partial x_3 } \frac{\partial n_3}{\partial x_1 }\\
&+ 2 n_2 \frac{\partial n_3}{\partial x_2 } \frac{\partial n_2}{\partial x_2 }
+ n_2 \frac{\partial n_3}{\partial x_1 } \frac{\partial n_1}{\partial x_2 }
+ n_2 \frac{\partial n_3}{\partial x_3 } \frac{\partial n_3}{\partial x_2 }\notag\\
&+ n_3 \frac{\partial n_3}{\partial x_1 } \frac{\partial n_1}{\partial x_3 }
+ n_3 \frac{\partial n_3}{\partial x_2 } \frac{\partial n_2}{\partial x_3 }
+ n_2 \frac{\partial n_3}{\partial x_2 } \frac{\partial n_1}{\partial x_1 }\notag\\
&+ n_3 \frac{\partial n_3}{\partial x_3 } \frac{\partial n_1}{\partial x_1 }
+ n_1 \frac{\partial n_3}{\partial x_1 } \frac{\partial n_2}{\partial x_2 }
+ n_3 \frac{\partial n_3}{\partial x_3 } \frac{\partial n_2}{\partial x_2 } \Bigg] \Bigg.  \notag\\
&-(K_{33}-K_{22})\Bigg[ \Bigg. n_3 \left(\frac{\partial n_1}{\partial x_3 }\right)^2 
+ n_3  \left(\frac{\partial n_2}{\partial x_3 }\right)^2
+ n_1 \frac{\partial n_1}{\partial x_3 } \frac{\partial n_1}{\partial x_1 } \notag\\
&+ n_1 \frac{\partial n_2}{\partial x_3 }\frac{\partial n_2}{\partial x_1 }
+ n_2 \frac{\partial n_1}{\partial x_3 }\frac{\partial n_1}{\partial x_2 }
+ n_2 \frac{\partial n_2}{\partial x_3 }\frac{\partial n_2}{\partial x_2 } \Bigg] \Bigg. \notag\\
&+ \varepsilon_0(\varepsilon_\mathrm{\parallel}-\varepsilon_\mathrm{\perp})E_3 \bigg(E_1n_1+E_2n_2+E_3n_3\bigg). \notag
\end{align}

Since the liquid crystals exhibit a dielectric anisotropy, one has to use Gauss' law in matter (no free charges)

\begin{align}\label{equation gauss law in matter}
\nabla \cdot \vec{D} = \nabla (\hat{\varepsilon} \cdot \vec{E}) = - \nabla (\hat{\varepsilon} \cdot \nabla \cdot \varphi) = 0
\end{align}
to calculate the electric field. In  \cref{equation gauss law in matter}, $\vec{D}$ denotes the dielectric displacement field, $\varphi$ the electric potential, and $\hat{\varepsilon}$ the dielectric tensor of the LCs, which has the form 
\begin{align}
\hat{\varepsilon} = \begin{pmatrix}
\varepsilon_\perp & 0 & 0 \\
0 & \varepsilon_\perp & 0 \\
0 & 0 & \varepsilon_\parallel
\end{pmatrix}.
\end{align}
The director $\vec{n}$, $\vec{E}$ and $\vec{D}$ are connected by the relation \cite{YangWu2014}

\begin{align}
\vec{D} = \varepsilon_0 \big(\varepsilon_\perp \vec{E} + \Delta \varepsilon (\vec{E} \cdot \vec{n}) \vec{n}\big).
\end{align}

 \cref{sec:numerical implementation of LC model} contains a description how to numerically solve the above equations. By modeling a SLM, the external electric field will be determined by the applied voltage over a pixel electrode. This voltage will represent the boundary conditions upon calculating the external electric field through Gauss Law in matter. 

If we numerically implement \cref{equation f1 explicit,equation f2 explicit,equation f3 explicit} (see \cref{sec:numerical implementation of LC model}), we will discretize our model using finite difference approximations of the form

\begin{align}
\dfrac{\partial^2 n_l}{\partial x_1^2} [i,j,k] = \dfrac{n_l[i+1,j,k] + n_l[i-1,j,k] - 2 n_l[i,j,k]}{(\Delta x_1 )^2}
\end{align}
which change if we swap the director on a gridpoint, e.g. $n_l[i+1,j,k]\rightarrow -n_l[i+1,j,k]$ \cite{Mori1999}. We will therefore take a look at an alternative formulation of \cref{equation f1 explicit,equation f2 explicit,equation f3 explicit}.
\FloatBarrier
\subsection{Tensor representation}\label{ssec:Tensor representation}

If we want to model a director distribution where the directors of two neighboring slices are oriented anti-parallel to one another, the model discussed in the previous section (\cref{ssec:theory vector representation}) yields an erroneous elastic energy \cite{YangWu2014,Anderson2001}. For the purpose of circumventing this problem, the tensor representation 

\begin{align}
\hat{Q} = \begin{pmatrix}
n_1^2 - \dfrac{1}{3} & n_1 n_2 & n_1 n_3 \\
n_2 n_1 & n_2^2 - \dfrac{1}{3} & n_2 n_3 \\
n_3 n_1 & n_3 n_2 & n_3^2 - \dfrac{1}{3}
\end{pmatrix}
\end{align}
can be used to calculate the Frank-Oseen free energy density

\begin{align}
\begin{split}
f =& \dfrac{1}{12} (K_{33} + 3 K_{22} - K_{11}) \sum_{j=1}^3 \sum_{k=1}^3 \sum_{l=1}^3 \dfrac{\partial Q_{jk}}{\partial x_l} \dfrac{\partial Q_{jk}}{\partial x_l} \\
&+ \dfrac{1}{2} (K_{11}-K_{22})  \sum_{j=1}^3 \sum_{k=1}^3 \sum_{l=1}^3 \dfrac{\partial Q_{jk}}{\partial x_k} \dfrac{\partial Q_{jl}}{\partial x_l} \\
&+ \dfrac{1}{2} (K_{33}-K_{11}) \sum_{j=1}^3 \sum_{k=1}^3 \sum_{l=1}^3 \sum_{m=1}^3 Q_{jk} \dfrac{\partial Q_{lm}}{\partial x_j} \dfrac{\partial Q_{lm}}{\partial x_l}.
\end{split}
\end{align}

The variations of $f$ with respect to the director $\delta f /\delta n_i$ can be expressed by the variation $\delta f / \delta Q_{ij}$ by

\begin{align}
\begin{split}
\dfrac{\delta f }{\delta n_i} = \sum_{j=1}^3 \sum_{k=1}^3 \dfrac{\delta f}{\delta Q_{jk}} \underbrace{\dfrac{\partial Q_{jk}}{\partial n_i}}_{\substack{n_j \delta_{ik} + n_k \delta_{ij}}} = \sum_{j=1}^3 \dfrac{\partial f}{\partial Q_{ji}}.
\end{split}
\end{align}
More explicitly, the variation can be written

\begin{align}\label{equation explizit variation energy tensor representation}
\begin{split}
\dfrac{\delta f }{\delta n_i} =& \dfrac{1}{3} (K_{33} + 3 K_{22} - K_{11}) \sum_{j=1}^3 \sum_{k=1}^3 n_j \dfrac{\partial^2 Q_{ji}}{\partial x_k^2} \\
& + (K_{11}-K_{22})  \sum_{j=1}^3 \sum_{k=1}^3 n_j \left( \dfrac{\partial^2 Q_{ik}}{\partial x_i \partial x_k} + \dfrac{\partial^2 Q_{ik}}{\partial x_j \partial x_k} \right) \\
& + \dfrac{1}{2}(K_{33} - K_{11})  \sum_{j=1}^3 \sum_{k=1}^3 \sum_{l=1}^3 n_j \left( 2 \dfrac{\partial Q_{lk}}{\partial x_l} \dfrac{\partial Q_{ji}}{\partial x_k} + 2 Q_{lk} \dfrac{\partial^2 Q_{ji}}{\partial x_k \partial x_l} - \dfrac{\partial Q_{lk}}{\partial x_i}\dfrac{\partial Q_{lk}}{\partial x_j} \right).
\end{split}
\end{align}

We will use this model in  \cref{ssec:diffraction efficiency hamamatsu slm} to model the director distribution of a SLM, where the directors of neighboring lattices will have anti-parallel orientation. Unfortunately, this model has the disadvantage of yielding non-physical numerical solutions if the angle of the directors between two adjacent lattices is greater than $90^\circ$, which could potentially be circumvented by increasing the number of gridpoints in the numerical implementation \cite{Anderson2001}. Additionally, the numerical implementation of the tensor representation (\cref{equation explizit variation energy tensor representation}) is more complex (and has a triple sum, which yields $81$ terms for the full $3$D implementation) and therefore slower than the vector method. Therefore, we will use the vector representation to simulate the director distribution for our SLM (see  \cref{sec:comparison of experiment with simulation}). 
\FloatBarrier
\section{Simplified 2D model}

If the applied electric field meets certain requirements the $3$D model can be simplified. 
The 3D model equations can be significantly simplified for the case that along the $x_2$ direction the applied voltage is constant, e.g. for a line grating along $x_1$. Here we assume that the alignment layer induces orientation along $x_1$.
In this case we have $\dfrac{\partial n_i}{\partial x_2} = 0$ and $n_2=0$, and  \cref{equation f1 explicit,equation f2 explicit,equation f3 explicit} reduce to the $2$D model

\begin{align}\label{equation stepsize director 2d binary grating 1}
F_1=&\bigg(K_{11}+(K_{33}-K_{22})n_1^2\bigg) \frac{\partial^2 n_1}{\partial x_1^2}\notag\\
&+ \bigg(K_{22}+(K_{33}-K_{22})n_1^2\bigg) \frac{\partial^2 n_1}{\partial x_3^2} \notag\\
&+ (K_{11}-K_{22})\frac{\partial^2 n_3}{\partial x_1 \partial x_3}\notag \\
&+ (K_{33}-K_{22})2n_1n_3\frac{\partial^2 n_1}{\partial x_1 \partial x_3}  \notag\\
&+ (K_{33}-K_{22})  \Bigg[ \Bigg. n_1 \left( \frac{\partial n_1}{\partial x_1 }\right)^2 
+ n_1 \frac{\partial n_1}{\partial x_3 } \frac{\partial n_3}{\partial x_1 }
+ n_3 \frac{\partial n_1}{\partial x_1 } \frac{\partial n_1}{\partial x_3 }\\
&+ n_3 \frac{\partial n_1}{\partial x_3 } \frac{\partial n_3}{\partial x_3 }
+ n_1 \frac{\partial n_1}{\partial x_1 } \frac{\partial n_3}{\partial x_3 } \Bigg] \Bigg.  \notag\\
&-(K_{33}-K_{22})\Bigg[ \Bigg.  
+ n_1  \left(\frac{\partial n_3}{\partial x_1 }\right)^2
+ n_3 \frac{\partial n_3}{\partial x_1 }\frac{\partial n_3}{\partial x_3 } \Bigg] \Bigg. \notag\\
&+ \varepsilon_0(\varepsilon_\mathrm{\parallel}-\varepsilon_\mathrm{\perp})E_1 \bigg(E_1n_1+E_3n_3\bigg),\notag
\end{align}
and
\begin{align}\label{equation stepsize director 2d binary grating 2}
F_3=& \bigg(K_{11}+(K_{33}-K_{22})n_1^2\bigg) \frac{\partial^2 n_3}{\partial x_3^2} \notag\\
&+ \bigg(K_{22}+(K_{33}-K_{22})n_1^2\bigg) \frac{\partial^2 n_3}{\partial x_1^2} \notag\\
&+ (K_{11}-K_{22})\frac{\partial^2 n_1}{\partial x_3 \partial x_1} \notag\\
&+ (K_{33}-K_{22}) 2n_1n_3\frac{\partial^2 n_3}{\partial x_1 \partial x_3} + \notag\\
&+ (K_{33}-K_{22})  \Bigg[ \Bigg. n_3 \left( \frac{\partial n_3}{\partial x_3 }\right)^2 
+ n_1 \frac{\partial n_3}{\partial x_1 } \frac{\partial n_1}{\partial x_1 }
+ n_1 \frac{\partial n_3}{\partial x_3 } \frac{\partial n_3}{\partial x_1 }\\
&+ n_3 \frac{\partial n_3}{\partial x_1 } \frac{\partial n_1}{\partial x_3 }
+ n_3 \frac{\partial n_3}{\partial x_3 } \frac{\partial n_1}{\partial x_1 } \Bigg] \Bigg.  \notag\\
&-(K_{33}-K_{22})\Bigg[ \Bigg. n_3 \left(\frac{\partial n_1}{\partial x_3 }\right)^2 
+ n_1 \frac{\partial n_1}{\partial x_3 } \frac{\partial n_1}{\partial x_1 } \Bigg] \Bigg. \notag\\
&+ \varepsilon_0(\varepsilon_\mathrm{\parallel}-\varepsilon_\mathrm{\perp})E_3 \bigg(E_1n_1+E_3n_3\bigg). \notag
\end{align}
 
Details how to numerically solve the $3$D and $2$D problems are given in  \cref{sec:numerical implementation of LC model}.

After discussing how to model the director distribution of a uniaxial nematic liquid crystal layer we will introduce a method with which we will propagate a plane wave through the LC layer.
\FloatBarrier
\chapter{Calculating the effects on light propagated through an LC layer by the Berreman $4\times 4$ matrix method}\label{sec:Berreman}

The Berreman method is a $4\times 4$ matrix formalism that considers the electric and magnetic field components in light propagation through stratified media, in which the dielectric tensor 

\begin{align}
\hat{\varepsilon} = \begin{pmatrix}
\varepsilon_{11} & \varepsilon_{12} & \varepsilon_{13}  \\
\varepsilon_{21} & \varepsilon_{22} & \varepsilon_{23}  \\
\varepsilon_{31} & \varepsilon_{32} & \varepsilon_{33}
\end{pmatrix}
\end{align}
only varies along $x_3$ \cite{Berreman1972,Eidner1989,WoehlerHaasFritschEtAl1988,Stallinga1999}. It yields results for changes to intensity, phase and polarization of the transmitted and reflected light. 

The Maxwell curl equations are

\begin{align}\label{equation maxwell curl 1}
\nabla \times \vec{E} &= -\frac{\partial \vec{B}}{\partial t}\\ \label{equation maxwell curl 2}
\nabla \times \vec{H} &= \frac{\partial \vec{D}}{\partial t}.
\end{align}
Considering an anisotropic dielectric medium with dielectric tensor $\hat{\varepsilon}$ without magnetization ($\vec{D} = \varepsilon_0 \hat{\varepsilon} \vec{E}$ and $\vec{H} = \frac{1}{\mu_0} \vec{B}$) we can write the electric and magnetic components for a monochromatic wave propagating in the $(x_1,x_3)$ plane as

\begin{align}
\vec{E}(x_1,x_3) &= \vec{E}_0(x_3)\mathrm{e}^{-i(k_{x_1} x_1 - \omega t)} \\
\vec{H}(x_1,x_3) &= \vec{H}_0(x_3)\mathrm{e}^{-i(k_{x_1} x_1 - \omega t)}.
\end{align}
Partial derivatives with respect to $x_1$ and $x_2$ therefore are

\begin{align}\label{ derivatives x1 x2 berreman}
\frac{\partial}{\partial x_1} = -ik_{x_1}, \qquad \frac{\partial}{\partial x_2} = 0.
\end{align}
With  \cref{ derivatives x1 x2 berreman} the two Maxwell equations \cref{equation maxwell curl 1,equation maxwell curl 2} reduce to 

\begin{align}
\begin{pmatrix}
-\dfrac{\partial E_2}{\partial x_3} \\
\dfrac{\partial E_1}{\partial x_3}+ ik_{x_1}E_3 \\
- i k_{x_1} H_2
\end{pmatrix} = i\mu_0 \omega
\begin{pmatrix}
H_1 \\
H_2 \\
H_3
\end{pmatrix}
\end{align}
and
\begin{align}
\begin{pmatrix}
- \dfrac{\partial H_2}{\partial x_3} \\
\dfrac{\partial H_1}{\partial x_3} + ik_{x_1} H_3 \\
-i k_{x_1} H_2
\end{pmatrix}=
\begin{pmatrix}
\varepsilon_{11} E_1 + \varepsilon_{12} E_2 + \varepsilon_{13} E_3 \\
\varepsilon_{21} E_1 + \varepsilon_{22} E_2 + \varepsilon_{23} E_3 \\
\varepsilon_{31} E_1 + \varepsilon_{32} E_2 + \varepsilon_{33} E_3 \\
\end{pmatrix}.
\end{align}

By expressing $H_3$ and $E_3$ in terms of $H_1$, $H_2$, $E_1$ and $E_2$, 

\begin{align}
H_3 &= \frac{k_{x_1}}{\mu_0 \omega} E_2 \\
E_3 &= \frac{\varepsilon_{31} E_1 + \varepsilon_{32} E_2 - \dfrac{k_{x_1}}{\varepsilon_0 \omega} H_2}{\varepsilon_{33}}
\end{align}
we get $4$ equations for the partial derivatives along $x_3$

\begin{align}\label{equation berreman derivatives along x3}
\begin{split}
\dfrac{\partial E_1}{\partial x_3} &= - i k_{x_1} \frac{\varepsilon_{31} E_1 + \varepsilon_{32} E_2 - \dfrac{k_{x_1}}{\varepsilon_0 \omega} H_2}{\varepsilon_{33}} + i\mu_0 \omega H_2 \\
\dfrac{\partial E_2}{\partial x_3} &= - i \mu_0 \omega H_1 \\
\dfrac{\partial H_1}{\partial x_3} &= \frac{-ik_{x_1}^2}{\mu_0 \omega} H_2 + i \varepsilon_0 \omega \left[ \varepsilon_{21} E_1 + \varepsilon_{22} E_2 + \varepsilon_{23} \left( \dfrac{\varepsilon_{31} E_1 + \varepsilon_{32} E_2 - \dfrac{k_{x_1}}{\varepsilon_0 \omega} H_2}{\varepsilon_{33}} \right) \right] \\
\dfrac{\partial H_2}{\partial x_3} &= -i \varepsilon_0 \omega \left[\varepsilon_{11} E_1 + \varepsilon_{12} E_2 + \varepsilon_{13} \left( \dfrac{\varepsilon_{31} E_1 + \varepsilon_{32} E_2 - \dfrac{k_{x_1}}{\varepsilon_0 \omega} H_2}{\varepsilon_{33}} \right) \right]. 
\end{split}
\end{align}
With $\dfrac{\omega}{k_0}=\dfrac{1}{\sqrt{\mu_0 \varepsilon_0}} = \dfrac{1}{\eta_0}$ we define the Berreman vector $\vec{\psi}$

\begin{align}
\vec{\psi} = \begin{pmatrix}
E_1 \\
\eta_0 H_2 \\
E_2 \\
-\eta_0 H_1
\end{pmatrix}.
\end{align}
Equations \cref{equation berreman derivatives along x3} can then be written

\begin{align}\label{equation berreman dgl}
\frac{\partial \vec{\psi}}{\partial x_3} = ik_0\hat{Q} \cdot \vec{\psi}
\end{align}
with the Berreman matrix

\begin{align}
\hat{Q} = \begin{pmatrix}
\dfrac{ - \chi_{x_1} \varepsilon_{13}}{\varepsilon_{33}} & \dfrac{ - \chi_{x_1}^2 }{\varepsilon_{33}} + 1 & \dfrac{ - \chi_{x_1} \varepsilon_{23}}{\varepsilon_{33}} & 0  \\
\dfrac{ - \varepsilon_{13}^2}{\varepsilon_{33}} + \varepsilon_{11} & \dfrac{ - \chi_{x_1} \varepsilon_{13}}{\varepsilon_{33}} & \dfrac{ - \varepsilon_{13} \varepsilon_{23}}{\varepsilon_{33}} + \varepsilon_{12} & 0 \\
0 & 0 & 0 & 1 \\
\dfrac{ - \varepsilon_{13} \varepsilon_{23}}{\varepsilon_{33}} + \varepsilon_{12} & \dfrac{ - \chi_{x_1} \varepsilon_{23}}{\varepsilon_{33}} &    \dfrac{- \chi_{x_1}^2 -  \varepsilon_{23}^2}{\varepsilon_{33}} + \varepsilon_{22} & 0
\end{pmatrix},
\end{align}
where $\chi_{x_1} = \dfrac{k_{x_1}}{k_0} = n \sin(\alpha)$ and $\alpha$ denotes the angle between $\vec{k}$ and $x_3$.

If $\hat{\varepsilon}$ is constant over a range $\Delta x_3$ the solution to  \cref{equation berreman dgl} is 

\begin{align}\label{equation solution berreman}
\vec{\psi}(x_3 + \Delta x_3) = \hat{P} \cdot \vec{\psi}(x_3)
\end{align}
with $\hat{P} = \mathrm{e}^{-ik_0\hat{Q}\Delta x_3}$.
 
For a liquid crystal layer divided into $N$ slabs, the overall propagator $\hat{B}$ is given by the matrix product of the propagators of the single slabs $\hat{P}_j$

\begin{align}
\hat{B} = \prod_{j=1}^N \hat{P}_j. 
\end{align}
The Berreman vectors before propagation, $\vec{\psi}_0$ and after propagation $\vec{\psi}_\mathrm{N}$ are then related by 

\begin{align}
\vec{\psi}_\mathrm{N} = \hat{B} \cdot \vec{\psi}_\mathrm{0}.
\end{align}



We can express the components of the dielectric tensor in terms of the director components or tilt angle $\theta$ and twist angle $\varphi$ by 

\begin{align}\label{equation epsilon tensor explizit with directors and angles}
\begin{split}
\varepsilon_{11} &= n_\mathrm{o}^2 + (n_\mathrm{e}^2 - n_\mathrm{o}^2) n_1^2 = \varepsilon_\perp + (\varepsilon_\parallel - \varepsilon_\perp) \cos^2 (\varphi) \cos^2 (\theta) \\
\varepsilon_{12} &= \varepsilon_{21} = (n_\mathrm{e}^2 - n_\mathrm{o}^2) n_1 n_2 = (\varepsilon_\parallel - \varepsilon_\perp)\sin (\varphi) \cos (\varphi) \cos^2 (\theta) \\
\varepsilon_{13} &= \varepsilon_{31} = (n_\mathrm{e}^2 - n_\mathrm{o}^2) n_1 n_3 = (\varepsilon_\parallel - \varepsilon_\perp) \cos (\varphi) \sin (\theta) \cos (\theta) \\
\varepsilon_{22} &= n_\mathrm{o}^2 + (n_\mathrm{e}^2 - n_\mathrm{o}^2) n_2^2 = \varepsilon_\perp + (\varepsilon_\parallel - \varepsilon_\perp) \sin^2 (\varphi) \cos^2 (\theta) \\
\varepsilon_{23} &= \varepsilon_{32} = (n_\mathrm{e}^2 - n_\mathrm{o}^2) n_2 n_3 = (\varepsilon_\parallel - \varepsilon_\perp) \sin(\varphi) \sin (\theta) \cos( \theta) \\
\varepsilon_{33} &= n_\mathrm{o}^2 + (n_\mathrm{e}^2 - n_\mathrm{o}^2) n_3^2 = \varepsilon_\perp + (\varepsilon_\parallel - \varepsilon_\perp)  \sin^2 (\theta). \\
\end{split}
\end{align}
$n_\mathrm{o}$ and $n_\mathrm{e}$ denote the ordinary and extraordinary refractive indices. In this case, $\hat{\varepsilon}$ is symmetric $\varepsilon_{ij} = \varepsilon_{ji}$.

To calculate the matrix exponential in  \cref{equation solution berreman} we use the Cayley-Hamilton theorem to express $\hat{P}$ with coefficients $\gamma_i$ \cite{WoehlerHaasFritschEtAl1988}

\begin{align}\label{equation cayley hamilton theorem}
\hat{P} = \gamma_1 \hat{I} + \gamma_2 (- i k_0 \Delta x_3) \hat{Q} + \gamma_3 (- i k_0 \Delta x_3)^2 \hat{Q}^2 + \gamma_4 (- i k_0 \Delta x_3)^3 \hat{Q}^3,
\end{align}
which are given by solving the linear equations

\begin{align}
\begin{pmatrix}
1 & (- i k_0 \Delta x_3) & (- i k_0 \Delta x_3)^2 & (- i k_0 \Delta x_3)^3 \\
1 & (- i k_0 \Delta x_3) & (- i k_0 \Delta x_3)^2 & (- i k_0 \Delta x_3)^3 \\
1 & (- i k_0 \Delta x_3) & (- i k_0 \Delta x_3)^2 & (- i k_0 \Delta x_3)^3 \\
1 & (- i k_0 \Delta x_3) & (- i k_0 \Delta x_3)^2 & (- i k_0 \Delta x_3)^3 
\end{pmatrix} \begin{pmatrix}
\gamma_1 \\
\gamma_2 \\
\gamma_3 \\
\gamma_4
\end{pmatrix} = \begin{pmatrix}
\mathrm{e}^{-ik_0 q_1 \Delta x_3} \\
\mathrm{e}^{-ik_0 q_2 \Delta x_3} \\
\mathrm{e}^{-ik_0 q_3 \Delta x_3} \\
\mathrm{e}^{-ik_0 q_4 \Delta x_3}
\end{pmatrix}
\end{align}
with eigenvalues $q_i$ of $\hat{Q}$

\begin{align}
q_1 &= \sqrt{n_\mathrm{o}^2 - \chi_{x_1}^2} \\
q_2 &= - \sqrt{n_\mathrm{o}^2 - \chi_{x_1}^2} \\
q_3 &= - \frac{\varepsilon_{13}}{\varepsilon_{33}} \chi_{x_1} + \frac{n_\mathrm{o}n_\mathrm{e}}{\varepsilon_{33}} \sqrt{\varepsilon_{33} - \left( 1 - \chi_{x_1}^2 \frac{n_\mathrm{e}^2 - n_\mathrm{o}^2}{n_\mathrm{e}^2}n_2^2\right)} \\
q_4 &= - \frac{\varepsilon_{13}}{\varepsilon_{33}} \chi_{x_1} - \frac{n_\mathrm{o}n_\mathrm{e}}{\varepsilon_{33}} \sqrt{\varepsilon_{33} - \left( 1 - \chi_{x_1}^2 \frac{n_\mathrm{e}^2 - n_\mathrm{o}^2}{n_\mathrm{e}^2}n_2^2\right)}.
\end{align}

The coefficients of the powers of $\hat{Q}$ in  \cref{equation cayley hamilton theorem} can be determined in closed form (for $q_i \neq q_j$, $i\neq j$) \cite{WoehlerHaasFritschEtAl1988,YangWu2014}:

\begin{align}
\begin{split}
\gamma_1  &= - \frac{q_2 q_3 q_4 \mathrm{e}^{-ik_0 q_1 \Delta x_3}}{(q_1 - q_2)(q_1 - q_3)(q_1 - q_4)} - \frac{q_1 q_3 q_4 \mathrm{e}^{-ik_0 q_2 \Delta x_3}}{(q_2 - q_1)(q_2 - q_3)(q_2 - q_4)} \\
& - \frac{q_1 q_2 q_4 \mathrm{e}^{-ik_0 q_3 \Delta x_3}}{(q_3 - q_1)(q_3 - q_2)(q_3 - q_4)} - \frac{q_1 q_2 q_3 \mathrm{e}^{-ik_0 q_4 \Delta x_3}}{(q_4 - q_1)(q_4 - q_2)(q_4 - q_3)}
\end{split} \\
\begin{split}
\gamma_2 (- i k_0 \Delta x_3) &= \frac{(q_2 q_3 + q_2 q_4 + q_3 q_4) \mathrm{e}^{-ik_0 q_1 \Delta x_3}}{(q_1 - q_2)(q_1 - q_3)(q_1 - q_4)} + \frac{(q_1 q_3 + q_1 q_4 + q_3 q_4) \mathrm{e}^{-ik_0 q_2 \Delta x_3}}{(q_2 - q_1)(q_2 - q_3)(q_2 - q_4)} \\
& + \frac{(q_1 q_2 + q_1 q_4 + q_2 q_4) \mathrm{e}^{-ik_0 q_3 \Delta x_3}}{(q_3 - q_1)(q_3 - q_2)(q_3 - q_4)} + \frac{(q_1 q_2 + q_1 q_3 + q_2 q_3) \mathrm{e}^{-ik_0 q_4 \Delta x_3}}{(q_4 - q_1)(q_4 - q_2)(q_4 - q_3)}
\end{split} \\
\begin{split}
\gamma_3 (- i k_0 \Delta x_3)^2 &= -\frac{(q_2 + q_3 + q_4) \mathrm{e}^{-ik_0 q_1 \Delta x_3}}{(q_1 - q_2)(q_1 - q_3)(q_1 - q_4)}- \frac{(q_1 + q_3 + q_4) \mathrm{e}^{-ik_0 q_2 \Delta x_3}}{(q_2 - q_1)(q_2 - q_3)(q_2 - q_4)} \\
& - \frac{(q_1 + q_2 + q_4 ) \mathrm{e}^{-ik_0 q_3 \Delta x_3}}{(q_3 - q_1)(q_3 - q_2)(q_3 - q_4)} - \frac{(q_1 + q_2  + q_3) \mathrm{e}^{-ik_0 q_4 \Delta x_3}}{(q_4 - q_1)(q_4 - q_2)(q_4 - q_3)}
\end{split} \\
\begin{split}
\gamma_4 (- i k_0 \Delta x_3)^3 &= \frac{\mathrm{e}^{-ik_0 q_1 \Delta x_3}}{(q_1 - q_2)(q_1 - q_3)(q_1 - q_4)} + \frac{\mathrm{e}^{-ik_0 q_2 \Delta x_3}}{(q_2 - q_1)(q_2 - q_3)(q_2 - q_4)} \\
& + \frac{\mathrm{e}^{-ik_0 q_3 \Delta x_3}}{(q_3 - q_1)(q_3 - q_2)(q_3 - q_4)} + \frac{\mathrm{e}^{-ik_0 q_4 \Delta x_3}}{(q_4 - q_1)(q_4 - q_2)(q_4 - q_3)}
\end{split}
\end{align}

In the case of a reflective SLM, the light travels twice through the LC layer after being reflected \cite{Stallinga1999}.  \cref{picture berreman lc propagation explanation} depicts the process schematically. The Berreman vector $\vec{\psi}_\mathrm{N}$ is the sum of the reflected and incident $H$ and $E$ fields, $\vec{\psi}_0$ is the Berreman vector at the (metallic) mirror with a vanishing electric field. $E_\mathrm{r\parallel}$ and $E_\mathrm{i\parallel}$ denote the reflected and incident components of the electric field parallel to the plane of incidence, $E_\mathrm{r\perp}$ and $E_\mathrm{i\perp}$ denote the reflected and incident components of the electric field perpendicular to the plane of incidence:

\begin{figure}[h!]
\centering
\includegraphics[width=6cm]{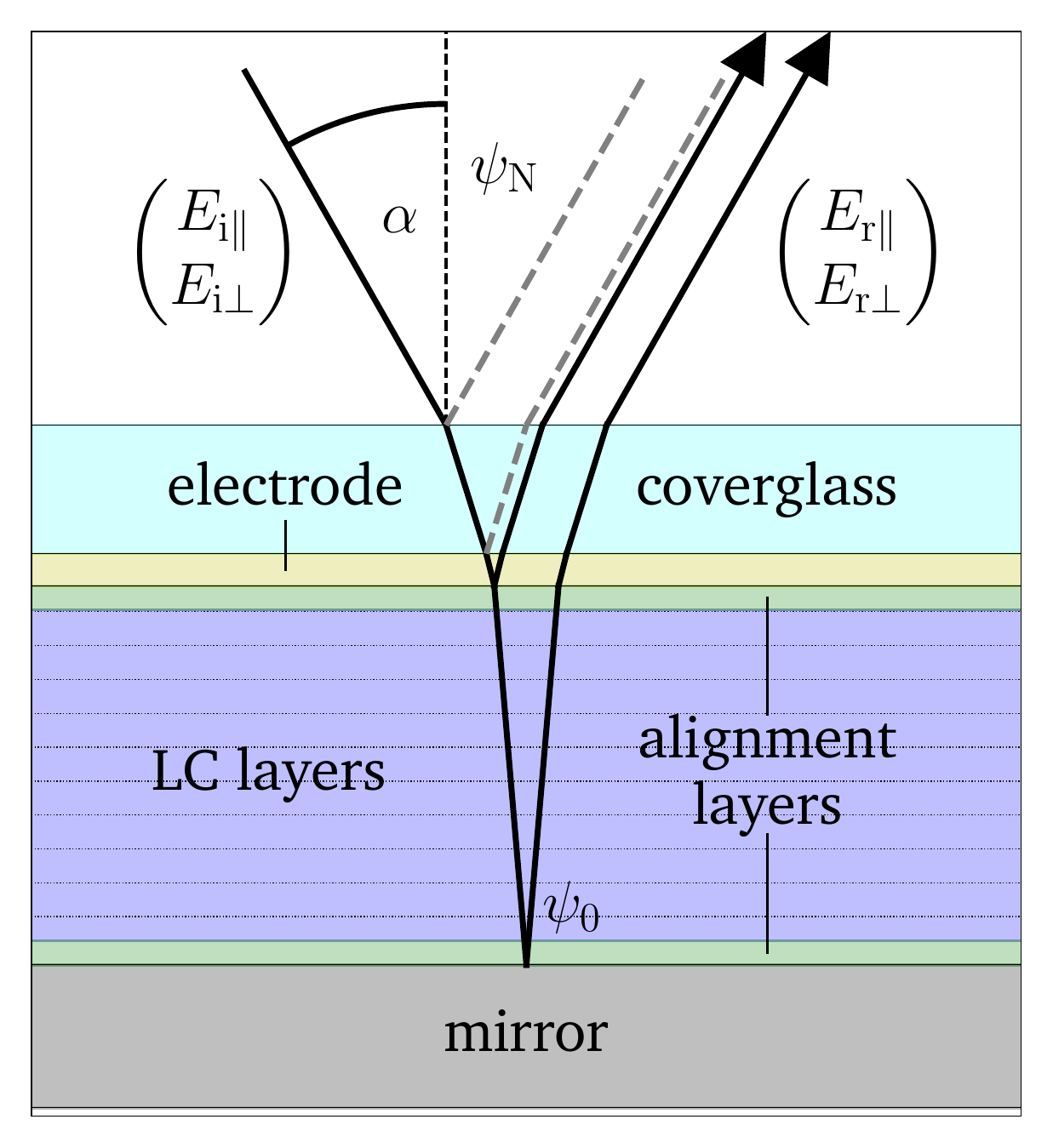}
\caption{Graphic representation of the propagation process using the Berreman matrix method.}
\label{picture berreman lc propagation explanation}
\end{figure} 

\begin{align}
\vec{\psi}_\mathrm{N} = \begin{pmatrix}
(E_\mathrm{r\parallel} + E_\mathrm{i\parallel} )/a \\
(E_\mathrm{r\parallel} - E_\mathrm{i\parallel}) a \\
(E_\mathrm{r\perp} + E_\mathrm{i\perp})/ b \\
(E_\mathrm{r\perp} - E_\mathrm{i\perp}) b
\end{pmatrix}\,\,\,\,\, 
\vec{\psi}_\mathrm{0} = \begin{pmatrix}
0 \\
B_\parallel \\
0 \\
B_\perp
\end{pmatrix} 
\end{align}
with 

\begin{align}
\begin{split}
a = \sqrt{n \cos (\alpha)}\\
b = \sqrt{\frac{n}{\cos (\alpha)}}.
\end{split}
\end{align}
With the propagator $\hat{B}$ we can solve for the reflected electric field components

\begin{align}\label{equation berreman vectors in and reflected}
\begin{pmatrix}
E_\mathrm{r\parallel} \\
E_\mathrm{r\perp}
\end{pmatrix} = -(C_+ C_-) \cdot \begin{pmatrix}
E_\mathrm{i\parallel} \\
E_\mathrm{i\perp}
\end{pmatrix} 
\end{align}
where

\begin{align}
\begin{split}
C_+ = \begin{pmatrix}
B_{11}^{-1} / a + a B_{12}^{-1} & B_{13}^{-1} / b + b B_{14}^{-1} \\
B_{31}^{-1} / a + a B_{32}^{-1} & B_{33}^{-1} / b + b B_{34}^{-1}
\end{pmatrix} \\
C_- = \begin{pmatrix}
B_{11}^{-1} / a - a B_{12}^{-1} & B_{13}^{-1} / b - b B_{14}^{-1} \\
B_{31}^{-1} / a - a B_{32}^{-1} & B_{33}^{-1} / b - b B_{34}^{-1}
\end{pmatrix}.
\end{split}
\end{align}
\FloatBarrier
\section{Slabs of isotropic media}

In the case of an isotropic medium with refractive index $n=\sqrt{\varepsilon}$ the Berreman matrix is reduced to \cite{YangWu2014}

\begin{align}
\hat{Q}_\mathrm{iso} =
\begin{pmatrix}
0 & -\dfrac{\chi_\mathrm{x_1}^2}{n^2}+1 & 0 & 0 \\
n^2 & 0 & 0 & 0 \\
0 & 0 & 0 & 1 \\
0 & 0 & -\chi_\mathrm{x_1}^2 + n^2
\end{pmatrix}.
\end{align}

The eigenvalues $q_{\mathrm{iso},i}$ of $\hat{Q}_\mathrm{iso}$ are degenerate

\begin{align}
\begin{split}
q_{\mathrm{iso},1/3} &= n \cos (\alpha) \\
q_{\mathrm{iso},2/4} &= - n \cos (\alpha)
\end{split}
\end{align}
which lead to 

\begin{align}
\begin{split}
\gamma_1 &= \cos ( n k_0 \cos (\alpha) \Delta x_3) \\
\gamma_2 (-ik_0 \Delta x_3) &= \dfrac{-i}{n \cos(\alpha)} \sin (n k_0 \cos (\alpha) \Delta x_3).
\end{split}
\end{align}
The propagation matrix $\hat{P}_\mathrm{iso} = \exp{ik_0 \Delta x_3 \hat{Q}_\mathrm{iso}}$ is then given by 

\begin{align}
\begin{split}
\hat{P}_\mathrm{iso} &= \gamma_1 \hat{I} + \gamma_2 (-i k_0 \Delta x_3 \hat{Q}_\mathrm{iso} )  \\
&=\left(\begin{matrix}
\cos( n k_0 \cos (\alpha) \Delta x_3) & i \dfrac{\cos(\alpha)}{n} \sin(n k_0 \cos (\alpha) \Delta x_3) \\  
- i \dfrac{n}{\cos(\alpha)} \sin(n k_0 \cos (\alpha) \Delta x_3) & \cos( n k_0 \cos (\alpha) \Delta x_3) \\ 
0 & 0 \\ 
0 & 0 
\end{matrix}\right.
\\
&\left. \begin{matrix}
 0 & 0 \\
 0 & 0 \\ 
 \cos( n k_0 \cos (\alpha) \Delta x_3) & -i \dfrac{1}{n \cos (\alpha) } \sin(n k_0 \cos (\alpha) \Delta x_3) \\
 - i n k_0 \cos(\alpha) \sin(n k_0 \cos (\alpha) \Delta x_3) & \cos( n k_0 \cos (\alpha) \Delta x_3)
\end{matrix}\right)
\end{split}
\end{align}

For $M$ slabs we have the overall propagation matrix

\begin{align}
\hat{B}_\mathrm{iso} = \prod_{i=1}^M \hat{P}_{\mathrm{iso},i}.
\end{align}

By introducing the Berreman vectors $\vec{\psi}_\mathrm{i} = (E_\mathrm{i\parallel},n/\cos(\alpha) E_\mathrm{i\parallel},E_\mathrm{i\perp},n\cos(\alpha)E_\mathrm{i\perp})$ and $\vec{\psi}_\mathrm{t,r} = (E_\mathrm{t\parallel},E_\mathrm{t\perp},E_\mathrm{r\parallel},E_\mathrm{r\perp})$ and matrices $\hat{A}_\mathrm{t}$, $\hat{A}_\mathrm{r}$

\begin{align}
\hat{A}_\mathrm{t} = 
\begin{pmatrix}1 & 0 & 0 & 0 \\
\dfrac{n}{cos(\alpha)} & 0 & 0 & 0 \\
0 & 1 & 0 & 0 \\
0 & n \cos(\alpha) & 0 & 0\end{pmatrix}
\end{align}

\begin{align}
\hat{A}_\mathrm{r} = \begin{pmatrix}
0 & 0 & 1 & 0 \\
0 & 0 & - \dfrac{n}{cos(\alpha)} & 0 \\
0 & 0 & 0 & 1 \\
0 & 0 & 0 & -n \cos(\alpha)\end{pmatrix}
\end{align}
the relation between $\vec{\psi}_\mathrm{i}$ and $\vec{\psi}_\mathrm{t,r}$ can be calculated to

\begin{align}
\vec{\psi}_\mathrm{t,r} = (\hat{A}_\mathrm{t} + \hat{B}_\mathrm{iso} \hat{A}_\mathrm{r})^{-1}\hat{B}_\mathrm{iso} \cdot \vec{\psi}_\mathrm{i}
\end{align}

In the isotropic case $E_1$ and $E_2$ decouples and no polarization conversion takes place.

The Berreman $4\times 4$ matrix method can be used to simulate phase-, polarization and amplitude of light propagated through slabs of anisotropic media. This method is more accurate than the extended Jones matrix method, because it additionally considers reflections at the interface between slabs. 

\begin{figure}[h!]
\centering
\includegraphics[width=12cm]{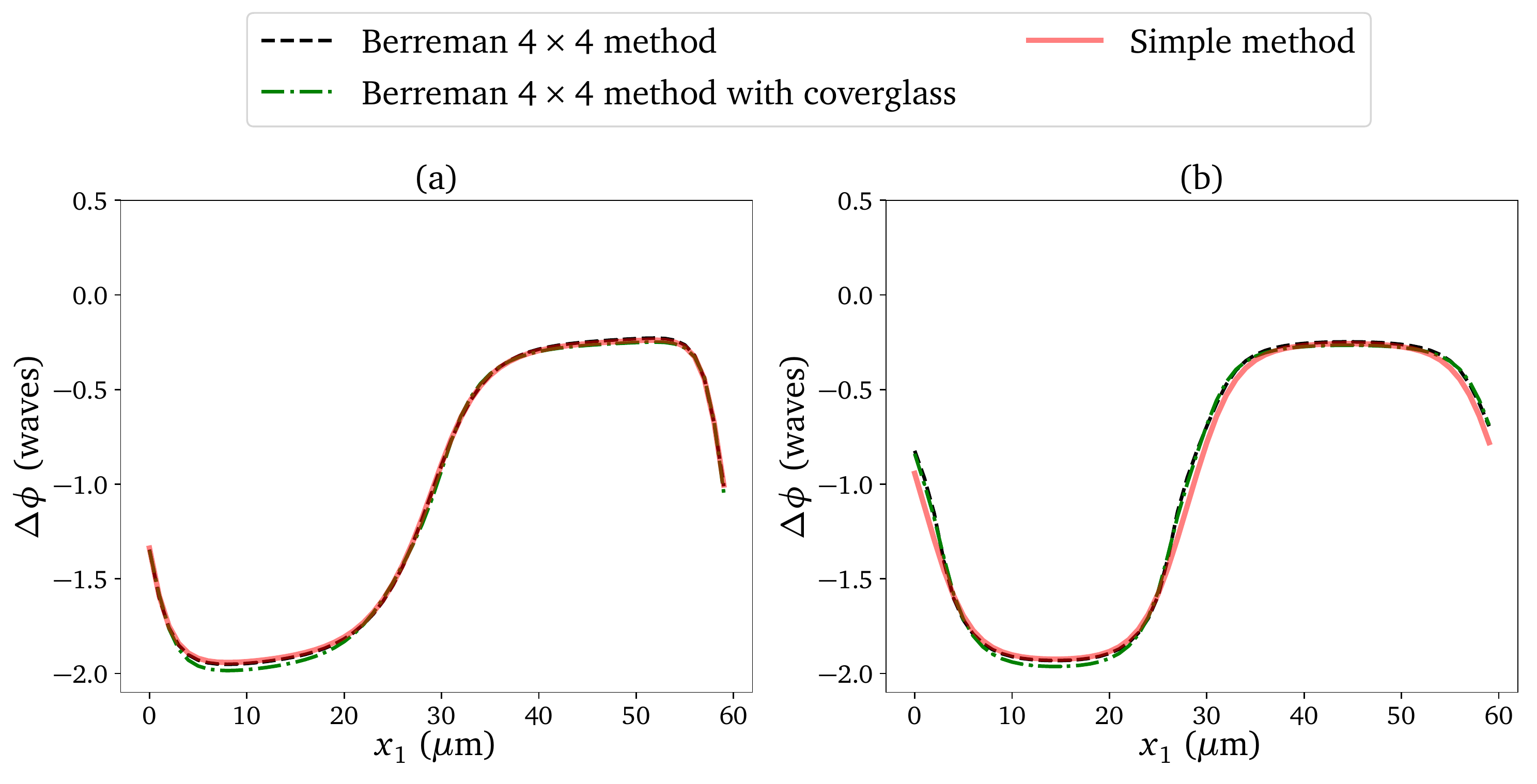}
\caption{Comparison of simulations of the phase shift done by the Berreman method (black and green) and done by the simple method (red), in (a) without and in (b) with polarization conversion.}
\label{picture comparison berreman simple}
\end{figure} 

\cref{picture comparison berreman simple} shows the calculated phase shifts done by the Berreman matrix method (black and green) and by the simple method (red). \cref{picture comparison berreman simple} (b) shows the phase profile for a LC layer, where no polarization conversion takes place and \cref{picture comparison berreman simple} (b) a layer where polarization conversion happens at the transition regions. Comparisons between simulations of the accumulated phase shift done with the Berreman method and the simple method (see \cref{equation delta phase}) show that the results are very similar in \cref{picture comparison berreman simple} (a). In \cref{picture comparison berreman simple} (b) the phase profile changes slightly. \cref{picture comparison berreman simple} (a) corresponds to a phase profile of a binary grating in the asymmetric direction and \cref{picture comparison berreman simple} (b) corresponds to a phase profile of a binary grating in the symmetric direction. For details, how the phase profiles were calculated, see \cref{sec:comparison of experiment with simulation}.

\FloatBarrier
\chapter{Numerical implementation}\label{sec:numerical implementation of LC model}
\section{3D model}

The aim in this section is show how to numerically calculate a stationary solution of  \cref{eulerlagrange ausgeschrieben,equation gauss law in matter} for given bounary conditions for the electric potential and the director distribution. To achieve this we will use the over-relaxation method with a central finite-difference approximation \cite{YangWu2014}. 

We will limit the region of the liquid crystals for the $3$D model to $4$ pixels with periodic boundary conditions to keep the computation time bearable. The director $\vec{n}$ and $\vec{E}$ of this region will be described by arrays of the size $3\times M \times N \times D$ and $3 \times M \times N \times D(1+\mathrm{d_{rel}})$, resp. Array entries $n[i,j,k]$ and $E[i,j,k]$ will represent the values of these quantities in space at position $(x_1,x_2,x_3) = (i\Delta x_1, j \Delta x_2, k \Delta x_3)$ with grid spacings $\Delta x_1$, $\Delta x_2$ and $\Delta x_3$.
We will also need arrays for the dielectric displacement field $\vec{D}$ (same sizes as $\vec{E}$) and the electric potential $\varphi$ (size $M \times N \times D(1+\mathrm{d_{rel}})$).  

The central finite-difference approximation to the $1^\mathrm{st}$ order partial derivatives are

\begin{align}\label{ersteabl}
\frac{\partial f}{\partial x_1}[i,j,k] &= \frac{f[i+1,j,k]-f[i-1,j,k]}{2\Delta x_1} \\
\frac{\partial f}{\partial x_2}{[i,j,k]} &= \frac{f[i,j+1,k]-f[i,j-1,k]}{2\Delta x_2} \\
\frac{\partial f}{\partial x_3}{[i,j,k]} &= \frac{f[i,j,k+1]-f[i,j,k-1]}{2\Delta x_3},
\end{align}
$2^\mathrm{nd}$ order partial derivatives of the form $\dfrac{\partial^2 f}{\partial x_i^2}$ can be approximated by

\begin{align}\label{zweiteabl}
\frac{\partial^2 f}{\partial^2 x_1}[i,j,k] \: &= \: \frac{f[i+1,j,k]+f[i-1,j,k]-2f[i,j,k]}{\Delta x_1^2} \\
\frac{\partial^2 f}{\partial^2 x_2}[i,j,k] \: &= \: \frac{f[i,j+1,k]+f[i,j-1,k]-2f[i,j,k]}{\Delta x_2^2} \\
\frac{\partial^2 f}{\partial^2 x_3}[i,j,k] \: &= \: \frac{f[i,j,k+1]+f[i,j,k-1]-2f[i,j,k]}{\Delta x_3^2}.
\end{align}
The derivatives for entries at the boundaries must be dealt with separately. For entries at the lateral boundaries in $x_1$ and $x_2$ periodic boundary conditions will be established,

\begin{align}\label{abl12randunten}
\frac{\partial f}{\partial x_1}[M,j,k] &\: = \: \frac{f[1,j,k]-f[M-1,j,k]}{2\Delta x_1} \\
\frac{\partial f}{\partial x_2}[i,N,k] &\: = \: \frac{f[i,1,k]-f[i,N-1,k]}{2\Delta x_2} \\
\frac{\partial^2 f}{\partial^2 x_1}[M,j,k] &\: = \: \frac{f[1,j,k]+f[M - 1,j,k]-2f[(M,j,k]}{\Delta x_1^2} \\
\frac{\partial^2 f}{\partial^2 x_2}[i,N,k] &\: = \: \frac{f[i,1,k]+f[i,N-1,k]-2f[i,N,k]}{\Delta x_2^2}
\end{align}
\begin{align}\label{abl12randoben}
\frac{\partial f}{\partial x_1}[1,j,k] &\: = \: \frac{f[2,j,k]-f[M,j,k]}{2\Delta x_1} \\
\frac{\partial f}{\partial x_2}[i,1,k] &\: = \: \frac{f[i,2,k]-f[i,N,k]}{2\Delta x_2} \\
\frac{\partial^2 f}{\partial^2 x_1}[1,j,k] &\: = \: \frac{f[2,j,k]+f[M,j,k]-2f[1,j,k]}{\Delta x_1^2} \\
\frac{\partial^2 f}{\partial^2 x_2}[i,1,k] &\: = \: \frac{f[i,2,k]+f[i,N,k]-2f[i,1,k]}{\Delta x_2^2}
\end{align}

For the electric potential $\varphi$ the boundary conditions at $[i,j,  D(1+d_\mathrm{rel}]$ are given by the electrode voltages $U_\mathrm{bcb}$ and $\varphi=0$ at $[i,j,1]$. The derivatives are then

\begin{align}\label{ablphirandunten}
\frac{\partial \varphi}{\partial x_3}[i,j,  D(1+d_\mathrm{rel})] \: &= \: \frac{U_\mathrm{bcb}[i,j]-\varphi[i,j, D(1+d_\mathrm{rel})-1]}{2\Delta x_3} \\
\frac{\partial^2 \varphi}{\partial^2 x_3}[i,j,  D(1+d_\mathrm{rel})] \: &= \: \frac{\varphi[i,j,D(1+d_\mathrm{rel})-1)]+U_\mathrm{bcb}[i,j]-2\varphi[i,j,D(1+d_\mathrm{rel})]}{\Delta x_3^2} 
\end{align}
and 

\begin{align}\label{ablphirandoben}
\frac{\partial \varphi}{\partial x_3}[i,j,  1] \: &= \: \frac{\varphi[i,j, 2]-0}{2\Delta x_3} \\
\frac{\partial^2 \varphi}{\partial^2 x_3}[i,j,  1] &\: = \: \frac{\varphi[i,j,2]+0-2\varphi[i,j,1]}{\Delta x_3^2}.
\end{align}

At the top and at the bottom surface of the LC-layer the molecules are anchored, which means that the angle between surface and director is constant. 
This angle is called the pretilt angle $\theta_\mathrm{p}$, and the derivatives can be written

\begin{align}\label{ablnunten}
\frac{\partial n_1}{\partial x_3}[i,j,1] &\: = \: \frac{n_1[i,j,2]-\cos\big(\theta_\mathrm{p}\big)}{2\Delta x_3} \\
\frac{\partial n_2}{\partial x_3}[i,j,1] &\: = \: \frac{n_2[i,j,2]-0}{2\Delta x_3} \\
\frac{\partial n_3}{\partial x_3}[i,j,1] &\: = \: \frac{n_3[i,j,2]-\sin\big(\theta_\mathrm{p}\big)}{2\Delta x_3} \\
\frac{\partial^2 n_1}{\partial^2 x_3}[i,j,1] &\: = \: \frac{n_1[i,j,2]+\cos\big(\theta_\mathrm{p}\big)-2n_1[i,j,1]}{\Delta x_3^2} \\
\frac{\partial^2 n_2}{\partial^2 x_3}[i,j,1] &\: = \: \frac{n_2[i,j,2]+0-2n_2[i,j,1]}{\Delta x_3^2} \\
\frac{\partial^2 n_3}{\partial^2 x_3}[i,j,1] &\: = \: \frac{n_3[i,j,2]+\sin\big(\theta_\mathrm{p}\big)-2n_3[i,j,1]}{\Delta x_3^2} 
\end{align}
and

\begin{align}\label{ablnoben}
\frac{\partial n_1}{\partial x_3}[i,j,D] &\: = \: \frac{\cos\big(\theta_\mathrm{p}\big)-n_1[i,j,D-1]}{2\Delta x_3} \\
\frac{\partial n_2}{\partial x_3}[i,j,D] &\: = \: \frac{0-n_2[i,j,D-1]}{2\Delta x_3} \\
\frac{\partial n_3}{\partial x_3}[i,j,D] &\: = \: \frac{\sin\big(\theta_\mathrm{p}\big)-n_3[i,j,D-1]}{2\Delta x_3} \\
\frac{\partial^2 n_1}{\partial^2 x_3}[i,j,D] &\: = \: \frac{n_1[i,j,D-1]+\cos\big(\theta_\mathrm{p}\big)-2n_1[i,j,D]}{\Delta x_3^2} \\
\frac{\partial^2 n_2}{\partial^2 x_3}[i,j,D] &\: = \: \frac{n_2[i,j,D-1]+0-2n_2[i,j,D]}{\Delta x_3^2} \\
\frac{\partial^2 n_3}{\partial^2 x_3}[i,j,D] &\: = \: \frac{n_3[i,j,D-1]+\sin\big(\theta_\mathrm{p}\big)-2n_3[i,j,D]}{\Delta x_3^2}.
\end{align}

The connection between the dielectric displacement field $\vec{D}$, electric field $\vec{E}$ and the director $\vec{n}$ is given by

\begin{align}\label{equation connection D E n}
\vec{D}=\varepsilon_0\bigg(\varepsilon_\mathrm{\perp}\vec{E}+\Delta\varepsilon(\vec{E}\cdot\vec{n})\vec{n}\bigg)
\end{align}
in the LC-layer. 
In the region between electrodes and LC-layer the medium is assumed to be isotropic with dielectric permittivity $\varepsilon_\mathrm{c}$. There, we can simply write

\begin{align}
\vec{D}=\varepsilon_0 \varepsilon_\mathrm{c}\vec{E}.
\end{align}

Using  \cref{equation connection D E n} the boundary conditions for $\vec{D}$ at the upper and lower end can be written in terms of $\varphi$ and $\vec{n}$.  
At the top electrode we have $\vec{E}=0$ and $\varphi=0$, so the derivatives at entries $[i,j,1]$ are

\begin{align}
\frac{\partial D_1}{\partial x_3}[i,j,1] &\: = \: \frac{D_1[i,j,2]-0}{2\Delta x_3} \\
\frac{\partial D_2}{\partial x_3}[i,j,1] &\: = \: \frac{D_2[i,j,2]-0}{2\Delta x_3} \\
\frac{\partial D_3}{\partial x_3}[i,j,1] &\: = \: \frac{D_3[i,j,2]-0}{2\Delta x_3} \\
\frac{\partial^2 D_1}{\partial^2 x_3}[i,j,1] &\: = \: \frac{D_1[i,j,2]+0-2D_1[i,j,1]}{\Delta x_3^2} \\
\frac{\partial^2 D_2}{\partial^2 x_3}[i,j,1] &\: = \: \frac{D_2[i,j,2]+0-2D_2[i,j,1]}{\Delta x_3^2} \\
\frac{\partial^2 D_3}{\partial^2 x_3}[i,j,1] &\: = \: \frac{D_3[i,j,2]+0-2D_3[i,j,1]}{\Delta x_3^2} 
\end{align}

At the bottom electrode we have $\varphi = U_\mathrm{bcb}$, but no electric field $\vec{E}$. The partial derivatives at
$[i,j,D(1+d_\mathrm{rel}]$ can be written

\begin{align}
\frac{\partial D_1}{\partial x_3}[i,j,D(1+d_\mathrm{rel})] &\: = \: \frac{0-D_1[i,j,D(1+d_\mathrm{rel})-1]}{2\Delta x_3} \\
\frac{\partial D_2}{\partial x_3}[i,j,D(1+d_\mathrm{rel})] &\: = \: \frac{0-D_2[i,j,D[1+d_\mathrm{rel})-1]}{2\Delta x_3} \\
\frac{\partial D_3}{\partial x_3}[i,j,D(1+d_\mathrm{rel})] &\: = \: \frac{0-D_3[i,j,D(1+d_\mathrm{rel})-1]}{2\Delta x_3} \\
\frac{\partial^2 D_1}{\partial^2 x_3}[i,j,D(1+d_\mathrm{rel})] &\: = \: \frac{D_1[i,j,D(1+d_\mathrm{rel})-1)]+0-2D_1[i,j,D(1+d_\mathrm{rel})]}{\Delta x_3^2} \\
\frac{\partial^2 D_2}{\partial^2 x_3}[i,j,D(1+d_\mathrm{rel})] &\: = \: \frac{D_2[i,j,D(1+d_\mathrm{rel})-1)]+0-2D_2[i,j,D(1+d_\mathrm{rel})]}{\Delta x_3^2} \\
\frac{\partial^2 D_3}{\partial^2 x_3}[i,j,D(1+d_\mathrm{rel})] &\: = \: \frac{D_3[i,j,D(1+d_\mathrm{rel})-1)]+0-2D_3[i,j,D(1+d_\mathrm{rel})]}{\Delta x_3^2} 
\end{align}


As initial values for the electric potential we choose

\begin{align}\label{equation starting phi}
\varphi^{(0)}[i,j,k] & = kU_\mathrm{bcb}[i,j]/D,
\end{align}
which simply corresponds to a constant electric field oriented along $x_3$. 

For $\vec{n}$ the initial values are chosen to be

\begin{align}
\theta_\mathrm{0} & = \frac{\pi}{180}\bigg(\theta_\mathrm{p}+\theta_\mathrm{max} \sin\big(k\pi\big)\bigg) \\
n_\mathrm{1}^{(0)}[i,j,k] & = \cos\big(\theta_\mathrm{0}\big) \\
n_\mathrm{2}^{(0)}[i,j,k] & = 0 \\
n_\mathrm{3}^{(0)}[i,j,k] & = \sin\big(\theta_\mathrm{0}\big).
\end{align}
This roughly approximates a solution for some intermediate voltage, see \cref{picture Residual and number of iterations required to satisfy the abortion condition.}. For a pretilt angle $\theta_\mathrm{p}=10^\circ$ we choose $\theta_\mathrm{max}=50^\circ$.

For the electric field, the initial values are simply calculated from $\varphi$,  \cref{equation starting phi},

\begin{align}\label{laplacegleichung}
\vec{E}=-\vec{\nabla}\varphi
\end{align}
by using finite differences and boundary conditions. To calculate the dielectric displacement $\vec{D}$ we define

\begin{align}
V[i,j,1:D]&=\varepsilon_\mathrm{\perp}-\varepsilon_\mathrm{c} \\
V[i,j,D+1:D(1+d_\mathrm{rel})]&=0.
\end{align}
The notation $V[i,j,1:D]$ denotes a sub-array of $V$ with indices $1\leq k \leq D$.

\FloatBarrier
\subsection*{Iterative algorithm}

The first step is to initialize the dielectric displacement field with

\begin{align}
D_1[i,j,k]&=V[i,j,k]E_1[i,j,k]\\
D_2[i,j,k]&=V[i,j,k]E_2[i,j,k]\\
D_3[i,j,k]&=V[i,j,k]E_3[i,j,k]
\end{align}
over the whole array $[1:M,1:M,1:D(1+d_\mathrm{rel})]$.

On the sub-array $[1:M,1:M,1:D]$,  the dielectric displacement field is updated by

\begin{align}
\begin{split}
D_i[i,j,k]=& D_i[i,j,k] + (\varepsilon_\mathrm{\parallel}-\varepsilon_\mathrm{\perp})\bigg(E_1[i,j,k]n_1[i,j,k]+E_2[i,j,k]n_2[i,j,k]\\
&+E_3[i,j,k]n_3[i,j,k]\bigg)n_i[i,j,k]
\end{split}
\end{align}
for $i=1,2,3$.

To ensure continuity of $\vec{D}$ at the LC/mirror interface ($[i,j,D+1]$), the components $D_1$ and $D_3$ are determined by

\begin{align}
\begin{split}
D_1[i,j,D+1]=&(\varepsilon_\mathrm{\perp}-\varepsilon_\mathrm{c})E_1[i,j,D+1]\\
&+ (\varepsilon_\mathrm{\parallel}-\varepsilon_\mathrm{\perp})\bigg(E_1[i,j,D+1]\cos\big(\theta_\mathrm{p}\frac{\pi}{180}\big)\\
&+E_3[i,j,D+1]\sin\big(\theta_\mathrm{p}\frac{\pi}{180}\big)\bigg)\cos\big(\theta_\mathrm{p}\frac{\pi}{180}\big)
\end{split}
\end{align}

\begin{align}
\begin{split}
D_3[i,j,D+1]=&(\varepsilon_\mathrm{\perp}-\varepsilon_\mathrm{c})E_3[i,j,D+1]\\
&+ (\varepsilon_\mathrm{\parallel}-\varepsilon_\mathrm{\perp})\bigg(E_1[i,j,D+1]\cos\big(\theta_\mathrm{p}\frac{\pi}{180}\big)\\
&+E_3[i,j,D+1]\sin\big(\theta_\mathrm{p}\frac{\pi}{180}\big)\bigg)\sin\big(\theta_\mathrm{p}\frac{\pi}{180}\big).
\end{split}
\end{align}
The $x_2$ component $D_2[i,j,D+1]$ stays unchanged.

To calculate the electrical potential we define
\begin{align}
F=&\varepsilon_\mathrm{c}\left( \frac{\partial^2 \varphi}{\partial x_1^2}+\frac{\partial^2 \varphi}{\partial x_2^2}+\frac{\partial^2 \varphi}{\partial x_3^2}\right)
+\frac{\partial D_1}{\partial x_1}+\frac{\partial D_2}{\partial x_2}+\frac{\partial D_3}{\partial x_3}.
\end{align}
The update $\Delta \varphi$ for the electric potential (not to be confused with the Laplace operator applied on $\varphi$) is then given by

\begin{align}
\Delta \varphi = F \cdot \Delta x_1 \cdot \Delta x_2 \cdot \Delta x_3
\end{align}
The potential at the step $\varphi^{(\tau +1)}$ is calculated from $\varphi^{(\tau)}$ simply by
 
\begin{align}
\varphi^{(\tau+1)}=\varphi^{(\tau)}+\Delta x_1 \Delta x_2 \Delta x_3 F.
\end{align}

To calculate the update for the director $\vec{n}$ we start by calculating the electric field from the updated potential

\begin{align}
E_1&=\frac{\partial \varphi}{\partial x_1}\\
E_2&=\frac{\partial \varphi}{\partial x_2}\\
E_3&=\frac{\partial \varphi}{\partial x_3}.
\end{align}

Using  \cref{equation f1 explicit,equation f2 explicit,equation f3 explicit} we then can calculate the update for the director components

\begin{align}
\Delta n_1 &= \Delta x_1 \Delta x_2 \Delta x_3 F_1\\
\Delta n_2 &= \Delta x_1 \Delta x_2 \Delta x_3 F_2\\
\Delta n_3 &= \Delta x_1 \Delta x_2 \Delta x_3 F_3.
\end{align}

The mixed second order partial derivatives in  \cref{equation f1 explicit,equation f2 explicit,equation f3 explicit} have to be calculated by first taking the derivative with respect to $x_3$, otherwise the boundary conditions from  \cref{ablnoben,ablnunten} no longer hold. Mixed second order partial derivatives with respect to $x_2$ and $x_1$ can be done either way. 
To ensure that $\vec{n}$ stays a unit vector, $\vec{n}$ is normalized after performing the update for time step at $\tau$, i.e.,

\begin{align}
n_i^{(\tau+1)} &= \dfrac{ \Delta n_i^{(\tau)} + n_i^{(\tau)}}{\sqrt{(\Delta n_1^{(\tau)} + n_1^{(\tau)})^2+(\Delta n_2^{(\tau)} + n_2^{(\tau)})^2+(\Delta n_3^{(\tau)} + n_3^{(\tau)})^2}} \\
\end{align}


\begin{figure}[h!]
\centering
\includegraphics[width=12cm]{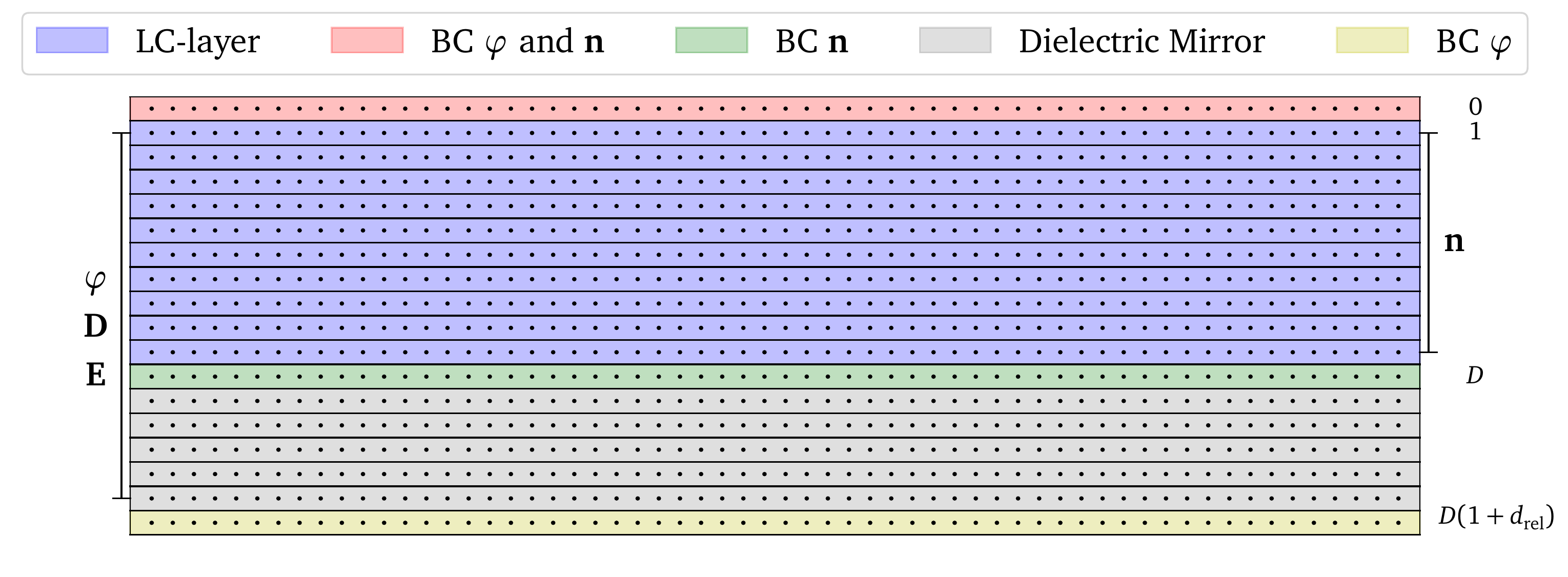}
\caption{$2$D slice of the computational space for $D=10$ and $M=N=60$.}
\label{picture computational space}
\end{figure}

\begin{figure}[h!]
\centering
\includegraphics[width=12cm]{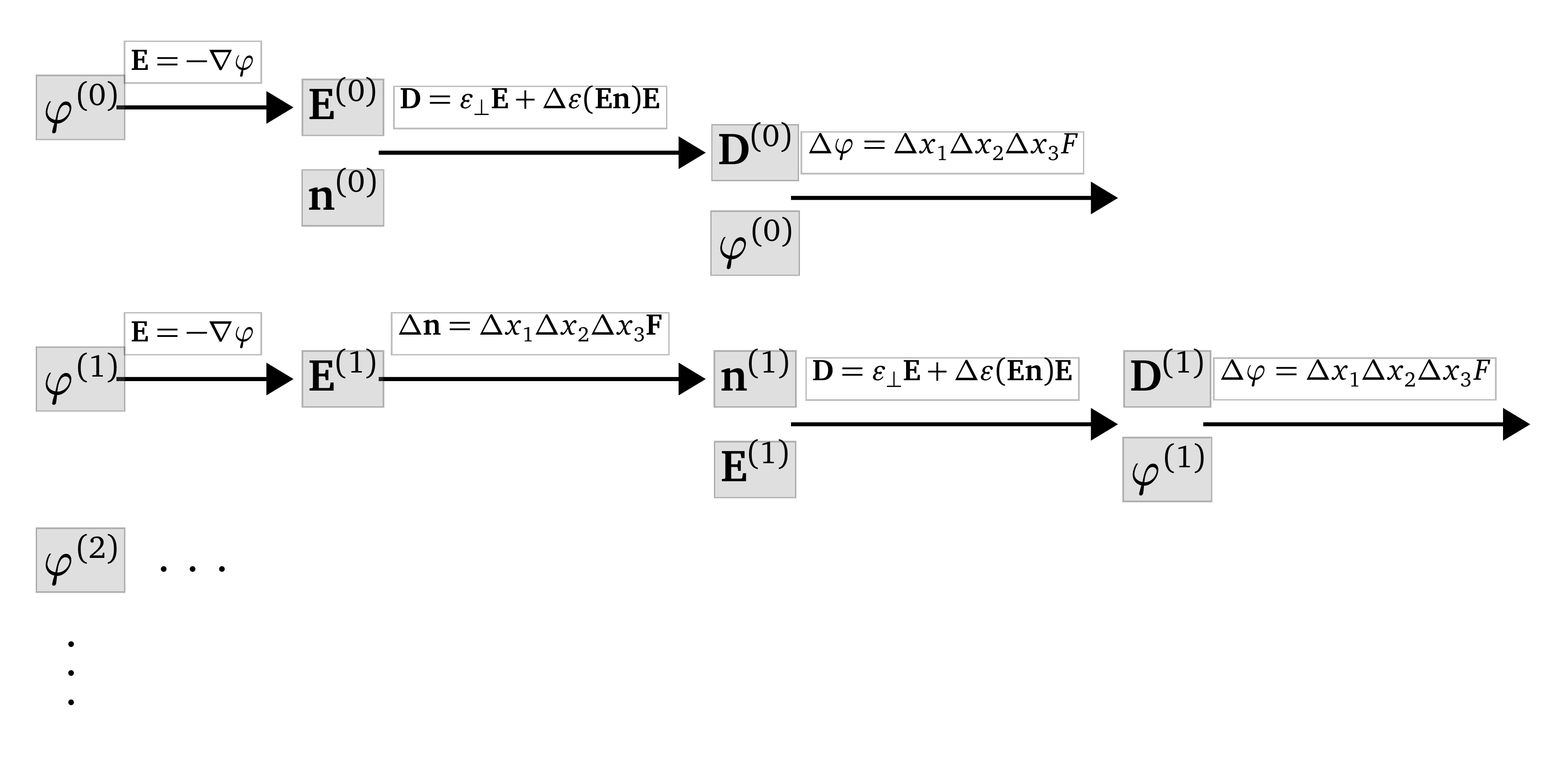}
\caption{Iteration process to calculate director $\vec{n}$ and electric field $\vec{E}$.}
\label{picture iteration process}
\end{figure}
\FloatBarrier
\section{2D model}
\label{sec:2dmodell}
\FloatBarrier
\subsection*{Binary gratings}

For the $2$D model (see  \cref{sec:theory}) the numerical solution is performed similar to the $3$D model. We only need much smaller arrays of size $2\times M\times D$ for the director $\vec{n}$ and $M\times D(1+d_\mathrm{rel})$ for the electric potential $\varphi$, the electric field $\vec{E}$, and the dielectric displacement field $\vec{D}$.

The initial values are then given by

\begin{align}\label{equation start values 2d binary grating}
\varphi^{(0)}[i,k] & = k/DU_\mathrm{bcb} \\
\theta_\mathrm{0} & = \frac{\pi}{180}\bigg(\theta_\mathrm{p}+\theta_\mathrm{max} \sin\big(k\pi\big)\bigg) \\
n_\mathrm{1}^{(0)}[i,k] & = \cos\big(\theta_\mathrm{0}\big) \\
n_\mathrm{3}^{(0)}[i,k] & = \sin\big(\theta_\mathrm{0}\big).
\end{align}
The quantity to compute the updates for $\varphi$ is then

\begin{align}\label{equation stepsize potential 2d binary grating}
F=&\varepsilon_\mathrm{c} \left( \frac{\partial^2 \varphi}{\partial x_1^2}+\frac{\partial^2 \varphi}{\partial x_3^2}\right)+\frac{\partial D_1}{\partial x_1}+\frac{\partial D_3}{\partial x_3}
\end{align}
with $\Delta \varphi = \Delta x_1 \Delta x_3 F$.

For the update of the director we get 
\begin{align}\label{equation upate director 2d binary grating}
 n_1^{(\tau+1)} &= \frac{ \Delta n_1^{(\tau)} + n_1^{(\tau)}}{\sqrt{(\Delta n_1^{(\tau)} + n_1^{(\tau)})^2+(\Delta n_3^{(\tau)} + n_3^{(\tau)})^2}} \\
 n_3^{(\tau+1)} &= \frac{ \Delta n_3^{(\tau)} + n_3^{(\tau)}}{\sqrt{(\Delta n_1^{(\tau)} + n_1^{(\tau)})^2+(\Delta n_3^{(\tau)} + n_3^{(\tau)})^2}}.
\end{align}

\begin{figure}[h!]
\centering
\includegraphics[width=12cm]{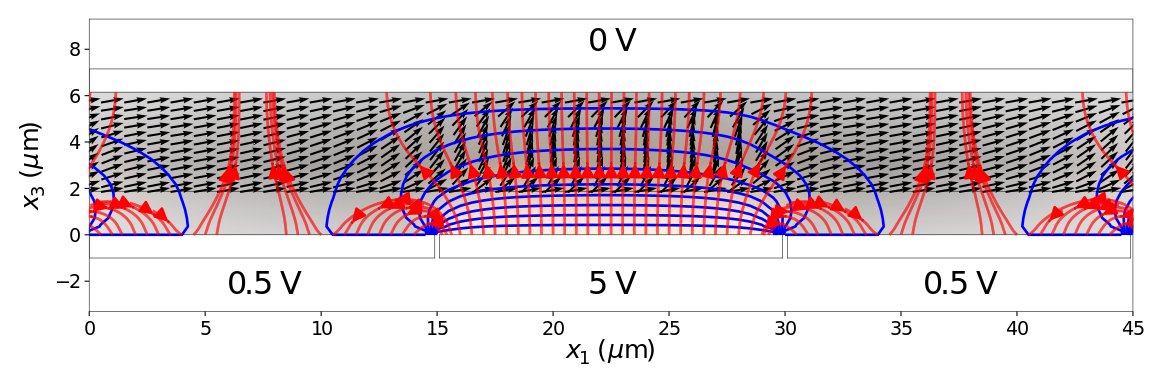}
\caption{Electric field lines (red), director distribution (black arrows), electric potential (background color) and equipotential contour lines (blue) of a binary grating along $x_1$.}
\label{picture potential director field x}
\end{figure}

 \cref{picture potential director field x} depicts the solutions for $\varphi$, $\vec{E}$ and $\vec{n}$ for parameters given in \cref{Table Simulation parameters theory 2d binary and blazed}.

\begin{table}
\begin{center}
\begin{tabular}[c]{r|r|r}
\hline
\multicolumn{3}{c}{Simulation Parameters}   \\
\thickhline
\rowcolor{bluetable1}
& $K_{11}$ & $19.41$ $\mathrm{pN}$ \\
\rowcolor{bluetable1}
&$K_{22}$ & $6.83$ $\mathrm{pN}$ \\
\rowcolor{bluetable1}
&$K_{33}$ & $9.61$ $\mathrm{pN}$ \\
\rowcolor{bluetable1}
&$\varepsilon_\parallel$ & $17.5$ \\
\rowcolor{bluetable1}
&$\varepsilon_\perp$ & $4.8$ \\
\rowcolor{bluetable1}
&$n_\mathrm{e}$ & $1.65$  \\
\rowcolor{bluetable1}
&$n_\mathrm{o}$ & $1.4$  \\
\rowcolor{bluetable1}
\multirow{-8}{*}{LC-Parameters}&$\theta_\mathrm{p}$ & $10^\circ$ \\ \hline
\rowcolor{bluetable2}
& $d$ & $4.25$ $\mu \mathrm{m}$ \\
\rowcolor{bluetable2}
&$d_\mathrm{rel}$ & $0.6$  \\
\rowcolor{bluetable2}
\multirow{-3}{*}{Geometry-Parameters}&$x$ & $30$ $\mu \mathrm{m}$ \\
\hline
\rowcolor{bluetable3}
&$\varepsilon_\mathrm{c}$ &$7$  \\
\hline
\end{tabular}
\end{center}
\caption{Simulation Parameters.}
\label{Table Simulation parameters theory 2d binary and blazed}
\end{table}

\FloatBarrier
\subsection*{Blazed gratings}
 
To calculate the solution for a blazed grating along $x_1$ of period $p$ we increase the number of grid points along $x_1$, so every pixel has $30$ entries, $M=p\cdot 30$. Consequently, we have $U_\mathrm{bcb}[1:M/p] = U_1$, ... , $U_\mathrm{bcb}[(p-1)M/p:M] = U_\mathrm{p}$. For the implementation equations \cref{equation start values 2d binary grating,equation stepsize potential 2d binary grating,equation stepsize director 2d binary grating 1,equation stepsize director 2d binary grating 2,equation upate director 2d binary grating} can be used.

Comparisons with measured data for blazed gratings are shown in \cref{sec:Blazed gratings}.

\begin{figure}[h!]
\centering
\includegraphics[width=12cm]{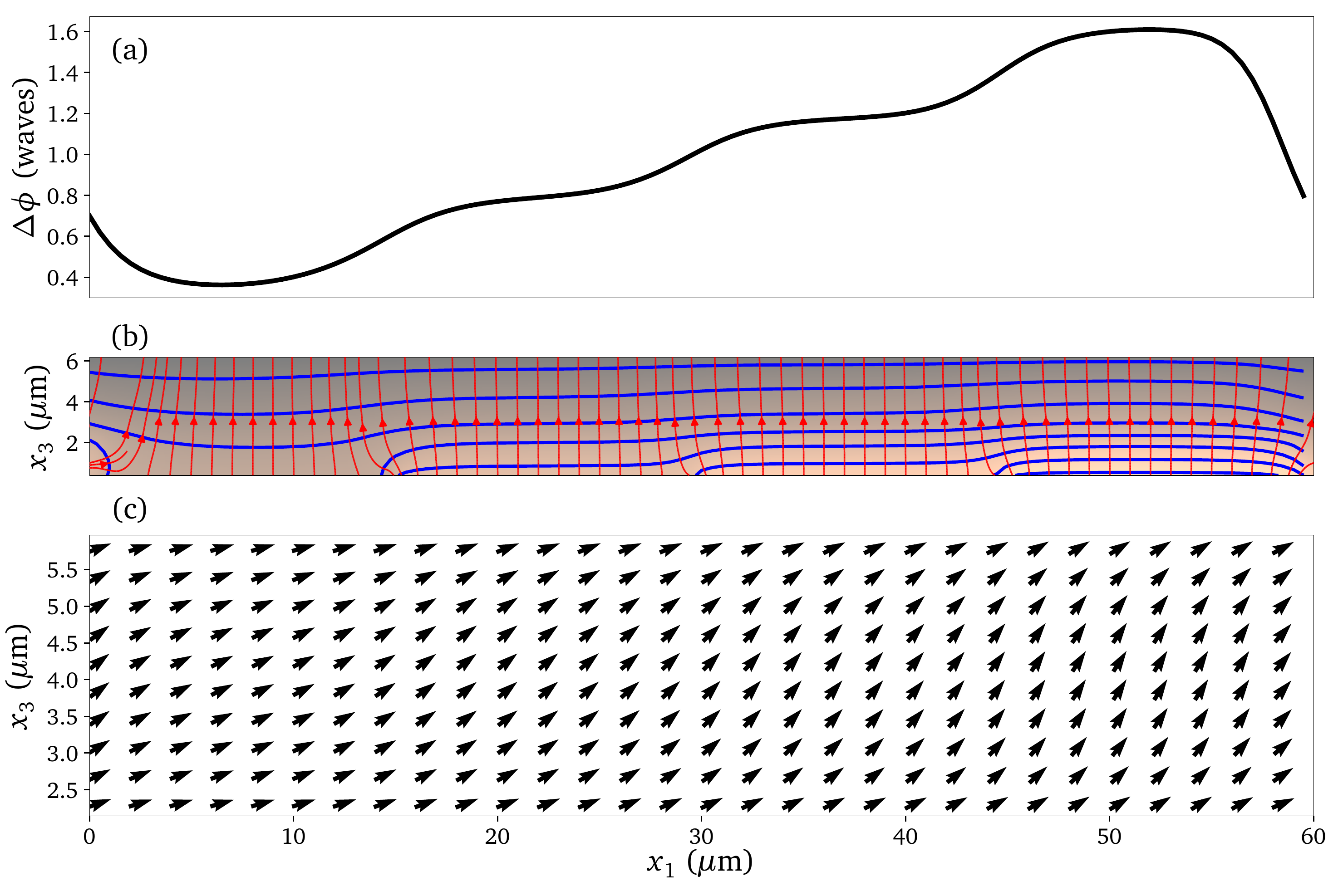}
\caption{Phase profile (a), Electric field lines (red), electric potential (background) and equipotential contour lines (blue) (b) and director distribution (c) of a blazed grating of period $4$ along $x_1$.}
\label{picture phase potential director blazed 4}
\end{figure}

\FloatBarrier
\chapter{Experimental setup}\label{sec:section experimental setup}

%
%

The fringing field effect modifies the realized phase pattern compared to the idealized behavior. We experimentally studied the response of our SLM, in particular we measured the diffraction efficiency for patterns with small periods, where fringing shows the largest effects.

To be able to compare experiment \& model calculations it is crucial to find values for the unknown SLM parameters, such as the Frank elastic coefficients ($K_{11}$, $K_{22}$, $K_{33}$), the dielectric anisotropy ($\Delta \varepsilon$) and the thickness of the LC-layer $d$ and dielectric mirror ($dd_\mathrm{rel}$). For this, several measurements are needed, in particular the measurement of the SLM response to a uniform pattern, which is used to determine the relation between control voltage and phase shift.

\begin{figure}[h!]
\centering
\includegraphics[width=12cm]{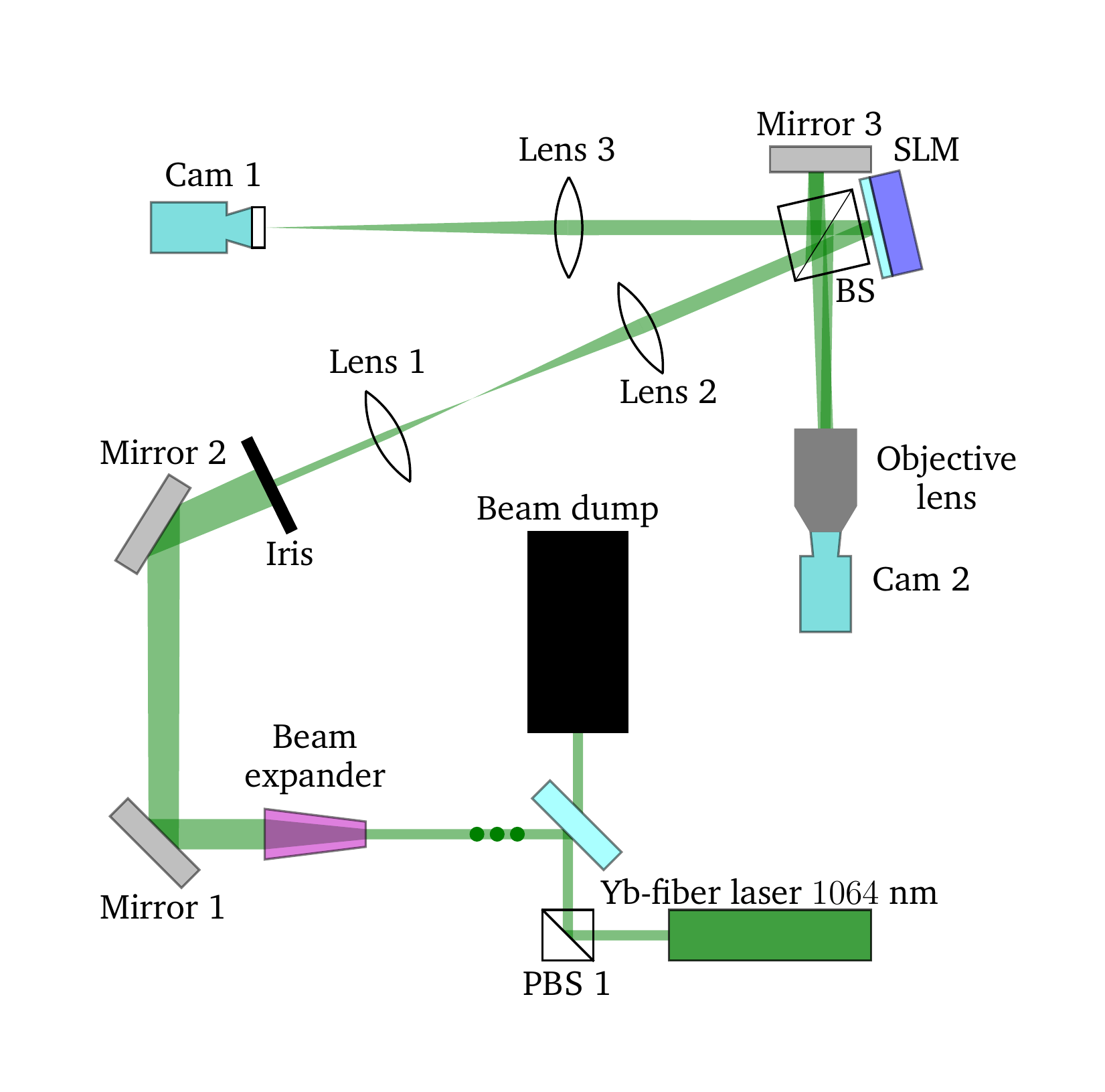}
\caption{Experimental setup for the calibration measurement.}
\label{picture setup LUT}
\end{figure}

 \cref{picture setup LUT} shows the experimental setup for the calibration measurement. A Yb fiber laser at $1064$ nm serves as coherent light source. By passing the laser beam through a polarizing beam splitter we ensure a clean polarization state. Subsequently, the intensity of the beam is reduced by a glass plate and a beam dump. The beam is then expanded and guided to the iris over two dielectric mirrors. Starting from the iris, the laser beam is passed through a $4$f setup with a magnification of $2$, that images the iris on the SLM, and after Lens $2$ optical attenuators are built into the setup. Afterwards, the beam is guided into an interferometer consisting of a beam splitter, another dielectric mirror, the SLM, and a camera with an objective lens (Cam 2). Between camera and objective lens a long pass filter is placed to suppress background light. For this measurement, the path of the interferometer leading to the other camera (Cam 1) is not significant and usually blocked. 
 
\begin{figure}[h!]
\centering
\includegraphics[width=12cm]{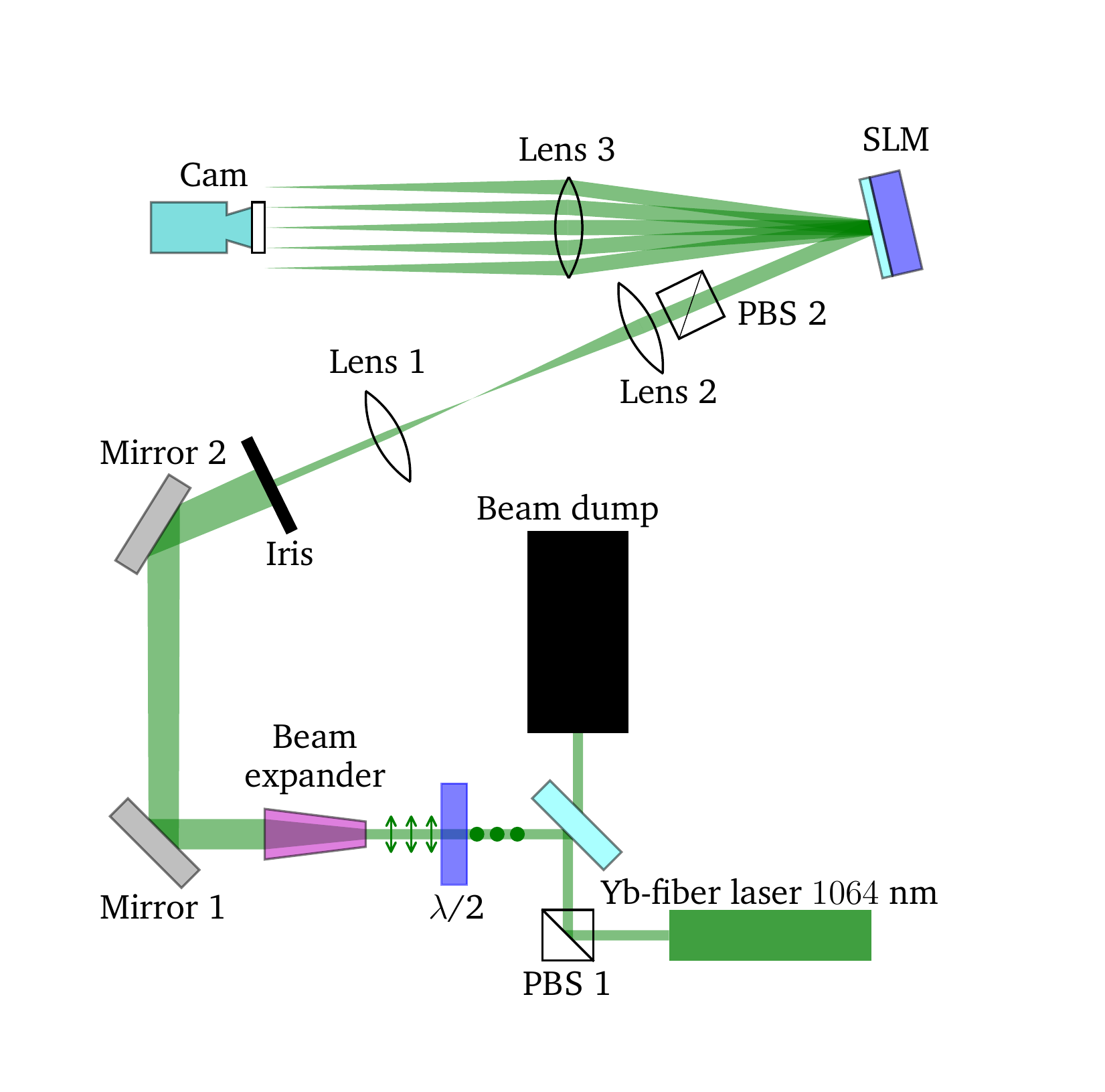}
\caption{Experimental setup for diffraction efficiency measurements.}
\label{picture setup gratings}
\end{figure}

In  \cref{picture setup gratings} we see the experimental setup for the diffraction efficiency measurements. As we recognized it is better to have the SLM oriented such that director and hence polarization are parallel to the optical table, we rotated the SLM by $90^\circ$ for these measurements, and we use a half-wave plate before the beam expander to change the polarization state of the light beam and an additional polarizing beam splitter after Lens $2$. The beam is then simply reflected by the SLM, collected by an additional Lens $3$ and a single diffraction order is selected by an iris and recorded by Cam $1$. Before Cam $1$ we place a low pass filter to suppress background light and a diffuser. The diffuser is used to circumvent interference fringes, which stem from the coverglass of the camera. The diffuser produces a speckle pattern, which averages those fringes out over the area of the camera.

The components used in the experimental setup:
\begin{itemize}
\item PBS $1$/$2$: Thorlabs PBS123, 
\item $\lambda/2$: Thorlabs AHWP$05$M-$950$ - Achromatic Half-Wave Plate, $690-1200$ nm, 
\item Beam expander: Thorlabs BE$02$M-A - $2$X Optical Beam Expander, AR Coated: $400-650$ nm, 
\item Lens $1$: Thorlabs AC$254$-$100$-B-ML - f=$100$ mm, $\varnothing 1"$ Achromatic Doublet, SM$1$-Threaded Mount, ARC: $650-1050$ nm, 
\item Lens $2$: Thorlabs AC$254$-$200$-B-ML - f=$200$ mm, $\varnothing 1"$ Achromatic Doublet, SM$1$-Threaded Mount, ARC: $650-1050$ nm, 
\item Lens $3$: Thorlabs AC$508$-$300$-B-ML - f=$300$ mm, $\varnothing 2"$ Achromatic Doublet, SM$2$-Threaded Mount, ARC: $650-1050$ nm,
\item BS: Thorlabs BS$014$ - $50$:$50$ Non-Polarizing Beamsplitter Cube, $700-1100$ nm, $1"$, 
\item Objective lens: Nikon Nikkor-P $55$mm f/$3.5$ micro macro, 
\item Beam dump: Thorlabs BT$600/$M,  
\item Camera: mvBlueFOX3, model $2024$G
\item Mirror $1,2,3$: Thorlabs BB$1$-E$03$ - $\varnothing1"$ Broadband Dielectric Mirror, $750$ - $1100$ nm
\item Optical Attenuators: various combinations of 
\begin{itemize}
\item Thorlabs ND$10$A - Reflective $\varnothing25$ mm ND Filter, SM1-Threaded Mount, Optical Density: 1.0 
\item Thorlabs NE$03$A-B - $\varnothing25$ mm AR-Coated Absorptive Neutral Density Filter, $650-1050$ nm, SM1-Threaded Mount, OD: $0.3$ 
\end{itemize}
\item Long Pass Filter: Thorlabs FGL$850$ - $\varnothing25$ mm RG$850$ Colored Glass Filter, $850$ nm Longpass      
\item Diffuser: Thorlabs DG$10$-$1500$-H$1$-MD - $\varnothing1"$ SM$1$-Mounted Frosted Glass Alignment Disk w/$\varnothing1$ mm Hole, mounted $5$ mm in front of camera sensor. 
\item SLMs
\begin{itemize}
\item BNS $512\times512$ XY Series 
\item Hamamatsu X$10468$-$07$
\end{itemize}
\end{itemize}
\FloatBarrier
\chapter{Comparison of experiment and simulation}\label{sec:comparison of experiment with simulation}

\begin{figure}[h!]
\centering
\includegraphics[width=12cm]{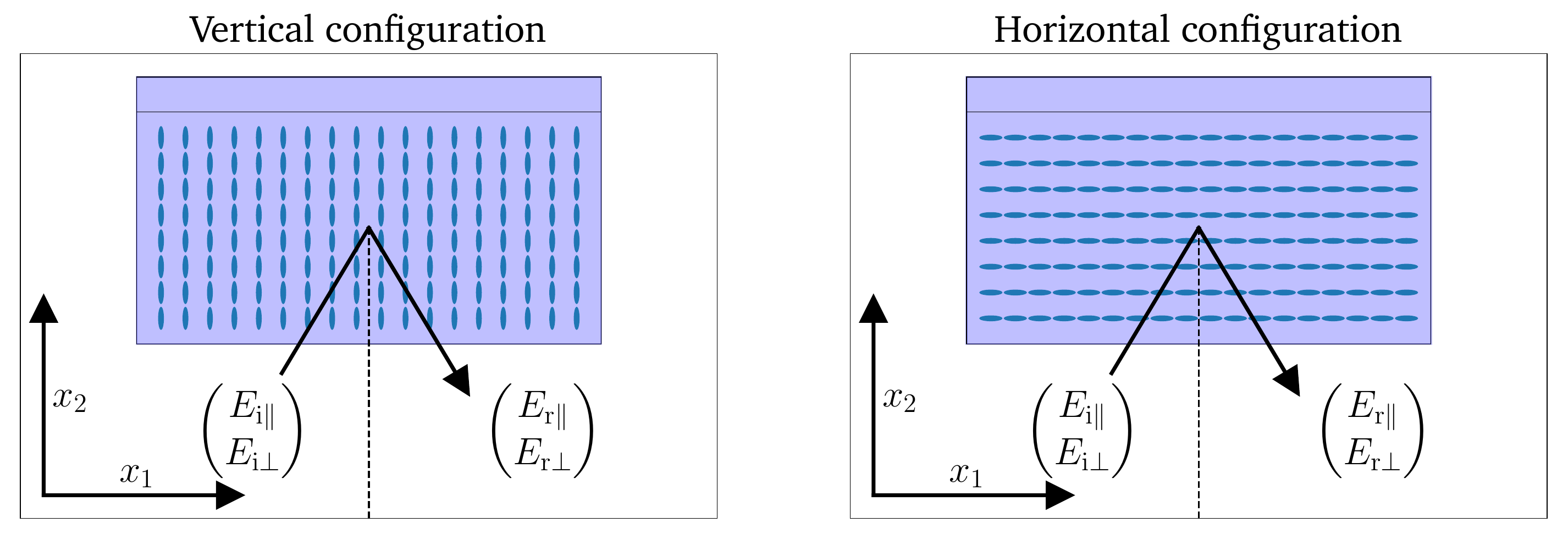}
\caption{The two basic SLM orientations in the experimental setup. Both pictures depict a front view on the SLM, where the plane of incidence is ($x_1,x_3$) and lies perpendicular to the picture plane ($x_1,x_2$).}
\label{picture horizontal vertical explanation}
\end{figure}

 \cref{picture horizontal vertical explanation} shows the two main orientations of the SLM in the experimental setup. If the LC director (at the alignment layer) lies in the plane of incidence of the incident light beam, we will refer to that configuration as horizontal. If the director lies perpendicular to the plane of incidence, we will call that configuration vertical. The vectors $\begin{pmatrix} E_\mathrm{i\parallel} \\E_\mathrm{i\perp}\end{pmatrix}$ and $\begin{pmatrix}E_\mathrm{r\parallel} \\E_\mathrm{r\perp}\end{pmatrix}$ refer to incident and reflected polarization components parallel and perpendicular to the plane of incidence (in correspondence with \cref{equation berreman vectors in and reflected})

\begin{align*}
\begin{pmatrix}
E_\mathrm{r\parallel} \\
E_\mathrm{r\perp}
\end{pmatrix} = -C_+ C_- \begin{pmatrix}
E_\mathrm{i\parallel} \\
E_\mathrm{i\perp}
\end{pmatrix} 
\end{align*}
in the Berreman $4\times4$ matrix formalism. 

This section includes simulations of the phase response vs. control voltage and diffraction efficiency simulations for constant, binary, checkerboard and blazed grating voltage patterns. For the phase/voltage calibration curve the director distribution was simulated with the $2$D model for a constant voltage pattern. The number of grid points along the $x_3$ axis of the director distribution was increased from $10$ to $111$ points through linear interpolation (alignment layers were also added). The extended director distribution was then subsequently fed to the Berreman matrix formalism through  \cref{equation epsilon tensor explizit with directors and angles}. To take into account the reflection from the coverglass-electrode-LC interfaces, the coverglass and electrode layer were implemented separately to determine the transmitted and reflected light from the intersections coverglass/electrode and electrode/LC. The field of the transmitted light $\begin{pmatrix} E_\mathrm{t\parallel,1} \\E_\mathrm{t\perp,1}\end{pmatrix}$ was then fed to the Berreman method for the LC layer, whereas the field reflected at the interface was added to the reflected field from the whole LC layer $\begin{pmatrix} E_\mathrm{r\parallel,2} \\E_\mathrm{r\perp,2}\end{pmatrix}$:

\begin{align}
\vec{E}_\mathrm{r} = \begin{pmatrix} E_\mathrm{r\parallel} \\E_\mathrm{r\perp}\end{pmatrix} = \begin{pmatrix} E_\mathrm{r\parallel,1} \\E_\mathrm{r\perp,1}\end{pmatrix} + \begin{pmatrix} E_\mathrm{r\parallel,2} \\E_\mathrm{r\perp,2}\end{pmatrix}
\end{align}

For the diffraction efficiency simulations the $2$D model was used for binary and blazed gratings in asymmetric direction, while for checkerboard and binary/blazed patterns in the symmetric direction the full $3$D simulation in vector representation was used. To determine the diffraction efficiency, a standard FFT algorithm was used.

The parameters shown in \cref{tabular parameters lut} are the parameters used for the simulations in this section, which are able to describe well the response of our SLM (BNS SLM). The parameters of the LC are $K_{11}$, $K_{22}$, $K_{33}$ (elastic coefficients), $\varepsilon_\parallel$, $\varepsilon_\perp$ (permittivity parallel and perpendicular to the long molecule axis), $n_\mathrm{e}$ and $n_\mathrm{o}$ (refractive indices parallel and perpendicular to the long molecule axis). For the LC parameters, the values were chosen similar to those of $4$-Cyano-$4'$-pentylbiphenyl ($5$CB) \cite{Bogi2001} and modified slightly, since the manufacturer noted the usage of a custom high birefringence LC mixture \cite{bns512}. The parameters for the pixel pitch $x/2$ and $y/2$ were known from the manufacturer \cite{bns512}. The thickness of the LC layer $d$ and the thickness of the dielectric mirror $dd_\mathrm{rel}$ are not published by the manufacturer and therefore tuned to fit measurements. The parameter $\varepsilon_\mathrm{c}$ denotes the (average) permittivity of the dielectric mirror, $n_\mathrm{coverglass}$ and $n_\mathrm{electrode}$ are the refractive indices of the coverglass and the transparent electrode, and $\alpha$ represents the angle of incidence of the laser beam on the coverglass. We assume that the maximum control voltage is $5$ V.

\begin{table}
\begin{center}
\begin{tabular}[c]{r|r|r}
\hline
\multicolumn{3}{c}{Simulation Parameters}   \\
\thickhline
\rowcolor{bluetable1}
& $K_{11}$ & $8.2$ $\mathrm{pN}$ \\
\rowcolor{bluetable1}
&$K_{22}$ & $3.9$ $\mathrm{pN}$ \\
\rowcolor{bluetable1}
&$K_{33}$ & $6.2$ $\mathrm{pN}$ \\
\rowcolor{bluetable1}
&$\varepsilon_\parallel$ & $14$ \\
\rowcolor{bluetable1}
&$\varepsilon_\perp$ & $8.5$ \\
\rowcolor{bluetable1}
&$n_\mathrm{e}$ & $1.9176$  \\
\rowcolor{bluetable1}
&$n_\mathrm{o}$ & $1.54$  \\
\rowcolor{bluetable1}
\multirow{-8}{*}{LC-Parameters}&$\theta_\mathrm{p}$ & $10^\circ$ \\ \hline
\rowcolor{bluetable2}
& $d$ & $3.98$ $\mu \mathrm{m}$ \\
\rowcolor{bluetable2}
&$d_\mathrm{rel}$ & $0.6$  \\
\rowcolor{bluetable2}
\multirow{-3}{*}{Geometry-Parameters}&$x$ & $30$ $\mu \mathrm{m}$ \\
\hline
\rowcolor{bluetable3}
&$\varepsilon_\mathrm{c}$ &$7$  \\
\rowcolor{bluetable3}
&$n_\mathrm{coverglass}$ & $1.575$  \\
\rowcolor{bluetable3}
&$n_\mathrm{electrode}$ & $1.575$  \\
\rowcolor{bluetable3}
&$\alpha$ & $4.2^\circ$ \\
\rowcolor{bluetable3}
&$\lambda$ & $1064$ nm \\
\hline
\end{tabular}
\end{center}
\caption{Simulation Parameters of the BNS SLM.}
\label{tabular parameters lut}
\end{table}




\FloatBarrier
\section{Uniform electric field}\label{ssec:uniform electric field}

\subsection{Phase response for uniform electric field}\label{ssec:LUT}

\begin{figure}[h!]
\centering
\includegraphics[width=12cm]{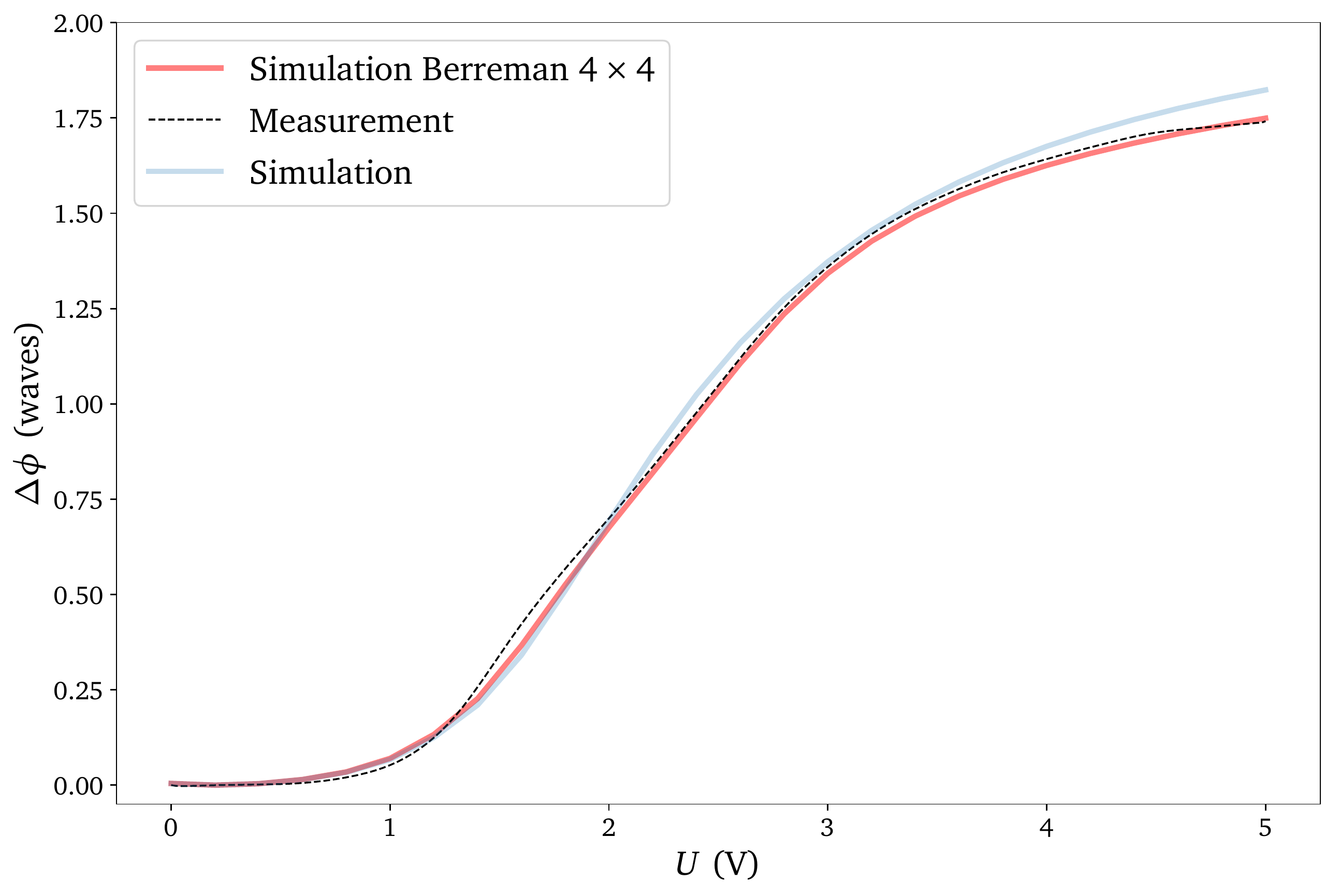}
\caption{Measurements (dashed black) and simulations (red and dashed grey) of the phase shift versus of the applied voltage.}
\label{picture compare measurement simulation lut}
\end{figure}

 \cref{picture compare measurement simulation lut} shows measurements (black dashed) and simulations (red and grey) of the accumulated phase shift of the light beam as a function of the applied voltage. The measurement was done with the experimental setup shown in \cref{picture setup LUT} with a period $32$ binary grating in the vertical configuration.
The grey line referred to as \glqq simple\grqq \, represents the calculation of the accumulated phase shift by

\begin{align}\label{equation delta phase 2}
\Delta \phi = 2 k \Delta x_3 \sum_{i=1}^{10} \left( n(\theta_i) - n(\theta_p)  \right)
\end{align}
with
\begin{align}\label{equation n eff 2}
n(\theta) = \frac{n_\mathrm{e} n_\mathrm{o}}{\sqrt{n_\mathrm{o}^2 + (n_\mathrm{e}^2 - n_\mathrm{o}^2)\sin ^2(\theta)}}.
\end{align}
The angle $\theta$ is calculated from the director component along $x_3$ by $\theta = \arcsin(n_3)$. By comparison with measurements in  \cref{picture compare measurement simulation lut}, this method (grey line) does fit well with measurements. So, even if the phase shift calculation with  \cref{equation delta phase 2} is a simple one, it is on par with the Berreman matrix calculation.
We can improve the simulated phase response slightly by using the Berreman $4 \times 4$ method to propagate the light beam through the LC layers, since it also includes the light reflected at the coverglass-LC interface. 

Since we have a uniform electric field, the LC director only varies along $x_3$ and so does the dielectric tensor. Therefore, it is justifiable to use the Berreman $4 \times 4$ method. Simulations with this method (red line in  \cref{picture compare measurement simulation lut}) stand in very good agreement with measurements in the operational range (approximately linear part from $\sim 1-3$ V) and in the saturated region ($\sim 3-5$ V). The threshold region of the LUT couldn't be resolved sharply by the utilized model. This discrepancy also arises in the simple simulation. This suggests that the cause for this error lies in the simulations of the director distribution. Apparently, the implemented method to determine the orientation of the director across the LC is not accurate at describing the Freedericksz transition near the threshold. 

In practice, the SLM has a spatially dependent LUT due to a curved silicon back plane. The LUT measurement shown in  \cref{picture compare measurement simulation lut} stems from a small region around the center of the SLM.

\FloatBarrier
\subsection{Polarization effects for uniform pattern}

\begin{figure}[h!]
\centering
\includegraphics[width=12cm]{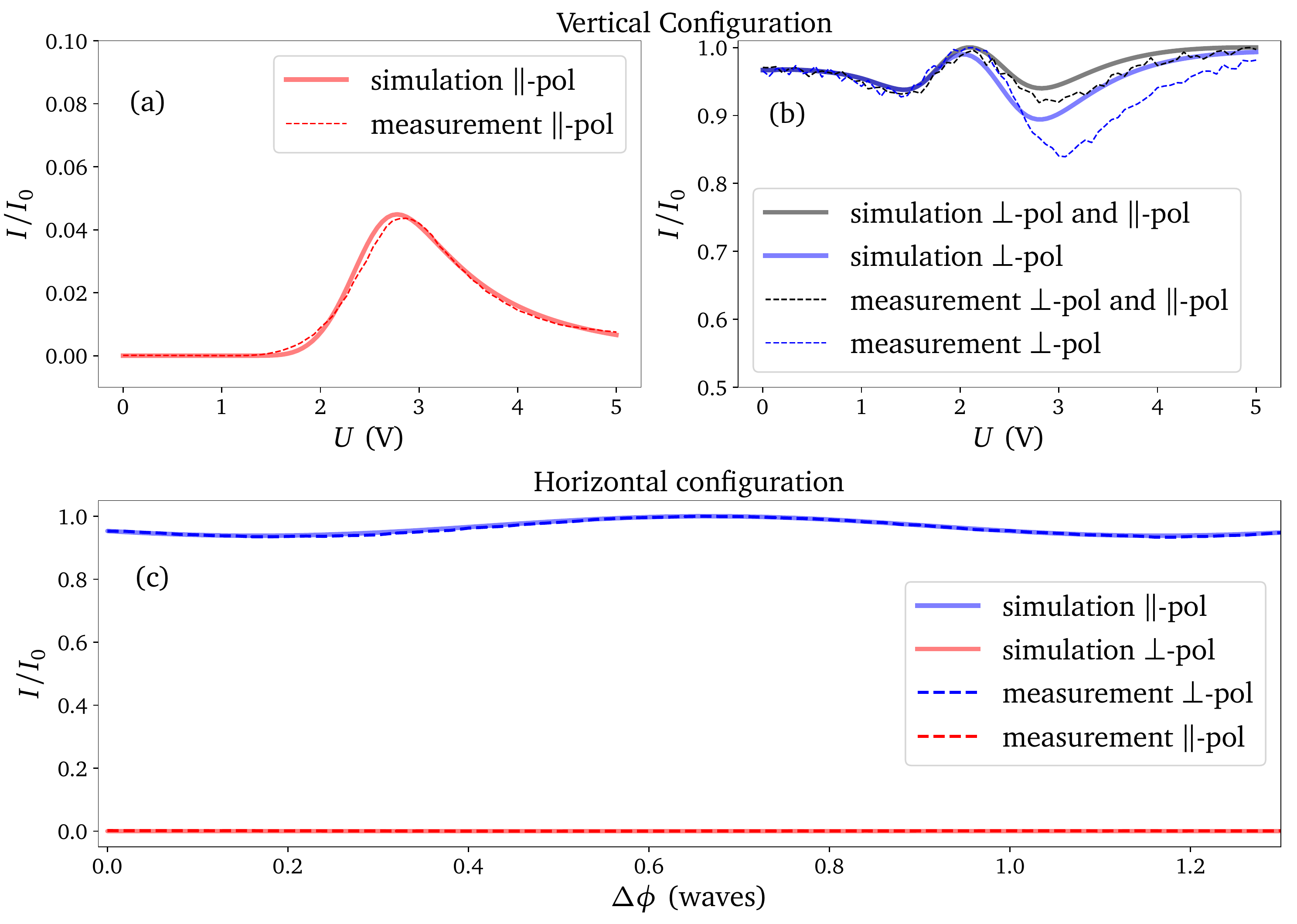}
\caption{Polarization conversion efficiencies for a uniform voltage pattern in vertical (a,b) and horizontal configuration (c).}
\label{picture compare measurement simulation const polx conv}
\end{figure}

In the vertical configuration, the incident light beam is polarized perpendicular ($\perp$-pol) with respect to the plane of incidence. We also have to account for polarization conversion if the incidence of the light beam is not orthogonal to the SLM surface. In  \cref{picture compare measurement simulation const polx conv} we see intensity measurements of the light beam after passing through the SLM for different polarization directions in the vertical configuration. In (a) simulations and measurements are depicted in red for the reflected $\parallel$-pol component. The red line (a) represents the amount of light converted from $\perp$-pol to $\parallel$-pol by the LC layer. At low voltages, no polarization change occurs whereas for $U>2$ V polarization conversion always occurs with a maximum of about $\sim 4\%$ of total intensity.  \cref{picture compare measurement simulation const polx conv} (b) shows the $\perp$-pol components and the total intensity. The total measured intensity in \cref{picture compare measurement simulation const polx conv} (b) is characterized by a modulation, which stems from interference between the partially reflected light beam at the interface between coverglass and liquid crystal layer and the light beam modulated by the LC layer. The modulation of the $\perp$-pol intensity in (b) is additionally characterized by the loss of light due to polarization conversion. In  \cref{picture compare measurement simulation const polx conv}, the simulation matches well with the measurement in \cref{picture compare measurement simulation const polx conv} (a), in \cref{picture compare measurement simulation const polx conv} (b) there is a discrepancy between simulation and measurement of the reflected $\perp$-pol. 

In the horizontal configuration (\cref{picture compare measurement simulation const polx conv} (c)) simulations and measurements show no polarization conversion, only a modulation, caused by interference of the partially and total reflected light beams in the $\parallel$-pol. These results suggest that the horizontal configuration is preferable to the vertical configuration, due to smaller (vanishing) polarization conversion efficiencies. 

\FloatBarrier
\section{Comparison of experiment and simulations for binary gratings}\label{ssec:comparison simulation experiment binary gratings}

\subsection{Validity of theoretical model}

In the previous chapter we discussed experiments and simulations of uniform electric fields applied on the SLM electrodes. This situation (uniform electric field) enabled us to make use of the Berreman $4\times 4$ method, since it assumes a variation of the dielectric tensor only along $x_3$. By applying a binary grating of some sort, the dielectric tensor also varies along $x_1$ or $x_2$ and the assumptions for the Berreman $4\times4$ method are not fully met.
  
\begin{figure}[h!]
\centering
\includegraphics[width=12cm]{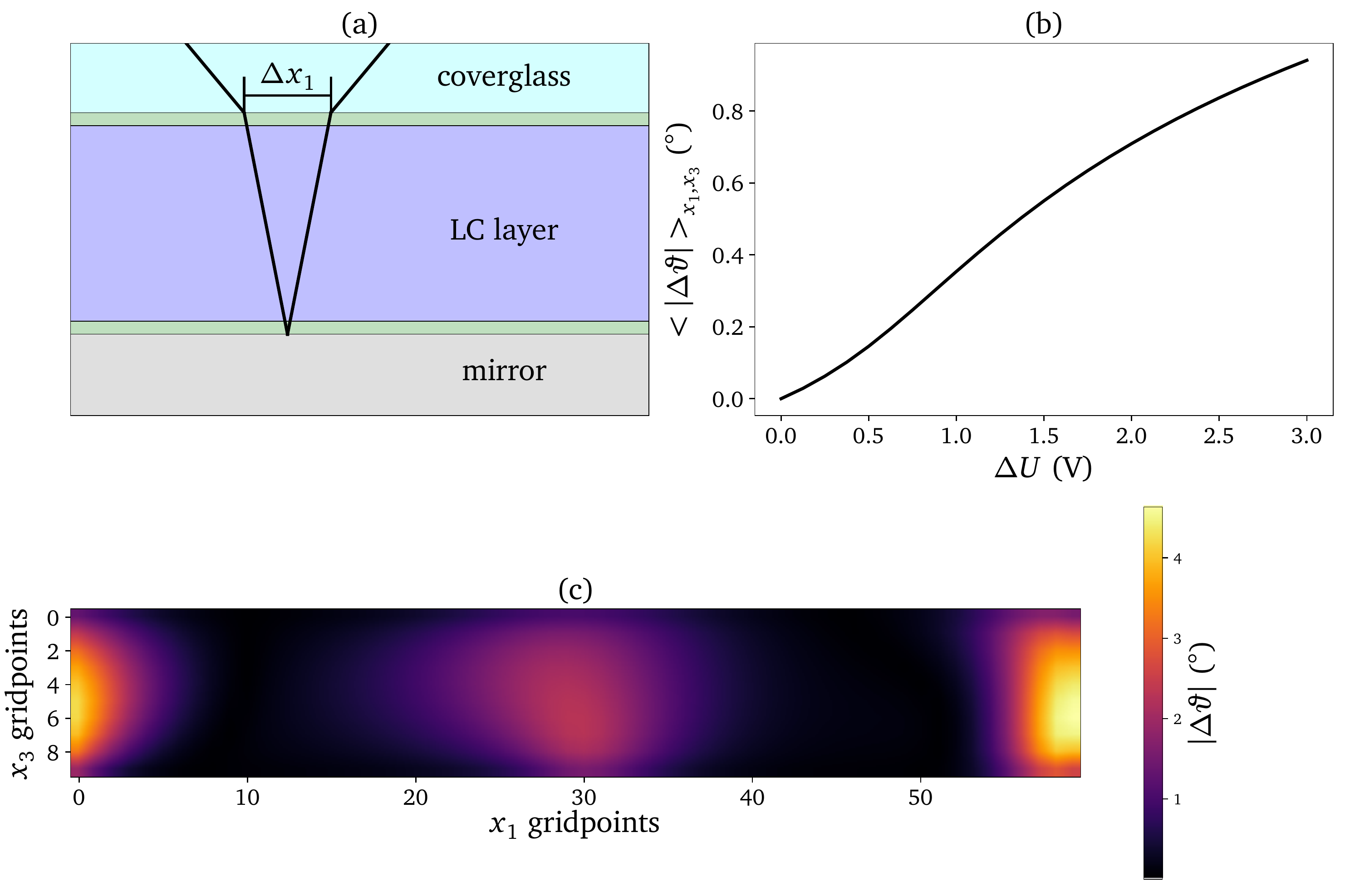}
\caption{Schematic representation of the approximate beam deviation (a), mean tilt angle deviation (b) and distribution of tilt angle deviation over the computational space (c).}
\label{picture justififaction berreman binary}
\end{figure}

The angle of the extraordinary light beam in the birefringent LC-layer is estimated to be $\sim 2.1^\circ$, which, assuming the light propagates at a straight line, causes a displacement of about $\sim 0.3$ $\mu$m, which is roughly $2\%$ of the size of a pixel, as shown in \cref{picture justififaction berreman binary} (a).
In \cref{picture justififaction berreman binary} (b) we see simulations for the mean tilt angle deviation over a range of $\Delta x_1 = 0.3$ $\mu$m, and \cref{picture justififaction berreman binary} (c) depicts the tilt angle deviations over a $2$D slice of the LC layer  for the voltage differences $U_1 = 1$ V and $U_2=4$ V for a binary grating in asymmetric direction.
 
The voltage range from $1$ V to $4$ V roughly represents the range at which the SLM is usually operated. Over the whole voltage range, the mean angle variation stays small. Therefore the effect of the displacement caused from the oblique incidence can be neglected. However, \cref{picture justififaction berreman binary} (c) shows the angle deviations for the biggest voltage difference, which reach $\sim 4.5^\circ$ at one point in the transition from high to low and represent the \glqq worst case scenario\grqq . The angle deviations of other patterns are smaller and we therefore neglect the effect of the $\Delta x_1=0.3$ $\mu$m displacement in the simulations in this \cref{sec:comparison of experiment with simulation}.


\FloatBarrier
\subsection{Sensitivity of model calculations to errors in parameter values}

We observe that the resulting simulation of the diffraction efficiency is robust against small changes in the transition regions of the phase and amplitude profiles. If we vary the splay and/or bend elastic coefficients ($K_{11}/K_{33}$) and therefore influence the phase profile, the change in the resulting diffraction efficiency vs. phase shift is not significant. On the other hand, parameters which change the geometry of the SLM setup such as the thickness of the LC layer $d$, thickness of the dielectric mirror or pixel pitch $x/y$ have a strong influence on the shape of the diffraction efficiency. Since the fringing field effect only becomes noticeable at the transition region between two pixels, the ratio of pixel pitch to thickness $x/d$ can be used to understand the fringing effect qualitatively. The bigger the ratio the smaller the total effect of fringing gets, since the shape of the electric field will match the voltage pattern on the electrodes more closely. Having big pixels reduces the fringing, but the downside is a loss in spatial resolution. A small thickness $d$ enables a high spatial resolution while keeping the effects of fringing small, but limits the maximum phase shift achievable by the SLM. However, this effect can be (partly) compensated by a preferably high birefringence $\Delta n$ of the LC material.

\FloatBarrier
\subsection{Binary grating along the asymmetric direction in horizontal configuration} \label{ssec:Binary grating along the asymmetric direction in horizontal configuration}

\begin{figure}[h!]
\centering
\includegraphics[width=12cm]{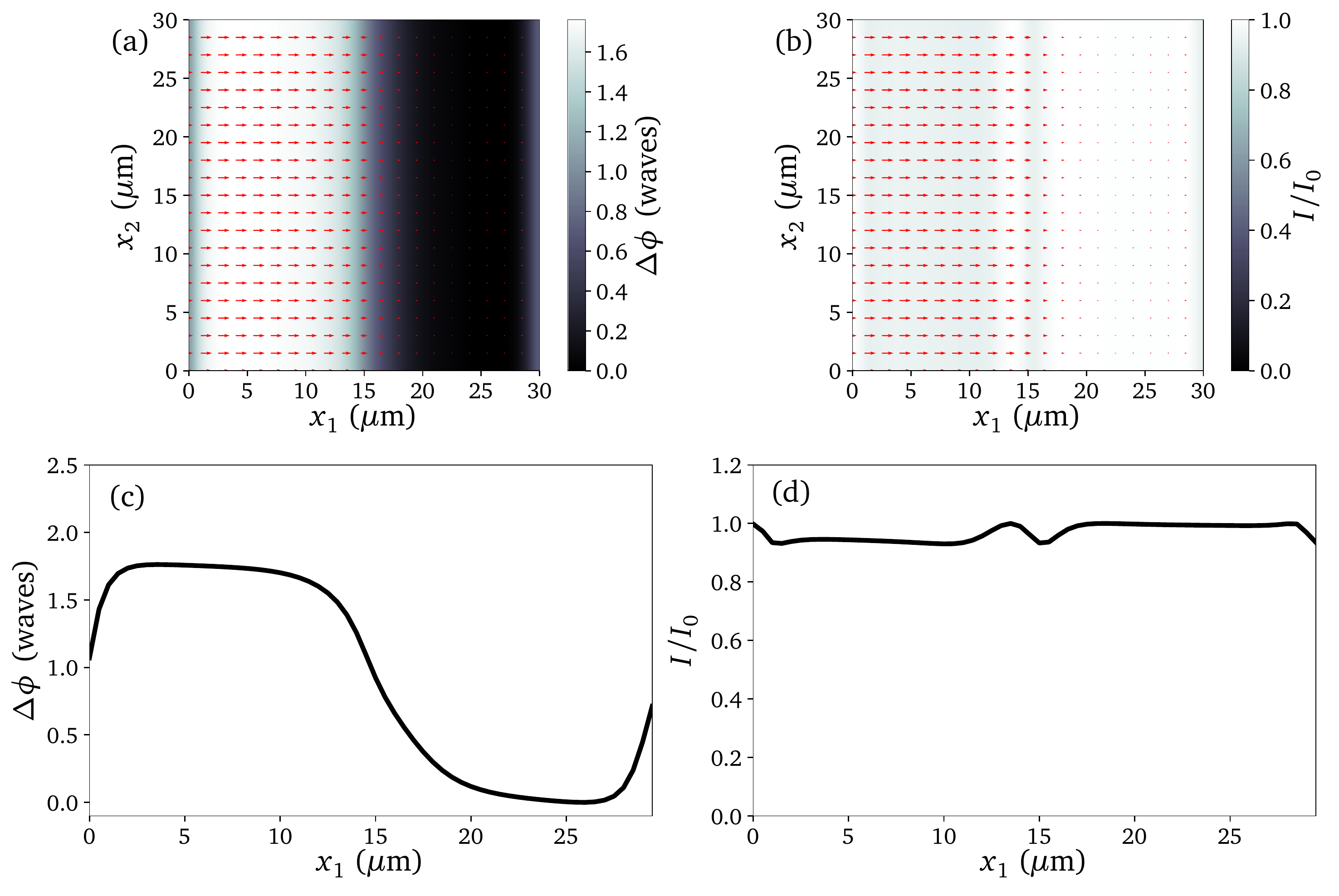}
\caption{Simulated $\parallel$-pol phase profiles (a,c) and intensity profiles (b,d) of a period $2$ binary grating in the asymmetric direction.}
\label{picture grx_lcdirh_polx_phase_int}
\end{figure}

 \cref{picture grx_lcdirh_polx_phase_int} shows simulations for the phase (a,c) and intensity profiles (b,d) for a period $2$ grating in the asymmetric direction in horizontal configuration for the $\parallel$-polarization.  \cref{picture grx_lcdirh_polx_phase_int} (a,b) shows the phase and intensity respectively over $4$ pixels in which the electrodes are driven by $0.5$ V and $5$ V. The red arrows depict the projection of the director along $x_1$ at the central ($6^\mathrm{th}$) layer.
In (c) and (d) slices through the phase and intensity profiles are shown. Since the incident light beam is polarized along $x_1$ and the director has no component along $x_2$ in this case, no polarization conversion is expected, therefore the intensity modulations are solely caused by interference. 


\begin{figure}[h!]
\centering
\includegraphics[width=12cm]{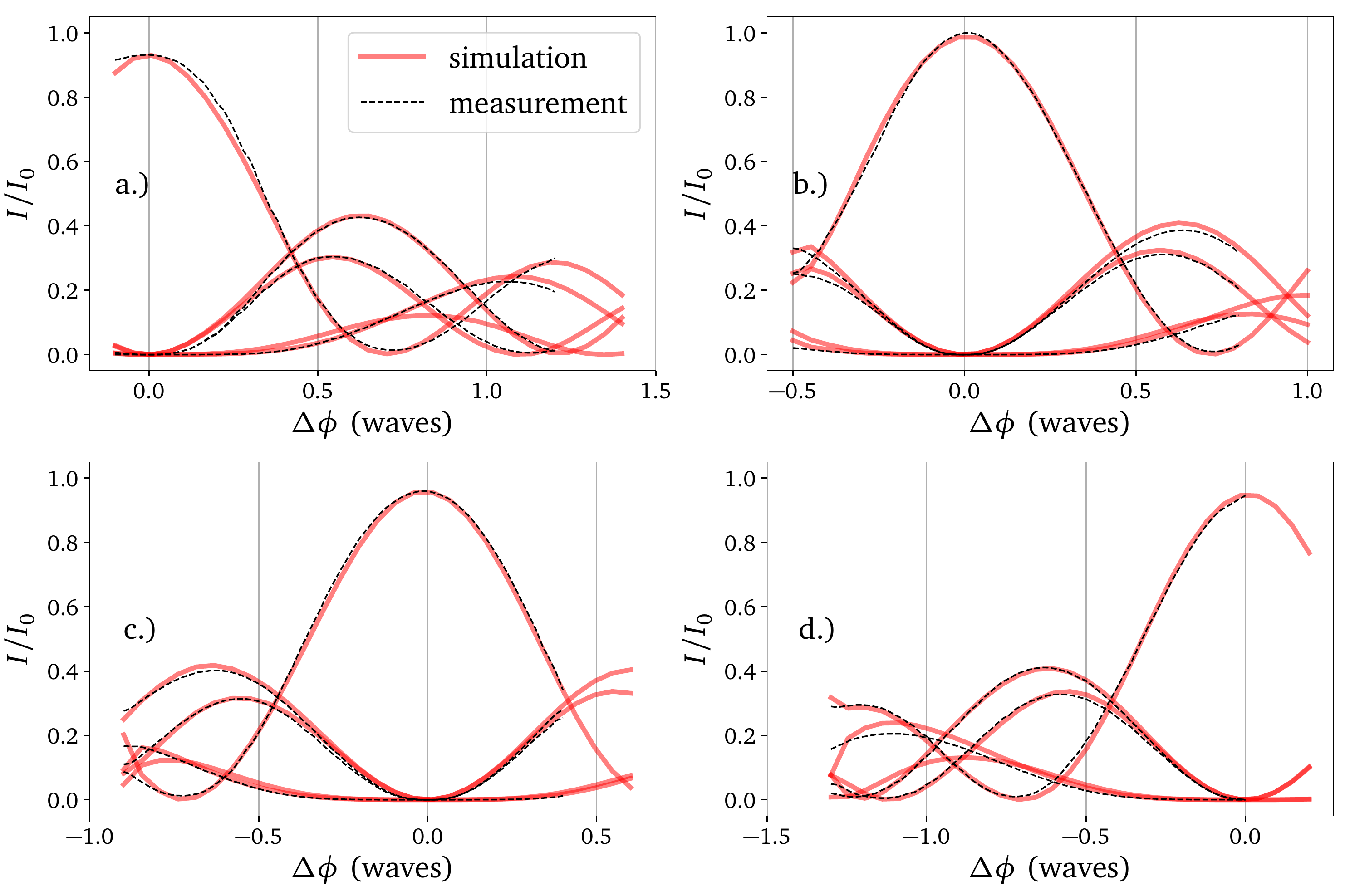}
\caption{Diffraction efficiency measurements and simulations for $\phi_\mathrm{ref} = 0.1$ (a), $\phi_\mathrm{ref} = 0.5$, (b), $\phi_\mathrm{ref} = 0.9$ (c) and $\phi_\mathrm{ref} = 1.3$ waves (d).}
\label{picture compare measurement simulation binary 2 x}
\end{figure}

The other polarization component ($\perp$-pol) vanishes in this situation.
%
%
Comparison of simulations and measurements for the diffraction efficiency is shown in  \cref{picture compare measurement simulation binary 2 x} (red and black respectively) for reference phases $\phi_\mathrm{ref} = 0.1$ (a), $\phi_\mathrm{ref} = 0.5$, (b), $\phi_\mathrm{ref} = 0.9$ (c) and $\phi_\mathrm{ref} = 1.3$ waves (d). In this case, only one $2^\mathrm{nd}$ diffraction order has been measured. The other is blocked by Lens $2$ (see  \cref{picture setup gratings} in  \cref{sec:section experimental setup}) of the $4$f setup. By comparing the $0^\mathrm{th}$ orders for different reference phases, one can see a modulation. This effect is due to the interference effect we discussed in the previous  \cref{ssec:uniform electric field} and only affects the $0^\mathrm{th}$ order. The simulations fit the measurements very well at low phase shifts $\Delta \phi$. The simulations for $2^\mathrm{nd}$ order deviate slightly from the measurements for high phase shifts in $(a)$ and $(d)$, as do the $0^\mathrm{th}$ orders. These deviations are possibly due to the violation of the preconditions for the applicability of the Berreman $4\times 4$ method or due to changes in the utilized LUT in the measurement as discussed in \cref{ssec:LUT}. 

\FloatBarrier
\subsection{Binary grating along the symmetric direction in horizontal configuration}
\label{ssec:Binary grating along the symmetric direction in horizontal configuration}

\begin{figure}[h!]
\centering
\includegraphics[width=12cm]{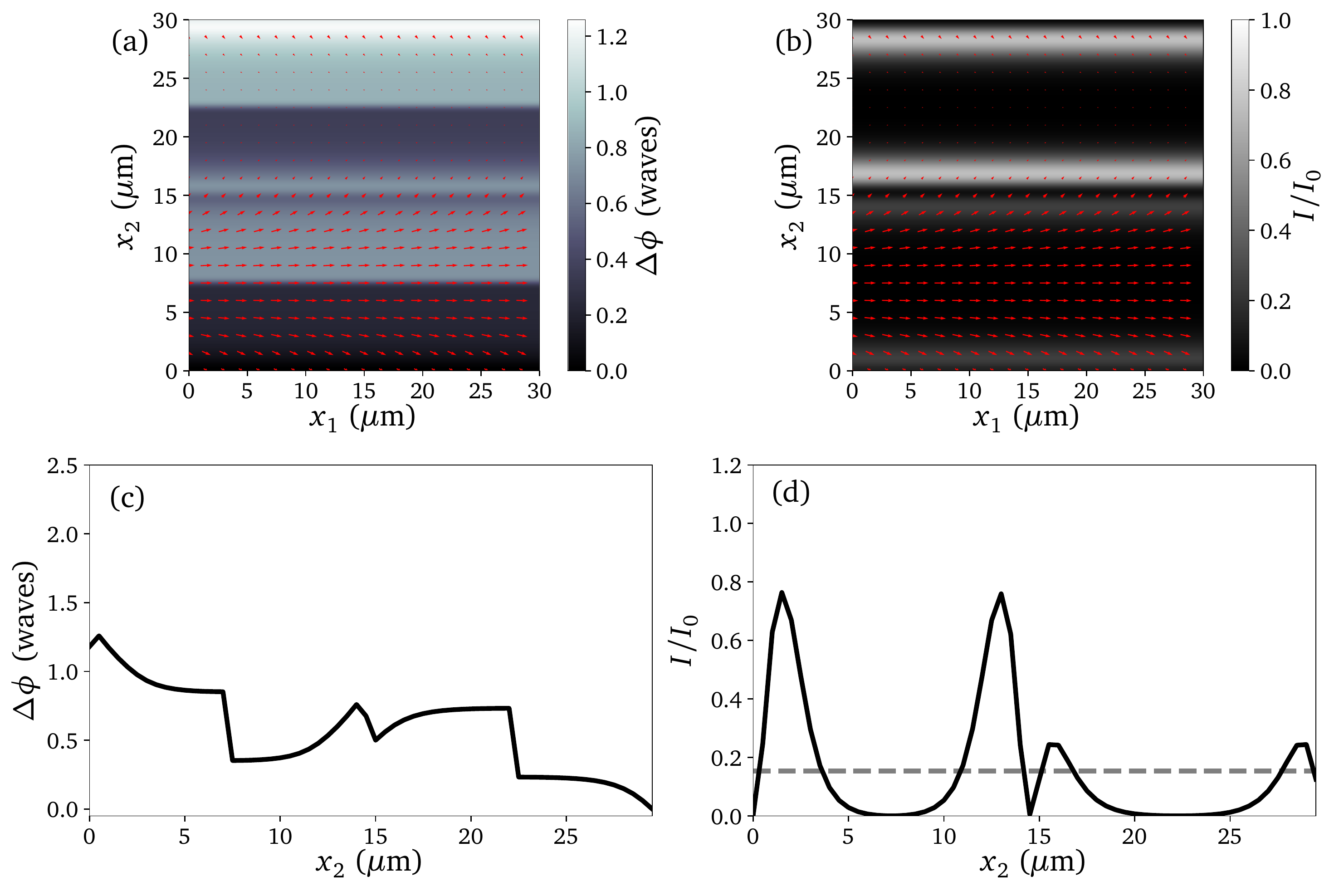}
\caption{Simulated $\parallel$-pol phase profiles (a,c) and intensity profiles (b,d) of a period $2$ binary grating in the symmetric direction.}
\label{picture gry_lcdirh_poly_phase_int}
\end{figure}

For gratings in the symmetric direction the applied electric field has also a component along $x_2$ in the horizontal configuration ($x_1$ in the vertical configuration) and therefore also the director. Due to a non-uniform twist angle $\varphi$ we expect polarization conversion effects in addition to the interference effect.  \cref{picture gry_lcdirh_poly_phase_int} shows correspondingly the simulated phase and intensity profiles for $\perp$-pol light. Looking at the intensity profiles \cref{picture gry_lcdirh_poly_phase_int} (b,d) one sees cave-ins  at the transition regions, which stem from polarization conversion. This is due to the shape of the electric field, which causes the director orientation to possess a twist angle $\varphi$. The dashed line in \cref{picture gry_lcdirh_poly_phase_int} (d) shows the mean intensity of the $\parallel$-pol component of the reflected light, which is at about $\sim 85.5\%$. The simulated intensity profiles suggest that polarization conversion produces strong variations in the field amplitude over a small region of $1-2$ $\mu$m. Since the used wavelength is $\lambda = 1.064$ $\mu$m, the Berreman matrix method is no longer a suitable method to simulate the phase and amplitude profiles accurately.

\begin{figure}[h!]
\centering
\includegraphics[width=12cm]{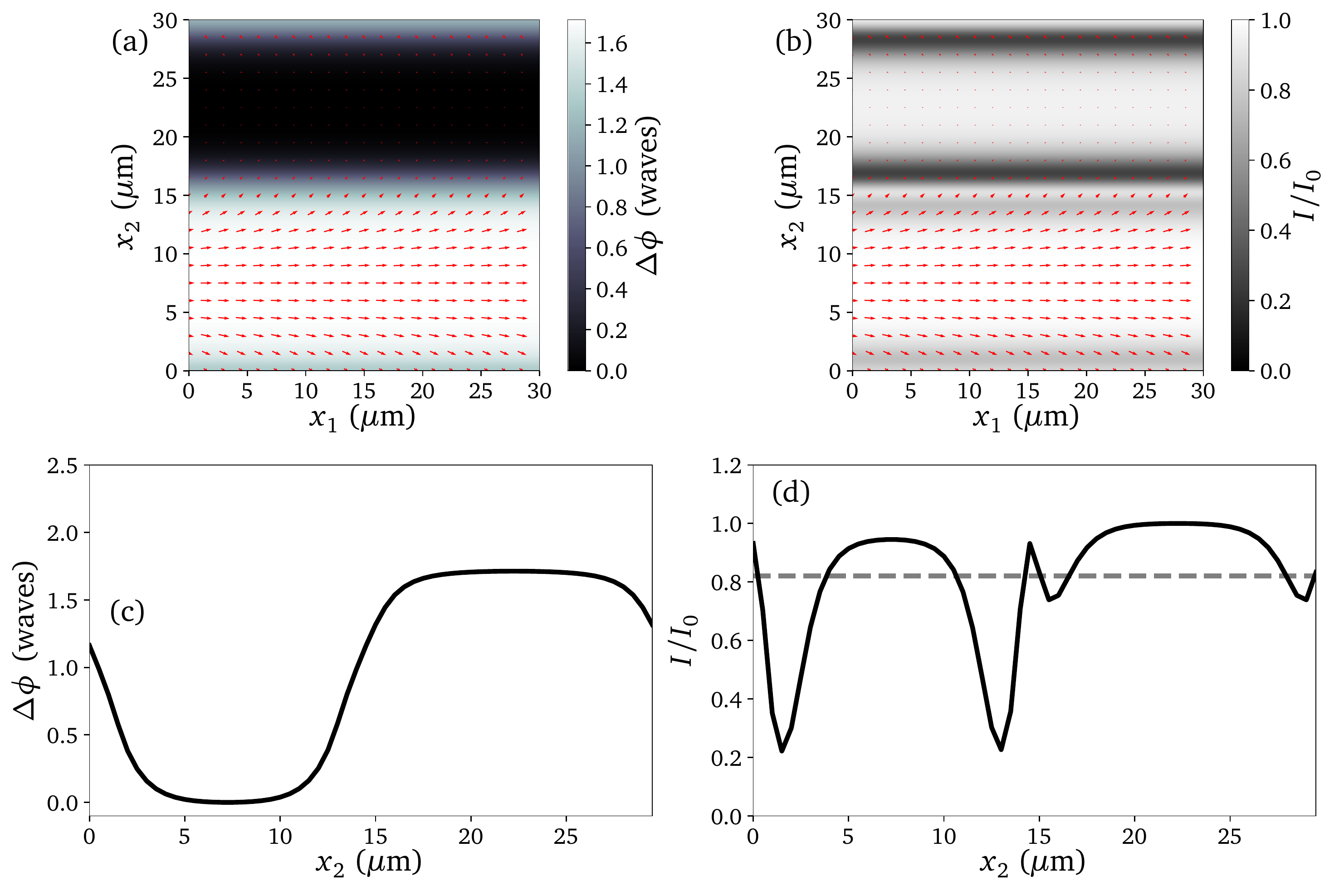}
\caption{Simulated $\perp$-pol phase profiles (a,c) and intensity profiles (b,d) of a period $2$ binary grating in the symmetric direction.}
\label{picture gry_lcdirh_polx_phase_int}
\end{figure}

 \cref{picture gry_lcdirh_poly_phase_int} shows the $\parallel$-pol component. The intensity profiles (\cref{picture gry_lcdirh_poly_phase_int} (b,d)) are complementary to those in  \cref{picture grx_lcdirh_polx_phase_int}. 

\begin{figure}[h!]
\centering
\includegraphics[width=12cm]{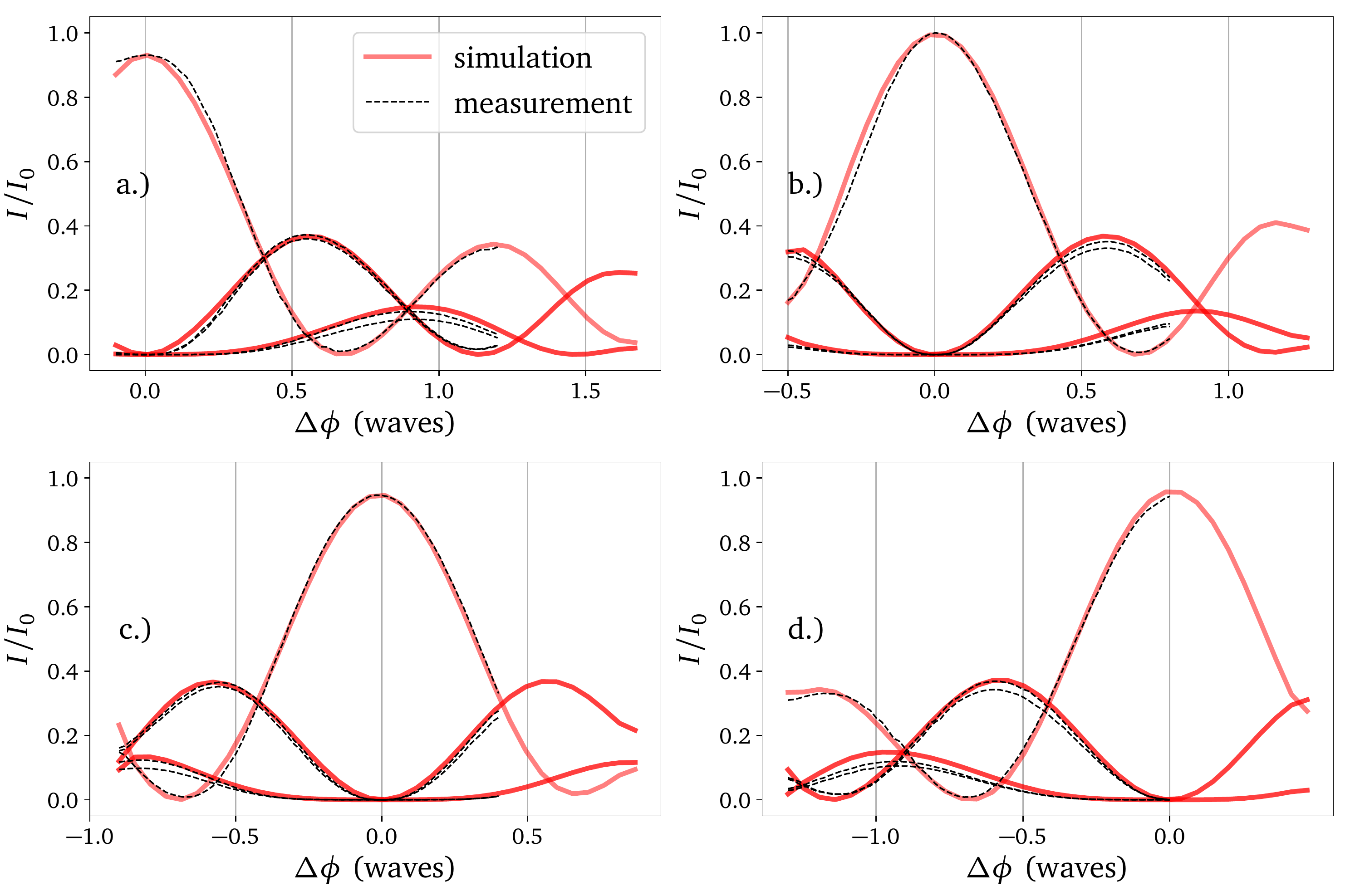}
\caption{Diffraction efficiency measurements and simulations for $\phi_\mathrm{ref} = 0.1$ (a), $\phi_\mathrm{ref} = 0.5$, (b), $\phi_\mathrm{ref} = 0.9$ (c) and $\phi_\mathrm{ref} = 1.3$ waves (d).}
\label{picture compare measurement simulation binary 2 y}
\end{figure}

 \cref{picture compare measurement simulation binary 2 y} shows the simulated and measured diffraction efficiency curves of the $0^\mathrm{th}$, $1^\mathrm{st}$ and $2^\mathrm{nd}$ orders. The simulations of the $1^\mathrm{st}$ diffraction orders match the measurements very well over the whole range of $\Delta \phi$. There are some deviations in the $0^\mathrm{th}$ and $1\mathrm{st}$ orders if one of the phase levels of the binary grating is small. This is due to the fact that close to the threshold region the LUT is different in simulations and experiments. In this range the conversion between phase and control voltage is sensitive to small errors, because of the small slope. Another factor is the spatial dependent LUT in measurements, in simulation we have only one LUT which corresponds to a position at the center of the SLM (measurement is done on an area of the SLM where multiple LUTs are being used).
 
The calculated $2^\mathrm{nd}$ orders have a systematically higher intensity in all cases. This discrepancy most probably stems from the experimental setup. The $2^\mathrm{nd}$ orders of a binary grating just fit through Lens $3$ (\cref{picture setup gratings}).

Other factors that influence the measurements are temperature, beam width and change of the SLM orientation between LUT and diffraction efficiency measurements. 

\FloatBarrier
\subsection{Binary grating along the asymmetric direction in vertical configuration}

\begin{figure}[h!]
\centering
\includegraphics[width=12cm]{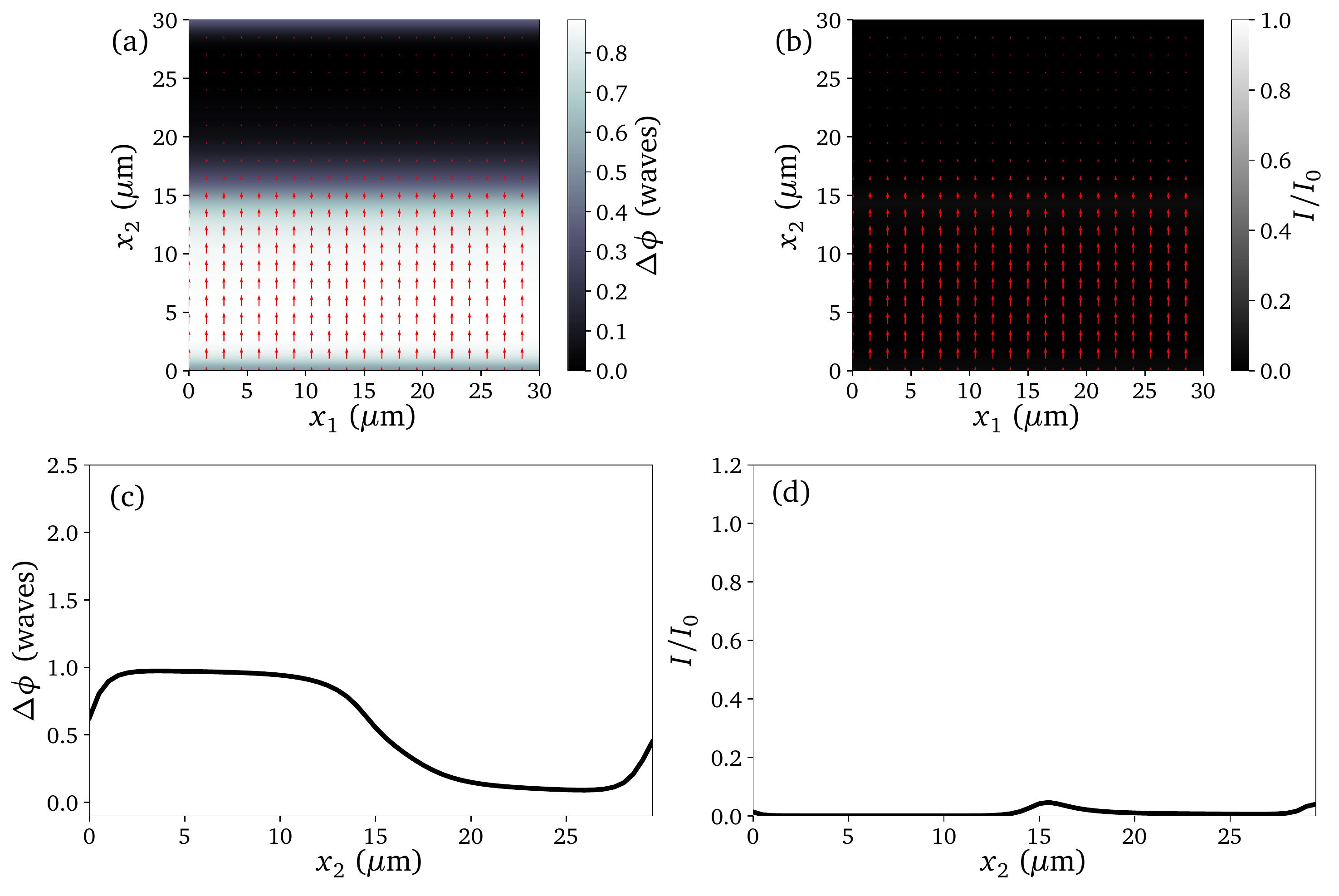}
\caption{Simulated $\parallel$-pol phase profiles (a,c) and intensity profiles (b,d) of a period $2$ binary grating in the asymmetric direction.}
\label{picture grx_lcdirv_polx_phase_int}
\end{figure}

\begin{figure}[h!]
\centering
\includegraphics[width=12cm]{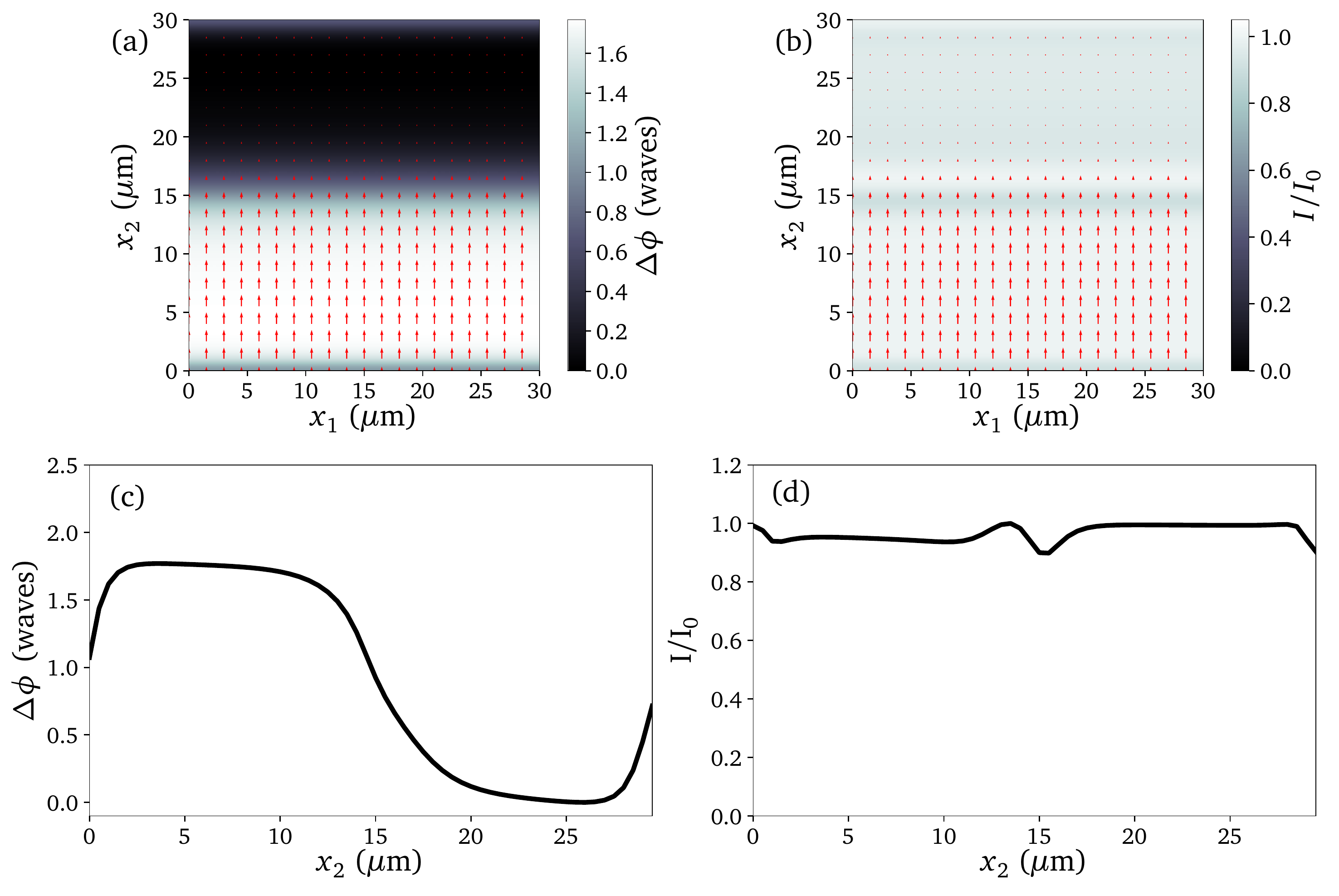}
\caption{Simulated $\perp$-pol phase profiles (a,c) and intensity profiles (b,d) of a period $2$ binary grating in the asymmetric direction.}
\label{picture grx_lcdirv_poly_phase_int}
\end{figure}

 \cref{picture grx_lcdirv_polx_phase_int,picture grx_lcdirv_poly_phase_int} show the simulated phase and intensity profiles for a grating in the asymmetric direction in the vertical configuration. This case is similar to the uniform pattern in the vertical direction. The applied voltages are $0.5$ V/$5$ V and we see almost no polarization conversion occurring there. At the transition regions we see some conversion happening, which is due to the tilt angle $\theta$ roughly around $45^\circ$, which maximizes the projection of the polarization vector on the \glqq wrong\grqq\, axis of the LC molecules.

 \cref{picture grx_lcdirv_poly_phase_int} shows the phase and intensity profiles of the $\perp$-pol contribution. By looking at both polarization directions, the effect of polarization conversion is expected to be small, around half the value of those for a uniform electric field.  

\begin{figure}[h!]
\centering
\includegraphics[width=8cm]{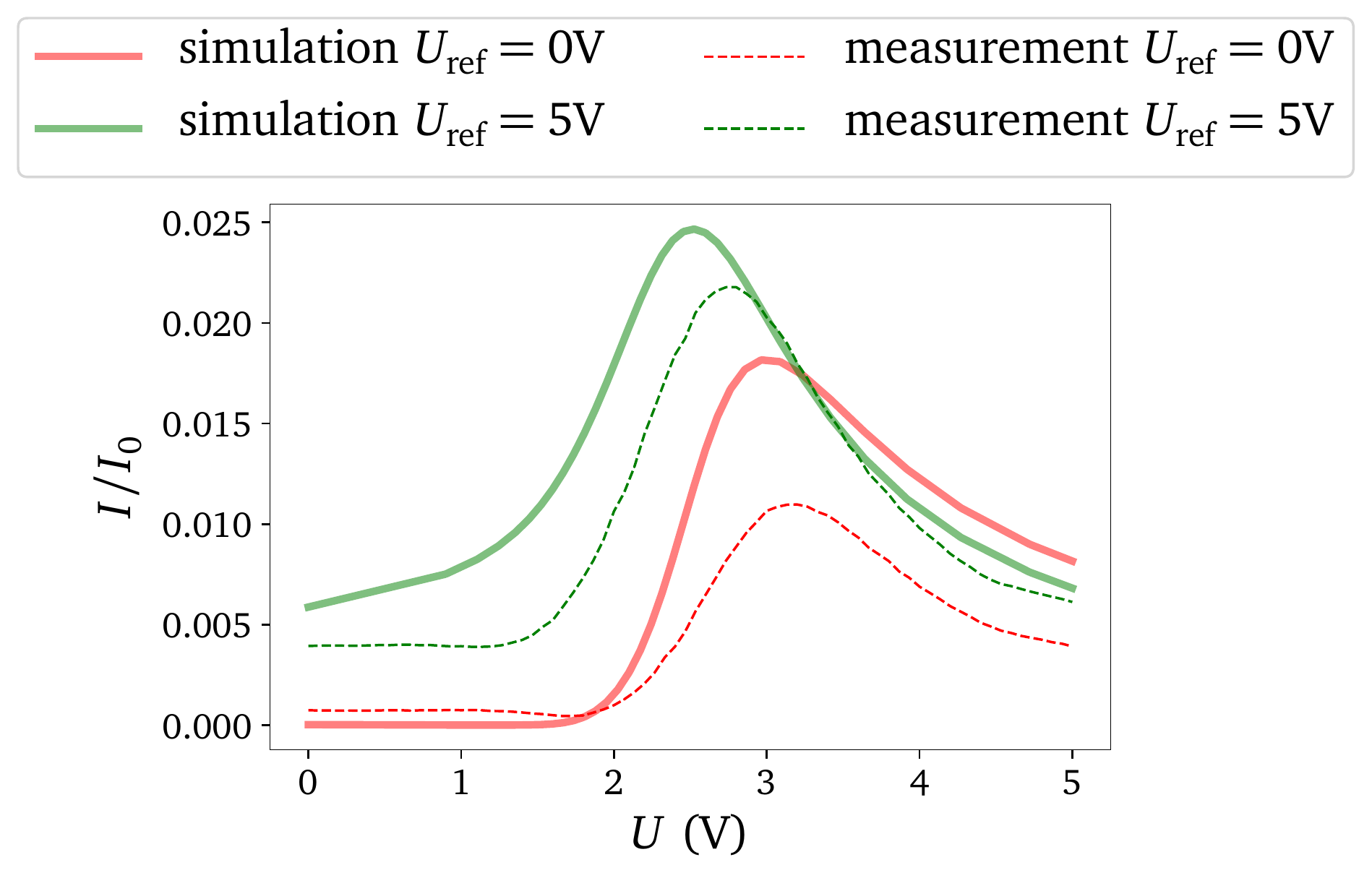}
\caption{Simulated $\parallel$-pol total intensity as a function of voltage for a period $2$ binary grating in the asymmetric direction for different reference voltages.}
\label{picture compare measurement simulation binary 2 x v polx}
\end{figure}

In  \cref{picture compare measurement simulation binary 2 x v polx} we see simulations and measurements of the $\parallel$-pol direction for the highest and lowest reference voltages $0$ V and $5$ V. These measurements were done with the experimental setup for the calibration (\cref{picture setup LUT}). As expected, the polarization conversion efficiencies are smaller than those in the uniform case and never exceed $2.5$\%. The simulated curves have systematically higher intensities. Especially the measurements can be subject to errors such as misalignment of the polarizer, resulting in transmitting some of the other, much stronger polarization components, the beam splitter not perfectly splitting $50$:$50$ and having a small dependence on polarization and camera sensitivity. Simulations and measurements have also been shown to be very sensitive to the angle of incidence, and this also poses a possible error source. Nonetheless, the calculations fit the measurements qualitatively well but seem to have some systematic error stemming from reasons stated before. 

\FloatBarrier
\subsection{Binary grating along the symmetric direction in vertical configuration}\label{ssec:Binary grating along the symmetric direction in vertical configuration}

\begin{figure}[h!]
\centering
\includegraphics[width=12cm]{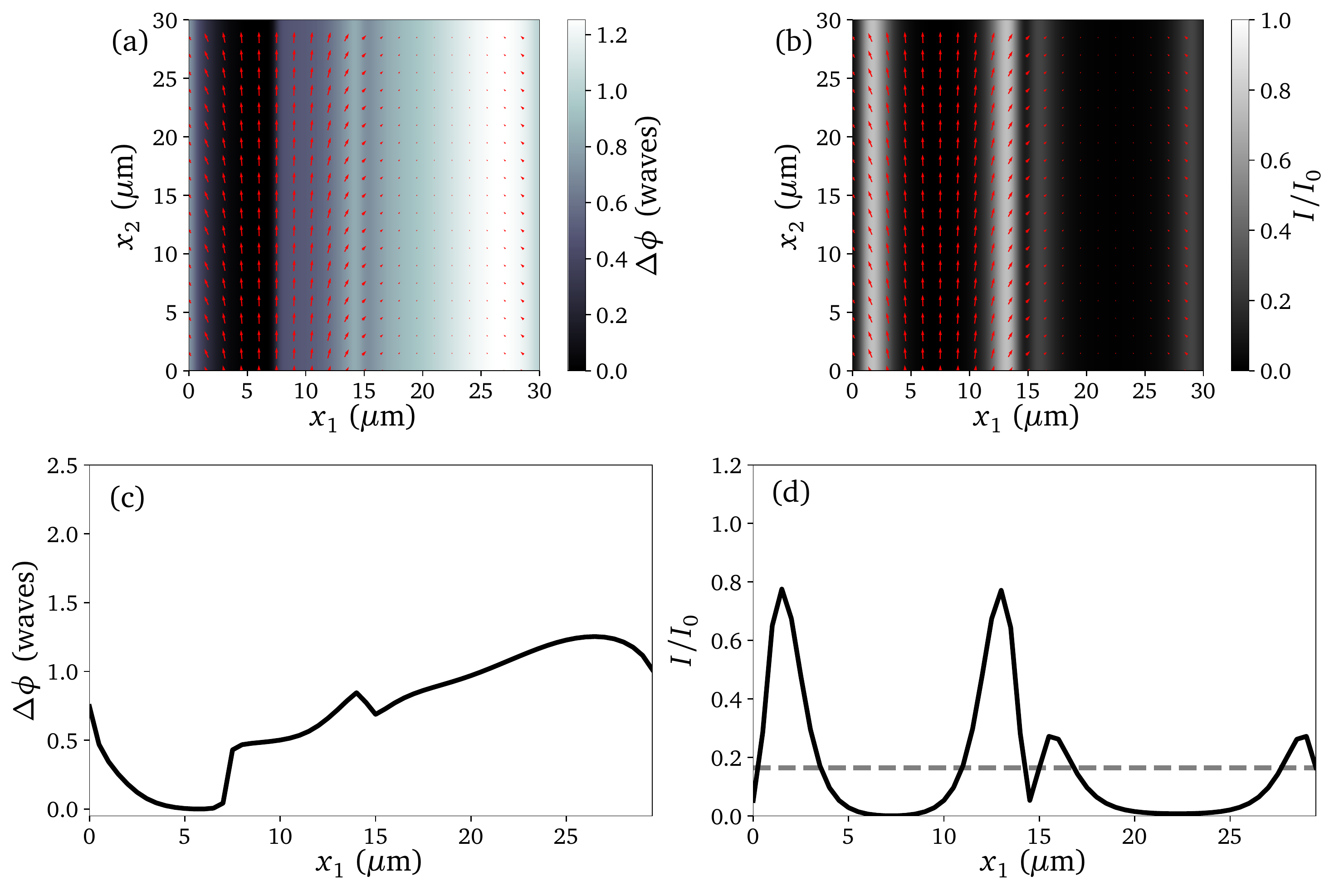}
\caption{Simulated $\parallel$-pol phase profiles (a,c) and intensity profiles (b,d) of a period $2$ binary grating in the symmetric direction.}
\label{picture gry_lcdiv_polx_phase_int}
\end{figure}
\begin{figure}[h!]
\centering
\includegraphics[width=12cm]{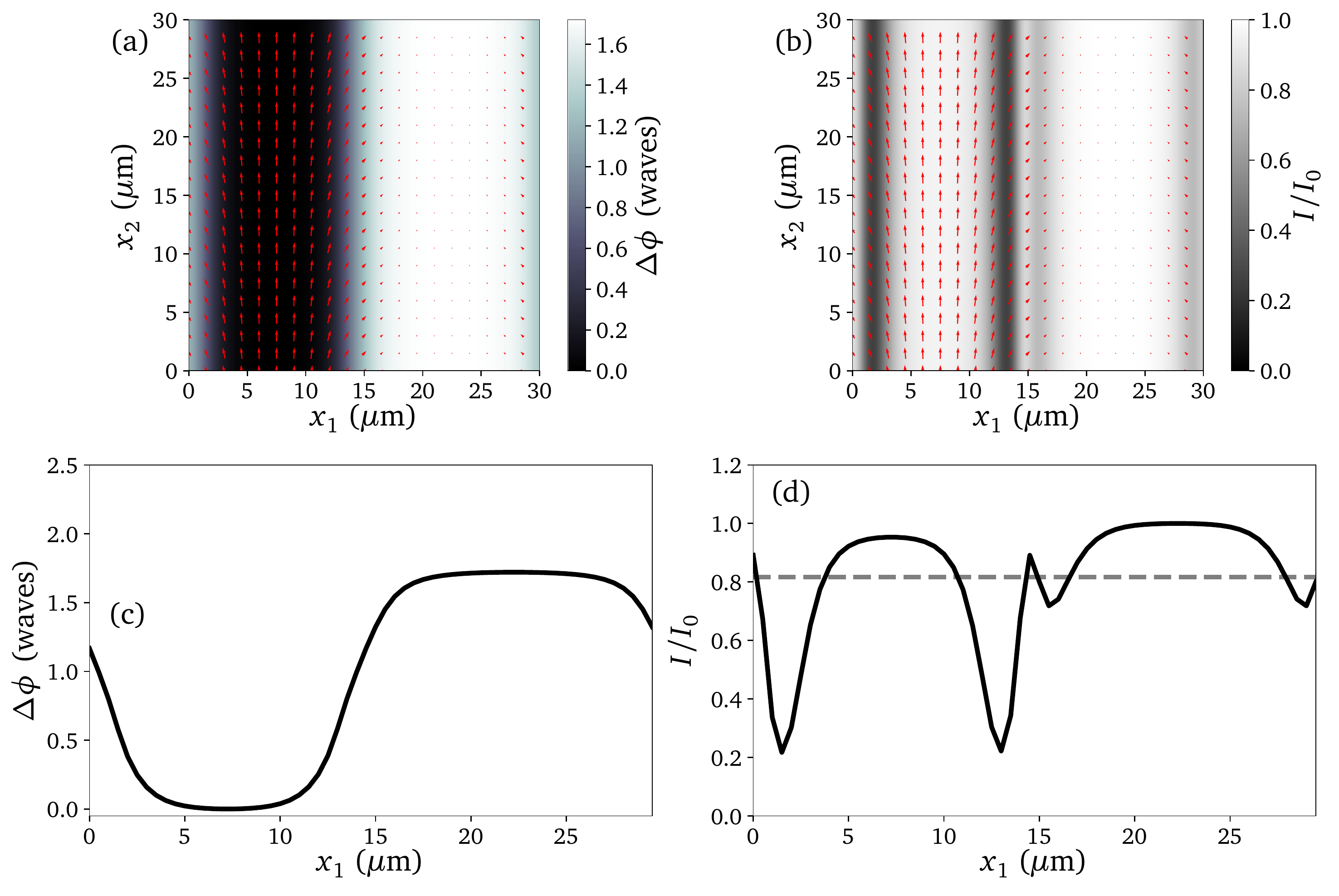}
\caption{Simulated $\perp$-pol phase profiles (a,c) and intensity profiles (b,d) of a period $2$ binary grating in the symmetric direction.}
\label{picture gry_lcdiv_poly_phase_int}
\end{figure}

In  \cref{picture gry_lcdiv_polx_phase_int,picture gry_lcdiv_poly_phase_int} we see the phase and intensity profiles for a binary grating in the symmetric direction in the vertical configuration. The profiles are similar to those in the horizontal configuration. The effect of polarization conversion is slightly higher in the vertical than in the horizontal configurations. The projection of the polarization vector onto the ordinary axes of the LCs are slightly higher because the directors never lie in the plane of incidence. Roughly speaking, the polarization conversion in this case has two components, one stemming from the same reason as in the uniform case and the other stemming from the twist angle caused by the shape of the electric field. 

\begin{figure}[h!]
\centering
\includegraphics[width=8cm]{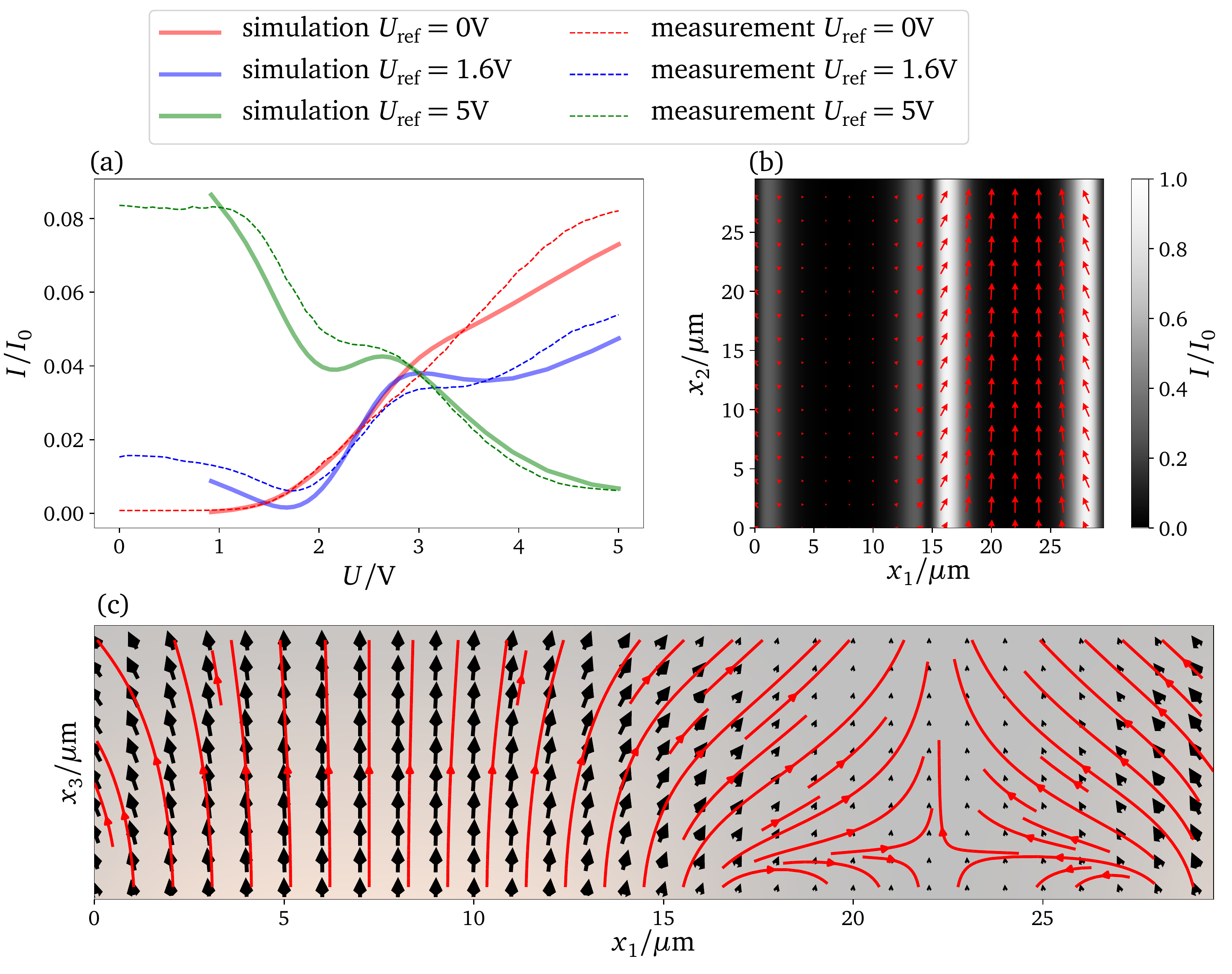}
\caption{Simulated and measured $\parallel$-pol total intensity as a function of voltage for a period $2$ binary grating in the symmetric direction for different reference voltages (a), intensity profile (b) and slice of the director distribution (c).}
\label{picture compare measurements simulation binary 2 y v polx}
\end{figure}

 \cref{picture compare measurements simulation binary 2 y v polx} shows intensity measurements and simulations for the $\parallel$-pol component at different reference voltages. The simulations fit the measurements qualitatively well, but both measurements and simulations are subject to the errors mentioned in the asymmetric case. The predicted maximum intensities from  \cref{picture gry_lcdiv_poly_phase_int} is around $18$\%, but the measurements only include the first $2-3$ diffraction orders. 
 
As mentioned in \cref{ssec:Binary grating along the symmetric direction in horizontal configuration}, the amplitude profiles simulated by the Berreman matrix method show structures at the scale of the used wavelength ($\sim 1$ $\mu$m) and therefore the simulation does not describe the profiles accurately. However, the intensity measurements in \cref{picture compare measurements simulation binary 2 y v polx} show, that the Berreman matrix method is able to describe the overall effect of polarization conversion.
 
\FloatBarrier
\section{Comparison of experiment and simulations for checkerboard patterns}



\FloatBarrier
\subsection{Horizontal configuration}

\begin{figure}[h!]
\centering
\includegraphics[width=12cm]{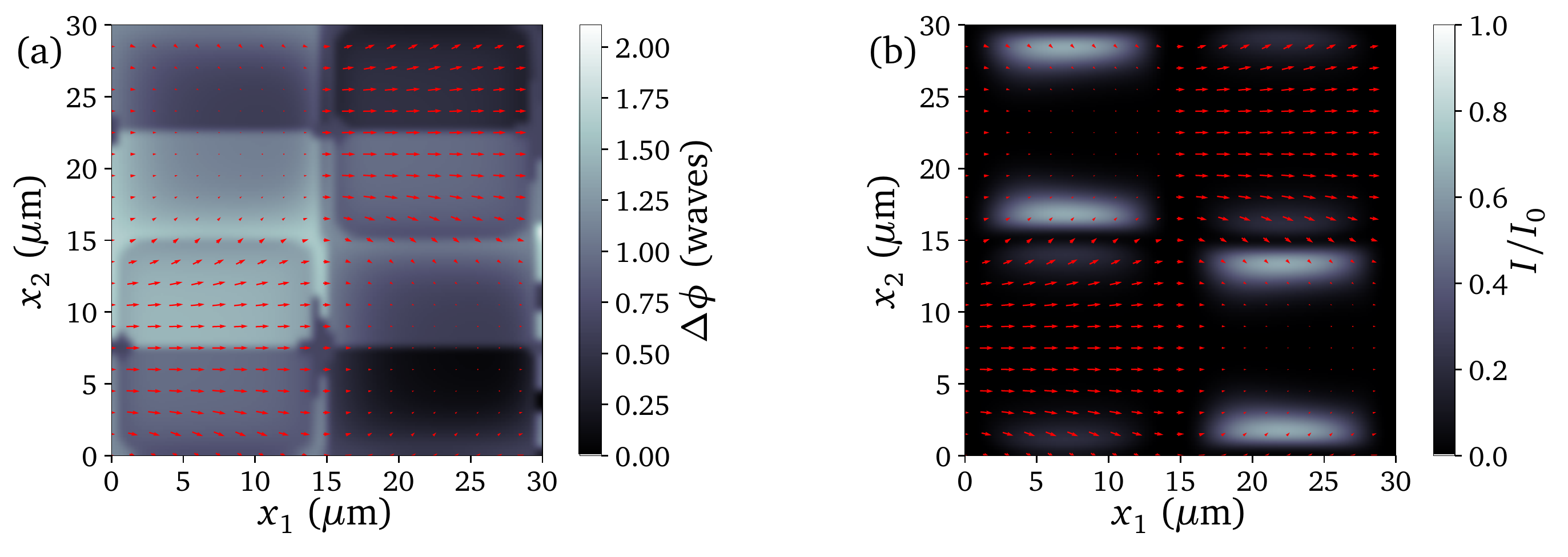}
\caption{Simulated $\perp$-pol phase profile (a) and intensity profile (b) of a checkerboard pattern in the horizontal configuration.}
\label{picture grsb_lcdih_poly_phase_int}
\end{figure}

\begin{figure}[h!]
\centering
\includegraphics[width=12cm]{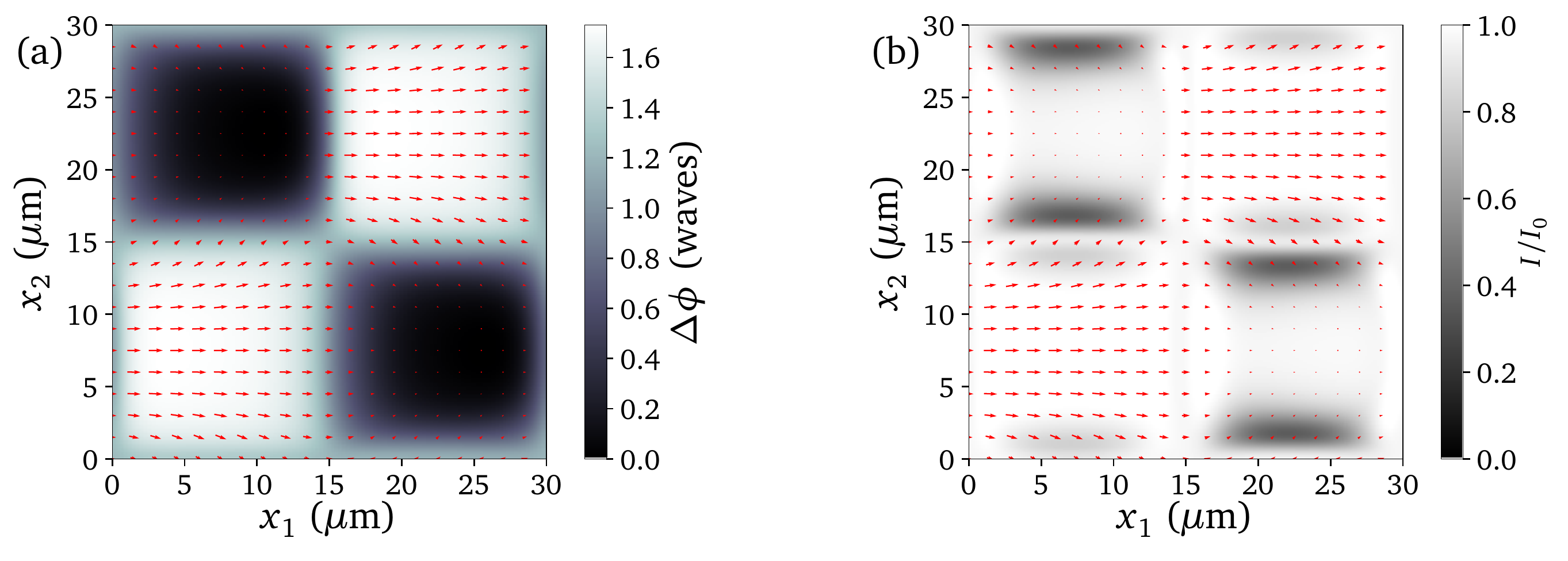}
\caption{Simulated $\parallel$-pol phase profile (a) and intensity profile (b) of a checkerboard pattern in the horizontal configuration.}
\label{picture grsb_lcdih_polx_phase_int}
\end{figure}

 \cref{picture grsb_lcdih_poly_phase_int,picture grsb_lcdih_polx_phase_int} show the phase and intensity profiles for a checkerboard pattern in the horizontal configuration (the phase profile in  \cref{picture grsb_lcdih_poly_phase_int} is mostly not relevant due to vanishing intensities). Speaking of overall intensity, the effects of polarization conversion for this pattern are smaller than in the binary symmetric case. The intensity profile of the $\perp$-pol shows $4$ strong intensity and $4$ weak intensity spots where polarization conversion is happening, similar to the case of binary patterns in the symmetric configuration. 

\begin{figure}[h!]
\centering
\includegraphics[width=12cm]{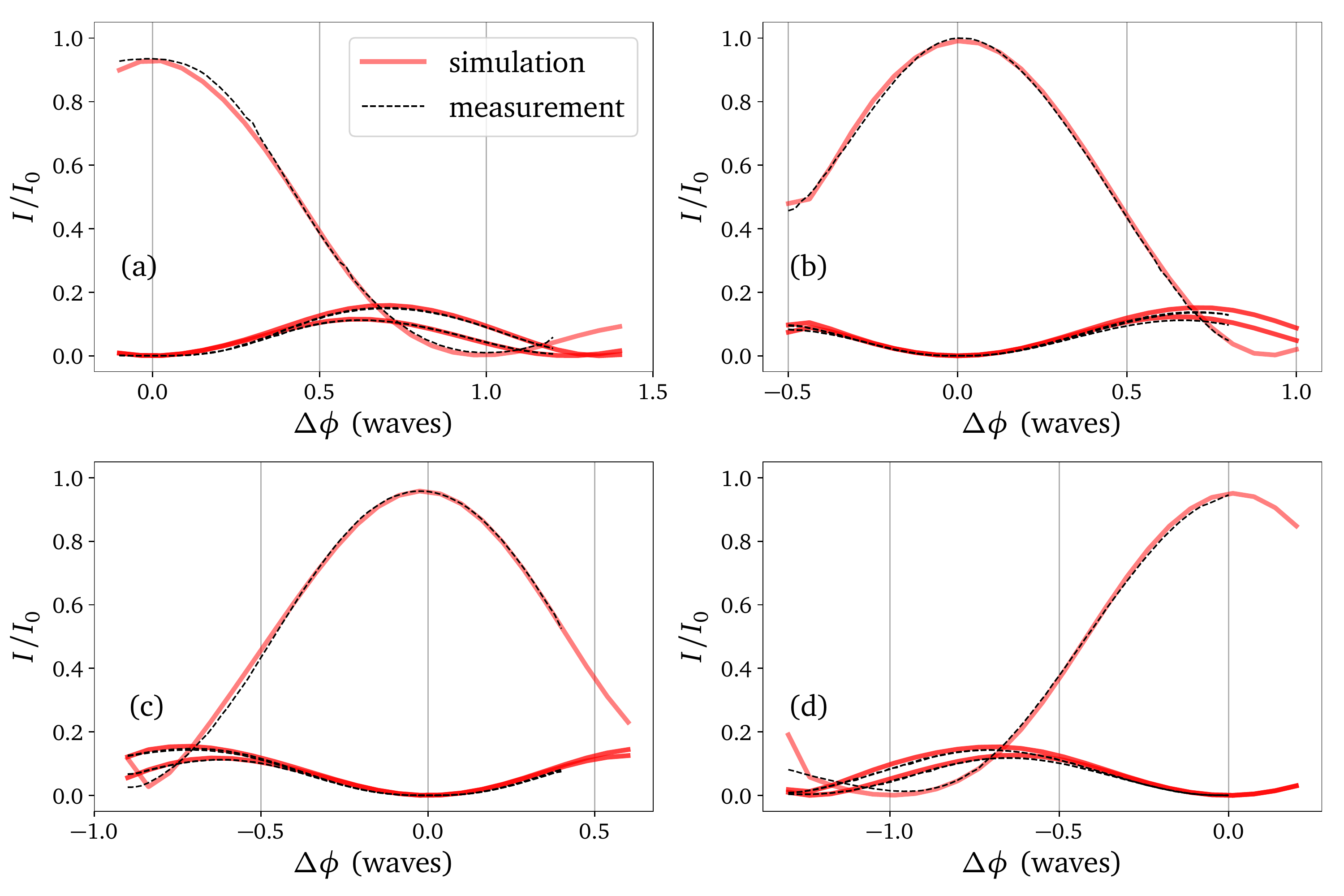}
\caption{Diffraction efficiency measurements and simulation for $\phi_\mathrm{ref} = 0.1$ (a), $\phi_\mathrm{ref} = 0.5$, (b), $\phi_\mathrm{ref} = 0.9$ (c) and $\phi_\mathrm{ref} = 1.3$ waves (d).}
\label{picture compare measurement simulation sb}
\end{figure}

In  \cref{picture compare measurement simulation sb} we see simulations and measurements for the diffraction efficiency of a checkerboard pattern. The measurements are matched very well by simulations. Only for $\Delta \phi$ at the lower margin there are some deviations, like in the case of binary gratings. The checkerboard pattern is most sensitive to fringing, since the minimum of the $0^\mathrm{th}$ order is located at $\Delta \phi \approx 2\pi$ (instead of at $\Delta \phi = \pi$ in the idealized case).


\FloatBarrier
\subsection{Vertical configuration (only simulations)}

\begin{figure}[h!]
\centering
\includegraphics[width=12cm]{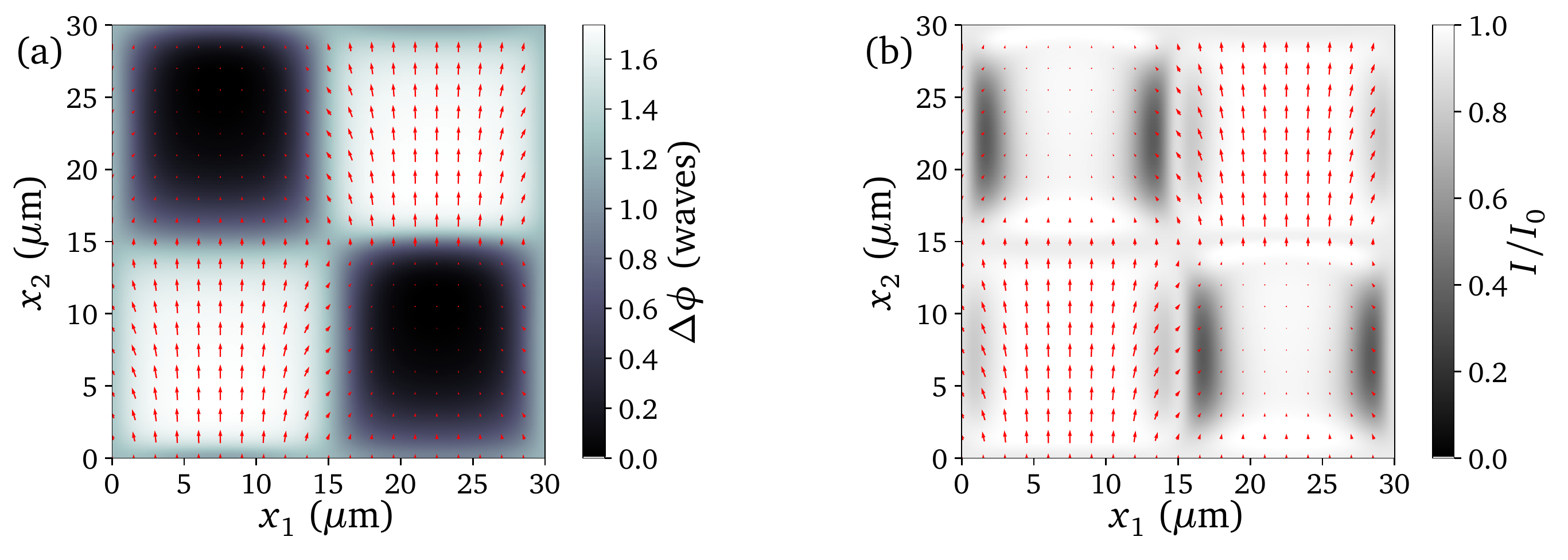}
\caption{Simulated $\perp$-pol phase profile (a) and intensity profile (b) of a checkerboard pattern.}
\label{picture grsb_lcdiv_poly_phase_int}
\end{figure}

\begin{figure}[h!]
\centering
\includegraphics[width=12cm]{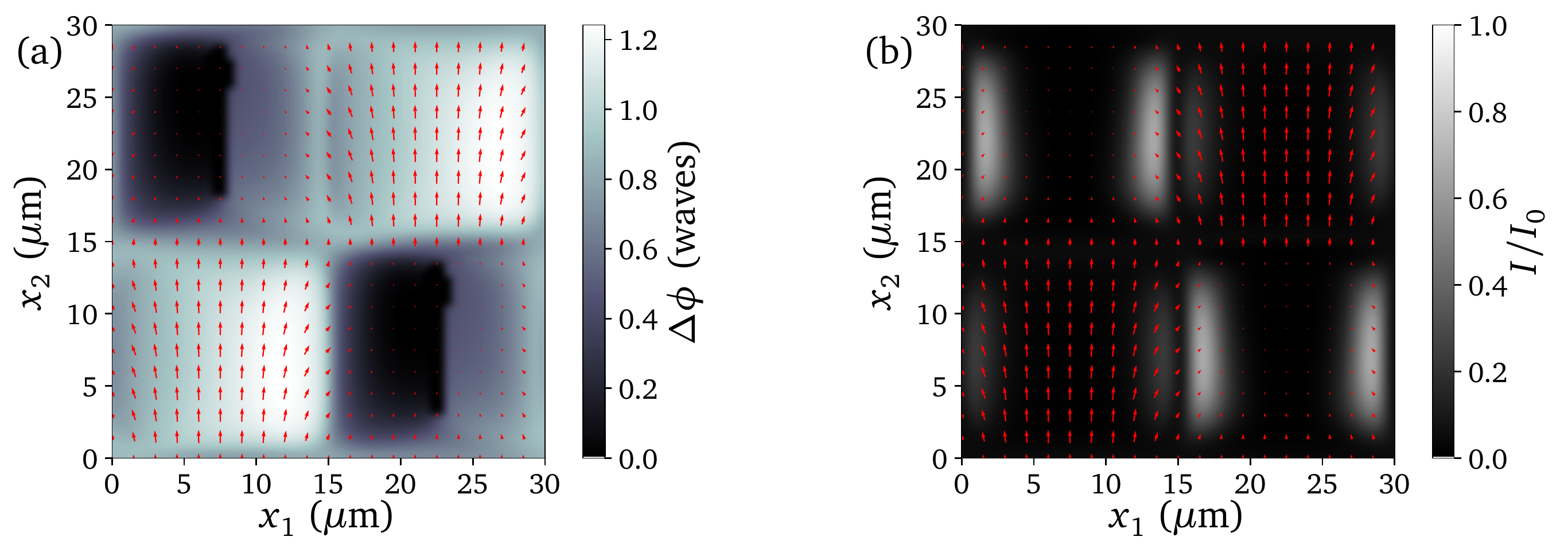}
\caption{Simulated $\parallel$-pol phase profile (a) and intensity profile (b) of a checkerboard pattern.}
\label{picture grsb_lcdiv_polx_phase_int}
\end{figure}

 \cref{picture grsb_lcdiv_poly_phase_int,picture grsb_lcdiv_polx_phase_int} show the profiles for the vertical configuration. We see the same characteristic of the $4$ weak and strong intensity spots as in the horizontal configuration but with additional polarization conversion happening at transition regions where the tilt angle is tilted by about $\sim 45^\circ$. 

\FloatBarrier
\section{Comparison of experiment and simulations for blazed gratings}
\label{sec:Blazed gratings}
\FloatBarrier
\subsection{Asymmetric direction in horizontal configuration}

\begin{figure}[h!]
\centering
\includegraphics[width=12cm]{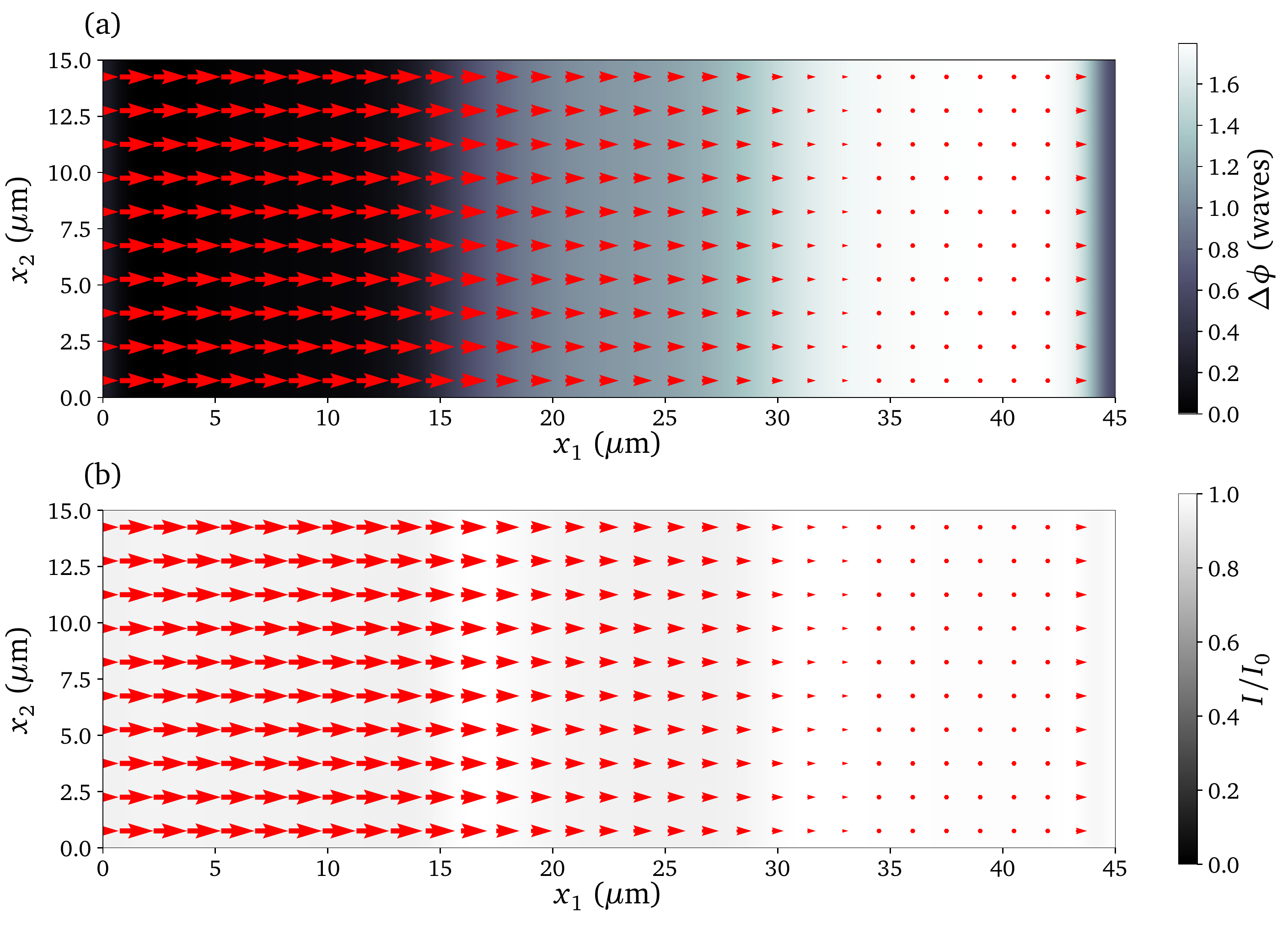}
\caption{Simulated $\parallel$-pol phase profile (a) and intensity profile (b) of a period $3$ blazed grating in the asymmetric direction.}
\label{picture grblx_lcdih_polx_phase_int_per3}
\end{figure}

\begin{figure}[h!]
\centering
\includegraphics[width=12cm]{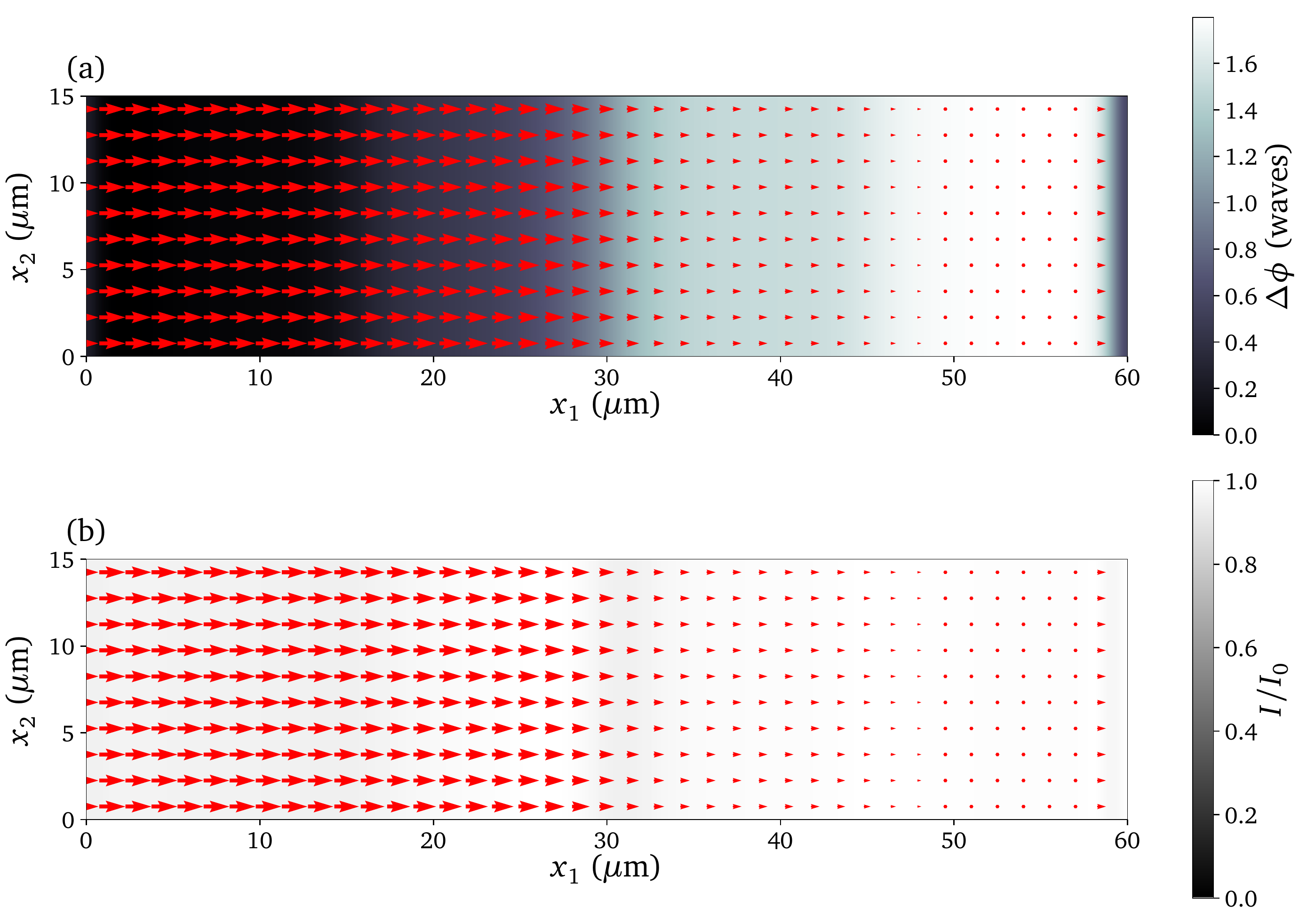}
\caption{Simulated $\parallel$-pol phase profile (a) and intensity profile (b) of a period $4$ blazed grating in the asymmetric direction.}
\label{picture grblx_lcdih_polx_phase_int_per4}
\end{figure}

\begin{figure}[h!]
\centering
\includegraphics[width=12cm]{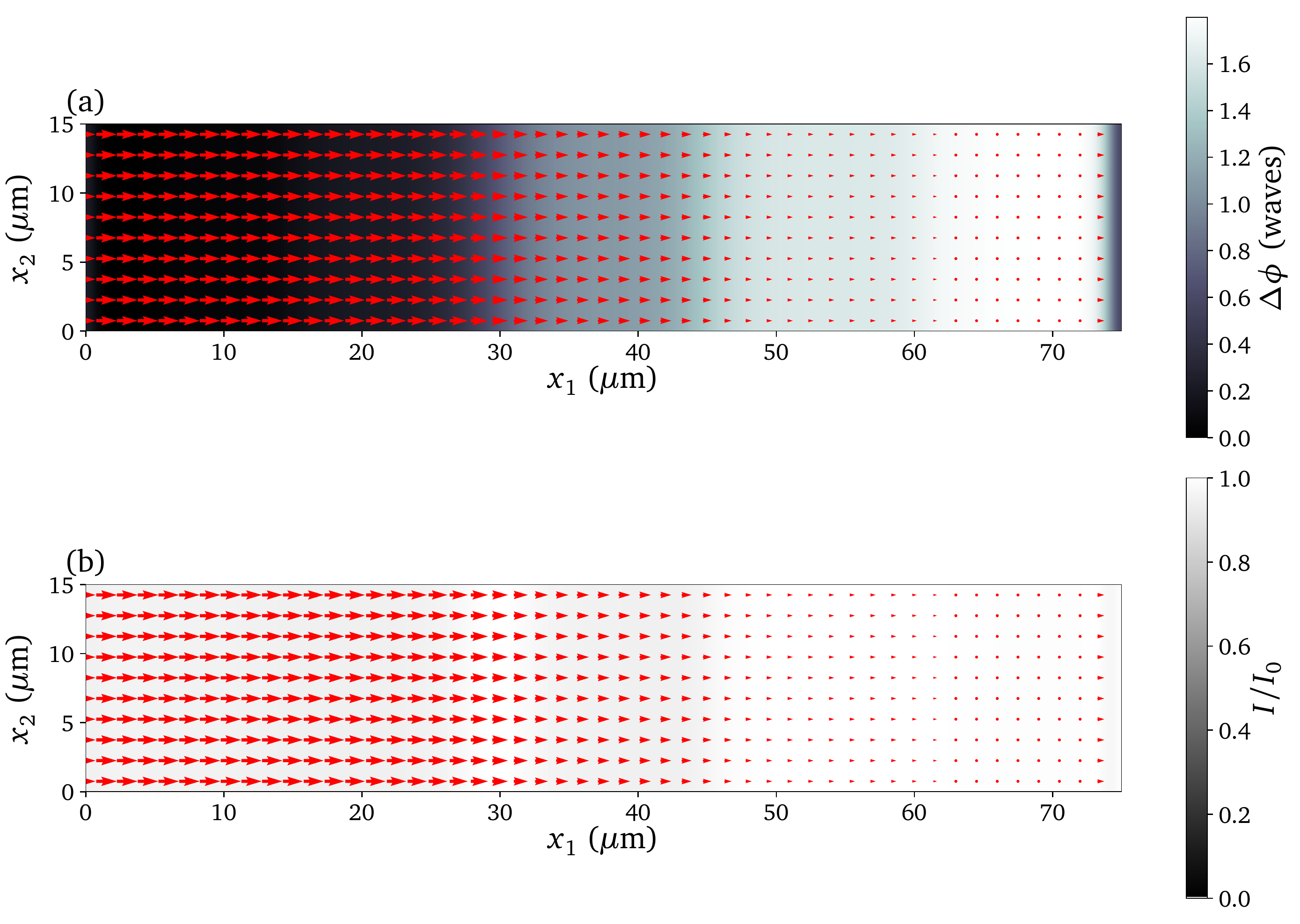}
\caption{Simulated $\parallel$-pol phase profile (a) and intensity profile (b) of a period $5$ blazed grating in the asymmetric direction.}
\label{picture grblx_lcdih_polx_phase_int_per5}
\end{figure}

\cref{picture grblx_lcdih_polx_phase_int_per3,picture grblx_lcdih_polx_phase_int_per4,picture grblx_lcdih_polx_phase_int_per5} show simulations of the phase and intensity profiles of blazed gratings with period $3$, $4$ and $5$ in horizontal configuration respectively. As in the asymmetric binary grating case along horizontal direction, the simulations predict that no polarization conversion occurs in this case. 

\begin{figure}[h!]
\centering
\includegraphics[width=6cm]{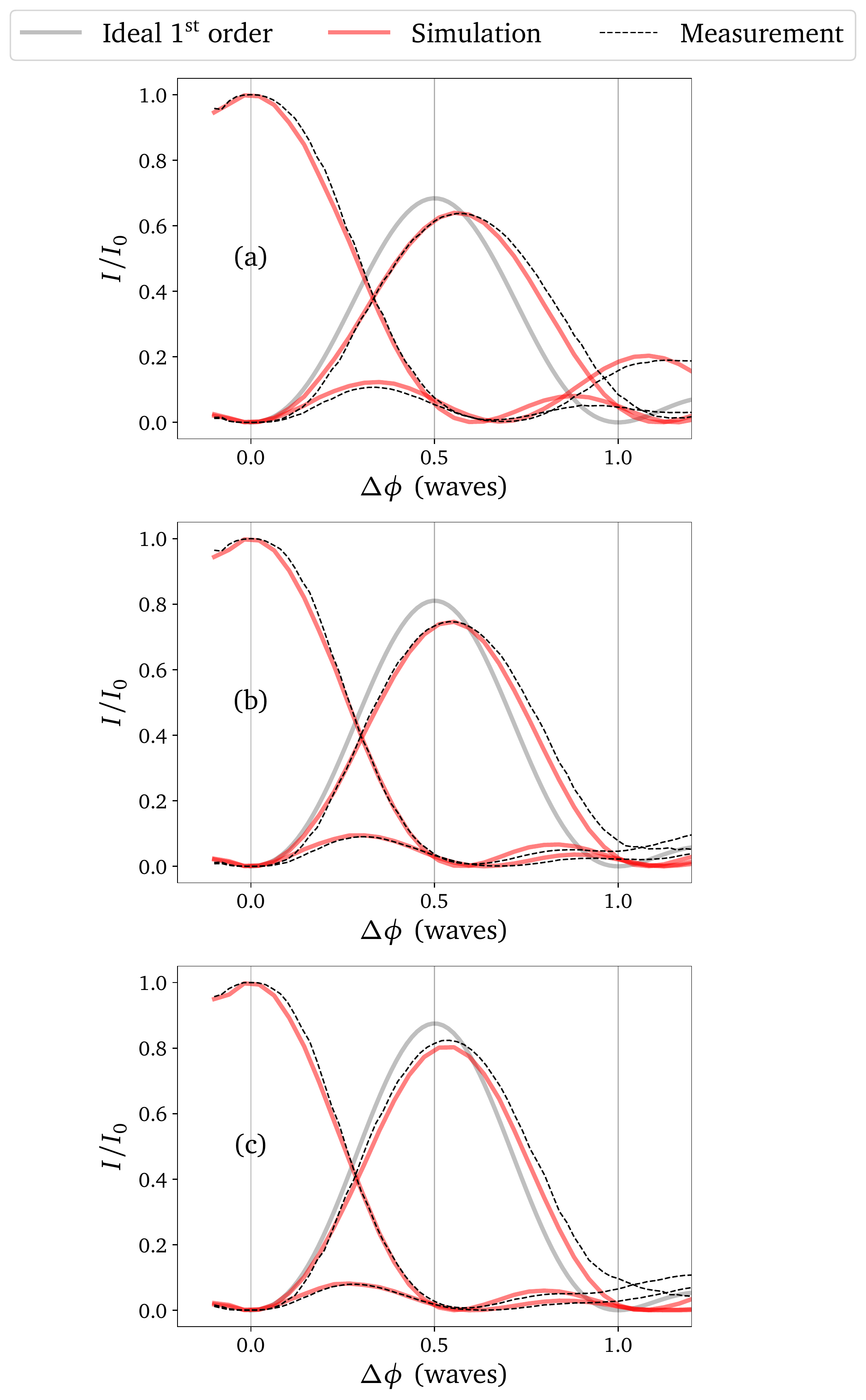}
\caption{Diffraction efficiency simulations and measurements for blazed gratings in the asymmetric direction with (a) period $3$, (b) period $4$ and (c) period $5$.}
\label{picture compare simulations with measurements blazed asymmetric}
\end{figure}

In  \cref{picture compare simulations with measurements blazed asymmetric} we see simulations and measurements for a blazed grating voltage pattern in the asymmetric direction for grating periods $3$, $4$ and $5$. The phase shifts in this figure have been calculated by the formula $\Delta \phi = 2\dfrac{(p-1)}{p} \Delta \tilde{\phi} + p_\mathrm{ref}$, where $\Delta \tilde{\phi}$ represents the (actually realized) phase shift and $p$ the period of the blazed grating. We see that the simulations match the measurements well for small phase shifts ($0-0.5\Delta \phi$). For higher phase shifts the efficiency curves of the measurement have a broader shape. The reasons for these discrepancies are the same as mentioned in  \cref{ssec:Binary grating along the asymmetric direction in horizontal configuration}, but the effects are stronger since the phase shifts here are much larger.

\FloatBarrier
\subsection{Symmetric direction in horizontal configuration}

\begin{figure}[h!]
\centering
\includegraphics[width=12cm]{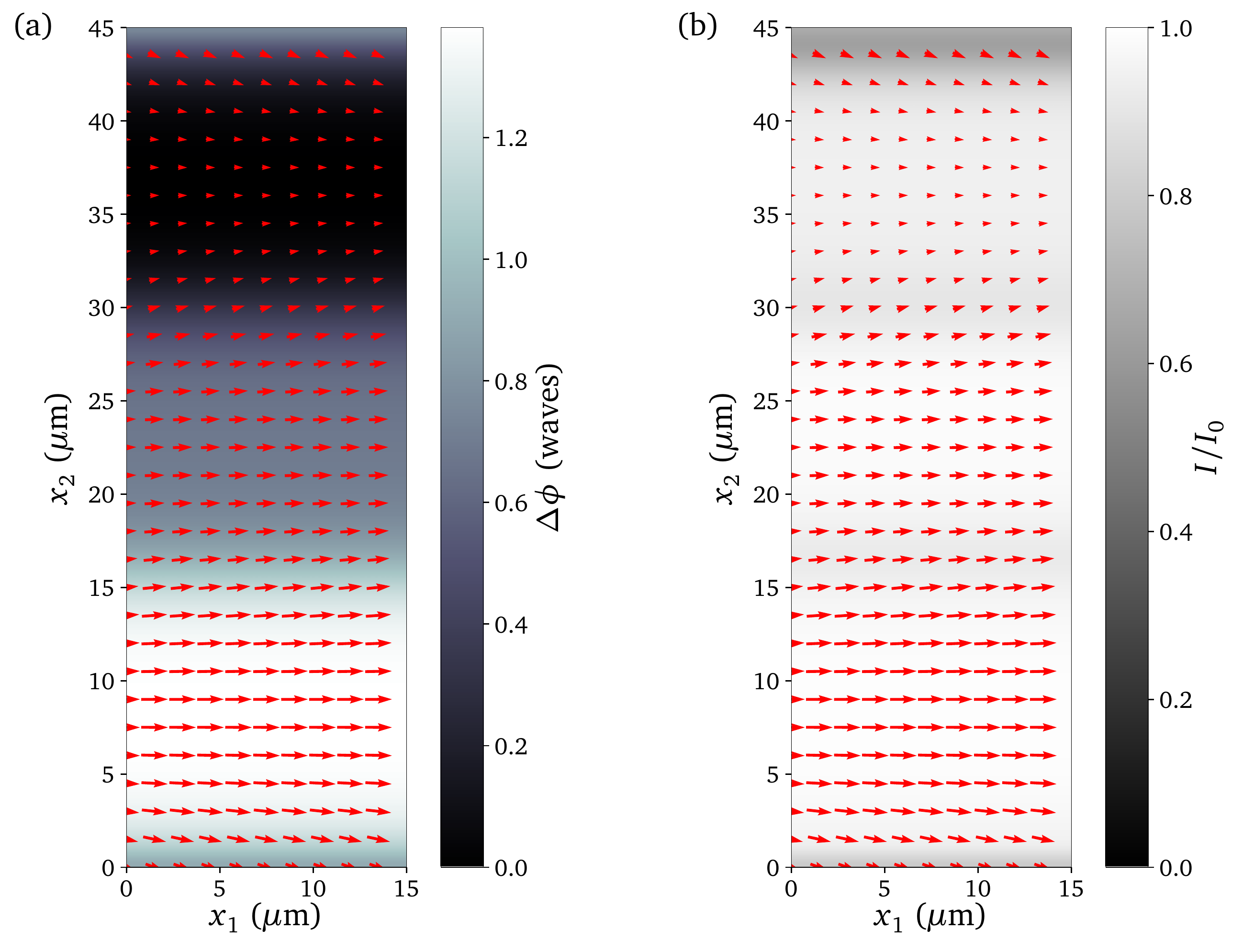}
\caption{Simulated $\parallel$-pol phase profile (a) and intensity profile (b) of a period $3$ blazed grating in the symmetric direction.}
\label{picture grbl_lcdih_polx_phase_int_per3}
\end{figure}

\begin{figure}[h!]
\centering
\includegraphics[width=12cm]{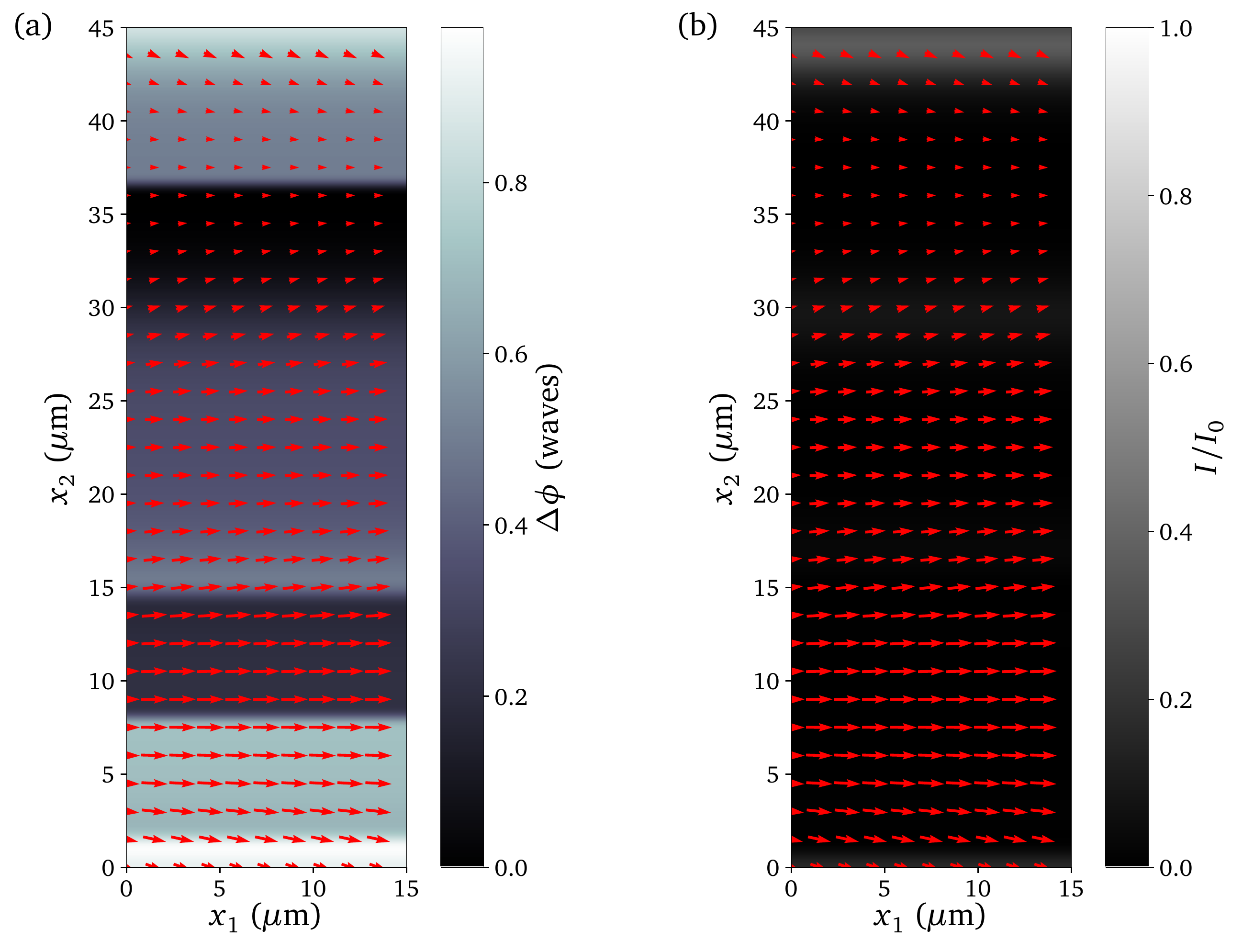}
\caption{Simulated $\perp$-pol phase profile (a) and intensity profile (b) of a period $3$ blazed grating in the symmetric direction.}
\label{picture grbl_lcdih_poly_phase_int_per3}
\end{figure}

\begin{figure}[h!]
\centering
\includegraphics[width=12cm]{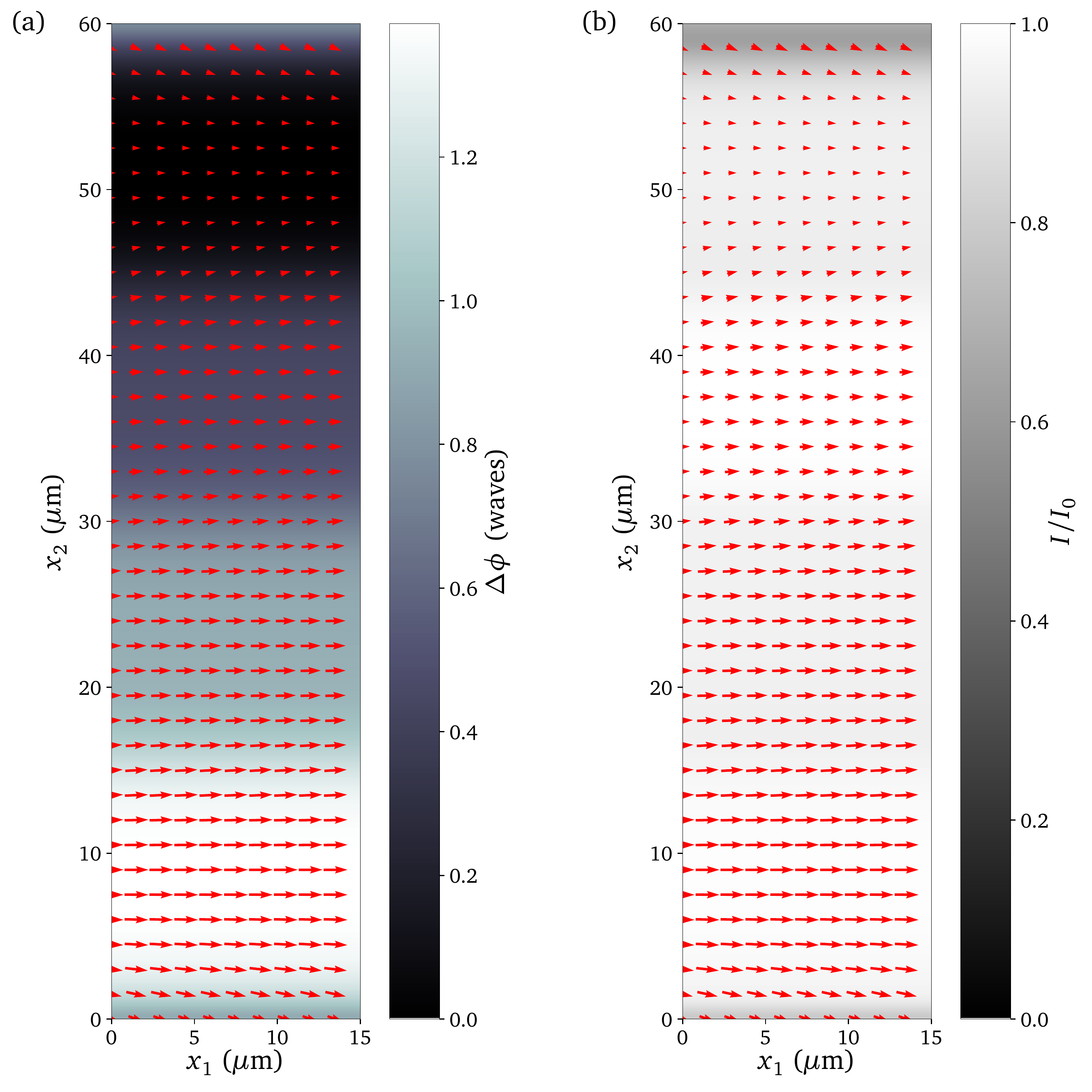}
\caption{Simulated $\parallel$-pol phase profile (a) and intensity profile (b) of a period $4$ blazed grating in the symmetric direction.}
\label{picture grbl_lcdih_polx_phase_int_per4}
\end{figure}

\begin{figure}[h!]
\centering
\includegraphics[width=12cm]{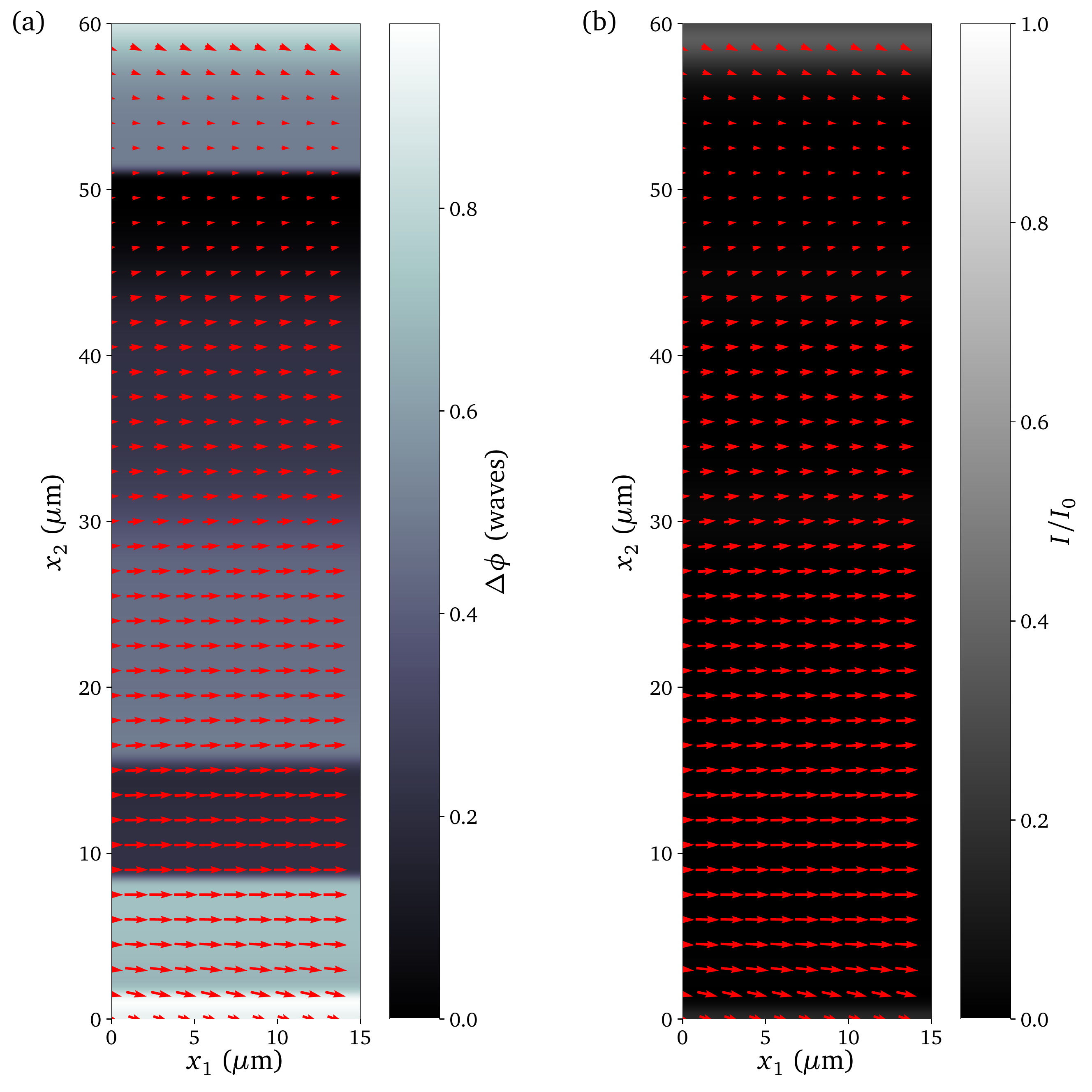}
\caption{Simulated $\perp$-pol phase profile (a) and intensity profile (b) of a period $4$ blazed grating in the symmetric direction.}
\label{picture grbl_lcdih_poly_phase_int_per4}
\end{figure}

\begin{figure}[h!]
\centering
\includegraphics[width=12cm]{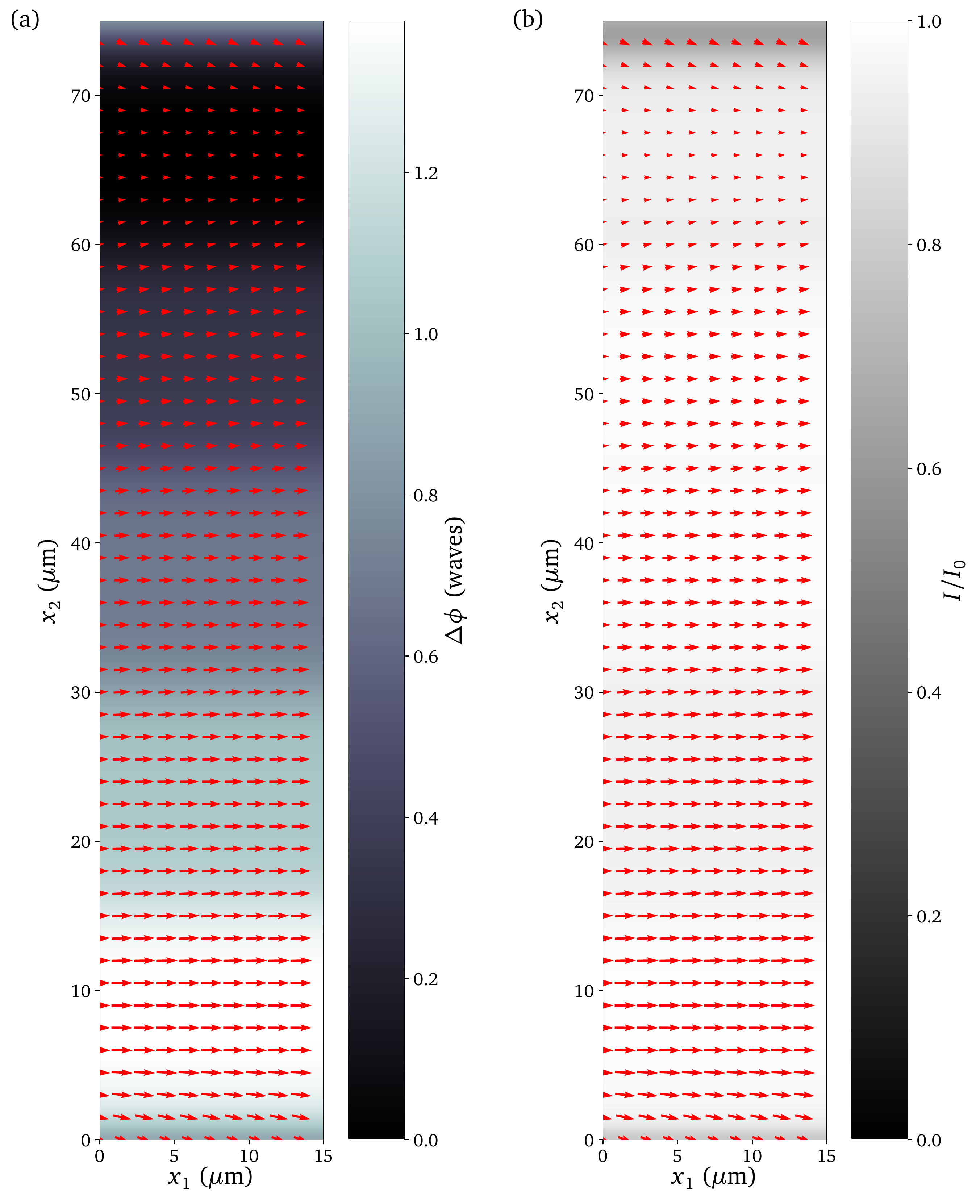}
\caption{Simulated $\parallel$-pol phase profile (a) and intensity profile (b) of a period $5$ blazed grating in the symmetric direction.}
\label{picture grbl_lcdih_polx_phase_int_per5}
\end{figure}

\begin{figure}[h!]
\centering
\includegraphics[width=12cm]{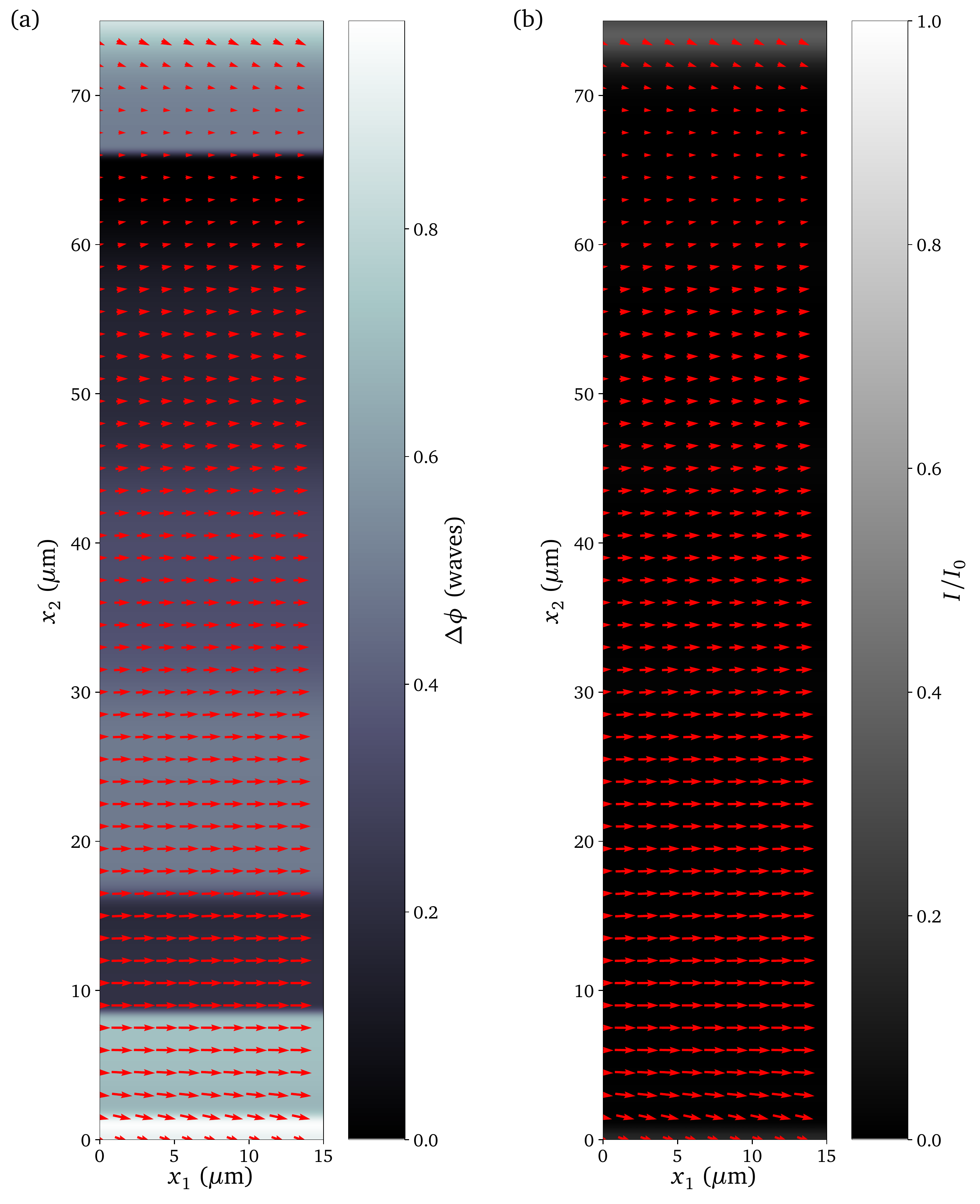}
\caption{Simulated $\perp$-pol phase profile (a) and intensity profile (b) of a period $5$ blazed grating in the symmetric direction.}
\label{picture grbl_lcdih_poly_phase_int_per5}
\end{figure}

\cref{picture grbl_lcdih_polx_phase_int_per3,picture grbl_lcdih_polx_phase_int_per4,picture grbl_lcdih_polx_phase_int_per5} show the phase and intensity profiles of the $\parallel$-pol and \cref{picture grbl_lcdih_poly_phase_int_per3,picture grbl_lcdih_poly_phase_int_per4,picture grbl_lcdih_poly_phase_int_per5} depict the $\perp$-pol of a period $3,4$ and $5$ blazed grating along the symmetric direction in horizontal configuration respectively. These simulations were done with the full $3$D model for a row of $3$, $4$ and $5$ pixels with each $30$ grid points per pixel. 

As in the case of a binary grating in the symmetric direction, here, we have to account for polarization effects. It happens mainly at the transition region between pixels with the highest and lowest voltage, everywhere else very little conversion occurs. The overall effect is smaller as in the binary grating case for all grating periods.

\begin{figure}[h!]
\centering
\includegraphics[width=6cm]{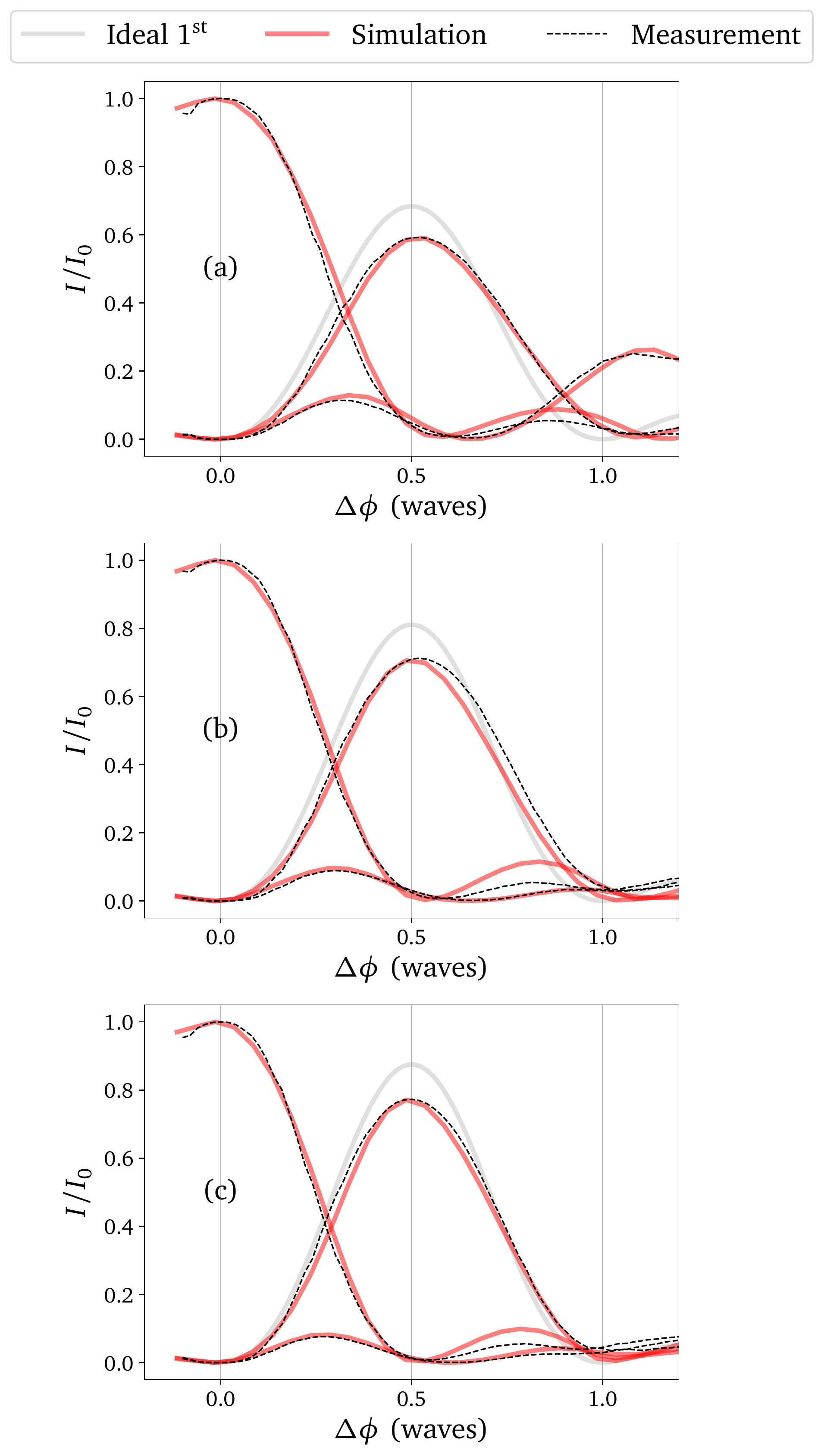}
\caption{Diffraction efficiency simulations and measurements for blazed gratings in the symmetric direction with (a) period $3$, (b) period $4$ and (c) period $5$.}
\label{picture compare simulations with measurements blazed symmetric}
\end{figure}

In \cref{picture compare simulations with measurements blazed symmetric} diffraction efficiency simulations and measurements for a blazed grating along the symmetric direction is shown for period $3$ (a), $4$ (b) and $5$ (c). As in the asymmetric case, the phase shift values were calculated by $\Delta \phi = 2\dfrac{(p-1)}{p} \Delta \tilde{\phi} + p_\mathrm{ref}$. 

The simulations fit the measurements very well for all periods. However, we see a small lateral displacement between simulated and measured diffraction efficiency curve for the period $4$ blazed grating in \cref{picture compare simulations with measurements blazed symmetric} (b), which does not appear in (a,c). 

\FloatBarrier
\section{Simulations for angle dependence of polarization conversion}

In  \cref{ssec:Binary grating along the symmetric direction in vertical configuration} we saw that the polarization conversion efficiency can reach up to $\sim 18\%$ for a binary grating in the symmetric direction. Now we will take a look at simulations of the angle dependence of said efficiency for different patterns in the vertical configuration.

\begin{figure}[h!]
\centering
\includegraphics[width=12cm]{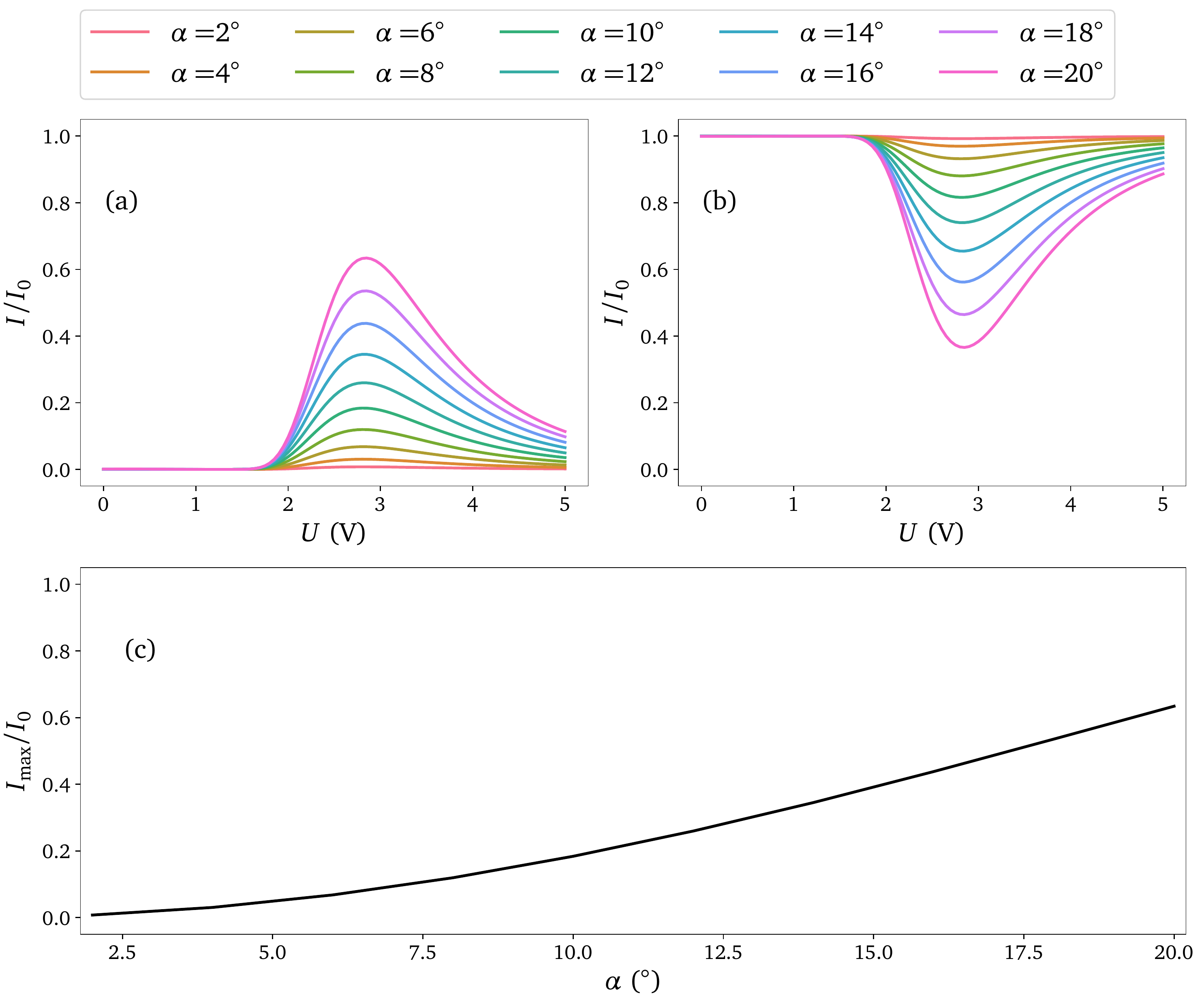}
\caption{Polarization conversion efficiencies for uniform electric field pattern at different angles in the vertical configuration. (a) shows the intensity of $\parallel$-pol, (b) $\perp$-pol and (c) maximum intensity of $\parallel$-pol.}
\label{picture simulation pol conv angle dependence const}
\end{figure}

 \cref{picture simulation pol conv angle dependence const} shows the conversion efficiencies of a constant voltage pattern for angles between $2^\circ$ and $20^\circ$, where (a) and (b) show the $\parallel$-pol and $\perp$-pol and (c) the maximum of the $\parallel$-pol intensity. The curves in (a) show a strict monotonous increase in efficiency upon increasing the angle of incidence, up to a maximum of $\sim 65\%$ for $\alpha=20^\circ$. Simulations show that in the horizontal configuration (not shown here) no polarization conversion takes place, independent from the angle.

\begin{figure}[h!]
\centering
\includegraphics[width=12cm]{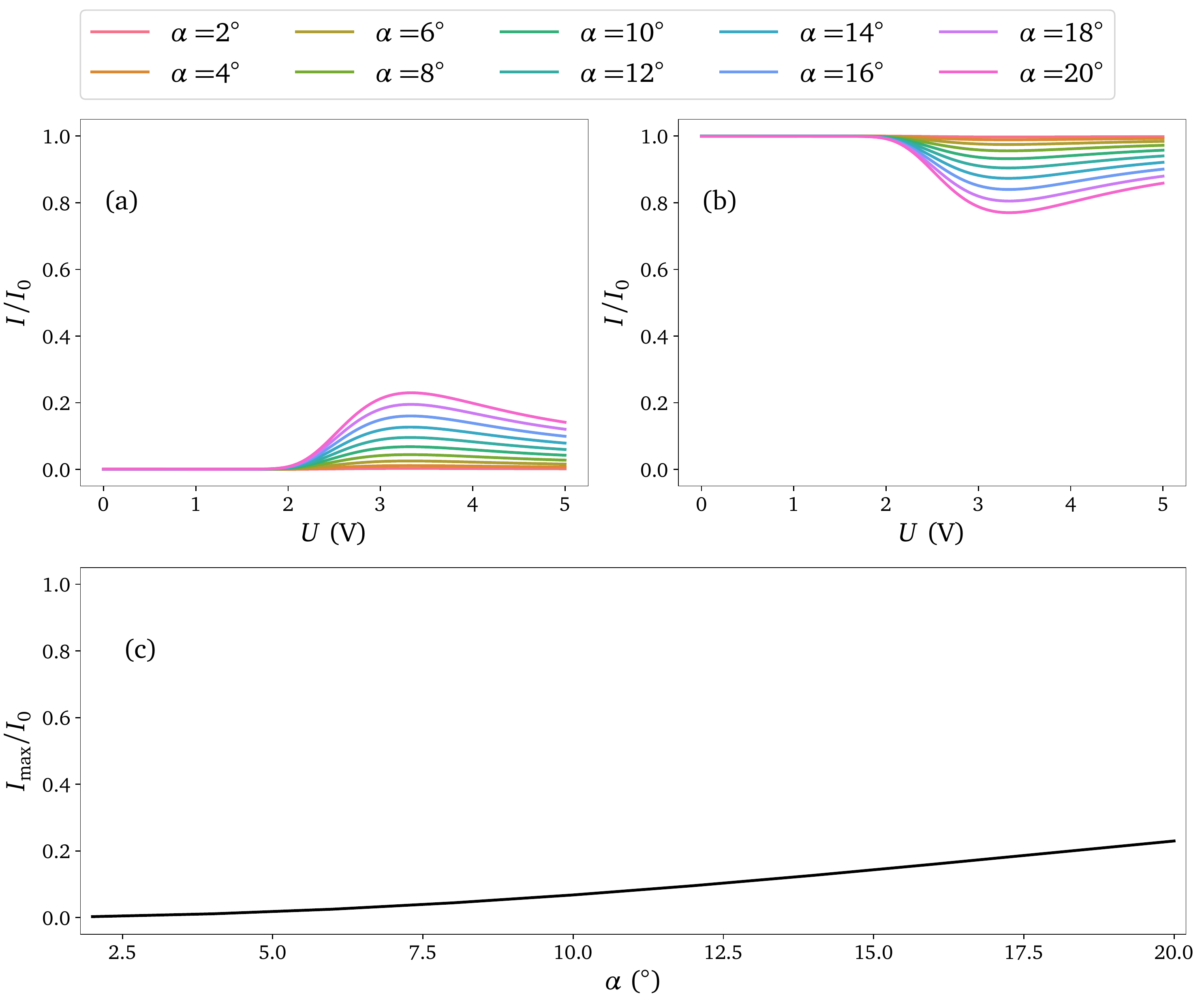}
\caption{Polarization conversion efficiencies for binary grating pattern in the asymmetric direction at different angles in the vertical configuration. (a) shows the intensity of $\parallel$-pol, (b) $\perp$-pol and (c) maximum intensity of $\parallel$-pol.}
\label{picture simulation pol conv angle dependence x}
\end{figure}

In  \cref{picture simulation pol conv angle dependence x} we see the conversion efficiencies for a binary grating in the asymmetric direction. The reference voltage in the simulations was set to $0.2$ V, which is below threshold. The effect is qualitatively similar to the uniform case, but quantitatively smaller. As in the uniform case, in the horizontal configuration no conversion takes place.

\begin{figure}[h!]
\centering
\includegraphics[width=12cm]{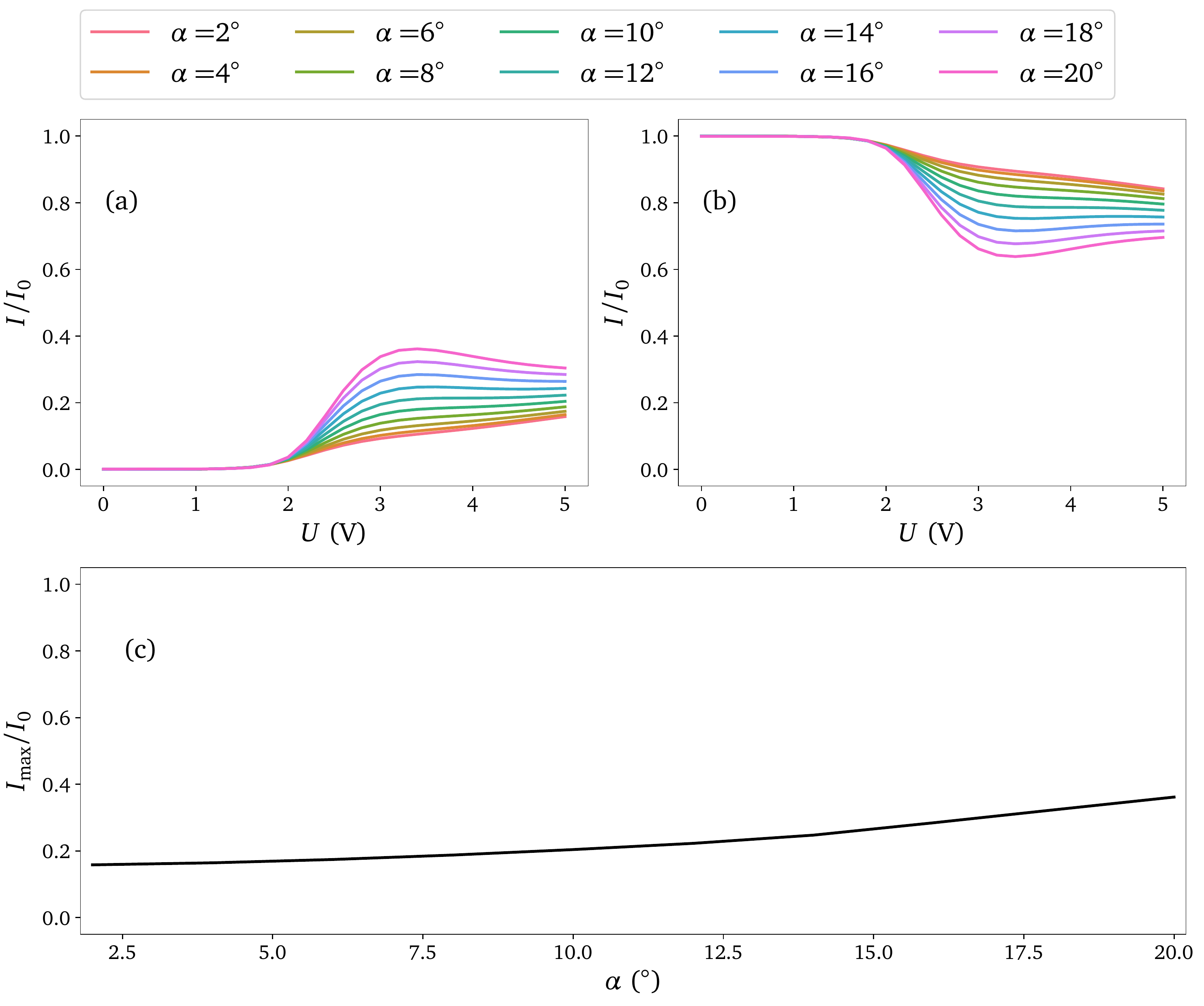}
\caption{Polarization conversion efficiencies for binary grating pattern in the symmetric direction at different angles in the vertical configuration. (a) shows the intensity of $\parallel$-pol, (b) $\perp$-pol and (c) maximum intensity of $\parallel$-pol.}
\label{picture simulation pol conv angle dependence y}
\end{figure}

\begin{figure}[h!]
\centering
\includegraphics[width=12cm]{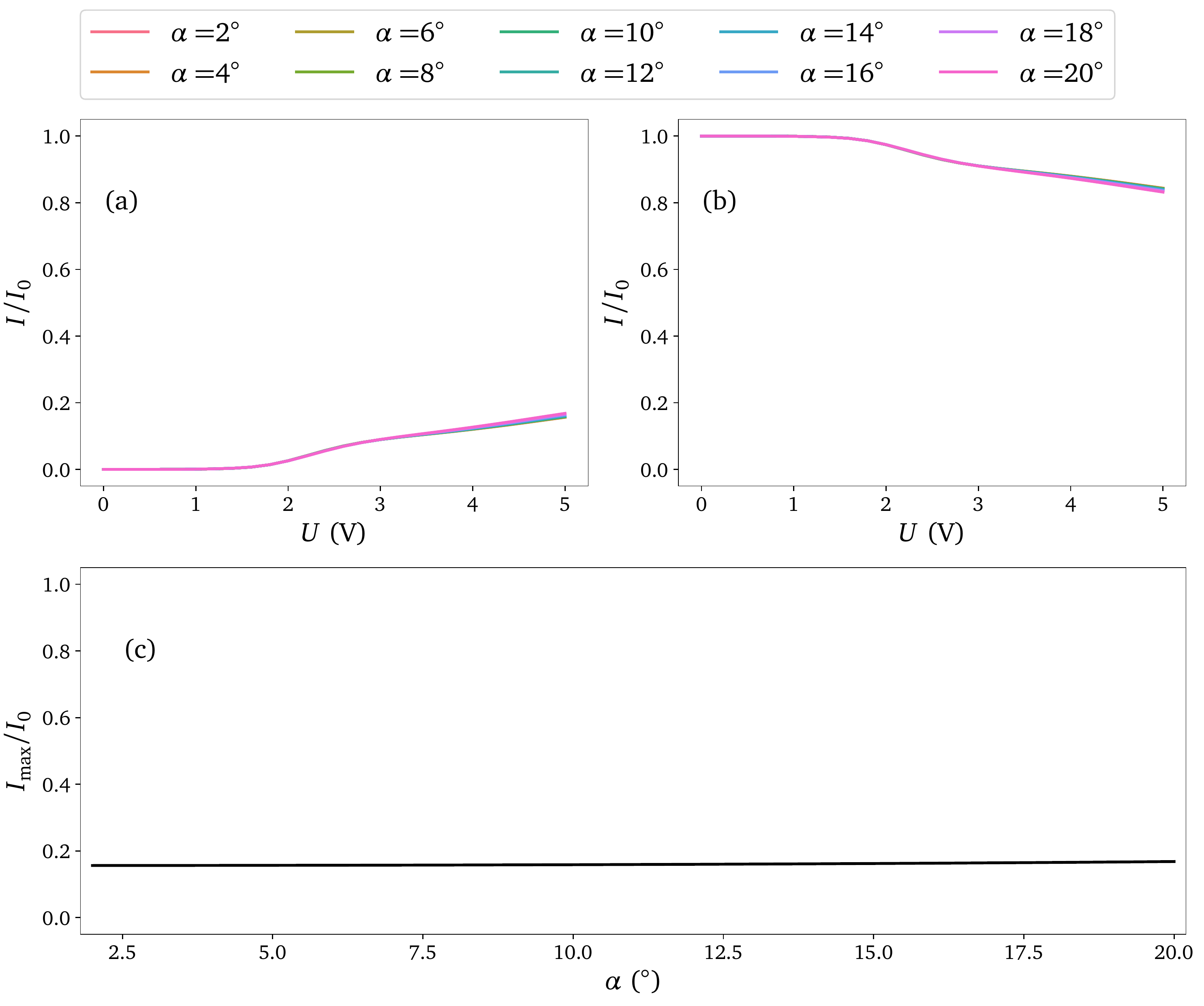}
\caption{Polarization conversion efficiencies for binary grating pattern in the symmetric direction at different angles in the horizontal configuration. (a) shows the intensity of $\perp$-pol, (b) $\parallel$-pol and (c) maximum intensity of $\perp$-pol.}
\label{picture simulation pol conv angle dependence y h configuration}
\end{figure}

 \cref{picture simulation pol conv angle dependence y,picture simulation pol conv angle dependence y h configuration} depict the conversion efficiencies of a binary grating in symmetric direction in the vertical and horizontal configuration. In the vertical case we see an angle dependence on the curves, whereas in the horizontal configuration there is almost no angle dependence. All curves in the horizontal case correspond to the $\alpha=0^\circ$ case in the vertical direction. Both cases show an offset in the maximum efficiency.

\begin{figure}[h!]
\centering
\includegraphics[width=12cm]{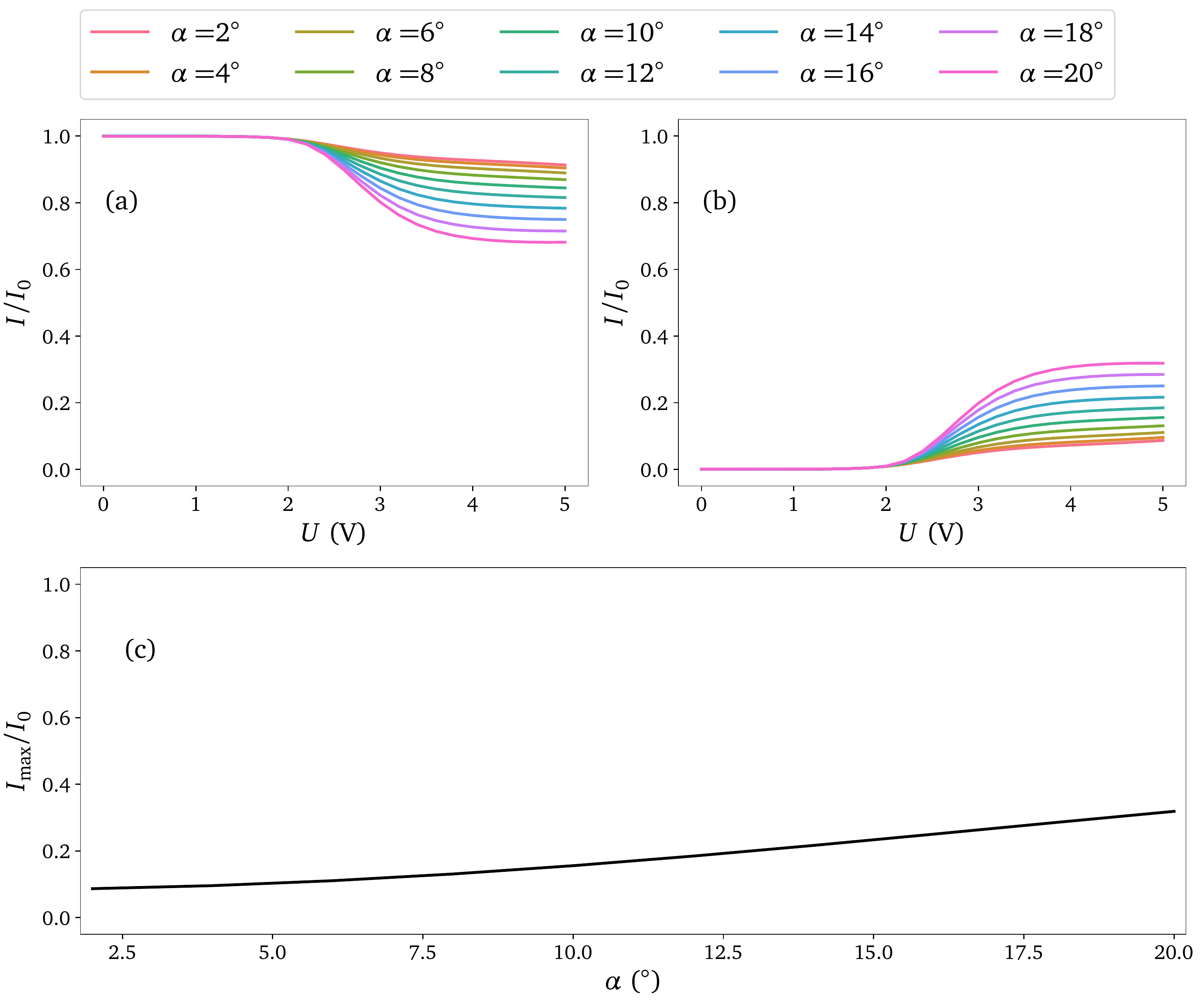}
\caption{Polarization conversion efficiencies for a checkerboard pattern at different angles in the vertical configuration. (a) shows the intensity of $\perp$-pol, (b) $\parallel$-pol and (c) maximum intensity of $\perp$-pol.}
\label{picture simulation pol conv angle dependence cb}
\end{figure}

\begin{figure}[h!]
\centering
\includegraphics[width=12cm]{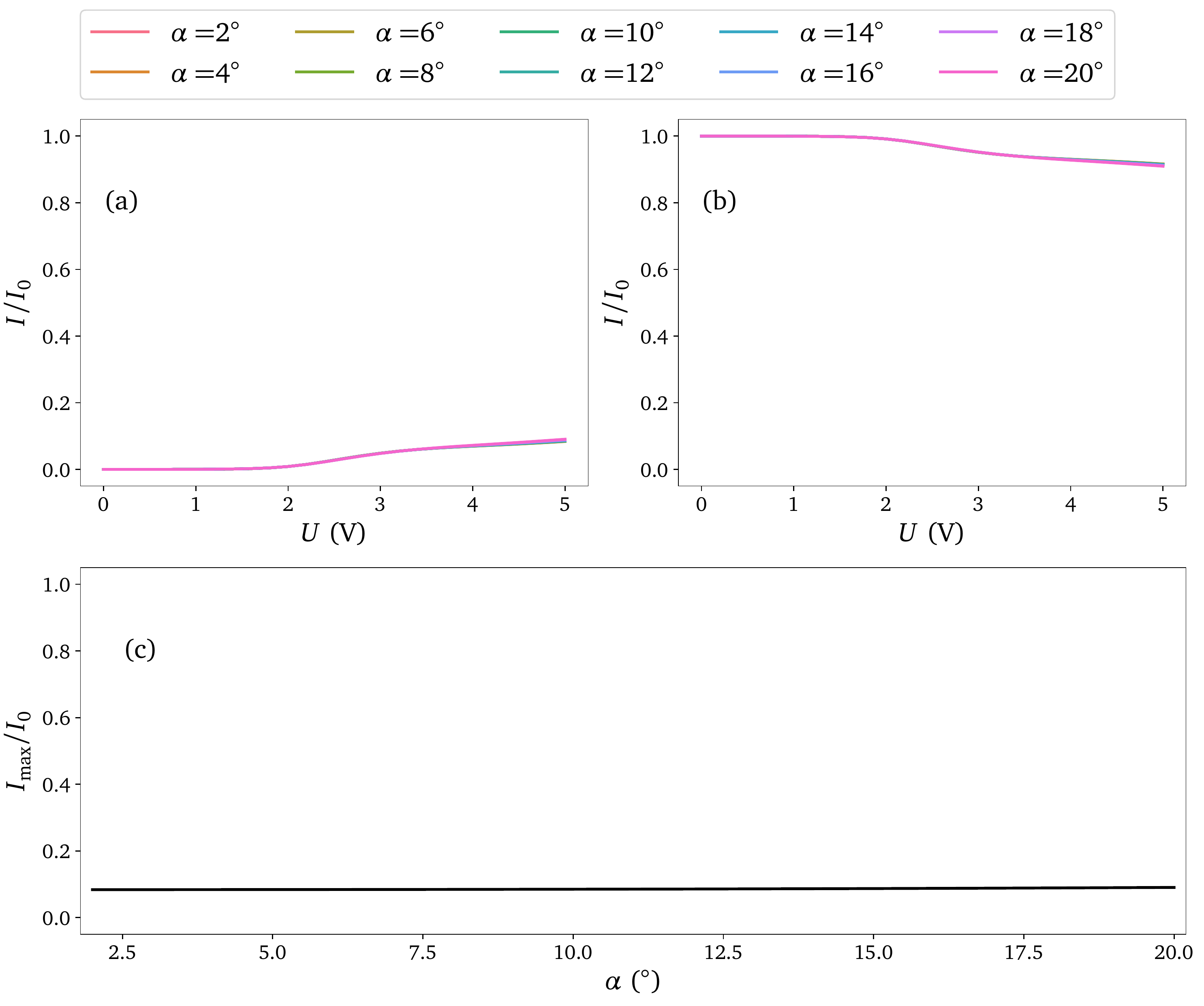}
\caption{Polarization conversion efficiencies for a checkerboard pattern at different angles in the horizontal configuration. (a) shows the intensity of $\perp$-pol, (b) $\parallel$-pol and (c) maximum intensity of $\perp$-pol.}
\label{picture simulation pol conv angle dependence cb h configuration}
\end{figure}

In  \cref{picture simulation pol conv angle dependence cb,picture simulation pol conv angle dependence cb h configuration} we see the polarization conversion efficiencies for a checkerboard pattern in the vertical and horizontal case. We see the same dependence on $\alpha$ as in the symmetric grating case in the vertical configuration and no influence in the horizontal configuration.

These simulations, together with the measurements done with an angle $\alpha=4.2^\circ$ suggest that the horizontal configuration is generally preferable, especially if operated at a large angle of incidence. The remaining effect of polarization conversion happening for the binary symmetric and checkerboard pattern can be minimized by using small voltage differences between pixels in the symmetric direction.
\FloatBarrier
\section{Diffraction efficiency of Hamamatsu SLM}\label{ssec:diffraction efficiency hamamatsu slm}

In this section we will take a look at a different SLM model, a Hamamatsu SLM without a built-in dielectric mirror (model X$10468$-$07$). In addition the driving voltage is inverted, which means that a control value of $0$ corresponds to the maximum applied voltage ($\sim 9-10$ V). The parameters used to simulate this SLM are shown in \cref{tabular parameters hamamatsu SLM}. 

\begin{table}
\begin{center}
\begin{tabular}[c]{r|r|r}
\hline
\multicolumn{3}{c}{Simulation Parameters}   \\
\thickhline
\rowcolor{bluetable1}
& $K_{11}$ & $19.41$ $\mathrm{pN}$ \\
\rowcolor{bluetable1}
&$K_{22}$ & $6.83$ $\mathrm{pN}$ \\
\rowcolor{bluetable1}
&$K_{33}$ & $9.61$ $\mathrm{pN}$ \\
\rowcolor{bluetable1}
&$\varepsilon_\parallel$ & $17.5$ \\
\rowcolor{bluetable1}
&$\varepsilon_\perp$ & $4.8$ \\
\rowcolor{bluetable1}
&$n_\mathrm{e}$ & $1.65$  \\
\rowcolor{bluetable1}
&$n_\mathrm{o}$ & $1.4$  \\
\rowcolor{bluetable1}
\multirow{-8}{*}{LC-Parameters}&$\theta_\mathrm{p}$ & $10^\circ$ \\ \hline
\rowcolor{bluetable2}
& $d$ & $8$ $\mu \mathrm{m}$ \\
\rowcolor{bluetable2}
&$d_\mathrm{rel}$ & $0.1$  \\
\rowcolor{bluetable2}
\multirow{-3}{*}{Geometry-Parameters}&$x$ & $40$ $\mu \mathrm{m}$ \\
\hline
\rowcolor{bluetable3}
&$\varepsilon_\mathrm{c}$ &$7$  \\
\rowcolor{bluetable3}
&$n_\mathrm{coverglass}$ & $1.575$  \\
\rowcolor{bluetable3}
&$n_\mathrm{electrode}$ & $1.575$  \\
\rowcolor{bluetable3}
&$\alpha$ & $4.2^\circ$ \\
\rowcolor{bluetable3}
&$\lambda$ & $633$ nm \\
\hline
\end{tabular}
\end{center}
\caption{Simulation Parameters for the Hamamatsu X$10468$-$07$ SLM.}
\label{tabular parameters hamamatsu SLM}
\end{table}

\begin{figure}[h!]
\centering
\includegraphics[width=12cm]{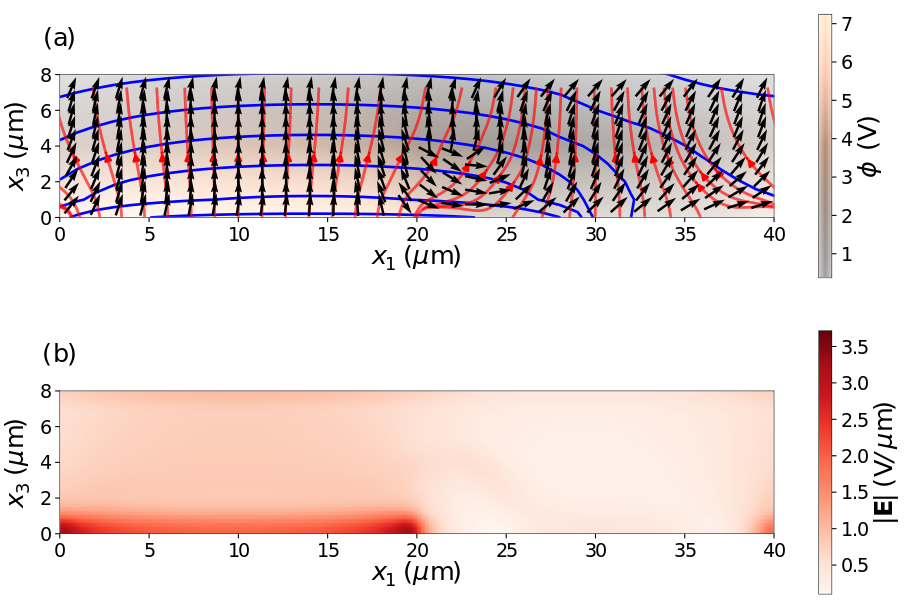}
\caption{(a) shows the director distribution (black arrows), electric field lines (red), electric potential contour lines (blue) and the electric potential itself (background color), (b) depicts $|\vec{E}|$.}
\label{picture director distribution thick hamamatsu}
\end{figure}

 \cref{picture director distribution thick hamamatsu} shows the simulated director distribution, electric field and the electric potential (a) and the absolute value of the electric field (b) for the Hamamatsu SLM for voltages $U_1 = 9$ V (left electrode) and $U_2 = 2.5$ V (right electrode). We see that the absence of space between electrodes and LC-layer causes a strong fringing field near the electrodes, where the electric field is strongest. The electric field near the electrodes forces the director to follow the electric field lines, which results in a peculiar orientation across the LC layer. This simulation was done with the vector representation. In the vertical slice $(20\, \mu \mathrm{m},x_3)$ we see that two adjacent directors are oriented anti-parallel. As discussed in the theory section (see  \cref{ssec:theory vector representation}) the vector representation yields an inaccurate free energy in this case.

\begin{figure}[h!]
\centering
\includegraphics[width=12cm]{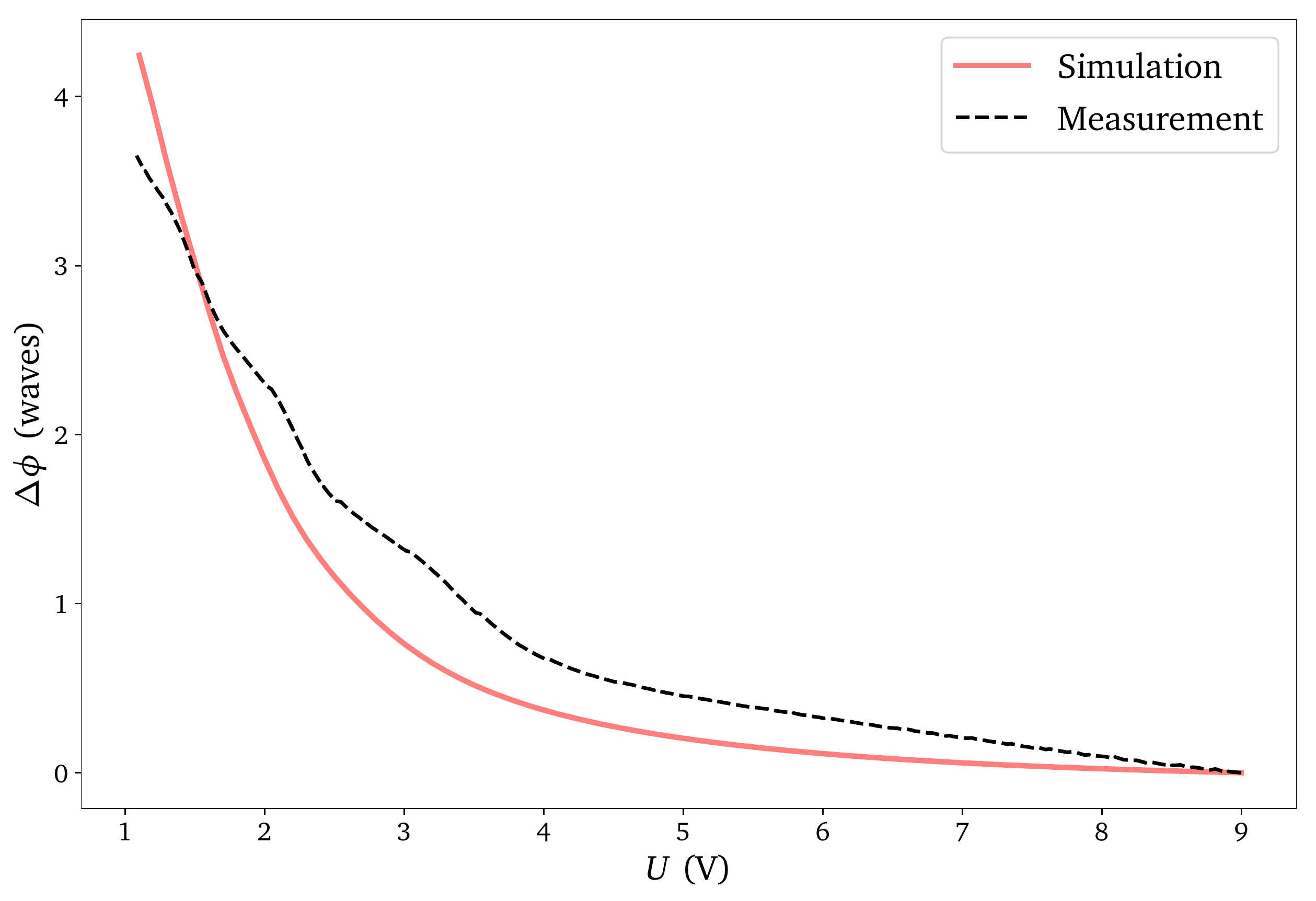}
\caption{Simulations (red) and measurements (black) of the phase shift vs. control voltage for the Hamamatsu SLM.}
\label{picture compare lut measurement thick hamamatsu}
\end{figure}

This SLM was delivered non calibrated. \cref{picture compare lut measurement thick hamamatsu} shows the simulation (red) and measurement (black) for for a uniform pattern of the Hamamatsu SLM. The simulation fits the measurement qualitatively and deviates quantitatively. 

\begin{figure}[h!]
\centering
\includegraphics[width=12cm]{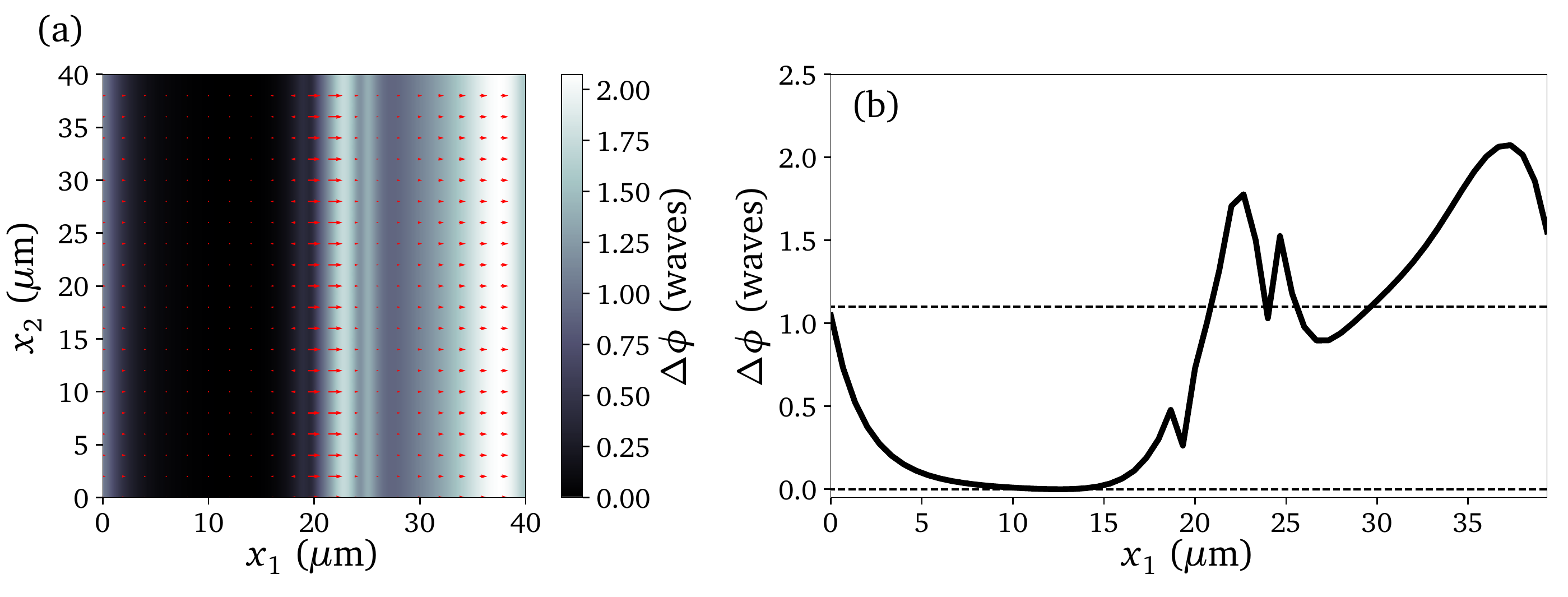}
\caption{(a) shows the phase profile over four pixels and a slice of $\vec{n}$ projected on the $(x_1,x_2)$ plane, at the center of the LC layer, (b) depicts a slice through the phase profile in (a).}
\label{picture phase profile thick hamamatsu}
\end{figure}

 \cref{picture phase profile thick hamamatsu} shows the resulting phase profile from the director distribution in  \cref{picture director distribution thick hamamatsu} in $2$D (a) and $1$D (b). The dashed grey lines in (b) represent the corresponding phase values $(p_1,p_2) = (0,1.1)$ waves for the voltages $(U_1,U_2) = (9,2.5)$ V in the LUT (see  \cref{picture compare lut measurement thick hamamatsu}. We see, that the phase values for the second pixel in  \cref{picture phase profile thick hamamatsu} (b) do not correspond to the expected values predicted by the calibration. This discrepancy is caused by the fringing field near the electrodes, which results in a small tilt angle $\theta$ in the transition region. This causes the phase profile to rise significantly above the expected phase value. 

\begin{figure}[h!]
\centering
\includegraphics[width=12cm]{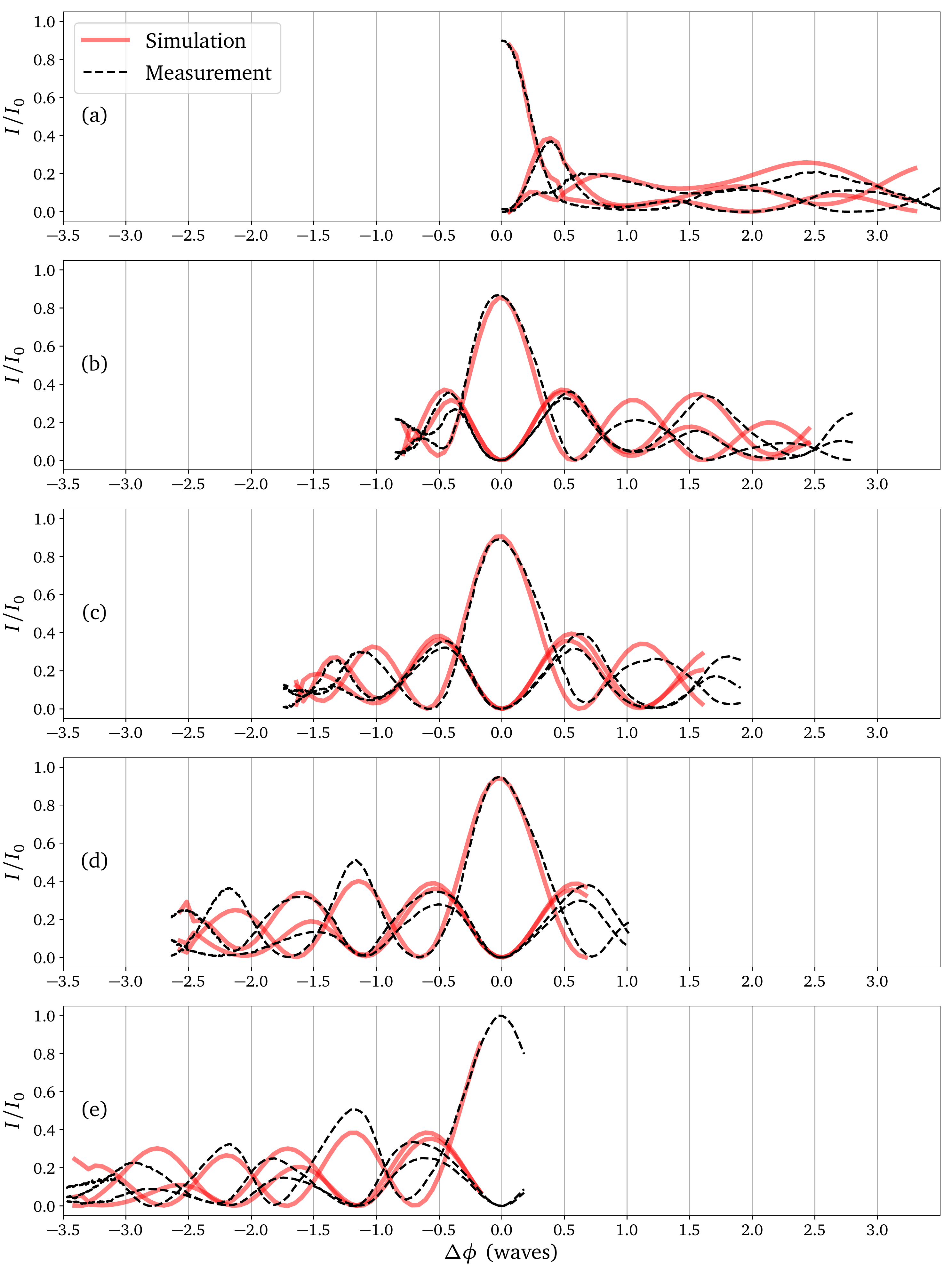}
\caption{Diffraction efficiency simulations (red) and measurements (black) of a binary grating in the asymmetric direction for reference phase values $\phi_\mathrm{ref} = 0.$ (a), $\phi_\mathrm{ref} = 0.85$ (b), $\phi_\mathrm{ref} = 1.7$ (c), $\phi_\mathrm{ref} = 2.63$ (d), $\phi_\mathrm{ref} = 3.47$ waves (e).}
\label{picture compare de measurement thick hamamatsu}
\end{figure}

 \cref{picture compare de measurement thick hamamatsu} shows diffraction efficiency simulations and measurements for a period $2$ binary grating in the asymmetric direction for five phase reference values $\phi_\mathrm{ref} = 0.$ (a), $\phi_\mathrm{ref} = 0.85$ (b), $\phi_\mathrm{ref} = 1.7$ (c), $\phi_\mathrm{ref} = 2.63$ (d) and $\phi_\mathrm{ref} = 3.47$ waves (e). The diffraction efficiency curves in (a) are of very unusual shape. The missing space between electrodes and LC-layer combined with a high reference voltage cause the SLM to be hardly usable in this region. At reference phase values corresponding to intermediate voltages (b,c,d) the diffraction efficiency curves behave similar to the curves of the BNS model (see  \cref{ssec:comparison simulation experiment binary gratings}). At very high phase reference levels (low voltages) (e) and low phase reference levels (a) this SLM deviates from the usual behavior. The deviations are more pronounced at low phase reference levels than at high phase reference levels.

The Simulations in  \cref{picture compare de measurement thick hamamatsu} fit the measurements qualitatively well, even at small phase reference values (a).

\begin{figure}[h!]
\centering
\includegraphics[width=12cm]{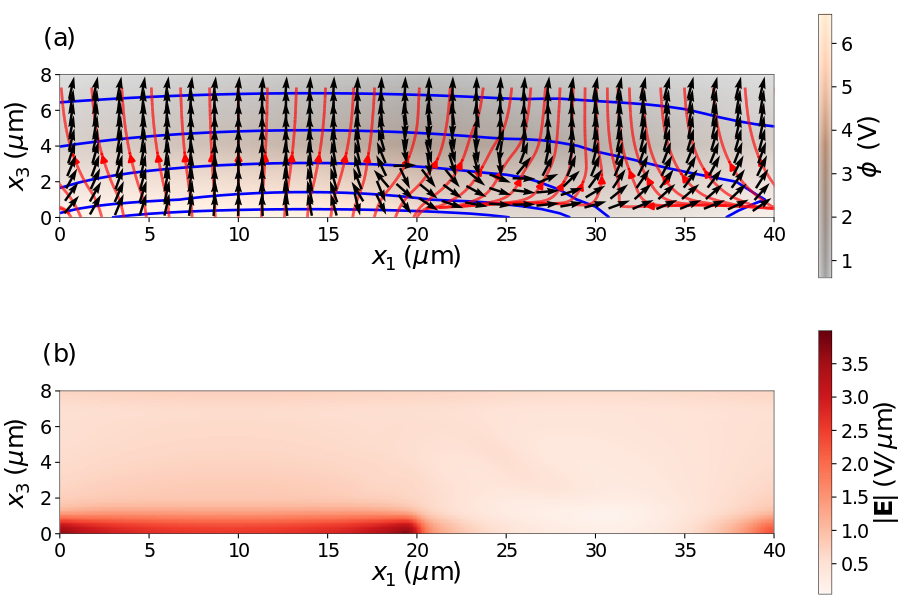}
\caption{(a) shows the director distribution calculated by the tensor representation of the director (black arrows), electric field lines (red), electric potential contour lines (blue) and the electric potential itself (background), (b) depicts $|\vec{E}|$.}
\label{picture director distribution tensor thick hamamatsu}
\end{figure}

 \cref{picture director distribution tensor thick hamamatsu} shows simulations for the director distribution done by the tensor method (see  \cref{ssec:theory vector representation}). By comparing  \cref{picture director distribution tensor thick hamamatsu} with  \cref{picture director distribution thick hamamatsu}, we see differences in the director distribution. Whereas in  \cref{picture director distribution thick hamamatsu} we saw only one pair of adjacent directors anti-parallel aligned, in  \cref{picture director distribution tensor thick hamamatsu} we see several pairs of directors which are oriented anti-parallel.  
However, the director distribution in  \cref{picture director distribution tensor thick hamamatsu} also shows multiple pairs of directors, where the included angle exceeds $90^\circ$. 

\begin{figure}[h!]
\centering
\includegraphics[width=12cm]{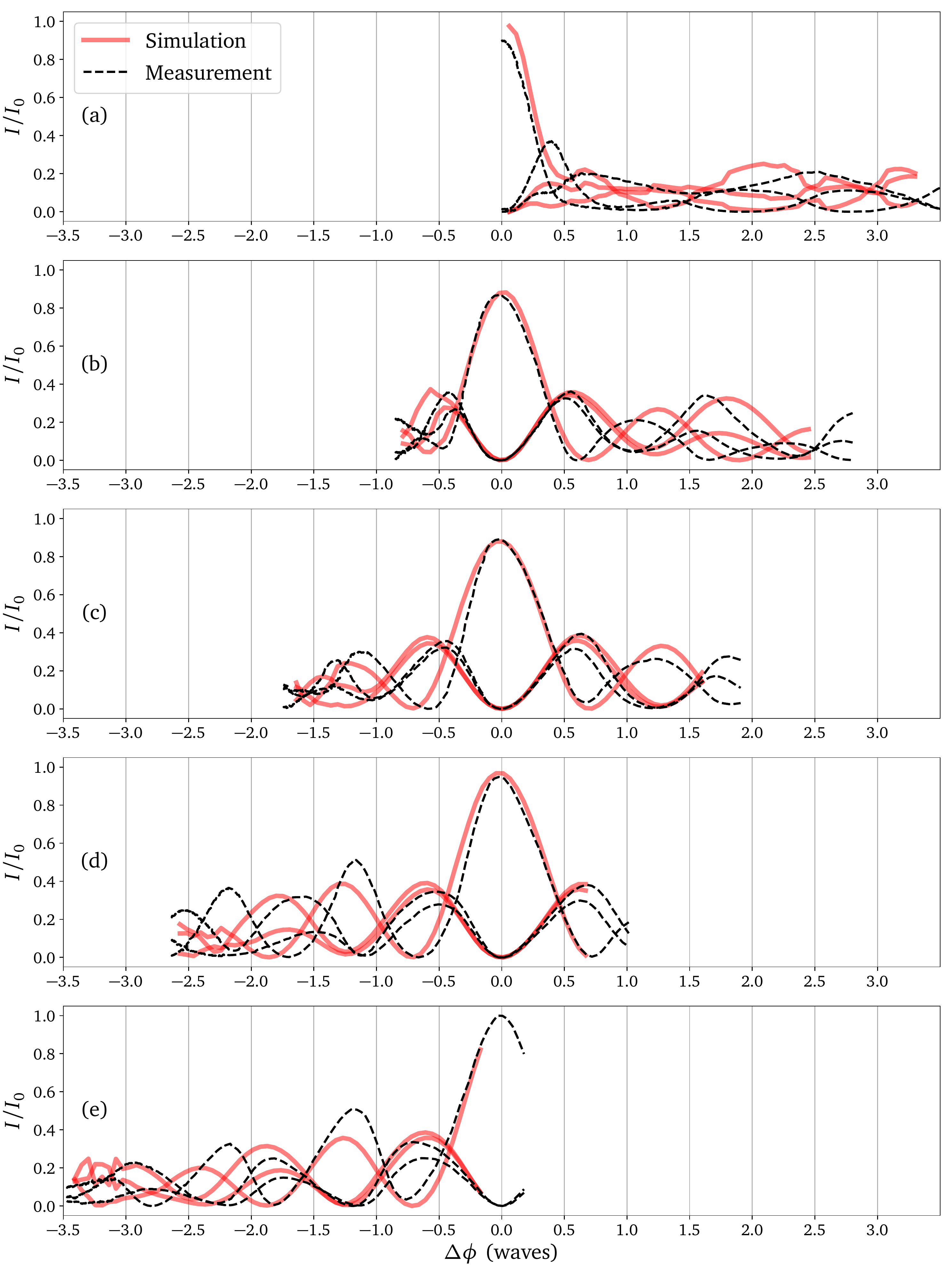}
\caption{Diffraction efficiency simulations done by the tensor method (red) and measurements (black) of a binary grating in the asymmetric direction for reference phase values $\phi_\mathrm{ref} = 0.$ (a), $\phi_\mathrm{ref} = 0.85$ (b), $\phi_\mathrm{ref} = 1.7$ (c), $\phi_\mathrm{ref} = 2.63$ (d), $\phi_\mathrm{ref} = 3.47$ waves (e).}
\label{picture compare de measurement thick hamamatsu tensor}
\end{figure}

In  \cref{picture compare de measurement thick hamamatsu tensor} we see the diffraction efficiency simulations done by the tensor method and measurements. The simulations were done with the same parameters and phase values as in  \cref{picture compare de measurement thick hamamatsu}, the measurements shown are the same in both pictures. We see that the simulations fit the measurements qualitatively in (b-e), but in (a) we see a jittery curve. This jittering of the curve is caused by the high voltage, which again causes the simulation to yield non-physical solutions for the director distribution.

We saw in  \cref{picture director distribution thick hamamatsu,picture director distribution tensor thick hamamatsu} that both methods used to simulate the director distribution (vector and tensor method) differ and from \cref{ssec:Tensor representation} we know that the vector method does not conserve the $n\rightarrow -n$ symmetry and the tensor method may yield non physical results. However, simulations done by the vector method were able to produce smooth diffraction efficiency curves (\cref{picture compare de measurement thick hamamatsu}) for all reference phase values and match experiments better, whereas simulations done with the tensor method did not yield smooth curves for low reference phase values (\cref{picture compare de measurement thick hamamatsu tensor} (a)).   

\FloatBarrier
\chapter{Fast $2$D model}\label{sec:Fringer}

The simulations to determine the director distribution from  \cref{sec:directormodeling3D} for a grid size $10\times 60 \times 60$ take about $\sim 2-5$min to converge to a solution. Since these simulations only include $4$ pixels, simulating the director distribution for an arbitrary voltage pattern over $500\times 500$ pixels would be off limits. To make the model useful for practical application, we use the information we have gathered in  \cref{sec:comparison of experiment with simulation} about the phase and amplitude profiles of simple voltage patterns to build an approximate, but much faster model. 

\begin{figure}[h!]
\centering
\includegraphics[width=12cm]{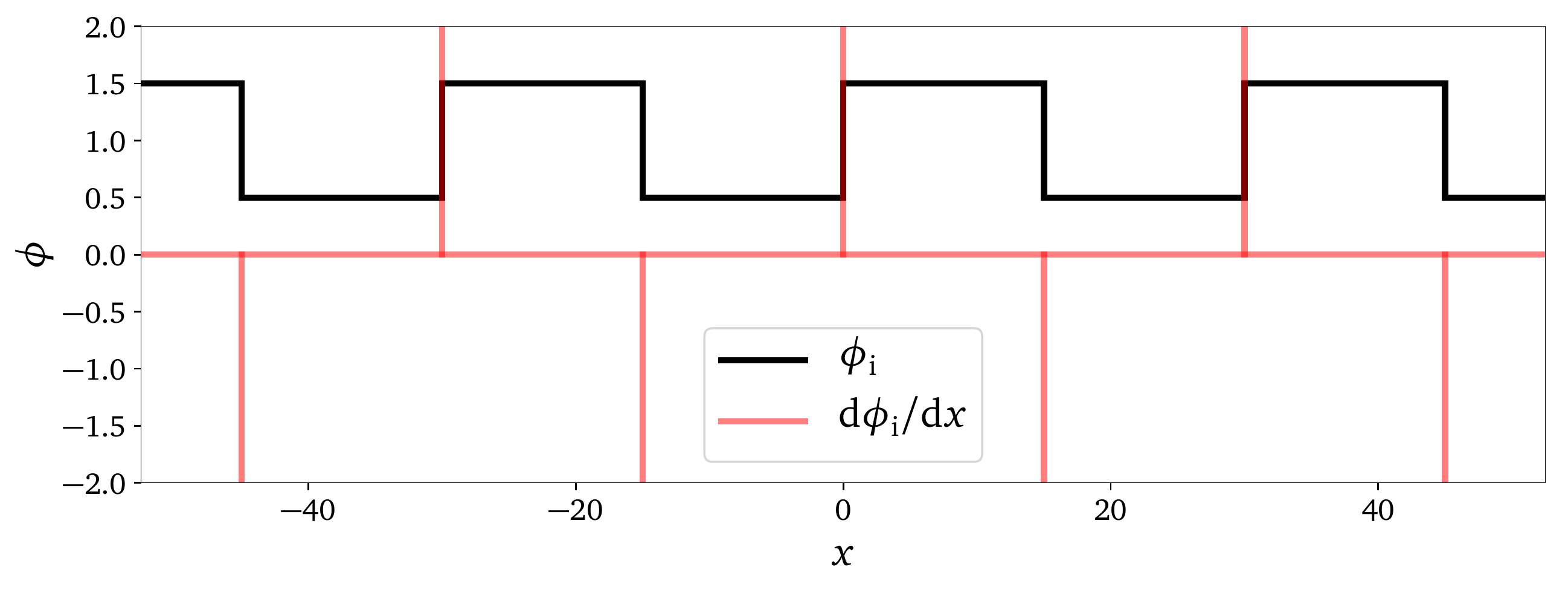}
\caption{Idealized phase profile $\phi_\mathrm{i}$ (black) and the derivative of $\phi_\mathrm{i}$ (red).}
\label{picture fringer ideal phase profile}
\end{figure}

For the moment we will restrict our view to a period $2$ binary grating. The phase response can be described approximately by a convolution of the ideal phase profile $\phi_\mathrm{i}$ with a kernel $k$ of gaussian \cite{PerssonEngstroemGoksoer2012,Haellstig2004} or exponential \cite{EfronApterBahat-Treidel2004} shape. 

\begin{align}\label{equation phi_r = phi_i * k}
\phi_\mathrm{r}(x) = (\phi_\mathrm{i} * k)(x)
\end{align} 
The ideal phase profile $\phi_\mathrm{i}$ (\cref{picture fringer ideal phase profile}) represents a step-like function with
\begin{align}
\dfrac{\mathrm{d} \phi_\mathrm{i}}{\mathrm{d} x}  = \sum_{j=-\infty}^\infty (p_2 - p_1) (-1)^j \delta(x-jx_\mathrm{pix}),
\end{align} 
where $x_\mathrm{pix}$ denotes the pixel pitch.

We observe that it is not possible to generate asymmetric diffraction efficiency curves with a linear convolution, e.g. \ref{equation phi_r = phi_i * k}.
Moreover, simulations of the phase response also show a nonlinear behavior in $p_1$ and $p_2$. The idea is therefore to make $k(x)$ depend on ($p_1,p_2$). This will be realized by introducing $4$ parameters, which depend on $(p_1,p_2)$. The goal then is to fit the phase profiles of the simulations, yielding a set of parameters for every $(p_1,p_2)$. To model the asymmetry properly, we will use two kernels, $k_1$ and $k_2$ to fit the simulations. Using the relation for differentiation for the convolution

\begin{align}
\dfrac{\mathrm{d}}{\mathrm{d}x} (f*g)(x) = \int_{-\infty}^\infty g(t) \dfrac{\partial}{\partial x} f(x-t) \mathrm{d}t
\end{align}
we can write the derivative of  \cref{equation phi_r = phi_i * k}
\begin{align}\label{equation d phase kernel faltung}
\begin{split}
\dfrac{\mathrm{d}}{\mathrm{d}x}\phi_\mathrm{r}(x) &= ((\dfrac{\mathrm{d}}{\mathrm{d}x}\phi_\mathrm{i}) * k)(x) \\
&= \Big( \Big. \sum_{j=-\infty}^{\infty} \Big[ \Big. ( p_2-p_1) \delta(t-j\mathrm{x_{pix}}) \\
&+ (p_1-p_2) \delta(t-(j+1)\mathrm{x_{pix}})\Big. \Big] * k(t)\Big. \Big)(x)  \\
&=  \sum_{j=-\infty}^{\infty} \Big[ \Big. ( p_2-p_1) \Big( \Big. \delta(t-j\mathrm{x_{pix}}) *  k(t) \Big. \Big)(x) + (p_1-p_2) \Big( \Big. \delta(t-(j+1)\mathrm{x_{pix}}) * k(t)\Big. \Big)(x)  \Big. \Big] \\
\end{split}.
\end{align}

Now we take  \cref{equation d phase kernel faltung} and write it with two separate kernels

\begin{align}
\begin{split}
(D\phi_\mathrm{r})(x) &=  \sum_{j=-\infty}^{\infty}  ( p_2-p_1) \left[ \underbrace{\Big( \Big. k_1(t) * \delta(t-j\mathrm{x_{pix}}) \Big. \Big)(x)}_{\substack{k_1(x-j\mathrm{x_{pix}})}} - \underbrace{\Big( \Big. k_2(t) * \delta(t-(j+1)\mathrm{x_{pix}})\Big. \Big) (x)}_{\substack{k_2(x-(j+1)\mathrm{x_{pix}})}}\right]\\
\end{split}
\end{align} 
and with $K_{1,2} = \int k_{1,2}(x) \mathrm{d}x$ we get

\begin{align}\label{equation approximation real phase profile herleitung}
\begin{split}
\phi_\mathrm{r}(x) = \int \sum_{j=-\infty}^{\infty} (p_2-p_1) \Big[ \Big. k_1 ( x - j\mathrm{x_{pix}}) - k_2 (x - (j+1) \mathrm{x_{pix}}) \Big. \Big] \mathrm{d}x \\
\approx \sum_{j=-m}^{m} (p_1-p_2) \Big[ K_1(x-j\mathrm{x_{pix}}) - K_2 (x-(j+1)\mathrm{x_{pix}})\Big] + C.
\end{split}
\end{align}

We can justify the approximation of the sum in  \cref{equation approximation real phase profile herleitung} by choosing a kernel with vanishing contribution if shifted by more than $m\mathrm{x_{pix}}$. Physically, this approximation means that for every pixel only the surrounding $m$ pixels influence the phase profile significantly. Simulations show, that almost all the contributions of the fringing field effect reside in adjacent pixels. Therefore, a value of $m=2$ (or $m=3$) suffices. 


\FloatBarrier
\section{Construction of the fit-function}

To model the simulated phase profiles, we choose an asymmetric kernel $k$ depending on parameters $[ \mathrm{x_0}, \mathrm{c_p},\mathrm{c_m},\mathrm{n}]$

\begin{align}
k(x)_{[\mathrm{x_0}, \mathrm{c_p},\mathrm{c_m},\mathrm{n}]}=
\begin{cases}
\mathrm{N} \exp(-|(x-\mathrm{x_0})/\mathrm{c_m}|^{\mathrm{n_m}}) & x < \mathrm{x_0}, \\
\mathrm{N} \exp(-|(x-\mathrm{x_0})/\mathrm{c_p}|^{\mathrm{n_p}}) & x \geq \mathrm{x_0},
\end{cases}
\end{align}
with $\mathrm{N} = \frac{2}{\mathrm{c_p} + \mathrm{c_m}}$. Similar kernels have been used by \cite{EfronApterBahat-Treidel2004,PerssonEngstroemGoksoer2012} to describe the fringing field effect by linear convolution. This kernel was chosen by combining the generalized Gaussian kernel in \cite{PerssonEngstroemGoksoer2012} and the (asymmetric) exponential kernel in \cite{EfronApterBahat-Treidel2004} in $1$D. 
With the integral

\begin{align}
E_\mathrm{n}(x) = \int_{-\infty}^x \exp(-|t|^\mathrm{n}) \mathrm{d}t 
\end{align}
the integrated kernel $K$ can then be written  

\begin{align}
\begin{split}
K_{[ \mathrm{x_0}, \mathrm{c_p},\mathrm{c_m},\mathrm{n}]}(x) &= \int_{-\infty}^x k_{[ \mathrm{x_0}, \mathrm{c_p},\mathrm{c_m},\mathrm{n}]}(x') \mathrm{d}x' \\
&=\begin{cases}
\mathrm{N} \mathrm{c_m} \dfrac{E_\mathrm{n}((x-\mathrm{x_0})/\mathrm{c_m})}{E_\mathrm{n}(\infty)} & x < \mathrm{x_0} \\
\mathrm{N} \left[(\mathrm{c_m} - \mathrm{c_p}) 0.5 + \mathrm{c_p} \dfrac{E_\mathrm{n}((x-\mathrm{x_0})/\mathrm{c_p})}{E_\mathrm{n}(\infty)}\right] & x \geq \mathrm{x_0}.
\end{cases}
\end{split}
\end{align}

\begin{figure}[h!]
\centering
\includegraphics[width=12cm]{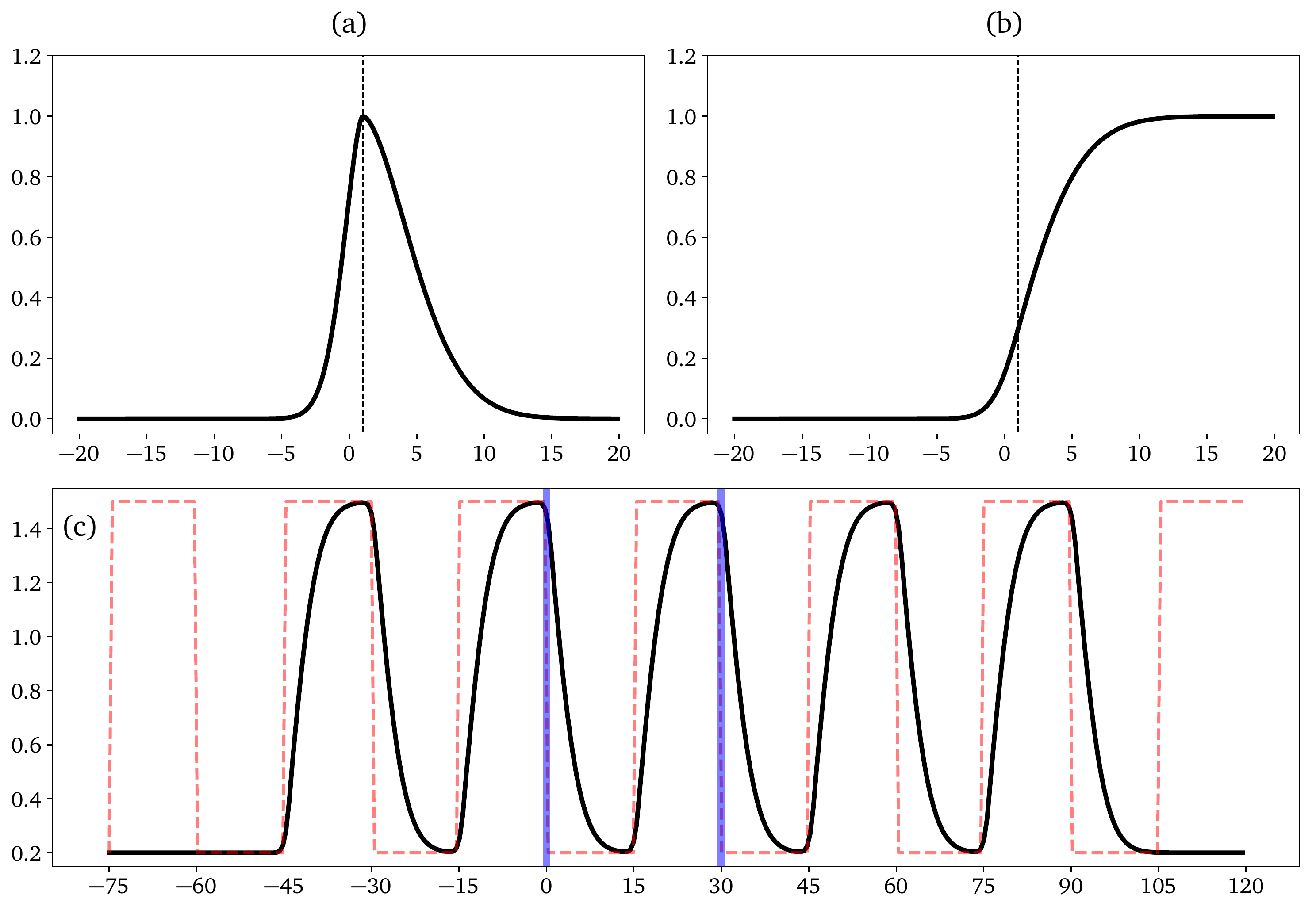}
\caption{Kernel $k$ (a), integrated kernel $K$ (b) and function for the phase profile $s_j$ (c).}
\label{picture kernel intkernel}
\end{figure}

The function $s_j$ to describe the simulated phase profile of binary gratings then is

\begin{align}\label{equation fitfunction}
\begin{split}
s_j(x) = \sum_{i=-2}^{2} (p_2-p_1) \Big[ \Big. & (-1)^i K_{[ \mathrm{x_{0_2}}+(2i+1)x_j/2 , \mathrm{c_{p_2}} , \mathrm{c_{m_2}}, \mathrm{n_2}]} (x) \\ &+ (-1)^{i+1} K_{[ \mathrm{x_{0_1}}+(2i+2)x_j/2 , \mathrm{c_{p_1}} , \mathrm{c_{m_1}}, \mathrm{n_1}]}(x) \Big. \Big] + p_1.
\end{split}
\end{align}

 \cref{picture kernel intkernel} (a) depicts typical examples for the kernel, (b) the integrated kernel and (c) the fit function $s_j$ (c) that match the simulations, where the blue lines isolate the relevant profile over two pixels. The fit function $s_j$ depends on a total of $8$ parameters for two kernels. 

\begin{figure}[h!]
\centering
\includegraphics[width=12cm]{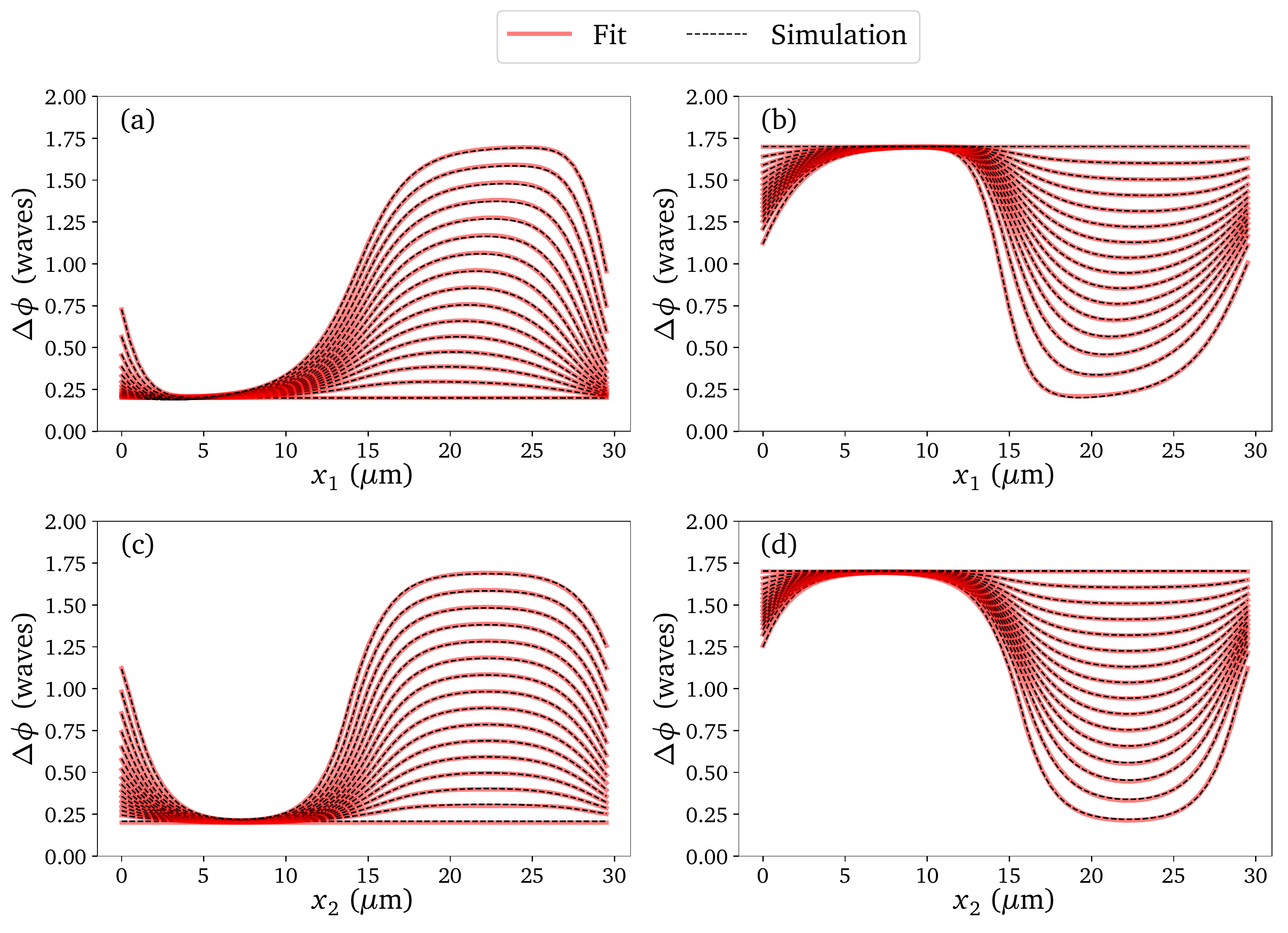}
\caption{Fit for a binary grating in the asymmetric direction (a,b) and symmetric direction (c,d).}
\label{picture compare fit with simulations}
\end{figure}

 \cref{picture compare fit with simulations} shows the fits (red) and simulations (black) for binary gratings in the asymmetric (a),(b) and symmetric (c),(d) direction. We see, that for several phase differences of the binary grating the chosen kernel with this amount of degrees of freedom is able to fit the simulations very well. 

At this point we emphasize again that our goal is to find a model for the kernel that describes the phase profiles of the simulations good enough to calculate an area over several hundred pixels. In order to describe the phase profiles over the whole phase-range of the SLM ($\sim 0-1.5$ waves), we found that $8$ parameters for the asymmetric and $4$ parameters for the symmetric direction are necessary. A kernel with $6$ and $3$ parameters, resp., was also implemented, which showed deviations at high and low phase values. 

We fit the profile function \cref{equation fitfunction} to simulations and determine $8$ parameters for each set of phase values ($p_1,p_2$) of the binary gratings on a grid of $31\times 31$ values.

\begin{figure}[h!]
\centering
\includegraphics[width=12cm]{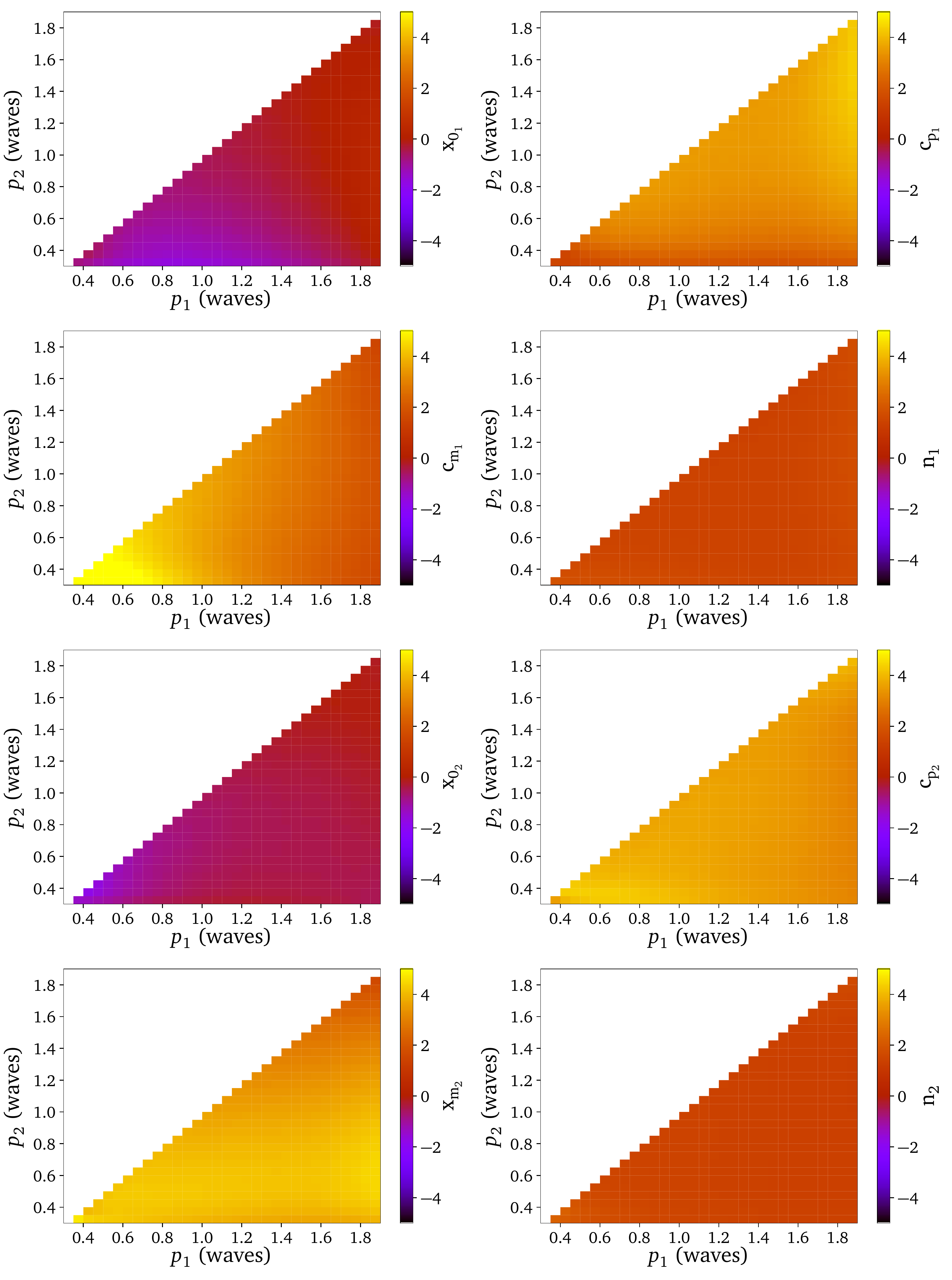}
\caption{Fit parameters for a binary grating in the asymmetric direction.}
\label{picture fit params x1}
\end{figure}

\begin{figure}[h!]
\centering
\includegraphics[width=12cm]{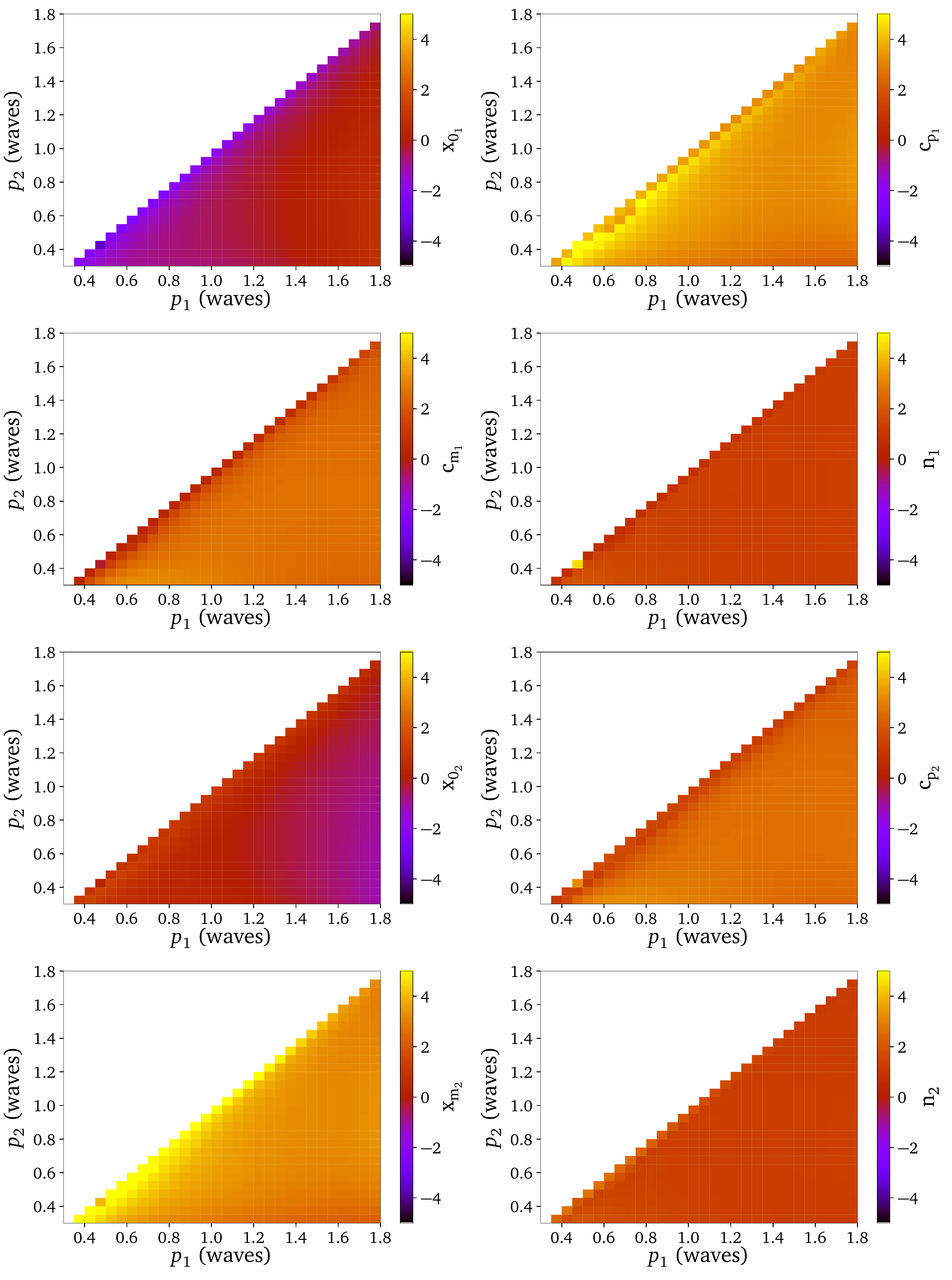}
\caption{Fit parameters for a binary grating in the symmetric direction.}
\label{picture fit params x2}
\end{figure}

\begin{figure}[h!]
\centering
\includegraphics[width=12cm]{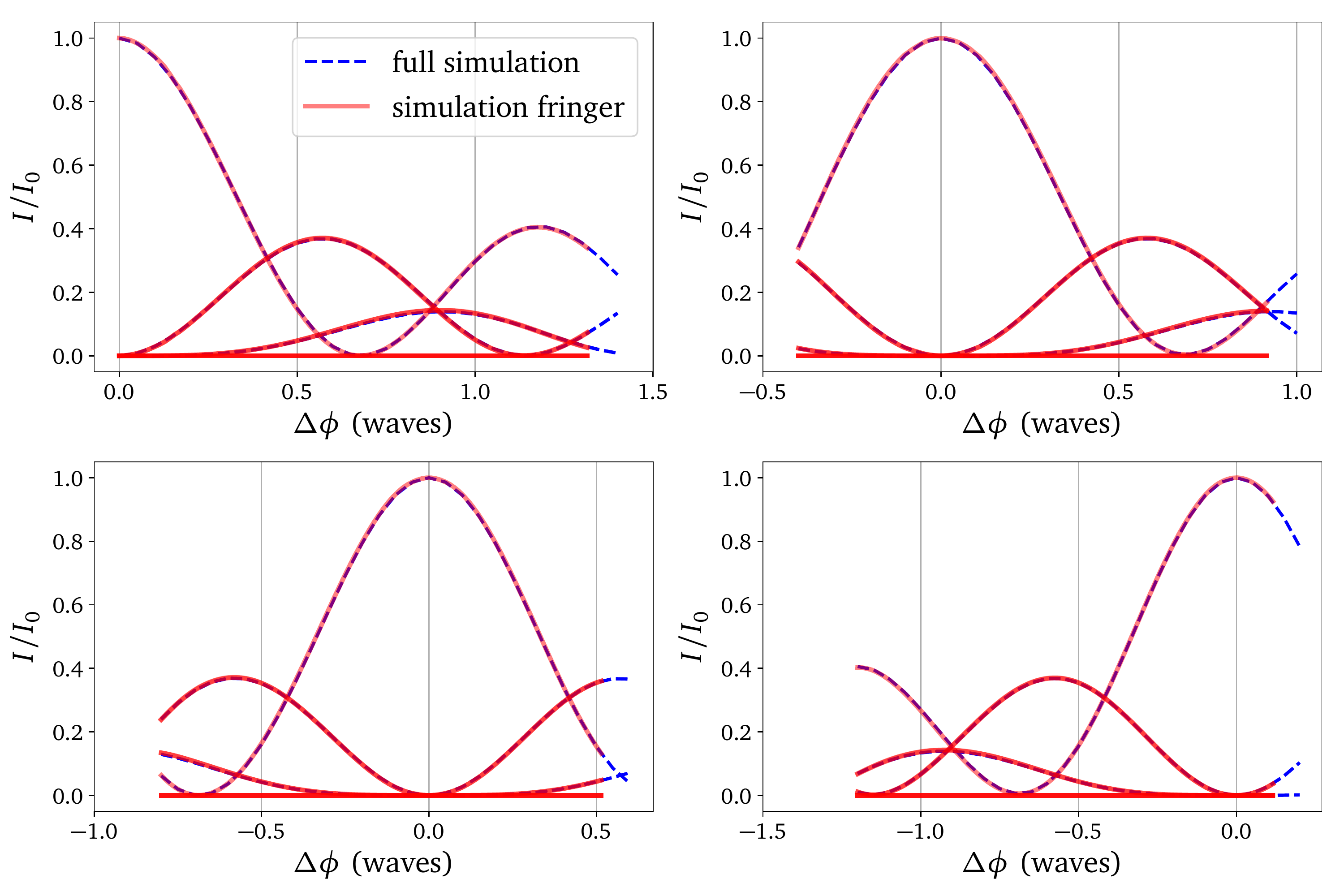}
\caption{Diffraction efficiency of a binary grating (period $2$) in the symmetric direction done by full simulation of the LC-directors and simulation done by fits for reference phases $\phi_\mathrm{ref}=0$ (top left), $\phi_\mathrm{ref}=0.4$ (top right), $\phi_\mathrm{ref}=0.8$ (bottom left), $\phi_\mathrm{ref}=1.2$ waves (bottom right).}
\label{picture compare fit with simulations de y}
\end{figure}

In  \cref{picture fit params x1} and \cref{picture fit params x2} parameters are shown for the asymmetric and symmetric direction, respectively. The graphs of the parameters in dependence of the phase values are mostly smoothly shaped, thus, it is justified to interpolate linearly between grid points. For phase values close to the diagonal ($p_1 \approx p_2$) in the symmetric direction (\cref{picture compare fit with simulations de y} some parameters show isolated jumps. These values can lead to erroneous parameters estimates when interpolating. However, in this case the pre-factor $(p_2 - p_1)$ in  \cref{equation fitfunction} is very small and the effect on the calculation of the phase profiles is negligible.

The fit parameters were calculated on a triangular grid, because for a change $(p_1,p_2)\rightarrow (p_2,p_1)$ also the m and p parameter values exchange.

\begin{figure}[h!]
\centering
\includegraphics[width=12cm]{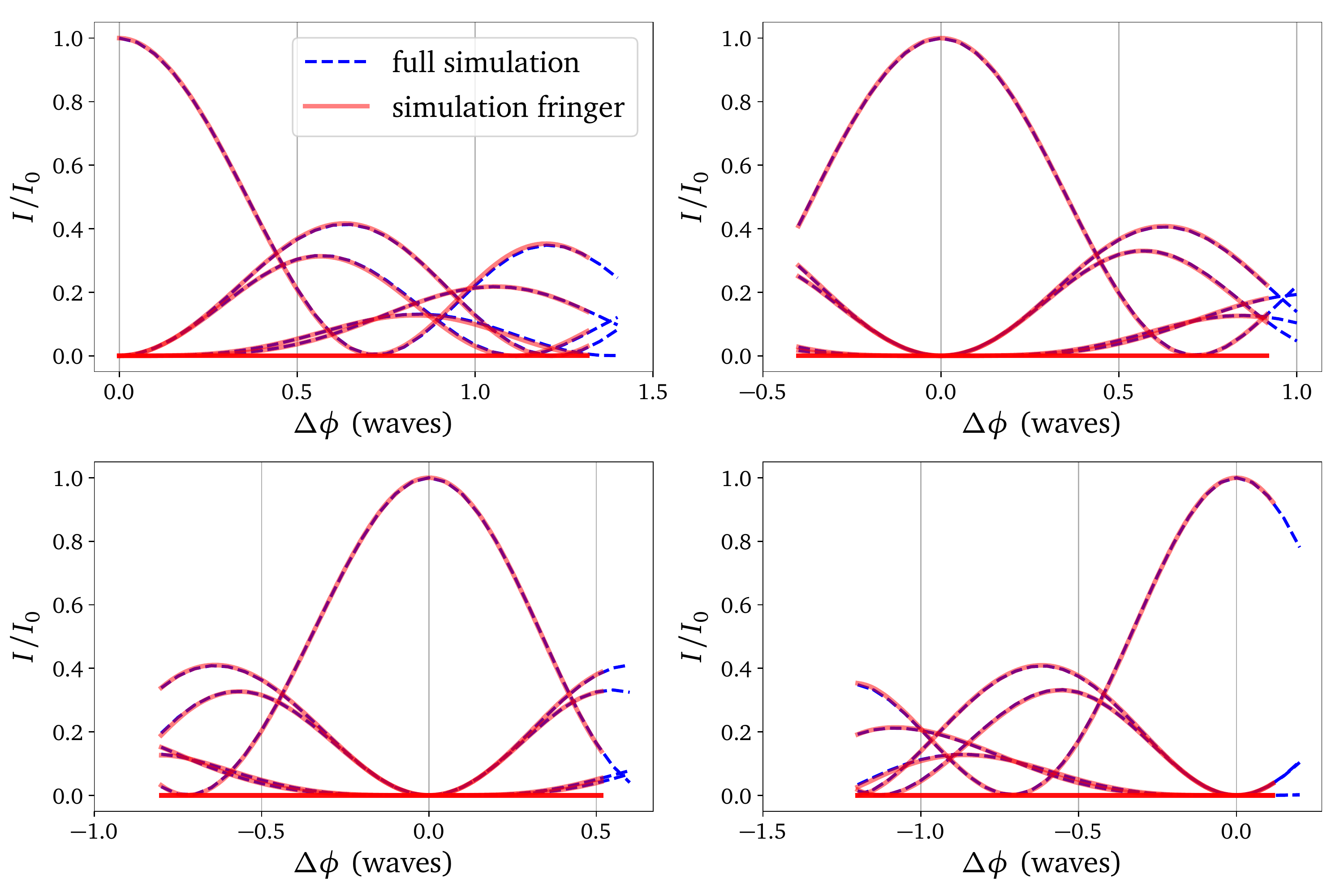}
\caption{Diffraction efficiency of a binary grating (period $2$) in the asymmetric direction done by full simulation of the LC-directors and simulation done by fits for reference phases $\phi_\mathrm{ref}=0$ (top left), $\phi_\mathrm{ref}=0.4$ (top right), $\phi_\mathrm{ref}=0.8$ (bottom left), $\phi_\mathrm{ref}=1.2$ waves (bottom right).}
\label{picture compare fit with simulations de x}
\end{figure}

To test our fit model, we compare the diffraction efficiency of binary gratings based either on simulating the fringing with fit-functions with interpolated parameters or based on a full simulation as shown in  \cref{picture compare fit with simulations de x} and \cref{picture compare fit with simulations de y}. Clearly, the fit-functions are also able to reproduce the desired diffraction efficiency curves of the full simulations for binary gratings. 

\FloatBarrier
\section{Fast 2D fringing model}\label{sec:fast 2d fringing model}

\begin{figure}[h!]
\centering
\includegraphics[width=10cm]{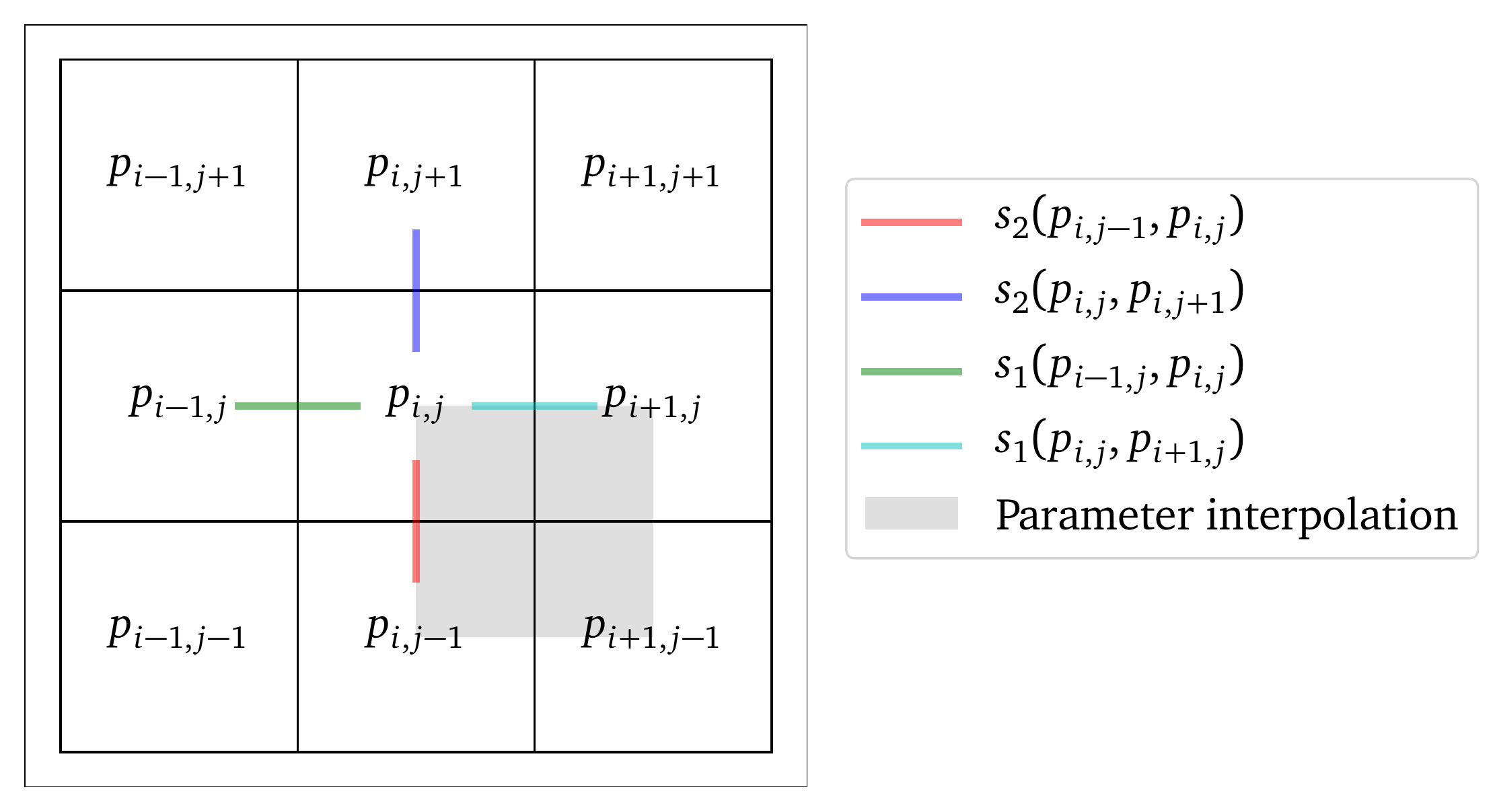}
\caption{Conceptual method, how to construct a $2$D model from $1$D transition curves for a single pixel.}
\label{picture fringer explanation pixels}
\end{figure}

Next, we want to formulate a model to describe the $2$D phase profiles of an arbitrary voltage pattern using our fit model with corresponding parameters in the symmetric and asymmetric direction. However, this poses a challenge since we cannot simply superimpose the phase profiles of the gratings in symmetric and asymmetric direction. This, in turn, has its cause in the nonlinear behavior of the LCs themselves.  \cref{picture fringer explanation pixels} presents the concept on which this $2$D model is drafted. We now assume, that the profile at the transition between two pixels with phase values $p_1$ and $p_2$ can be approximated by our integrated kernel $K$ with corresponding parameters. 

For an array of pixels with values $p(i,j)$ at discrete pixel coordinated $(i,j)$ we now define a continuous phase function $P(x_1,x_2)$ which describes the phase response of the SLM over the area of one pixel

\begin{align}\label{equation fringer pixel formula}
\begin{split}
P(x_1,x_2) &= p_{i,j} \\
      &+ (p_{i+1,j} - p_{i,j}) \cdot K_{\mathrm{a},(i,j)\rightarrow (i+1,j)}(x_1) \\
      &+ (p_{i,j} - p_{i-1,j}) \cdot K_{\mathrm{a},(i-1,j)\rightarrow (i,j)}(x_1) \\
      &+ (p_{i,j+1} - p_{i,j}) \cdot K_{\mathrm{s},(i,j)\rightarrow (i,j+1)}(x_2) \\
      &+ (p_{i,j} - p_{i,j-1}) \cdot K_{\mathrm{s},(i,j-1)\rightarrow (i,j)}(x_2) \\
      &+ (p_{i+1,j+1} - p_{i+1,j} - p_{i,j+1} + p_{i,j}) \cdot K_{\mathrm{a},(i,j)\rightarrow (i+1,j)}(x_1) \cdot  K_{\mathrm{s},(i,j)\rightarrow (i,j+1)}(x_2) \\
      &+ (p_{i+1,j} - p_{i+1,j-1} - p_{i,j} + p_{i,j-1}) \cdot  K_{\mathrm{a},(i,j)\rightarrow (i+1,j)}(x_1) \cdot K_{\mathrm{s},(i,j-1)\rightarrow (i,j)}(x_2) \\
      &+ (p_{i,j+1} - p_{i-1,j+1} - p_{i,j} + p_{i-1,j}) \cdot  K_{\mathrm{a},(i-1,j)\rightarrow (i,j)}(x_1) \cdot K_{\mathrm{s},(i,j)\rightarrow (i,j+1)}(x_2) \\
      &+ (p_{i,j} - p_{i-1,j} - p_{i,j-1} + p_{i-1,j-1}) \cdot  K_{\mathrm{a},(i-1,j)\rightarrow (i,j)}(x_1) \cdot K_{\mathrm{s},(i,j-1)\rightarrow (i,j)}(x_2) 
\end{split}
\end{align}
where $K_\mathrm{a}$ and $K_\mathrm{s}$ describe the integrated kernels for parameters in the asymmetric and symmetric direction and

\begin{align}
K_{\mathrm{m},(i,j)\rightarrow (k,l)}(x) := K_{[ \mathrm{x_0}(i,j\rightarrow k,l), \mathrm{c_p}(i,j\rightarrow k,l),\mathrm{c_m}(i,j\rightarrow k,l),\mathrm{n}(i,j\rightarrow k,l)]}(x)\qquad m=(a,s)
\end{align}
defines the integrated kernel dependent on parameters $\mathrm{x_0}$, $\mathrm{c_p}$, $\mathrm{c_m}$ and $\mathrm{n}$ which in turn depend on phase values $p_{i,j}$ and $p_{k,l}$. The first $5$ terms in  \cref{equation fringer pixel formula} include the transitions along the center of a pixel in the symmetric and asymmetric direction, while the last $4$ summands take the influence of the adjacent diagonal pixels into account. In the numerical implementation of this model, the parameters were also interpolated linearly from the center to the edge of a pixel. So the parameters of the transition $(i,j)\rightarrow(i+1,j)$ were mixed with the parameters of the adjacent transition $(i,j+1)\rightarrow(i+1,j+1)$ from the center to the upper part of the pixel (see grey sector in  \cref{picture fringer explanation pixels}. In the lower part of the pixel the transition parameters $(i,j)\rightarrow(i+1,j)$ were mixed with $(i,j-1)\rightarrow(i+1,j-1)$. The profiles of the transitions in the symmetric and asymmetric direction were treated equally. In this mixing process, the parameters are interpolated linearly with $p = w p_1 + (1-w) p_2$ with the weight $w(t) = 1.5(t-0.5)$, where $t$ is the $x_1$ or $x_2$ and assumes values $t\in [0,1]$.

\begin{figure}[h!]
\centering
\includegraphics[width=12cm]{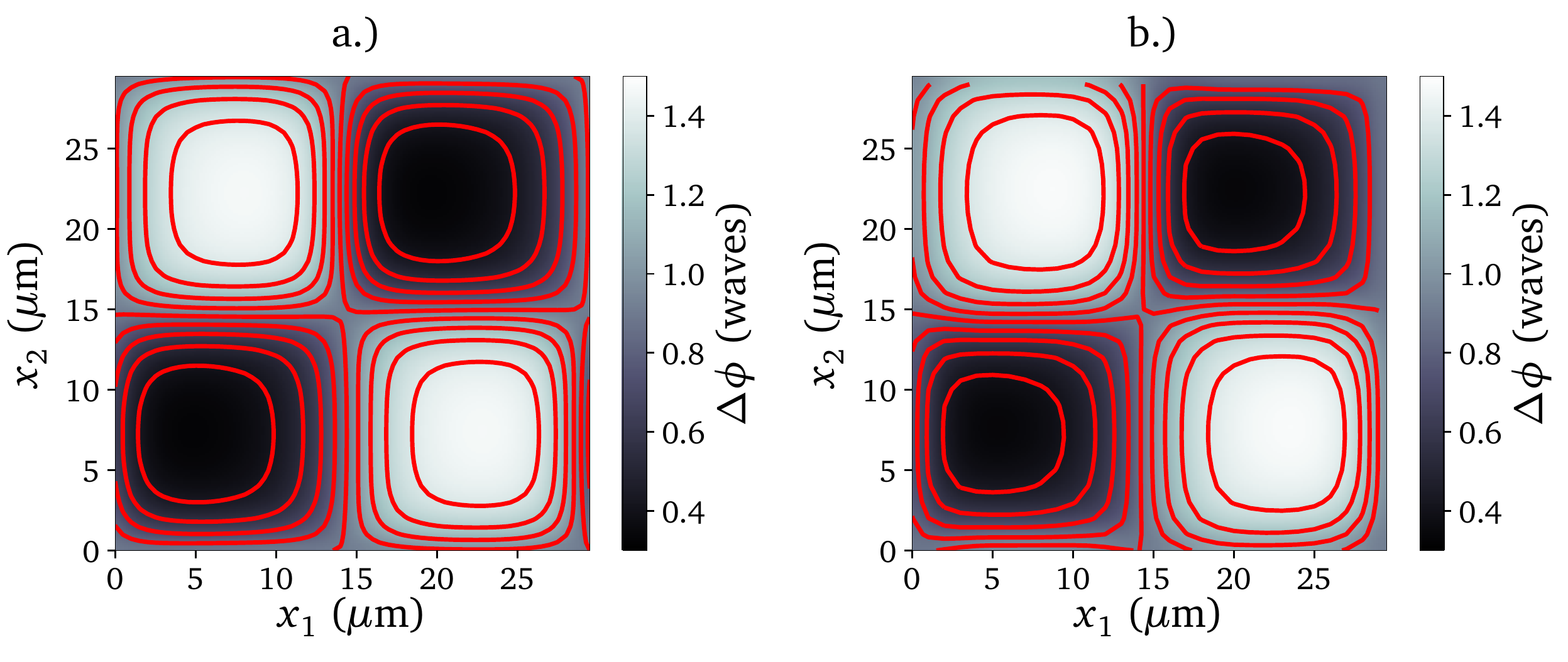}
\caption{Phase profile of a checkerboard pattern by full simulation (a) and the fast $2$D model (b).}
\label{picture compare phase profile fringer 3d}
\end{figure}

To test this $2$D model for the phase profile we compared it to a full simulation of a checkerboard pattern.  \cref{picture compare phase profile fringer 3d} shows the comparison of the phase profiles with corresponding contour line values where (a) represents the phase profile of the fast model and (b) depicts the phase profile of the full simulation. We see differences mainly at the lowest contour lines and at the center of the $4$ pixels, where the fast model shows contorted contour lines in comparison to the full simulation.

\begin{figure}[h!]
\centering
\includegraphics[width=12cm]{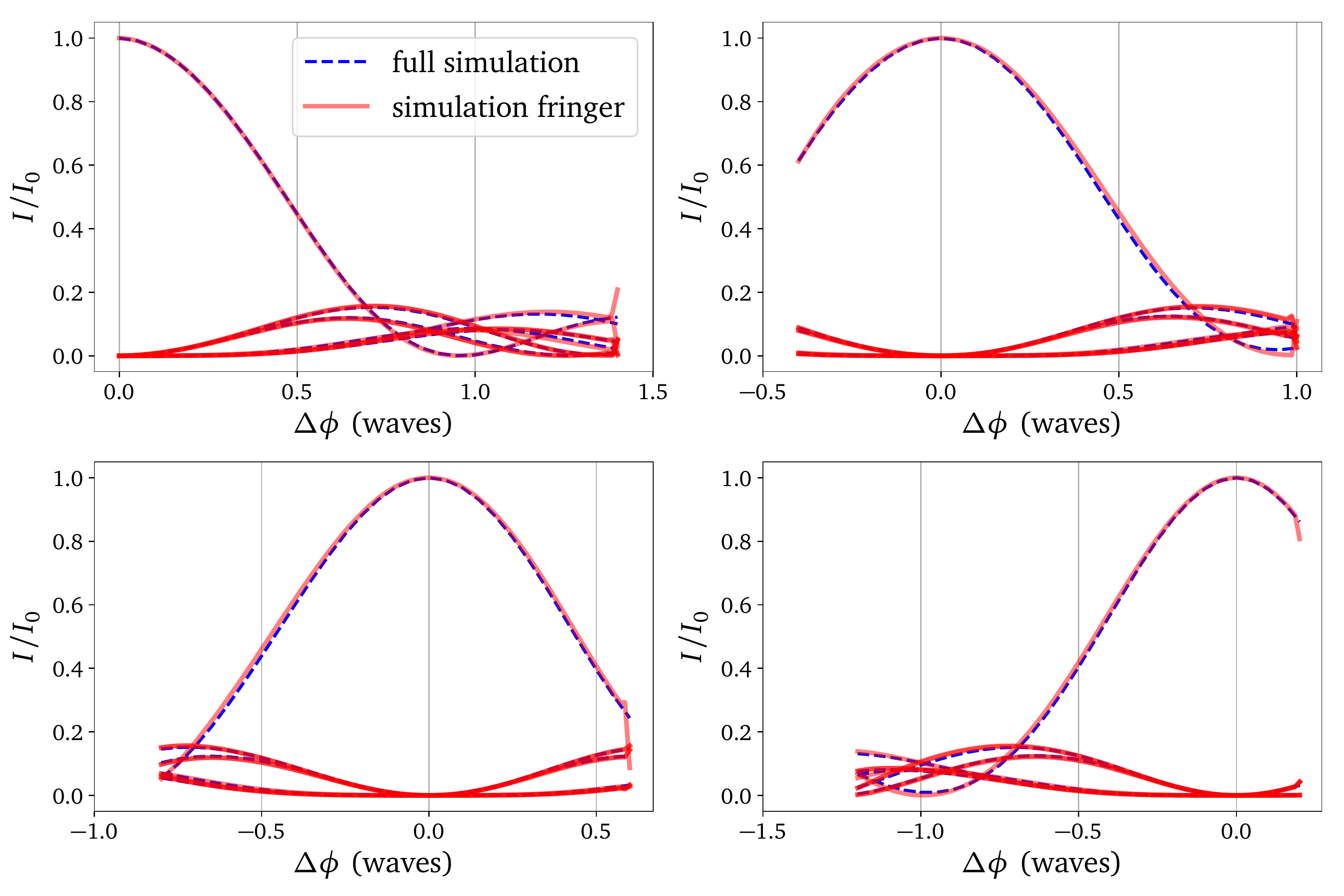}
\caption{Diffraction efficiency simulation of a checkerboard pattern done by the fast $2$D model and full simulation for reference phases $\phi_\mathrm{ref}=0$ (top left), $\phi_\mathrm{ref}=0.4$ (top right), $\phi_\mathrm{ref}=0.8$ (bottom left), $\phi_\mathrm{ref}=1.2$ waves (bottom right).}
\label{picture compare fringer lcdir checkerboard de}
\end{figure}

Another way of testing, and the more relevant one, is the comparison of the diffraction efficiencies for the full $3$D model and the fast fringing model for a checkerboard pattern, shown in \cref{picture compare fringer lcdir checkerboard de} for different reference phase values. The fast fringing model agrees very well with the full $3$D simulations for all phase shifts and reference phase values.

\begin{figure}[h!]
\centering
\includegraphics[width=6cm]{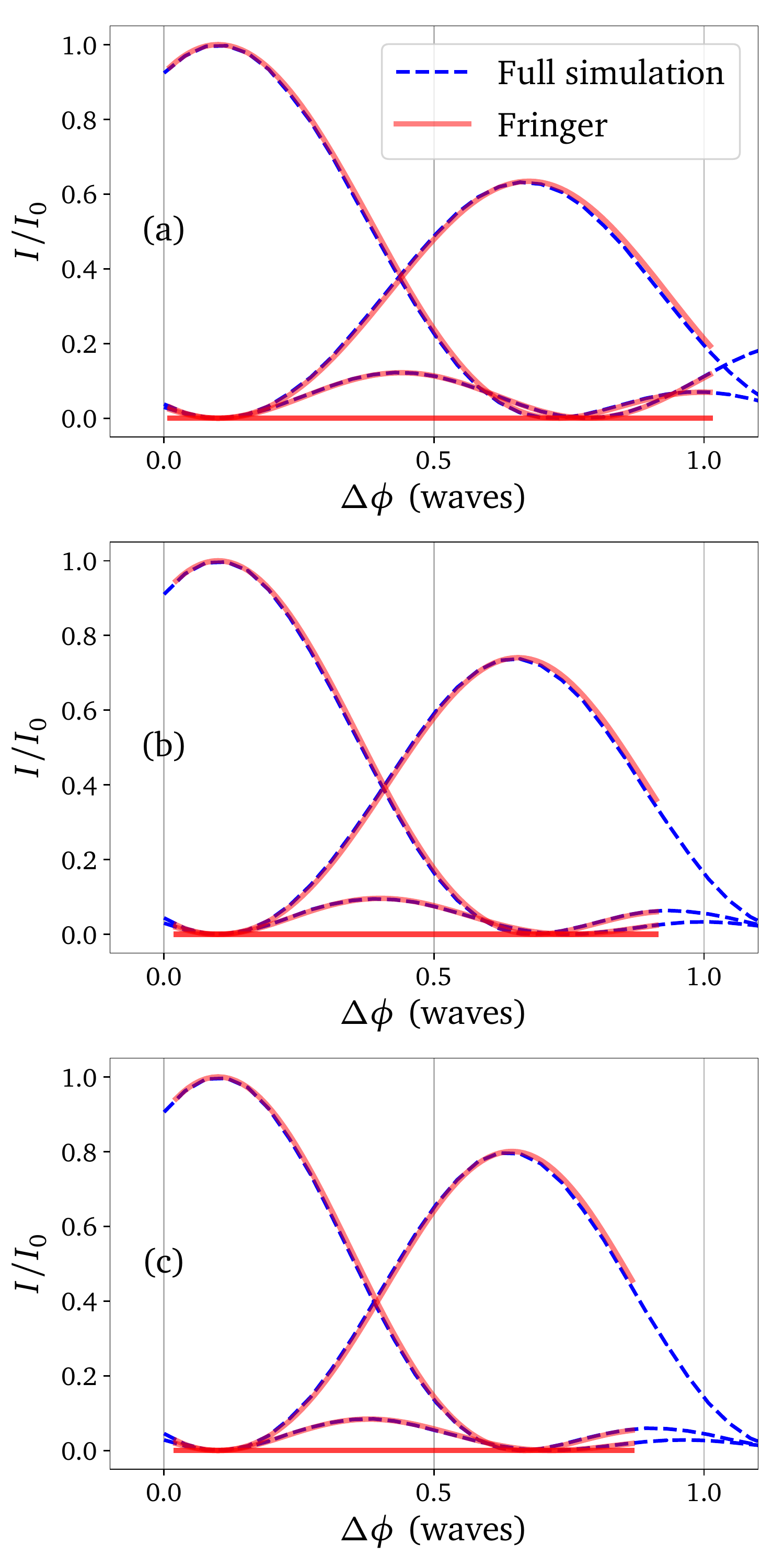}
\caption{Simulations done by the fast $2$D model (red) and full simulations (blue) for blazed gratings, a.) period $3$ vertical, b.) period $4$ vertical, c.) period $5$ vertical.}
\label{picture compare blazed lcdir}
\end{figure}

In  \cref{picture compare blazed lcdir} we see a comparison between simulated diffraction efficiencies for a blazed grating with the fast $2$D model (Fringer) and full simulation. 

\begin{figure}[h!]
\centering
\includegraphics[width=12cm]{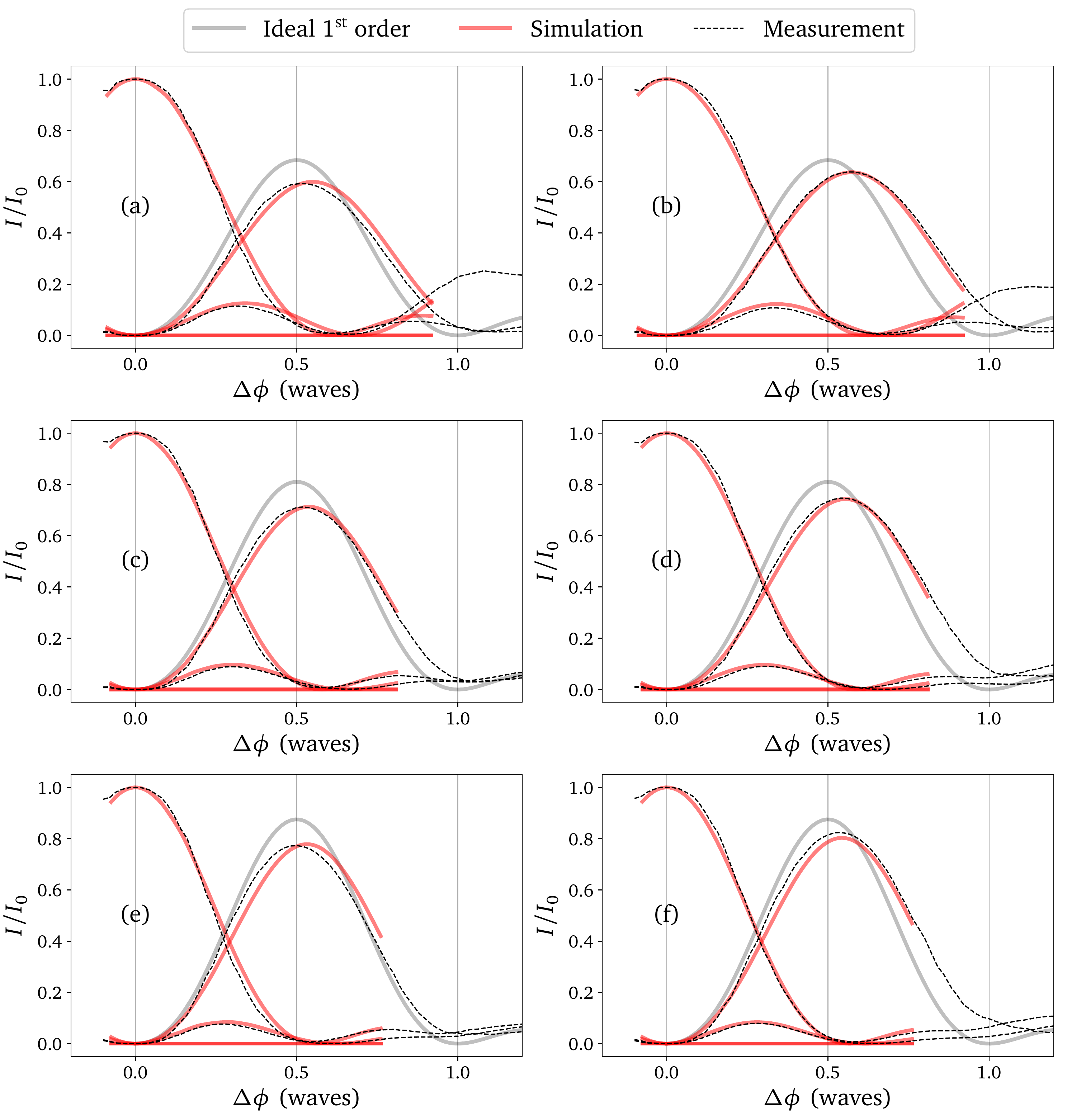}
\caption{Simulation (Fringer) and measurements for blazed gratings. (a) period $3$ symmetric, (b) period $3$ asymmetric, (c) period $4$ symmetric, (d) period $4$ asymmetric, (e) period $5$ symmetric, (f) period $5$ asymmetric.}
\label{picture compare blazed fringer}
\end{figure}

 \cref{picture compare blazed fringer} shows simulations and measurements for blazed gratings in the symmetric (a,c,e) and asymmetric direction (b,d,f). (a,b) have grating period $3$, (c,d) have period $4$ and (e,f) have period $5$. 

In all cases we observe a very good agreement between full simulations and the fast $2$D fringing model. As a huge improvement it allows us to calculate the effect of fringing much faster. Using a GPU it only takes a few ms for a $512\times 512$ SLM pattern.

\FloatBarrier

\section{Compensation of pixel crosstalk}

In this section we will use the fast $2$D model to calculate a phase pattern designed to create a test pattern, a regular spot pattern in the far field, where we compensate the detrimental effects of fringing on the spot uniformity. Specifically, we consider a $15 \times 19$ rectangular spot pattern, where the spots at the edge map to a binary grating of period $2$. This pattern was chosen as a test pattern, since in this configuration the spot uniformity is very sensitive to the fringing field effect \cite{Leonardo2007}. We will therefore test our fast $2$D model on such a spot pattern by evaluating the spot intensity modulations.

Upon considering such a spot pattern we meet a limitation regarding efficiency, since the diffraction efficiency of the SLM depends strongly on the spatial frequency. This means that we have to sacrifice light efficiency to gain a uniform spot pattern and vice versa.


Now we want to take the effects of the fringing field into consideration in the calculation of the phase pattern by implementing our fast $2$D model in the phase retrieval algorithms, namely a weighted Gerchberg Saxton (wGS) and a Nesterov accelerated gradient descent (Nagd) algorithm. We will start from a random phase pattern and use a weighted Gerchberg Saxton algorithm (without considering the fringing field effect) to find a starting value for further optimization. We then use the resulting phase pattern and feed it into a Nesterov accelerated gradient descent algorithm, where we minimize the mean square difference of the simulated spot intensities (now including fringing) from the target value. We decided to optimize our phase pattern with respect to the light efficiency, therefore, we choose target spot intensities in the shape of a $\mathrm{sinc} (\xi_1) \cdot \mathrm{sinc} (\xi_2)$ function, where $\xi_1$ and $\xi_2$ represent the coordinates in the Fourier plane. This target intensity profile corresponds roughly to the maximum diffraction efficiency ($1^\mathrm{st}$ order) of blazed gratings as shown in \cref{ssec:comparison simulation experiment binary gratings,sec:Blazed gratings}. This target intensity profile was chosen to maximize the diffraction efficiency of the SLM. All target spot intensities are additionally reduced by $\sim 20\%$ to ensure that the SLM is able to reach the desired diffraction efficiency. The error metric of our gradient descent algorithm also restricts the phase values to $0.2-1.5\cdot$ waves by penalizing values outside the interval. This is done to not exceed the phase range of our SLM. 

 

\begin{figure}[h!]
\centering
\includegraphics[width=12cm]{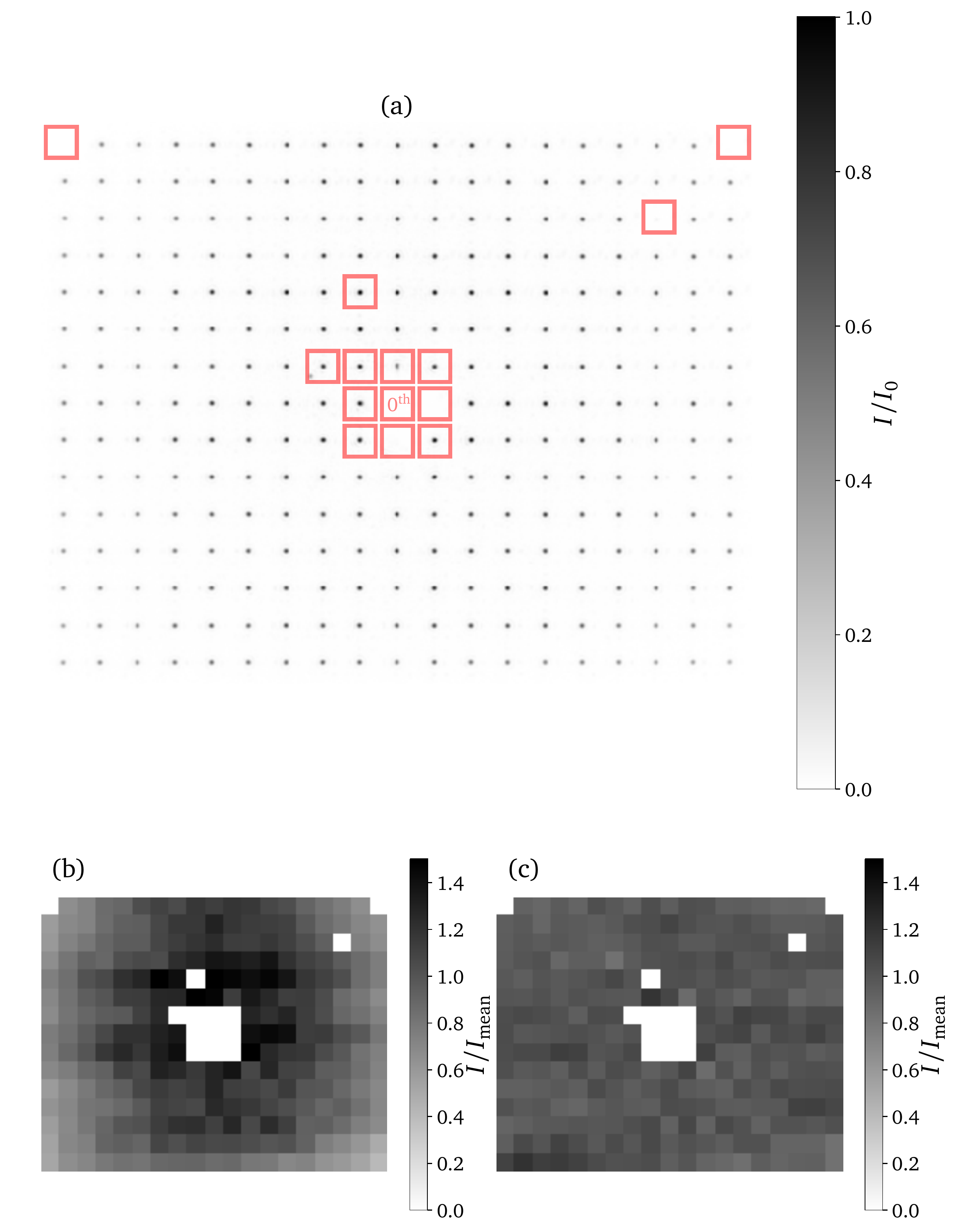}
\caption{Raw measurement data of a $15\times 19$ spot pattern (with fringing compensation) (a) (the red squares represent spots, which have been excluded), spot intensities (b) and corrected array of spot intensities (c).}
\label{picture spot uniformity measurement and correction}
\end{figure}

\cref{picture spot uniformity measurement and correction} (a) shows the measurement of a $15\times 19$ spot pattern (with some missing spots), which stems from a phase pattern calculated considering the effects of fringing. The $0^\mathrm{th}$ order was blocked during the measurement, which masks a few additional spots at the center. We see that the spot pattern has the shape of the product of two sinc-functions, as discussed above. The spots in the red squares have been excluded in the following evaluation. The spots at the center were excluded since the blockage of the $0^\mathrm{th}$ order also affected surrounding spots in the measurement process. An additional spot in this central region was excluded since it overlapped with a back reflection spot. Other empty spot locations were omitted by purpose in the target test pattern.
 
In \cref{picture spot uniformity measurement and correction} (b) we see the first step of the evaluation of the measurement in (a). The red squares represent excluded spots.  The squares in \cref{picture spot uniformity measurement and correction} (b) correspond to the sum of a square region around a spot in (a). We can clearly see in \cref{picture spot uniformity measurement and correction} (b) that spots at the center have more intensity than spots at the edges. 

\cref{picture spot uniformity measurement and correction} (c) depicts the second step in the evaluation. Here, we divide the spot rows and columns through the respective mean of the rows and columns to get rid of slowly varying interference fringes, which are visible in (b). This also removes the difference in the diffraction efficiency between center and border from the data. 

\begin{figure}[h!]
\centering
\includegraphics[width=12cm]{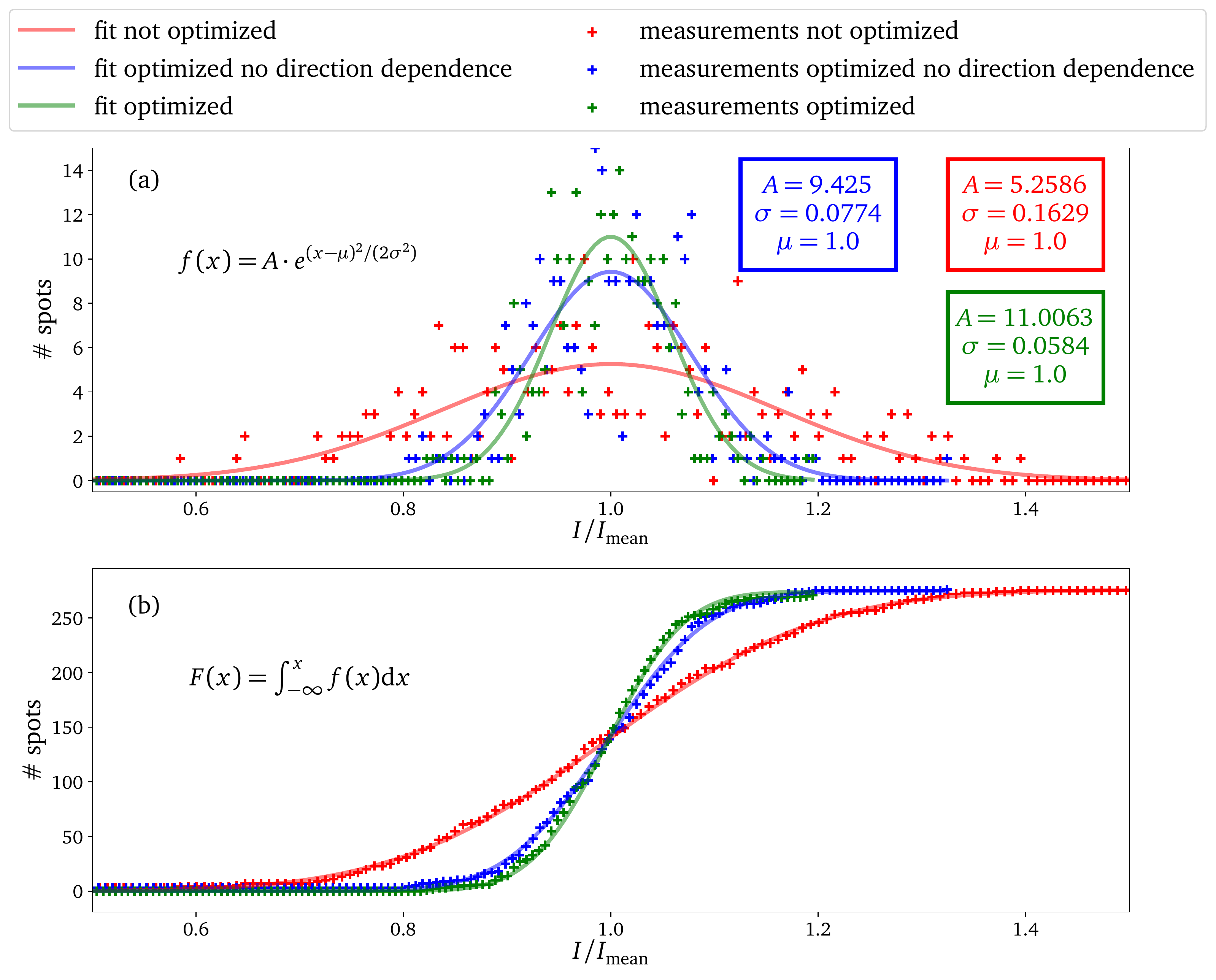}
\caption{Histogram of the spot intensities with corresponding Gaussian fits and fit parameter are shown in (a). Cumulative histograms with integrated fits are shown in (b). The red spots correspond to intensity measurements corresponding to the calculated phase pattern without the consideration of the fringing field effect. In the measurements of the blue spots, the fringing effect has been compensated, but with the direction dependence swapped. The green spots represent measurements, where the fast $2$D model has been used to compensate the fringing field effect.}
\label{picture spot uniformity fits}
\end{figure}

\cref{picture spot uniformity fits} (a) shows histograms of evaluated spot intensity measurements of a phase pattern without compensation (red), compensated (green) and compensated but with symmetric and asymmetric directions exchanged (blue). Additionally, Gaussian fits with corresponding fit parameters are shown. In \cref{picture spot uniformity fits} (b) we see the cumulative (integrated) histograms from (a) with corresponding fits, which are less noisy. We see from the values of the width $\sigma$ that without optimization the fringing field effect strongly reduces the spot uniformity (\cref{picture spot uniformity fits} red). With proper modeling of the fringing field effect (green line), the spot uniformity is strongly increased. Even by modeling the fringing field with the wrong parameters (\cref{picture spot uniformity fits} blue) we can increase the spot uniformity significantly compared to the not optimized case.


\chapter{Conclusion and Outlook}

In this thesis we have examined the fringing field effect in LC based SLMs closely and we were able to model the SLM response precisely, which allows us to achieve greater accuracy in generating complex light fields. 

First, we analyzed the diffraction efficiency measurements of period $2$ binary gratings of the BNS $512\times512$ XY Series SLM. We saw a distinctive behavior of the diffraction efficiency curves depending on the orientation of the grating with respect to the easy axis of the LC molecules, resulting in a symmetric and asymmetric diffraction. By modeling the LC director distribution, we were able to reproduce the diffraction efficiency measurements of the BNS $512\times512$ XY Series SLM for binary, blazed and checkerboard patterns. Additionally, the polarization conversion efficiencies of binary gratings were measured and simulated for two different SLM orientations (horizontal and vertical), finding lower efficiencies for the SLM in the horizontal orientation (with the easy axis of the LC molecules lying in the plane of incidence of the light beam).

Furthermore, we compared simulations done with the tensor and vector representation of the Hamamatsu X$10468$-$07$ SLM with corresponding diffraction efficiency measurements and concluded that for this specific case the simulations using the vector representation yields more reliable and physical plausible solutions than the tensor representation.

Using simulations of the phase profile, a fast and precise model was formulated and programmed on the GPU, the model being able to calculate the phase profiles of a $500 \times 500$ pixel region within a time frame of $<10$ ms.

Generally, the fast $2$D model can be used to model the phase response of a variety of nematic SLMs with similar composition like the SLM studied in this thesis. Specifically, the model will be used in torque measurements in holographic optical trapping, where a precise knowledge of the phase response is crucial. 

We implemented the fast $2$D model using two phase retrieval algorithms (WGS and NAGD) to calculate a phase profile corresponding to a regular spot pattern. In the experiment, the calculated pattern was displayed on the SLM and measurements of the spot intensity showed a significant improvement in spot uniformity compared to measurements, where the phase patterns were calculated without compensation or by ignoring the direction-dependence of the fringing field effect.  

The fast $2$D model could be improved by further investigation of the parameter interpolation from the $1$D fit functions to the $2$D model. Additionally, one could also take the effect of polarization conversion into account and thus develop a model which calculates the phase and amplitude response of the SLM.

Regular spot patterns with high uniformity can be used in parallelized material processing to increase the efficiency in the treatment (e.g. welding, cutting, etc.) of a variety of different materials (metal, plasic, organic materials, etc.), in microscopy to parallelize point scanning (e.g. confocal microscopy) and in synthetic holography to suppress artifacts.








\nocite{*}

\bibliography{}
\bibliographystyle{plain}
\end{document}